\documentclass[12pt,a4paper]{book}

\usepackage[hcentering,vcentering,a4paper]{geometry}
\usepackage{fancyhdr}

\usepackage[pdfborder={0 0 0}]{hyperref}
\usepackage{lmodern}

\usepackage{cite}
\usepackage{color}
\usepackage{axodraw}
\usepackage{graphicx}
\usepackage{enumerate}
\usepackage{subfigure}
\usepackage{psfrag}
\usepackage{multirow}
\usepackage{amsmath}
\usepackage{amsfonts}
\usepackage{amssymb}

\newcommand{\be}{\begin{equation}}
\newcommand{\ee}{\end{equation}}

\newcommand{\bea}{\begin{eqnarray}}
\newcommand{\eea}{\end{eqnarray}}

\newcommand{\nn}{\nonumber}

\newcommand{\diag}{\operatorname{diag}}

\newcommand{\h}{\mathfrak h}

\newcommand{\M}{\mathcal M}

\newcommand{\tr}{\operatorname{tr}}

\newcommand{\parasep}{\begin{center}*\hspace{2em}*\hspace{2em}*\end{center}}

\begin{document}

\title{General Warped Five-Dimensional Models} %TODO
\author{Joan Antoni Cabrer Rubert}
\date{\today}

\thispagestyle{empty}

\vspace{1.0cm}

\noindent {\Huge {\textbf{Studies on Generalized\vspace{0.4cm}\\Warped Five-Dimensional Models}}}

\vspace{1.8cm}

\noindent {\LARGE \bf  Joan Antoni Cabrer Rubert}

\vspace{3.0cm}

\noindent {\Large Mem\`oria de recerca presentada per a l'obtenci\'o del t\'itol de
\vspace{0.1cm}\\Doctor en F\'isica}
\\
\ 
\\
\noindent {\Large Director de tesi: 
\vspace{0.2cm}
 \\
\Large \bf Dr. Mariano Quir\'os Carcel\'en}

\vfill

\noindent {\Large Institut de F\'isica d'Altes Energies
\vspace{0.1cm}
\\
Grup de F\'isica Te\`orica
\vspace{0.1cm}
\\
Departament de F\'isica -- Facultat de Ci\`encies
\vspace{0.1cm}
\\
\bf Universitat Aut\`onoma de Barcelona
}
\\ 
\ 
\\
\ 
\\
\noindent{\Large Desembre de 2011}

\newpage
\thispagestyle{empty}
\ 

\newpage
\thispagestyle{empty}
\

%%%%%%%%%%%%%%%%%%%%%%%%%%%%%%%%%%%%%%%%
%%%%%%%%%%%%%%%%%%%%%%%%%%%%%%%%%%%%%%%%
\newpage
\thispagestyle{plain}
\section*{\Huge Abstract}

Models of warped extra-dimensions have been studied over the last decade as candidates to complete the Standard Model (SM) of particle physics, for they provide a natural mechanism to address its hierarchy problem. In this thesis we study a number of aspects of the five-dimensional warped models, and in particular the possibility of generalizing the well-known Randall-Sundrum (RS) solution, which is based on the Anti-de Sitter metric (AdS). 

We first discuss on the construction of soft-wall models, which are a modification of RS where the infrared brane is substituted by a naked singularity in the metric. We provide recipes for constructing consistent models of this kind and address the issue of how the length of the extra dimension can be stabilized. We also discuss on the spectrum of fluctuations that arise in soft-wall models, finding that it can range from a continuous spectrum above a mass gap to a discrete spectrum with a variable level spacing. We discuss on the possible applications of soft-wall models, and finally present a concrete model where a large ultraviolet/infrared hierarchy can be generated without any fine-tuning.

Next, we return to the original two-brane setup to study how the electroweak symmetry can be broken in warped models with generalized metrics when the Higgs boson propagates in the bulk. We show how the bounds on the Kaluza-Klein (KK) scale that arise from electroweak precision observables can be alleviated when the Higgs is localized towards the infrared brane. We apply our results to a minimal 5D extension of the SM and consider the AdS geometry and a deformation of it inspired by soft-walls. We find that the deformed geometry greatly reduces the bounds on the KK scale, to a point where the KK states can be within the range of the LHC and the little hierarchy problem can be removed without requiring the introduction of any custodial symmetry.

Finally, we study the propagation of all SM fermions in the bulk of the extra dimension, which we use to address the flavor puzzle of the SM. We find general explicit expressions for oblique and non-oblique electroweak observables, as well as flavor and CP violating operators. We apply these results to the RS model and the model with deformed geometry, for which we perform a statistical analysis departing from a random set of 5D Yukawa couplings. The comparison of the predictions with the current experimental data exhibits an improvement of the bounds in our model with respect to the RS model.

\newpage
\thispagestyle{plain}
\section*{\Huge Resum}

%{\huge Estudis sobre models generalitzats amb una dimensi\'{o} extra corbada}

Durant la passada d\`ecada, els models de dimensions extra corbades han estat estudiats com a candidats per a completar el Model Est\`{a}ndard (ME) de la f\'{i}sica de part\'{i}cules. En aquesta tesi estudiarem una s\`{e}rie d'aspectes dels models amb una dimensi\'{o} extra corbada; en particular, la possibilitat de generalitzar la ben coneguda soluci\'{o} de Randall-Sundrum (RS), la qual es basa en la m\`{e}trica Anti-de Sitter (AdS). 

Primer, discutim la construcci\'{o} dels models de \emph{soft-wall}, que s\'on una modificaci\'o de RS on la brana infraroja  ha sigut substitu\"ida per una singularitat nua a la m\`etrica. Donem receptes per a construir models consistents d'aquests tipus i estudiem com la longitud de la dimensi\'o extra pot ser estabilitzada. Tamb\'e estudiem l'espectre de les fluctuacions que apareixen en els models de \emph{soft-wall}, i trobem que podem obtenir des d'un espectre continu a partir d'una certa massa fins a un espectre discret amb un espaiament variable. Discutim les possibles aplicacions dels models de \emph{soft-wall}, i finalment presentem un model concret on es pot generar una jerarquia  ultravioleta/infraroja prou gran sense necessitat de cap ajust fi.

Despr\'es, retornem a la construcci\'o original amb dues branes per tal d'estudiar com la simetria electrod\`ebil pot ser trencada en models amb m\`etriques generalitzades quan el bos\'o de Higgs es propaga a l'engr\'os de la dimensi\'o extra. Veiem com les cotes sobre l'escala dels modes de Kaluza-Klein (KK), que apareixen a causa dels observables electrod\`ebils de precisi\'o, poden ser redu\"ides quan el Higgs est\`a localitzat a prop de la brana infraroja. Apliquem els nostres resultats a una extensi\'o m\'inima del ME en 5D, i considerem la geometria AdS i una deformaci\'o d'aquesta inspirada pels models de \emph{soft-wall}. Trobem que la geometria deformada redueix enormement les cotes sobre l'escala de KK, fins al punt en qu\`e els estats de KK es poden trobar dins del rang de l'LHC i el problema de la petita jerarquia pot ser eliminat sense requerir la introducci\'o de cap simetria custodial.

Finalment, estudiem la propagaci\'o de tots els fermions del ME al llarg de la dimensi\'o extra, la qual cosa fem servir per tractar el problema del sabor en el ME. Trobem expressions generals i expl\'icites per als observables electrod\`ebils oblics i no oblics, aix\'i com per als operadors que violen sabor i la simetria CP. Apliquem aquest resultat al model RS i al model amb geometria deformada, per la qual cosa fem un estudi estad\'istic a partir d'un conjunt aleatori d'acoblaments de Yukawa en 5D. La comparaci\'o de les prediccions amb les dades experimentals actuals mostren una millora de les cotes en el nostre model en comparaci\'o amb RS.

%%%%%%%%%%%%%%%%%%%%%%%%%%%%%%%%%%%%%%%%

\tableofcontents

%%%%%%%%%%%%%%%%%%%%%%%%%%%%%%%%%%%%%%%%

\chapter*{Preface}
\markboth{Preface}{}
\addcontentsline{toc}{chapter}{Preface}

% The Standard Model... Completion... 

%%% TODO: Afegir 1-2 paràgrafs sobre la motivació d'aquests models, la importància d'aquesta tesi, etc...?

The Standard Model of particle physics (SM), which describes three of the four fundamental interactions, is one of the most successful theories in the history of science, measured in terms of the agreement of its predictions with the results of many experiments. However, there are a number of reasons that make most physicists believe it is an incomplete theory. One of the most important reasons is the hierarchy problem, related to the question of why the gravitational force is much weaker than the nuclear and electroweak forces.

The quest of solving the problems of the SM has led vast amounts of research during the past decades, and several models have been proposed as candidates to extend the SM. Among them, we will choose to study models of extra dimensions, which attempt a geometrical explanation of the hierarchy problem. More concretely, we will focus on models of warped extra dimensions, which have been able to elegantly provide solutions to some of the SM's problems, while providing interesting possibilities for model building.

In this thesis, we will study several aspects of models of warped five-dimensional spaces. The intention is to present the most general results, so that they can be applied to a wide variety of geometries. Later, we will apply them to more concrete models that provide some additional interest.

The contents of this thesis are the fruit of a collaboration with Dr. Mariano Quir\'os and Dr. Gero von Gersdorff, whom I have had the pleasure to work with and learn from. Previously, our work was published in Refs.~\cite{Cabrer:2009we,Cabrer:2010si,Cabrer:2011fb,Cabrer:2011mw,Cabrer:2011vu,Cabrer:2011qb}. In this thesis we will review the results first presented in these publications, with the aim of providing a global approach to them.

The structure of this thesis is as follows:

In \textbf{Chap.~\ref{chap:introduction}} we will very briefly review some of the basics of the Standard Model (SM) of particle physics, in order to understand its theoretical problems and motivate the need for New Physics. In particular, we will pay attention to the Higgs mechanism, as it gives rise to the Hierarchy Problem, arguably the major motivation for New Physics. Afterwards, we will introduce the Randall-Sundrum model (RS), which represents the original and most simple formulation of models with a warped extra dimension. We will show how this model can deal with the Hierarchy Problem and also how it can be used to explain the hierarchy between the fermion masses in the SM.

\textbf{Chap.~\ref{chap:softwalls}} will be devoted to introducing the so-called Soft-Wall models, a class of warped 5D spaces which feature a naked singularity located at a finite distance from a (UV) brane. This constitutes an alternative to the hard-wall models such as RS, where two branes are used to set a finite length for the extra dimension. We will discuss on how to construct soft-wall models, on the conditions required for their consistency, and on how to stabilize the length of the extra dimension using a bulk scalar field. We will also classify these models in function of the mass spectrum of Kaluza-Klein states, which can range from a continuum spectrum with a mass gap to a discrete spectrum with a non-trivial spacing. 

In \textbf{Chap.~\ref{chap:ewsbbulkhiggs}} we will consider the question of how Electroweak Symmetry Breaking (EWSB) can be modeled in general 5D Warped models with two branes, where we will allow the Higgs boson to propagate in the bulk of the extra dimension. This will be used to describe a minimal 5D extension of the SM. We will provide expressions for electroweak precision observables in a simplified setup where all fermions are located on the UV brane (the discussion of fermions in the bulk is delayed to Chap.~\ref{chap:fermionsbulk}). Finally, we will apply our results to the RS model, and we will see how the experimental bounds on the KK scale can be relaxed by considering a heavy bulk Higgs, without requiring the presence of an additional Custodial Symmetry.

In \textbf{Chap.~\ref{chap:noncustodial}} we will meet a model of warped EWSB that does not require a Custodial Symmetry, even with a light Higgs boson. The model is based on a deformed Anti-de Sitter metric and a bulk Higgs boson. In fact, our metric is inspired by the soft-wall models of Chap.~\ref{chap:softwalls}, although here we are considering models with two branes. We will see how we can obtain low bounds on the KK scale, that can be less than $1~\mathrm{TeV}$, using a minimal 5D extension of the SM.

\textbf{Chap.~\ref{chap:fermionsbulk}} is about the propagation of fermions in the bulk, something that is not considered before for simplicity. We will generalize the results of the previous chapters to this case, and see how a theory of flavor can be constructed in general warped models with a bulk Higgs. We will discuss on the main sources of experimental constraints, namely the apparition of flavor and $CP$ violating processes and anomalous contributions to some electroweak observables. These results will be applied to RS and to the non-custodial model described in Chap.~\ref{chap:noncustodial}, and we will see how the former is also useful to relax the experimental constraints from flavor.

Finally, \textbf{Chap.~\ref{chap:conclusion}} is devoted to some concluding remarks. In particular, we will review the big picture of our work, we will discuss the consequences of soft-wall and non-custodial models and the research paths they might open, and we will identify some of the topics not covered in this thesis that are worth of future research. 

This thesis also includes five appendices where more technical details are provided that can be useful to follow our results. In \textbf{App.~\ref{fluctuationsgauge}} the propagation of gauge bosons in the bulk of a 5D warped space is discussed. In \textbf{App.~\ref{app:propagators}} and \textbf{App.~\ref{app:fermions}} we derive the propagators for bulk gauge bosons and fermions, respectively. \textbf{App.~\ref{4f}} contains results about the four-fermion terms that appear from the exchange of KK gauge bosons. Finally, in \textbf{App.~\ref{RH}} we extend the analysis done in Chap.~\ref{chap:fermionsbulk} to the case where a right-handed hierarchy is followed by the 5D Yukawa matrices.

\newpage

\section*{Acknowledgements}
\markboth{Preface}{Acknowledgements}
\addcontentsline{toc}{section}{Acknowledgements}

First and foremost I would like to show my most sincere gratitude to my supervisor, Mariano Quir\'os. 
I want to thank him for his guidance and support, for his friendliness, for his endless patience and understanding, and for treating me as a collaborator from the very beginning, which allowed me to witness how ideas are fostered in science, certainly the best way to learn. I have been deeply inspired by his invaluable scientific intuition, his intellectual courage and his passion for research. I just cannot emphasize enough how lucky I feel about having had the opportunity to work with Mariano and learn from him.

I am also extremely grateful to Gero von Gersdorff, with whom I have had the pleasure to work during all these years. Gero has always amazed me for his knowledge and inspiration for physics, which he has always been willing to share. I want to thank him for the many interesting discussions, for always helping me with my doubts and for his support. 

I am indebted to Jos\'e Ram\'on Espinosa, who first taught me about this amazing field and introduced me to Mariano. Also many thanks to \`Alex Pomarol, Christophe Grojean and Jos\'e Santiago for being in my thesis committee and for their insightful comments about this work.

I have been lucky enough to share my daily work at IFAE with a cheerful group of fellows, now turned into friends. I want to thank them for the good times, and for the many interesting and fun conversations. I certainly could have not wished for better work colleagues. To all of them I wish the best of luck, and I hope that we will meet many more times in the future. Thanks to Marc~R. and Marc~M. for the many chats about physics and life in general and for the often awaited snack breaks; to Javi and Oriol for the discussions and journeys we shared; to my office mates Carles, Simone and Joan (plus of course those mentioned above), with whom spending the largest part of my time was a pleasure;  to Germano, Diogo, Clara, Alvise, Thomas, Sebastian, Juanjo, Volker, Nikos, Felix, Pere, Max... for the great conversations, lunches and parties.

During these years I had the opportunity to get out of Barcelona many times and meet amazing people from all over the world. I would like to thank the CERN theory group for their hospitality during the three months I stayed there, which was truly inspiring because of the many great people I had the chance to meet. I am also indebted to everyone at the theory department at Columbia U. for the great month I spent there and the many interesting discussions. Getting to know the people of the Theorie des Cordes group in \'{E}cole Polytechnique was a true pleasure, and I really want to thank them all for those four months.  I would also like to thank the friends I have met along my journeys, and who I expect to meet many more times: Bryan, Pantelis, Sommath, Luis, Juanjo, Josemi, Roberto... 
  
Many thanks to the amazing musicians of the UAB Combo for the groove that has accompanied me during the last three years. Playing with them was a boost of energy that kept my batteries charged during every week.

Thanks to all my old and new friends from Barcelona and Mallorca for the great times during these years. To my friends in (non-particle) physics for the many interesting discussions.  Many thanks to my flatmates for making getting back to home a pleasure every day. Very special thanks to Rebecca, whose support and advice  has been a blessing. 

And last, but most important, I would like to thank my parents for always being there to help me in every possible way. It is their unconditional support what has always given me the confidence and courage to continue learning and pursuing my goals.

%%%%%%%%%%%%%%%%%%%%%%%%%%%%%%%%%%%%%%%%

\chapter{Introduction}
\label{chap:introduction}

The Standard Model (SM) is one of the most successful theories in science to date. It describes three of the four known fundamental interactions (electromagnetism and the strong and weak nuclear forces)  in the unifying framework of the gauge principle. This model divides the most fundamental entities of Nature in two blocks: the matter particles and the force carriers. The matter particles are fermions, with spin-$1/2$, and are classified in two groups as quarks (the constituents of protons and neutrons) or leptons (which include the electron). The force carriers are vector bosons, with spin-$1$, and include the photons (which mediate electromagnetism), the $W$ and $Z$ bosons (responsible for weak interactions) and eight gluons  (that mediate strong interactions). The interactions between matter particles and force carriers are dictated by a gauge symmetry, described by the group $SU(3)_c \times SU(2)_L \times U(1)_Y$, which provides a predictive and elegant mathematical structure from which the precise form of the interactions arises.

The key to the SM's success is that it has been able to explain a wide range of microscopic phenomena and to pass several experimental tests over the past decades, many of them within extraordinary levels of precision. Although some observables have been measured to be slightly deviating from the {SM} prediction\footnote{One example is the anomalous magnetic moment of the muon, which features a $3.2\sigma$ deviation from the {SM} prediction.\cite{Hagiwara:2006jt}}, the difficulty associated with their calculation and measurement have not affected the general consensus that the {SM} is still unchallenged experimentally. 

However, one of the key ingredients of the {SM} is yet to be discovered: the Higgs boson, a spin-$0$ particle that has eluded experimental detection to date. The importance of the Higgs relies in that its Vacuum Expectation Value ({VEV}) is responsible for breaking the electroweak symmetry $SU(2)_L \times U(1)_Y$ down to the $U(1)_{EM}$ of quantum electrodynamics, and in this process it provides masses to the rest of (massive) elementary particles. The confirmation of this {EWSB} mechanism, which would require finding the Higgs and analyzing its properties, remains the major cornerstone for the validation of the {SM}.

In fact, the {EWSB} mechanism of the {SM}, although theoretically self-consistent, features an unnatural property called the hierarchy problem \cite{PhysRevD974,PhysRevD1667,PhysRevD2619}. This problem is related to the question of why the weak force is $10^{32}$ times stronger than gravity. More precisely, for the {EWSB} mechanism to work, the Higgs boson is required to have a mass of the order of the weak scale ($\sim 100$~GeV). On the other hand, the Higgs mass receives quantum corrections that are quadratically sensitive to any new physics that might appear above that scale. Therefore, there should be a contribution of the order of the Planck mass ($\sim 10^{19}$~GeV) squared, where new physics is expected to appear to describe gravity. One would thus expect that these corrections make the Higgs several orders of magnitude heavier than required by {EWSB}, unless there is a huge fine-tuning cancellation between the quantum corrections and the bare mass. Or if some kind of new physics appears a little above the weak scale to counteract these corrections.

In addition to the hierarchy problem, there are a number of theoretical issues that lead us to think that the {SM} is not a complete theory. Of course, the {SM} does not describe gravity, and it will need to be extended at the Planck scale when gravity becomes important compared to the other three interactions. But there are still some issues relevant at much lower scales. To begin with, the {SM} does not have a candidate for a Dark Matter particle, although its existence has been inferred by astrophysical arguments. Another issue is the \emph{strong CP problem}, related to the difficulty of explaining why the {SM} does not seem to violate the CP-symmetry. Another example is the \emph{flavor puzzle}, or the fact that the {SM} lacks of any structure to explain the pattern of fermion masses, which span 12 orders of magnitude with no apparent relation between them.

The quest for a solution to some of these issues, and in particular to the hierarchy problem, has led to vast amounts of research for constructing new models of physics beyond the {SM}. Most of the research has been conducted along three major directions: \emph{supersymmetry}, consisting of the introduction of an extra symmetry that protects the quadratic contributions to the Higgs mass; \emph{Higgsless} models, which replace a fundamental Higgs by another entity capable of mediating {EWSB}, thereby the problem of quadratic corrections; and models of \emph{extra dimensions}, that rely on the introduction of extra spatial dimensions in order to account for the scale between gravity and weak interactions. This last direction is the one we will choose to explore in this thesis.

The question of which (if any) low-energy extension of the {SM} solves the hierarchy problem (and the other theoretical issues) might in fact be solved soon. At the moment of writing this thesis, the Large Hadron Collider (LHC), a particle accelerator with a potential collision energy of $14$~TeV, is completing its first year of operation. The data collected by the particle detectors is now above $1$~fb$^{-1}$, although no conclusive hints of new physics or the Higgs boson have been found. However, given the energy range of the experiment, the question of which is the true {EWSB} mechanism will hopefully be solved soon.

In this thesis, we will be studying one of these classes of models proposed to solve the Hierarchy problem: models of warped extra dimensions. A very well-known example of them is the {RS} model \cite{Randall:1999ee, Randall:1999vf}, which features just one extra dimension. In this thesis we will consider generalizations of the {RS} model, to study their phenomenology and confront them with the currently available experimental data. In particular, in Chapter~\ref{chap:softwalls} we will consider a class of models with naked singularities, and in Chapter~\ref{chap:ewsbbulkhiggs} we will study general warped models with a Higgs propagating in the bulk. 

One of the key phenomenological predictions of the {RS}-like models is that, for every {SM} particle that propagates in the bulk of the extra dimensions, a tower of particles appears which have the same quantum numbers. These excitations are referred to as {KK} modes. If we expect to solve the fine-tuning problem in this framework, the {KK} modes should appear at the {EWSB} scale, not too far from the weak scale, what would mean that, in general, they should be within the range of the LHC. However, the extremely precise measurement of some {SM} observables are pushing lower bounds on the masses of the {KK} modes. The construction of warped 5D models that are able to alleviate these bounds is an ongoing area of research, and in fact we will present a proposal in this direction in Chapter~\ref{chap:noncustodial}.

In the remaining of this chapter we will briefly present some basic facts about the {SM}, with the intention to understand the hierarchy problem that arguably motivates most of the research in physics beyond the {SM}. We will afterwards give a brief review about the construction of 5D warped models, and in particular about the {RS} model, which will serve as a starting point for the rest of this thesis.

\section{The Standard Model and its hierarchy problem}

Let us now present a very brief review about the {SM}. The intention of this section is only to present some of the basic facts about its {EWSB} mechanism in order to understand the hierarchy problem. A complete general introduction to the {SM} can be found in  Refs.~\cite{Donoghue:1992dd,Peskin:1995ev}, among many others.  

\subsection{Elementary particles and force carriers}
Subatomic matter is made of spin-$1/2$ particles (fermions). The {SM} classifies fermions in two groups: quarks and leptons. In total, the {SM} distinguishes 24 different fermions: 6 quarks and 6 leptons, each with its corresponding anti-particle. Moreover, the elementary fermions are grouped into three generations, each comprising two quarks and two leptons (plus their antiparticles). This classification is represented in Tab.~\ref{tab:fermions}, where the mass and charge of each particle is also shown.

\begin{figure}
\centering
\includegraphics[width=0.7\textwidth]{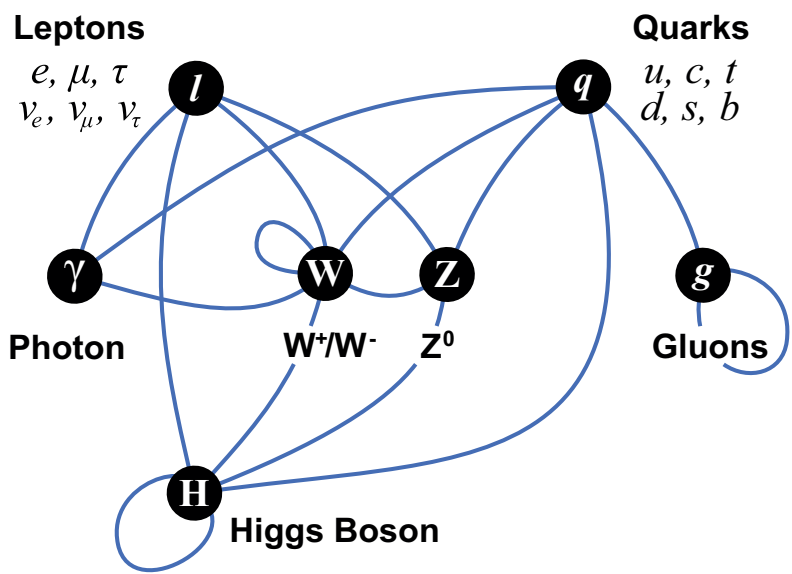}
 \caption{\it A diagram summarizing the tree-level interactions between elementary particles described in the Standard Model. The black circles represent different kinds of particles, and the blue lines connecting them represent interactions that can take place. The diagram is organized so that the matter particles are in the top row, the force carriers are in the middle row, and the Higgs boson is found at the bottom \cite{standardmodeldiagram}.
}
\label{fig:standardmodel}
\end{figure}

\begin{table}
\centering
\begin{tabular}{c|c|c|c|}
 \multicolumn{1}{c}{} & \multicolumn{1}{c}{I} & \multicolumn{1}{c}{II} & \multicolumn{1}{c}{III} 
\\ \cline{2-4}
\multirow{4}{*}{Quarks} & up ({\bf u}) & charm  ({\bf c}) & top  ({\bf t}) 
\\
& {\footnotesize $2.4$~MeV ~~~~ $2/3$} & {\footnotesize $1.3$~GeV ~~~~ $2/3$} & {\footnotesize $171$~GeV ~~~~ $2/3$} 
\\ \cline{2-4}
 & down  ({\bf d}) & strange  ({\bf s}) & bottom   ({\bf b})
 \\
 & {\footnotesize $4.8$~MeV ~~~~ $-1/3$} & {\footnotesize $104$~MeV ~~~~ $-1/3$} & {\footnotesize $4.2$~GeV ~~~~ $-1/3$} 
\\ \cline{2-4} 
\multicolumn{4}{c}{\vspace{-0.3cm}}
\\ \cline{2-4}
\multirow{4}{*}{Leptons}  & electron  ({\bf e}) & muon  ({\boldmath${\mu}$}) & tau  ({\boldmath${\tau}$})
\\
& {\footnotesize $0.51$~MeV ~~~~ $-1$} & {\footnotesize $106$~MeV ~~~~ $-1$} & {\footnotesize $1.8$~GeV ~~~~ $-1$} 
\\ \cline{2-4}
& e-neutrino ({\boldmath${\nu_e}$})& $\mu$-neutrino ({\boldmath${\nu_\mu}$})& $\tau$-neutrino ({\boldmath${\nu_\tau}$})
\\
& {\footnotesize $\lesssim $~$\mathcal{O}($eV$)$ ~~~~ $0$} & {\footnotesize $\lesssim $~$\mathcal{O}($eV$)$ ~~~~ $0$} & {\footnotesize $\lesssim $~$\mathcal{O}($eV$)$ ~~~~ $0$} 
\\ \cline{2-4}
\end{tabular}
 \caption{\it Fermionic content of the Standard Model. Along with every element we show its mass (left) and its electric charge (right). For each of the fermions shown here there is a corresponding anti-particle with equal mass and opposite electric charge.}
\label{tab:fermions}
\end{table}
Quarks and leptons interact by exchanging force carriers, which are spin-$1$ bosons responsible for the electromagnetic, weak and strong interactions. Electromagnetism is mediated by photons ($\gamma$), it has an infinite interaction range and affects all particles with an electric charge. The weak force affects all fermion and is mediated by $W^{\pm}$ and $Z^{0}$ bosons. Finally, the strong force, is mediated by gluons ($g$) and acts only on quarks. In Fig.~\ref{fig:standardmodel} a schematic view of the {SM} particles and their interactions can be found. 

 The strength of the interactions between particles is predicted by the {SM} from a simple and elegant mathematical principle: by postulating the existence of a gauge symmetry. The Lagrangian of a physical system is said to be gauge invariant, or equivalently to have a gauge symmetry, if it remains constant under a continuous phase transformation. The Lagrangian of the {SM} originates from imposing symmetry under the gauge group $SU(3)_c \times SU(2)_L \times U(1)_Y$, where $SU(3)_c$ is the group responsible for the strong interaction, and $SU(2)_L \times U(1)_Y$ is responsible for the weak and electromagnetic interactions unified into the electroweak theory.

However,  this gauge symmetry requires all gauge bosons to be massless, which is not the case as the $W^{\pm}$ and $Z^{0}$ are massive (with masses $80.4$~GeV and $91.2$~GeV respectively). In order to solve this, it is required to break the $SU(2)_L \times U(1)_Y$ subgroup into the $U(1)_{em}$ of electromagnetism. The Higgs field is the responsible for triggering this {EWSB} mechanism. This constitutes a central, and yet unconfirmed, point of the {SM}, so it is worth to study it with a bit of detail.

\subsection{The Higgs mechanism}
Let us have a look at the electroweak part of the {SM} Lagrangian, dictated by the gauge symmetry $SU(2)_L \times U(1)_Y$, with respective gauge bosons $W^{i}_\mu$ ($i=1,2,3$) and $B_\mu$ and couplings $g$ and $g'$. It can be written as
\begin{equation}
\mathcal{L}_{\mathrm{EW}} = - \frac14 F^{i\mu\nu} F^{i}_{\mu\nu} - \frac14 B^{\mu\nu}B_{\mu\nu}  + \sum_k i \bar{\psi}_k \gamma_\mu D^\mu \psi_k \,,
\label{eq:smlag}
\end{equation}
where 
\begin{gather}
F^{i}_{\mu\nu}  = \partial_\mu W_\nu^i  - \partial_\nu W_\mu^i + g \varepsilon^{ijk} W_\mu^{j} W_\nu^{k} 	\,,
\\
B_{\mu\nu} = \partial_\mu B_\nu - \partial_\nu B_\mu \,,
\\
D_\mu = \partial_\mu - i g \frac{\vec\sigma}{2} \vec{W} - i g' \frac{Y}{2} B_\mu \,.
\end{gather}
$\psi$ represent the fermions, which under $SU(2)_L$ are organized in multiplets as
\begin{equation}
\psi_L = \left( \begin{array}{c} f_L \\ f'_L \end{array} \right) \,,~~ \psi_R = f_R \,, ~~ \psi'_R = f'_R \,,
\end{equation}
where for quarks $f=u, c, t$ and $f' = d, s, b$, and for leptons $f=\nu_e, \nu_\mu, \nu_\tau$ and $f'=e, \mu, \tau$, respectively for each generation.

Since we know that the $W^{\pm}$ and $Z^0$ gauge bosons are massive, we would like to consider additional terms in the Lagrangian \eqref{eq:smlag}. One might naively add mass terms of the form
\begin{equation}
m^2_a W_\mu^a W^{a\,\mu} \,,
\end{equation}
but in fact these terms violate the gauge symmetry, and are thus forbidden. Therefore, we need to find a mechanism by which these masses can be accounted for. 

The {SM} solves this by introducing the Higgs, a scalar (spin-$0$) field that transforms as a doublet under $SU(2)$:
\begin{equation}
H = \left( \begin{array}{c} H^+ \\ H^0 \end{array} \right) \,.
\end{equation}
The introduction of this field leads to an additional piece in the Lagrangian 
\begin{equation}
\mathcal{L}_{\mathrm{Higgs}} = \left( D_\mu H \right)^\dagger \left( D^\mu H \right) - V(H^\dagger H ) \,,
\label{eq:HiggsL}
\end{equation}
where $V$ is the Higgs potential and is given by 
\begin{equation}
V(H^\dagger H) =  \mu^2 H^\dagger H + \lambda ( H^\dagger H)^2 \,.
\label{eq:VHHsm}
\end{equation}
The quartic coupling $\lambda$ in this potential needs to be positive, so that $V$ is bounded from below. As for the mass term $\mu^2$, there is no restriction as for its sign. However, if we choose it to be negative, we can see that this potential has a minimum which is not at $\langle H \rangle = 0$, and therefore the Higgs acquires a non-trivial {VEV}. This means that the vacuum of the theory breaks spontaneously the gauge invariance, and this process is known as \emph{spontaneous symmetry breaking}. More explicitly, the $SU(2)_L \times U(1)_Y$ is broken down to $U(1)_{em}$.

Let us see how the fact that the Higgs acquires a {VEV} leads to masses for the gauge bosons. We can always use a gauge transformation to write the {VEV} of the Higgs as
\begin{equation}
\langle H \rangle =  \frac{1}{\sqrt{2}} \left( \begin{array}{c}0 \\v \end{array} \right) \,,
\end{equation}
where $v = \sqrt{- \mu^2 / (2 \lambda)}$ as can be obtained from Eq.~\eqref{eq:VHHsm}. When substituting this {VEV} into the Lagrangian \eqref{eq:HiggsL}, the following terms arise
\begin{equation}
\left( D_\mu H \right)^\dagger \left( D^\mu H \right) \supset \frac{g^2 v^2}{4} W^+_\mu W^{+\,\mu} + \frac{g^2 v^2}{4}  W^-_\mu W^{-\,\mu}  +\frac{ v^2 (g^2 + g'^2) }{4} Z_\mu Z^\mu \,,
\label{eq:DmuHcon}
\end{equation}
where we have expressed the original gauge boson degrees of freedom in terms of the so-called \emph{weak basis}, as
\begin{align}
W^+_\mu  &= \frac{1}{\sqrt{2}} \left( W^1_\mu  - i W^2_\mu \right)  \,,
\\
W^-_\mu  &= \frac{1}{\sqrt{2}} \left( W^1_\mu  + i W^2_\mu \right)   \,,
\\
Z_\mu  &= \displaystyle \frac{1}{\sqrt{g^2 + g'^2}} \left( g W_\mu^3 - g' B_\mu \right)  \,,
\\
A^{\gamma}_\mu  &= \displaystyle \frac{1}{\sqrt{g^2 + g'^2}} \left( g' W_\mu^3 + g B_\mu \right)  \,.
\end{align}
Notice how in Eq.~\eqref{eq:DmuHcon} we can find mass terms for the first three of these fields. Therefore $W^{\pm}_\mu$ and $Z_\mu$ can be readily identified as the charged and massive neutral gauge bosons, with masses $m_W = v g/2 $ and $m_Z = v \sqrt{g^2 + g'^2}/2$. The measured values for these masses and the couplings fix the Higgs {VEV} to $v= 246$~GeV. The remaining degree of freedom, $A^{\gamma}_\mu$ features no mass term, and is identified as the photon. 

The Higgs mechanism is also used to generate fermion masses, which are also protected by chiral symmetry. In order to do so, we need to include in the Lagrangian couplings between the fermions and the Higgs, the so-called Yukawa couplings:
\begin{equation}
 \mathcal{L}_\mathrm{Y} = \overline{q}_L \tilde{H} Y_u^\dagger u_R + \overline{q}_L  H {Y}_d^\dagger d_R + \overline{l}_L  H {Y}_e^\dagger e_R + \hbox{h.c.},
 \label{eq:LyukawaSM}
\end{equation}
 where $\tilde{H} = i \sigma^2 H^*$ and $Y_i$ are complex $3\times3$ matrices in the generation space (we have ignored the generation indexes here). After the Higgs gets a {VEV}, the fermions acquire a mass proportional to their Yukawa couplings.

\subsection{The hierarchy problem}
After the Higgs obtains its {VEV}, in the scalar sector of the theory there is left one massive degree of freedom: the physical Higgs field, an excitation over the vacuum Higgs value (i.e., a particle). Expressing the physical Higgs as $h(x)$, the Higgs doublet after acquiring a {VEV} reads
\begin{equation}
H(x) = \frac{1}{\sqrt{2}} \left( \begin{array}{c}0 \\v + h(x) \end{array} \right) \,.
\end{equation}
Substituting this expression in the Lagrangian \eqref{eq:HiggsL}, we find a mass term for the Higgs, from which we read the tree-level mass $m_H^2 = - \mu^2  = 2 v^2 \lambda$.

\begin{figure}
\begin{center}
\begin{picture}(108,75)(-54,-18)
\SetWidth{0.9}
\DashLine(-55,0)(-22,0){4}
\DashLine(55,0)(22,0){4}
\CArc(0,0)(22,0,360)
\Text(-50,7)[c]{$h$}
\Text(0,32)[c]{$\psi$}
\end{picture}
\hspace{1.4cm}
\begin{picture}(108,75)(-54,-18)
\SetWidth{0.9}
\Text(0,49)[c]{$W,Z$}
\Text(-41,7)[c]{$h$}
\DashLine(-46,0)(46,0){4}
\SetWidth{0.9}
\PhotonArc(0,22)(17,-18,342){5}{9}
\end{picture}
\end{center}
\vspace{-0.3cm}
 \caption{\it Feynman diagrams corresponding to one-loop corrections to the Higgs mass from fermions (left) and gauge bosons (right).}
\label{fig:hloops}
\end{figure}
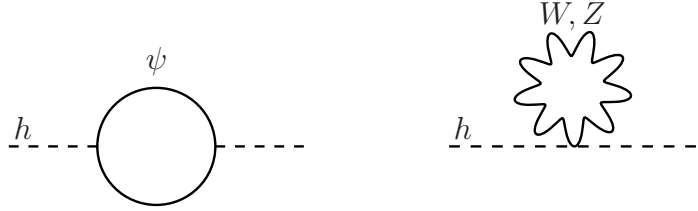
From measurements of the properties of the electroweak interaction we know that the Higgs mass should be around $~\mathcal{O}(100~\mathrm{GeV})$. However, a problem arises when we consider the quantum corrections to its tree-level mass. Because of the Higgs being an scalar field, every particle that couples to it will yield huge corrections to its mass, which are quadratic on the cutoff of the theory (i.e.~the scale at which the theory is no longer valid). These corrections arise from diagrams such as the ones shown in Fig.~\ref{fig:hloops}. The major contribution comes in fact from the most massive particle coupling to the Higgs, the top quark, and it is given at one-loop level by
\begin{equation}
\Delta m_H^2 = - \frac{Y_t}{16 \pi^2} \Lambda^2_{\mathrm{UV}} + \cdots   \,,
\end{equation}
where $Y_t \approx 1$ is the Yukawa coupling of the top quark. $\Lambda_{\mathrm{UV}}$ is the cutoff scale of the theory, and should be interpreted as the energy scale at which new physics enter to alter the behavior of the theory. Since the {SM} needs to be completed at the scale at which gravity effects become important, the Planck scale $M_{Pl} \sim 10^{19}$~GeV, the cutoff should be of this order. The fact that the corrections to the Higgs mass are so large compared with its expected value is what is known as the hierarchy problem. Reconciling the equation
\begin{equation}
\underbrace{m_H^2}_{\sim 10^4~\mathrm{GeV}^2} = {m_H^{(0)}}^2 
+ 
 \underbrace{\Delta m_H^2}_{\sim 10^{37}~\mathrm{GeV}^2}
\end{equation} 
would require an enormous amount of fine-tuning between the bare mass and the radiative corrections, of more than $30$ orders of magnitude.  

Implicit in the formulation of the hierarchy problem lies the assumption that no new physics appear between the weak scale and the Planck scale. Although, to date, the successes of the {SM} when confronted to experiment do not lead us to a need of extending the model for phenomenological reasons, the hierarchy problem itself is a good motivation to study the introduction of new physics that are able to remove this problem, and has in fact motivated huge amounts of research. One of the first proposals in this direction are models of \emph{supersymmetry}, which introduce a symmetry between fermions and bosons that cancels the quadratic contributions to the Higgs mass \cite{Martin:1997ns}. Other proposals have been in the direction of replacing a fundamental scalar Higgs by other entities, such as new composite states that play the role of the {SM} Higgs \cite{Dimopoulos:1979es,Csaki:2003zu}. In this thesis we will study the option of introducing extra spatial dimensions to address the hierarchy problem, and we will in short see how it can be done.  

Whatever the model chosen of physics beyond the {SM}, one has to be careful not to enter in conflict with the very precise measurement of observables, to which the {SM} agrees with extraordinary accuracy. The  increasing amount of precision that experiments can reach when measuring these observables pushes bounds on the scale where new physics can manifest, for a given model. This might lead to the so-called \emph{little hierarchy problem}, which would result from a model where new physics appears at a scale more than one order of magnitude above the weak scale. In this case there would still be a fine-tuning between the Higgs mass and its corrections, indeed much smaller than the original hierarchy problem but still not negligible. For example, for a theory where new physics appear at $1$~TeV, around a $1\%$ fine-tuning would be necessary. Constructing models that do not feature this little hierarchy problem without increasing their complexity is a difficult problem and a very active line of research, and the question will hopefully be illuminated soon by the LHC.

%%%%
%%%%
%%%%

\section[5D warped models: the RS model]{Five-dimensional warped models:\\ the Randall-Sundrum model}
Of the many different possibilities that can be explored to solve the hierarchy problem, in this thesis we will explore the so-called 5D warped models. In order to introduce them, let us very briefly explore here the first proposal of this kind of models: the Randall-Sundrum (RS) model \cite{Randall:1999ee, Randall:1999vf}. Complete reviews on this subject can be found in, e.g., Refs.~\cite{Davoudiasl:2009cd,Csaki:2004ay,Sundrum:2005jf}.

The RS model proposes that spacetime is described by a 5D Anti-de Sitter (AdS) metric, which is given in the proper coordinate system by
\begin{equation}
ds^2 = e^{-2ky} \eta_{\mu\nu} dx^\mu dx^\nu + dy^2 \,,
\end{equation}
where $k$ is the curvature scale of the warped extra dimension and $\eta_{\mu\nu}$ is the usual 4D Minkowski metric. We can also express this metric in conformal coordinates, defined by $z=e^{ky}$, as
\begin{equation}
ds^2 = \frac{1}{(k z)^2} \left(  \eta_{\mu\nu} dx^\mu dx^\nu + dz^2 \right) \,.
\end{equation}
This spacetime is a solution to the 5D action 
\begin{equation}
S =  \int d^4x d y \sqrt{-g} \left( M_5^3 R  + \Lambda \right)  \,,
\end{equation}
where $M_5$ is the 5D Planck scale. The constant $\Lambda$ is given by
\begin{equation}
\Lambda = - 12 M_5^3 k^2 \,,
\end{equation}
and is, in fact, a 5D cosmological constant with negative value.

In the original formulation of the RS model, often referred to as RS1 (see Ref.\cite{Randall:1999ee}), the spacetime is cut by two flat 4D boundaries, referred to as the UV (at $y=0$) and IR ($y=y_1$) branes,  and a $S^1/\mathbb{Z}_2$ symmetry is imposed that relates $y \leftrightarrow -y $. In the original formulation, all SM fields were localized on the IR brane, although it was afterwards realized that the resolving the hierarchy only required the Higgs to be IR-localized \cite{Goldberger:1999wh}.\footnote{There are a number of motivations to place the rest of SM fields in the bulk. One of them is to address the flavor puzzle of the SM, which can be explained if fermions are located at different distances from the IR brane, so that their physical masses can arise from Yukawa couplings of order one. Another reason is to exploit the AdS/CFT interpretation of the model, since all fields on the IR brane are interpreted as composite states, which is obviously not the case for SM particles. 
} The most distinct signature of this setup is the appearance of a tower of Kaluza-Klein (KK) modes for each field propagating in the bulk, with masses of the order
\begin{equation}
m_1 \sim k e^{-ky_1} \,, ~~~~ m_n  \sim n \, m_1 \,,
\label{eq:mKKRS}
\end{equation}
where $n=1$ represents the lightest new mode of the tower. This approximation is very rough and only gives us an idea of the order of magnitude. Finding the precise values for the KK masses requires solving the 5D equations of motion for each particular field. 

\subsection{Solving the hierarchy problem}

From Eq.~\ref{eq:mKKRS} we can already see how, if the distance between branes is large enough, we can obtain masses for the first KK modes which are suppressed with respect to the curvature scale $k$. In fact, something similar will happen to the Higgs VEV. Let us have a look at the piece to be added to the action for an IR localized Higgs
\begin{equation}
S_{\mathrm{H}} = \int d^4x dy \sqrt{-g} \delta(y-y_1) \left( \vert D_\mu H \vert^2 -  \lambda \left[ (H^\dagger H) - v_{5D}^2 \right]^2 \right) \,,
\end{equation}
where a generalization of the 4D Higgs potential (after {EWSB}) has been considered. By integrating over the fifth dimension 
%and reescaling $\tilde H = e^{-kb} H$ to obtain a properly normalized field, we obtain
% \begin{equation}
%S_{\mathrm{H}} = \int d^4x \vert \partial_\mu \tilde{H} \vert^2   -  \lambda \left( (H^\dagger H) - %e^{-2kb} v^2 \right)^2  \,,
%\end{equation}
we find that the 4D VEV is \emph{warped down} to 
\begin{equation}
v_{4D} = e^{-ky_1} v_{5D} \,.
\label{eq:vhierarchy}
\end{equation}
This means that, if we introduce a 5D Higgs VEV of the order of $k$, the \emph{physical} Higgs VEV will be exponentially suppressed in function of the distance between branes. Before claiming any relevance, we need to find where $M_5$ is compared to the 4D Planck scale. In fact, we find
\begin{equation}
M_{Pl}^2 = M_5^2 2 \int_0^{y_1} e^{-2ky} dy = \frac{M_5^3}{k} \left( 1 - e^{-2 k y_1} \right) \,.
\end{equation}
This shows that, naturally, one would expect $M_5$ and $k$ to be at the Planck scale. Since $v_{5D}$ is naturally expected to be of the same order, Eq.~\eqref{eq:vhierarchy} is telling us that we can induce the $\mathcal{O}(16)$ orders of magnitude between the gravitational and weak scales with a distance between branes $k y_1 \approx 35$, which is only moderately large. Here we have considered a IR localized Higgs, but we will see in Chapter~\ref{chap:ewsbbulkhiggs} how a bulk Higgs can also solve the hierarchy problem, provided it is localized towards the IR brane (see also Ref.~\cite{Huber:2000fh}).

We have therefore seen how the RS model solves the hierarchy problem in a original and elegant way. In fact, there are other ways by which a warped model can be exploited to address the hierarchy. One of them is by invoking AdS/CFT conjecture \cite{Maldacena:1997re,Gubser:1998bc,Witten:1998qj}. By this correspondence, the RS model is dual to a Conformal Field Theory (CFT) to which the fields living in the bulk are coupled. At the TeV scale the conformal symmetry is broken, producing a composite Higgs particle, which will in turn trigger {EWSB} \cite{Contino:2003ve}. A different possibility is to consider a Higgsless model, where the brane boundary conditions are used to trigger {EWSB}. 

\parasep

In the following chapters we will discuss some aspects of models with warped extra dimensions, generalizing the RS model to use an arbitrary metric. Whenever possible, we will provide the most general expressions and consider the RS limit of our models. Therefore more details about the RS models can be easily extracted from the remaining of this thesis.

%%%%%%%%%%%%%%%%%%%%%%%%%%%%%%%%%%%%%%%%

\chapter{Soft Walls}

\label{chap:softwalls}

The most common approach to constructing 5D warped models is by considering two branes, {UV} and {IR}, that define the length of the extra dimension. In this chapter we will consider the so-called \emph{soft-wall} models, in which one of the branes ({IR}) is replaced by a spacetime singularity, effectively setting a finite length of the extra dimension in the proper coordinate system. The term \emph{soft wall} emphasizes the fact that the {IR} brane (a \emph{hard wall}) is substituted by a singular solution which is reached \emph{softly} due to a smoothly vanishing metric.

In {RS} models with two branes, the distance between those needs to be stabilized by a certain mechanism, the most usual one being the Goldberger-Wise ({GW}) mechanism~\cite{Goldberger:1999uk,Goldberger:1999un}. 
This mechanism consists in the introduction of a background scalar field propagating in the bulk of the extra dimension which, after acquiring a coordinate-dependent vacuum expectation value, triggers a 4D effective potential for the radion field with a minimum, stabilizing the distance between the two branes at a given distance. On the other hand, this background scalar also generates a deviation from the {AdS} metric near the {IR}, although the mechanism is usually constructed so that this deformation is negligible. In this chapter we will see how, using this same mechanism, we can push further the modification of {AdS} and invoke a spacetime singularity at a finite proper distance, which will be automatically stabilized.

Warped models with singularities at a finite proper distance were first introduced\footnote{For an earlier analysis of models with exponential dilatonic brane potentials see Ref.~\cite{Chamblin:1999ya}.} to address the cosmological constant problem by self-tuning  \cite{ArkaniHamed:2000eg,Kachru:2000hf,Gubser:2000nd}, although later it was shown that the cosmological constant fine-tuning could not be removed this way \cite{Forste:2000ps,Csaki:2000wz,Forste:2000ft}. However, it was later found that the linear spectrum of mesons in QCD could be described by invoking the AdS/CFT correspondence in soft-wall models (AdS/QCD) \cite{Karch:2006pv,Gursoy:2007cb}.
 The name soft-wall was not introduced until later in Ref.~\cite{Batell:2008zm}. Other applications of soft-wall models that have been proposed include an holographic description \cite{Cacciapaglia:2008ns,Falkowski:2008yr} of the theory of unparticles \cite{Georgi:2007ek}, or as an alternative to {RS} to describe {EWSB} \cite{Falkowski:2008fz}. 

Another interesting feature of soft-wall models was described in Ref.~\cite{vonGersdorff:2010ht}, where it is shown that the interesting properties of soft-wall models can be equivalently described by means of an effective infrared brane, hiding the singularity and simplifying the physical understanding of these models. In this chapter, however, we will describe the full theory and discuss about the physical feasibility of the singularity.

In this chapter, which reviews the results first presented in \cite{Cabrer:2009we}, we will show how we can construct soft-wall models that include a built-in stabilization mechanism, which provides a new, natural setup to address the hierarchy problem using warped extra dimensions. We will describe the spectra of {KK} excitations and their particularities for soft-wall models. We will finally discuss on the possible applications of soft-wall models, and on the feasibility of applying them to {EWSB}. For this latter application, we will see that we need to reintroduce an {IR} brane, losing the ``soft wall'', in order to address the Higgs hierarchy problem correctly. However, the soft-wall models remain a consistent, interesting class of models to be studied and, as we will see later on in this thesis, some of their properties can be exploited even after inserting this {IR} brane. 

\section{The scalar-gravity background}
\label{sec:scalargravitybackground}

We will begin by reviewing the construction of backgrounds with 4D Poincar\'e invariance in order to construct models with spacetime singularities. We  will consider a scalar field $\phi(y)$ propagating in 5D gravity, described by the metric, in proper coordinates, 
\begin{equation}
ds^2 = e^{-2A(y)} dx^\mu dx^\nu \eta_{\mu \nu} + dy^2 \,,
\label{eq:metricfirstdefinition}
\end{equation}
where $\eta_{\mu \nu} = \mathrm{diag}(-, +, +, +)$ is the flat Minkowski metric and $A(y)$ is an arbitrary function. We will refer to the exponential $e^{-2A(y)}$ as the \emph{warp factor}. We will also consider a brane at $y=0$ and impose the orbifold $\mathbb{Z}_2$ symmetry $y \rightarrow -y$ under which $A(y)$ and $\phi(y)$ are even. 

It will also be useful to define the metric in conformally flat coordinates as
\be
ds^2=e^{-2A(z)}(dx^\mu dx^\nu\eta_{\mu\nu}+dz^2) .
\label{eq:conformal}
\ee
where $A(z)\equiv A[y(z)]$, and the relationship between $z$ and $y$ coordinates is given by 
\begin{equation}
\frac{dz}{dy} = e^{A(y)}
\,.
\label{eq:relationzy}
\end{equation}

Our setup is described, in proper coordinates, by the 5D action
\begin{equation}
S = \int d^4 x dy \sqrt{-g}  \left[ R - \frac12 (\partial \phi)^2 - V(\phi) \right] - \int d^4 x dy \sqrt{-g} \lambda(\phi) \delta(y) \,,
\label{eq:actiongphi}
\end{equation}
where we have introduced arbitrary bulk and brane potentials $V(\phi)$ and $\lambda(\phi)$, and where we have set the Planck mass to unity. The bulk {EOM}s that follow from \eqref{eq:actiongphi} read
\begin{align}
\phi'' - 6 A' \phi' - \partial_\phi V(\phi) &=0 \,, \label{eq:eom1}
\\
6 A'' - \phi'^2 &= 0 \,, \label{eq:eom2}
\\
24 A'^2 - \phi'^2 + 2 V(\phi) &= 0 \,. \label{eq:eom3}
\end{align}
Eq.~\eqref{eq:eom3} is the usual zero-energy condition arising from general covariance. After differentiating this equation with respect to $y$, it vanishes identically when Eqs.~(\ref{eq:eom1}--\ref{eq:eom2}) are satisfied. The system is then first order in $\phi$ and second order in $A$, and it has three integration constants. One of them is $A(0)$, that remains totally free, and the other two can be fixed from the {BC}s that follow from the boundary pieces of the {EOM}s,
\begin{align}
A'(0_+)&=\frac{2}{3}\lambda(\phi_0)\,,\label{bc1}\\
\phi'(0_+)&=\partial_\phi\lambda(\phi_0)\,,
\label{bc2}
\end{align}
where $\phi_0=\phi(0)$.
Using Eqs.~(\ref{bc1}--\ref{bc2}) in Eq.~(\ref{eq:eom3}) determines $\phi_0$,
\begin{equation}
\frac{1}{2}[\partial_\phi \lambda(\phi_0)]^2-\frac{1}{3}\lambda(\phi_0)^2=V(\phi_0)\,,
\end{equation}
which can be used to replace Eq.~(\ref{bc2}). 

%\subsection{The Superpotential Method}
In order to solve the system of Eqs.~(\ref{eq:eom1}--\ref{eq:eom3}), we will make use of the superpotential method, as introduced in Ref.~\cite{DeWolfe:1999cp}. This method consists in introducing an auxiliary function $W(\phi)$ related to the scalar potential by
\begin{equation}
V(\phi) \equiv \frac12 \left( \frac{\partial W(\phi)}{\partial \phi} \right)^2  - \frac13 W(\phi)^2 \,.
\label{eq:superpotential1}
\end{equation}
Using this ansatz, the bulk {EOM}s can be written as a simple system of first-order differential equations
\begin{align}
A'(y) &=  \frac16 W(\phi(y))\,, \label{eq:EOMsuperpotential1}
\\
\phi'(y) &= \partial_\phi W(\phi) \,, \label{eq:EOMsuperpotential2}
\end{align}
and the {BC}s are satisfied when
\begin{align}
W(\phi_0) &= \frac16 \lambda(\phi_0) \,,  \label{eq:BC1}
\\ 
\partial_\phi W(\phi_0) &= \frac16 \partial_\phi \lambda(\phi_0) \,. \label{eq:BC2}
\end{align}
Again, the system of Eqs.~(\ref{eq:EOMsuperpotential1}--\ref{eq:EOMsuperpotential2}) has three integration
constants and in principle every solution to Eqs.~(\ref{eq:eom1}--\ref{eq:eom3}) can be constructed in this way.  One
integration constant is the trivial additive constant $A(0)$ that does
not enter in the new set of equations. We are left with the integration
constant in Eq.~(\ref{eq:superpotential1}) and the value $\phi_0$ to fix the two
constraints Eq.~(\ref{eq:BC1}--\ref{eq:BC2}).  The equation for $W(\phi)$ is a complicated
non-linear differential equation, and in practice it is often easier to
start with a particular superpotential satisfying the {BC}s and deduce the potential needed to reproduce it.

\subsection{Solutions with spacetime singularities}
\label{subsec:solutionswithspacetimesingularities}
A particularity of the scalar-gravity system with one brane is the possibility of having naked curvature singularities at a finite proper distance. From Eq.~\eqref{eq:EOMsuperpotential2} we can easily find that, if the superpotential $W$ grows faster than $\phi^2$ for large values of $\phi$, the profile $\phi(y)$ diverges a finite value of $y \equiv y_s$. Moreover, the curvature scalar along the fifth dimension is given by
\begin{equation}
R(y) =  8  A''(y) - 20 A'(y)^2  = \frac43  \left( \frac{\partial W(\phi[y])}{\partial\phi} \right)^2  - \frac{10}{13} W ( \phi[y] )^2  \,,
 \label{eq:curvature}
\end{equation}
so that the curvature, in general, diverges at $y=y_s$. The interpretation is that the spacetime ends at $y_s$, and the region $y>y_s$ does not have any physical meaning. We are therefore using the naked singularity to cut the space instead of a brane.

Since we only are introducing one brane, it seems that we only have two constraints for the three integration constants of the {EOM}s. However, having dynamically generated a new boundary at the singularity, we must ensure that the boundary pieces of the {EOM}s vanish at $y=y_s$. Otherwise, the proposed solution would not extremize the action, resulting in a non-zero cosmological constant. These boundary pieces can easily read from the action \eqref{eq:actiongphi} and lead to the condition
\begin{equation}
\frac13 \lambda(\phi_0) - 2 W(\phi_0) + 2 \lim_{y\to y_s} e^{-4A(y)} W(\phi[y]) = 0 \,,
\label{eq:boundarypiece}
\end{equation}
where $\phi_0 \equiv \phi(0)$. The two first terms of this equation cancel when Eq.~\eqref{eq:BC1} is satisfied, while the last one depends on the behavior of the superpotential near the singularity. Let us now find under which conditions does the last term in \eqref{eq:boundarypiece} vanish. Using $\phi$ as a coordinate in Eqs.~(\ref{eq:EOMsuperpotential1}~--~\ref{eq:EOMsuperpotential2}), we find
\begin{equation}
\frac{dA}{d\phi} = \frac16 \frac{W(\phi)}{\partial_\phi W(\phi)} \,,
\end{equation}
from where we can see that, in order for the last term of \eqref{eq:boundarypiece} to vanish, W needs to grow more slowly than $e^{2\phi/\sqrt{6}}$ at large $\phi$. We thus arrive at a simple criterion for the existence of consistent, singular solutions:
\begin{equation}
\parbox{0.66\textwidth}{\it A singularity  with  $\phi(y_s)\to\infty$
is allowed if, and only if, $W(\phi)$  grows asymptotically more
slowly than $ e^{2\phi/\sqrt{6}}$. 
}\label{criterion}
\end{equation}
From this criterion follows that the potential $V$ needs to grow more slowly than $e^{2\phi/3}$, although this is not a sufficient  condition (consider e.g.~the trivial example $V=0$ which has the general solution $W \propto e^{2\phi/\sqrt{6}}$). 

It is instructive to compare our criterion with the one found in Ref.~\cite{Gubser:2000nd} where AdS-CFT duality was used to classify physical singularities. According to Ref.~\cite{Gubser:2000nd} admissible singularities are those whose potential is bounded above in the solution. Inspection of Eq.~(\ref{eq:superpotential1}) shows that singularities fulfilling (\ref{criterion}) have a potential that goes to $-\infty$, while those that fail (\ref{criterion}) go to $+\infty$. Although we here employ a much more basic condition (a consistent solution to the Einstein equations), which in particular can be applied to theories without any field theory dual, it is good to know that our allowed solutions have potentially consistent interpretations as 4D gauge theories at finite temperature. 

Furthermore, a detailed analysis about the feasibility of soft-wall singularities as the ones we will cover in the remaining of this chapter is conducted in Ref.~\cite{George:2011gs}, where they show that, in the cases where \eqref{criterion} is satisfied, the singularity satisfies the necessary unitary boundary conditions.

\subsection{On the cosmological constant fine-tuning}
So far, it might seem that one can obtain flat 4D solutions with fairly generic brane and bulk potentials without fine-tuning. This apparent self-tuning property was pointed out in \cite{ArkaniHamed:2000eg,Kachru:2000hf,Gubser:2000nd}, and was studied as a possible solution to the cosmological constant problem. However, as we will right now see, there is a hidden fine-tuning that uncovers again the cosmological constant problem \cite{Forste:2000ps,Csaki:2000wz,Forste:2000ft}. 

In fact, achieving a superpotential that grows more slowly than $e^{2\phi/\sqrt{6}}$ requires a hidden fine-tuning of the cosmological constant. 
To see this, it suffices to consider a potential that behaves asymptotically as
\begin{equation}
V(\phi) \sim  b e^{2\nu\phi/\sqrt{6}}\,.
\end{equation}
Writing $W(\phi)$ as
\be
W(\phi)=6 w(\phi) e^{\nu\phi/\sqrt{6}}\,,
\ee
we can express the solutions for $w(\phi)$ as the roots of
\be
e^{(4-\nu^2)(\phi-c)/\sqrt{6}}=
(2 w+\sqrt{b+4 w^2})^{\pm 2}(\nu w\mp\sqrt{b+4 w^2})^\nu\,,
\ee
where $c$ is an integration constant. For $\nu>2$ this implies that
$w(\phi)$ tends asymptotically to the constant
\be
w \approx \pm \sqrt{\frac{b}{\nu^2-4}}\,
\label{wconst}
\ee
at large $\phi$ and for $b>0$.  However, for $0<\nu<2$, $w$
generically behaves as
\be
w(\phi) \sim e^{(2-\nu)\phi/\sqrt{6}}\,.
\ee
Only if we adjust $c\to\infty$ we can achieve that $w$ behaves as in
Eq.~(\ref{wconst}).  In this case, $b$ has to be negative in order to
have a real solution for $W$.

The generic solution to Eq.~(\ref{eq:superpotential1}) thus grows as $W\sim
e^{\nu\phi/\sqrt{6}}$ for $\nu\geq 2$ and $W\sim e^{2\phi/\sqrt{6}}$ for $\nu\leq
2$. However, it is possible to arrange for $W\sim e^{\nu\phi/\sqrt{6}}$ in the
latter case by picking a particular value for the integration constant
in Eq.~(\ref{eq:superpotential1})~\footnote{Similar reasonings apply to potentials
that grow even more slowly, e.g., as a power. The generic solution
behaves as $e^{2\phi/\sqrt{6}}$, but particular solutions may exist that behave
as $\sqrt{V}$ and hence allow for consistent, yet fine-tuned,
flat backgrounds.}.  We are then left in two possible scenarios:

\begin{enumerate}[(a)]
\item The superpotential $W$ grows as $e^{2\phi/\sqrt{6}}$ or faster, and the {EOM}s are not satisfied at the singularity. The only consistent way-out is to resolve the singularity, for instance by introducing a second brane located at $y_s$ (or at $y < y_s$). In this case, the fine-tuning of the cosmological constant is reintroduced, as two additional {BC}s appear [the {IR} equivalents to Eqs.~(\ref{bc1}--\ref{bc2})] without increasing the number of free parameters \cite{Forste:2000ps,Csaki:2000wz,Forste:2000ft}.

\item The superpotential $W$ grows as $e^{\nu \phi/\sqrt{6}}$, with $\nu < 2$, or slower. The {EOM}s are satisfied at the singularity, and there is no need to resolve it. However, we need to adjust the integration constant of Eq.~\eqref{eq:superpotential1}, losing one of our parameters needed to satisfy the {BC}s of Eqs.~(\ref{eq:BC1}--\ref{eq:BC2}) and resulting in a fine-tuning of the brane tension. 
\end{enumerate}

It is important to realize that either fine-tuning precisely corresponds to the fine-tuning of the cosmological constant. In the second possibility above this is particularly obvious: the superpotential is completely specified by the bulk potential and the {BC} at $\phi \rightarrow \infty$. Eqs.~(\ref{eq:BC1}--\ref{eq:BC2}) are then simply the minimization of the 4D potential
\begin{equation}
V_4(\phi) = \lambda(\phi) - 6 W(\phi)
\label{eq:V4}
\end{equation}
under the condition that $V_4(\phi)$ vanishes at the minimum $\phi = \phi_0$. In fact, the brane potential $\lambda(\phi)$ should be determined by physics localized at the {UV} brane interacting with the scalar field $\phi$. For example, if the {SM} Higgs was localized at the {UV} brane it would generate a brane potential as $\lambda(\phi,H)$ which would in turn provide the effective brane potential $\lambda(\phi, \langle H \rangle)$ after {EWSB}. Hence, after the electroweak phase transition, there will be a $\phi$-dependent vacuum energy which will require re-tuning the cosmological constant to zero and possibly a shift in the minimum of Eq.~\eqref{eq:V4}.

As we have seen, there exist consistent solutions to the {EOM}s in the full closed interval $[0,y_s]$ that do not demand the introduction of a second brane or other means of resolving the singularity. However, this setup does not solve the cosmological constant problem. 

\section{The 4D spectrum}
\label{sec:the4dspectrum}

In this section we will study the fluctuation of the metric and the scalar around the classical background solutions for general metrics, and classify them according to the asymptotic behavior of the superpotential. A general ansatz to describe all gravitational excitations of the model is, using an appropriate gauge choice \cite{Csaki:2000zn}, 
\begin{gather}
 \phi(x,y) = \phi(y) + \varphi(x,y) , \\
 ds^2 = e^{-2A(y) - 2F(x,y)}(\eta_{\mu\nu} + h^{\mathrm{TT}}_{\mu\nu}(x,y)) dx^\mu dx^\nu + (1 + G(x,y))^2 dy^2 ,
\end{gather}
where $\phi(y)$ is the background scalar, solution of Eq.~\eqref{eq:EOMsuperpotential2}, and $h^{TT}_{\mu\nu}$ are the transverse traceless fluctuations of the metric. The Einstein equations that arise from this ansatz have the spin-two fluctuations decoupled from the spin-zero fluctuations, so we can proceed to study them independently.
\subsection{The graviton}

Let us first consider the graviton as the transverse traceless fluctuations of the metric
\begin{equation}
 ds^2 = e^{-2A(y)} (\eta_{\mu\nu} + h_{\mu\nu} (x,y) ) dx^2 + dy^2,
\end{equation}
where $h_{\mu}^{\ \mu} = \partial_\mu h^{\mu\nu} = 0$. In order to respect the orbifold symmetry and to keep the possibility of a constant profile zero mode, we will consider $h(y) = h(-y)$ which leads to the {BC} at the brane $h'(0)=0$. The part of the action quadratic in the  graviton fluctuations becomes
\begin{multline}
 S = \int d^4 x\, d y\, \sqrt{-g} R\\  \rightarrow -\frac14 \int d^4 x\, d y\, e^{-2 A(y)} \left(  \partial_\rho h_{\mu\nu} \partial^\rho h^{\mu\nu} + e^{-2A(y)}  \partial_y h_{\mu\nu} \partial_y h^{\mu\nu}\right) .
 \label{actionh}
\end{multline}
Using the ansatz
\begin{equation}
h_{\mu\nu} (x,y) =  h^{}_{\mu\nu} (x) h^{}(y) ,
\end{equation}
one can obtain the {EOM} for the wavefunctions $h(y)$, which is given by
\begin{equation}
 h''(y) - 4 A'(y) h'(y) + e^{2A(y)} m^2 h(y) = 0  .
\label{GEOMy}
\end{equation}
After an integration by parts in \eqref{actionh}, one finds an additional equation due to boundary terms at $y=y_s$,  
\begin{equation}
 e^{-4 A(y_s)} h'(y_s)  = 0 .
\label{GBCy}
\end{equation}
In addition, one has to impose that the solutions are normalizable, i.e.
\begin{equation}
 \int_{0}^{y_s} dy \, e^{-2A(y)} h^2(y) < \infty .
\end{equation}

It is now convenient to change to conformally flat coordinates, as defined in~\eqref{rel}. In this frame, rescaling the field by $ \tilde h(z) = e^{-3 A(z)/2} h(z)$, Eq.~\eqref{GEOMy} can be written as a Schroedinger-like equation, 
\begin{equation}
 - \ddot{\tilde{ h}}(z) +  V_{h}(z) \tilde h(z) = m^2 \tilde h(z) ,
\label{GSchroedinger}
\end{equation}
where the dots ($\cdot{}$) represent derivatives with respect to $z$ and the potential is given by
\begin{equation}
  V_{h} (z) = \frac{9}{4} \dot{A} (z)^2 - \frac{3}{2} \ddot{A}(z) 
  \label{eq:Vhschroedinger}
\end{equation}
The boundary equations are written in the $z$-frame as
\begin{equation}
 \left.e^{-3 A(z)} \dot{h}(z) \right\vert_{z_0,z_s} 
 = 
 \left.e^{-3 A(z)/2} \left( \dot{\tilde h} (z) + \frac32 \dot{A}(z) \tilde h(z) \right) \right\vert_{z_0,z_s} 
 = 0 \,,
 \label{Gconditionszs}
\end{equation}
and the normalizability condition is
\begin{equation}
 \int_{z_0}^{z_s} dz \, e^{-3A(z)} h^2(z) = \int_{z_0}^{z_s} dz \, \tilde{h}^2(z)  < \infty
 .
 \label{Gnormalizzs}
\end{equation}

\subsection{The radion-scalar system}
Now we consider the spin-zero fluctuations of the system. This is
\begin{gather}
 \phi(x,y) = \phi(y) + \varphi(x,y) , \\
 ds^2 = e^{-2A(y) - 2F(x,y)}\eta_{\mu\nu} dx^\mu dx^\nu + (1 + J(x,y))^2 dy^2 .
\end{gather}
With an appropriate gauge choice, the {EOM}s for the $y$-dependent part of the {KK} modes form a coupled system with only one degree of freedom. The derivation of the equations is given with detail in \cite{Csaki:2000zn}, and the result is
\begin{gather}
 F'' - 2 A' F' - 4 A'' F - 2 \frac{\phi''}{\phi'} F' + 4 A' \frac{\phi''}{\phi'} F = - m^2 e^{2A} F ,\\
 \phi'(y) \varphi(y) =  F'(y) - 2 A'(y) F(y) , \\
 J(y) = 2F(y) .
 \label{Feqmotion}
\end{gather}
The boundary equations on the brane depend on the brane tension $\lambda(\phi)$. The precise form of the dependence can be found in \cite{Csaki:2000zn}. At the singularity, similarly to the graviton case, one gets the boundary equation
\begin{equation}
 e^{-4A(y)} \varphi'(y)\vert_{y_s} = 0 ,
\label{Fboundary}
\end{equation}
and the normalizability condition
\begin{equation}
 \int_{0}^{y_s} dy e^{-2A(y)} \varphi^2 (y) = \int_{z_0}^{z_s} dz \tilde{\varphi}^2(z)  < \infty ,
\label{Fnorm}
\end{equation}
where the field has been rescaled by $\tilde \varphi (z) \equiv e^{-3A/2} \varphi (z)$.

It is convenient, as for the graviton,  to use conformally flat coordinates. Rescaling the field by $\tilde F(z) = e^{-3A(z)/2}  F(z)/{\dot\phi(z)}$, Eq.~\eqref{Feqmotion} can be written as the Schroedinger equation
\begin{equation}
 -\ddot{\tilde F}(z) +  V_{F} (z) \tilde F(z) = m^2 \tilde F(z) ,
 \label{Fschroedinger}
\end{equation}
where
\begin{equation}
 V_{F} (z) = \frac94 \dot{A}(z) ^2 + \frac52 \ddot{A}(z) - \dot{A}(z) \frac{\ddot{\phi}(z)}{\dot{\phi}(z)} - \frac{\dddot{\phi}(z)}{\dot{\phi}(z)}+ 2 \left( \frac{\ddot{\phi}(z)}{\dot{\phi}(z)}\right)^2 \ .
 \label{FVhatz}
\end{equation}
The relation between the rescaled field $\tilde F$ and the rescaled scalar field $\tilde\varphi$ is 
\begin{equation}
  \tilde\varphi (z) =  \dot{\tilde{F}}(z) + \left( \frac{\phi''(z)}{\phi'(z)} - \frac12 \dot{A}(z) \right) \tilde F(z) .
 \label{Ftildevarphi}
\end{equation}

\subsection{Classification of spectra}
\label{subsec:classificationofspectra}
Let us now try to classify the different mass spectra for the Kaluza-Klein modes that can be obtained from different superpotentials \cite{Cabrer:2009we} \cite{Gursoy:2007cb, Gursoy:2007er}. As we will shortly see, the mass spectrum of a soft-wall model depends basically on the asymptotic behavior of the superpotential $W(\phi)$ near the singularity. 

In general, all fields propagating in the bulk will have the same kind of spectrum, in terms of the dependence of the $n$-th {KK} mode mass to $n$. Hence, it will be sufficient to consider the case of one field propagating in the bulk, and it can be checked that the conclusions we are going to extract in this section apply to all kind of bulk fields. We will then consider the case of the graviton. Recall that the {EOM} for the fluctuations can be written, in conformal coordinates and after the redefinition $\tilde h(z) = e^{-3A(z)/2} h(z) $, as \eqref{GSchroedinger}
\begin{equation}
 - {\tilde{ h}}''(z) +  V_{h}(z) \tilde h(z) = m^2 \tilde h(z) ,
 ~~~~
  V_{h} (z) = \frac{9}{4} \dot{A} (z)^2 - \frac{3}{2} \ddot{A}(z) ,
  \label{eq:GSchroedinger2}
\end{equation}
and so we will only need to analyze the behavior of $A(z)$ for large values of $z$. 

Our aim is to classify different soft-wall models in terms of the asymptotic behavior of $A(z)$. In order to do that, we will consider two classes of superpotentials, following the notation of Ref.~\cite{vonGersdorff:2010ht}. Type-1 models (SW1) will be defined by superpotentials with an asymptotic behavior given by
\begin{equation}
W(\phi) \sim e^{\nu \phi/\sqrt{6}} \,, ~~~ \nu < 2 \,,
\label{eq:WSW1}
\end{equation}
where the condition on $\nu$ follows from the consistency requirement \eqref{criterion}. As we will shortly see, in the case $\nu=1$ the subleading behavior will become important, and so it will be convenient to define a second category of backgrounds (SW2) defined as
\begin{equation}
W(\phi) \sim e^{\phi/\sqrt{6}} \phi^\beta \,, ~~~ \beta > 0 \,. 
\label{eq:WSW2}
\end{equation}
With this classification we will cover all different possible kinds of spectra that can be obtained using simple soft-wall models.  

\subsubsection{SW1}
Let us first consider the SW1 class \eqref{eq:WSW1}. From Eqs.~(\ref{eq:EOMsuperpotential1}--\ref{eq:EOMsuperpotential2}) we know that the asymptotic behavior of the metric is given by
\begin{equation}
A(y) \sim - \frac{1}{\nu^2} \log \left( 1 - \frac{y}{y_s} \right) \,,
\end{equation}
where $y_s$ marks the position of the singularity. Since our Schroedinger-like potential is expressed in terms of the conformal coordinate $z$, we need to find $A(z)$ using the coordinate change of Eq.~\eqref{eq:relationzy}. The relation is given, at large $z$, by
\begin{equation}
z \sim 
\left\lbrace 
\begin{array}{ll}
- \left( y_s - y \right)^{1-1/\nu^2} \,,  ~~&\nu \neq 1 
\\
- y_s \log \left( 1 - \frac{y}{y_s} \right) \,,  &\nu = 1 
\end{array}
\right.
\,,
\end{equation}
which allows us to write
\begin{equation}
A(z) \sim 
\left\lbrace 
\begin{array}{ll}
\log(\rho z) \,,  ~~&\nu \neq 1 
\\
\rho z \,,  &\nu = 1 
\end{array}
\right.
\,,
\end{equation}
where $\rho = \mathcal{O}(y_s^{-1})$ is a constant of energy dimension, which will set the energy scale of the {KK} excitations. Finally the Schroedinger-like potential of Eq.~\eqref{eq:GSchroedinger2} behaves asymptotically as 
\begin{equation}
V_h(z) \sim 
\left\lbrace 
\begin{array}{ll}
1/z^2 \,,  ~~&\nu \neq 1 
\\
(9/4) \rho^2 \,,  &\nu = 1 
\end{array}
\right.
\,.
\end{equation}
We then find the following possibilities:
\begin{itemize}
\item When $\nu<1$ the singularity is located at $z_s \rightarrow \infty$, and the potential is bounded below by $0$. This means that there will not be confinement, and the lightest state will have mass $m=0$. In other words, we have a continuous mass spectrum without a mass gap. 
\item When $\nu=1$ the singularity is located at $z_s \rightarrow \infty$, and the potential is bounded below by $(9/4)\rho^2$. Again there will not be confinement, but this time the lightest state will have a mass $m > \sqrt{9/4} \rho$. In this case the spectrum is continuum but with a mass gap.
\item When $\nu>1$ the singularity is now at a finite $z_s$, which will lead to confinement and a discrete spectrum. We can approximate the behavior of the heavy mass modes by making use of an WKB approximation, which tells us that the $n$-th mass mode is given by
\begin{equation}
\int_{z_a}^{z_b} \sqrt{m_n^2 - V_h(z)}  \approx n \pi + \theta \,,
\label{eq:WKBapprox}
\end{equation}
where $z_a$ and $z_b$ correspond to the values for which $V_h(z_1) = V_h(z_2) = m^2$ (i.e.~the classical turning points) and $\theta$ is a constant that will be given by the {BC}s at the brane and the singularity (that we do not need to care about at this point). Taking the approximation $m_n^2 \gg V_h$ and $z_b = 1/m_n \gg z_a$ we obtain
\begin{equation}
m_n \sim n \,,
\end{equation}
that is, a linear mass spectrum. 
\end{itemize}

With these approximations we have obtained information about the general behavior of the {KK} spectrum for SW1 models. In Section~\ref{sec:aselfstabilizedsoftwallmodel} we will explicitly solve a particular example of SW1 models covering the three possibilities described above, and we will confirm the validity of this approximation in that case.

\subsubsection{SW2}
Now let us move on to SW2 models \eqref{eq:WSW2}. As in SW1 models, near the singularity the metric behaves as 
\begin{equation}
A(y) \sim - \log\left( 1 - \frac{y}{y_s} \right) \,,
\end{equation}
while in terms of the conformal coordinates it reads
\begin{equation}
\displaystyle
A(z)
\sim
\left\lbrace
\begin{array}{lll}
(\rho z)^{1/(1-2\beta)}  & , & 0 < \beta < 1/2
\\
e^{\rho z} & , &\beta = 1/2
\\ 
\left[\rho (z_s - z) \right]^{1/(1-2\beta)} & , & \beta > 1/2
\end{array}
\right.
\,,
\end{equation}
where $z_s$ is the position of the singularity in conformal coordinates $z_s = z(y_s)$. We can see that the singularity is located at $z_s = \infty$ when $\beta \leq 1/2$, and at a finite conformal length $z_s < \infty$ when $\beta > 1/2$. The potential \eqref{eq:GSchroedinger2} behaves, also near the singularity, as
\begin{equation}
V_h(z)/\rho^2 \sim 
\left\lbrace 
\begin{array}{ll}
(\rho z)^{4\beta/(1-2\beta)} \,,  ~~&0 < \beta < 1/2
\\
e^{2\rho z} \,,  &\beta = 1/2 
\\
\left[ \rho (z_s - z)\right]^{4\beta/(1-2\beta)} \,,  ~~& \beta > 1/2
\end{array}
\right.
\,.
\end{equation}
 Let us now explore the three cases separately 
\begin{itemize}
\item For $\beta < 1/2$, we can see that even if the position of the singularity is at $z_s = \infty$, the potential diverges there. Therefore, we can expect a discrete spectrum. Making a WKB approximation as in Eq.~\eqref{eq:WKBapprox}, for $m_n \gg V_h$ and $z_b = m_n^{1/(2\beta) - 1} \gg z_a$ we get that
\begin{equation}
m_n \sim n^{2\beta} \,,
\label{eq:spectrumSW1beta12}
\end{equation}
so that the spacing of the spectrum can be controlled by varying the parameter $\beta$. 
\item For $\beta = 1/2$ the arguments are the same as in the case we just explored, and it can easily be checked that the WKB approximation yields
\begin{equation}
m_n \sim n \,,
\end{equation}
which coincides with taking the appropriate limit in Eq.~\eqref{eq:spectrumSW1beta12}. 
\item For $\beta > 1/2$ the WKB approximation of Eq.~\eqref{eq:WKBapprox}, for $m_n \gg V_h$ and $z_b = z_s \gg z_a$, tells us that the spectrum is approximated by
\begin{equation}
m_n \sim n  \,,
\end{equation}
recovering again the linear result. In fact, this could have been expected, since it is known that 5D models with a finite conformal length, as in this case, always produce a linear spectrum.
\end{itemize}

\parasep

To conclude this section, let us recapitulate and summarize the classification of spectra we just obtained in Tab.~\ref{default}, where we also show some of the results we obtained in previous sections. We can see that we can obtain a broad range of spectra, ranging from a continuous tower of excitations to a linear spacing between masses, with all possibilities in between covered. This makes soft-wall models a powerful tool to describe phenomenologically a large set of physical situations. We will comment on the possible applications of the different kinds of soft-wall models later on in Sec.~\ref{sec:applicationsofsoftwallmodels}.

\begin{table}[h]
\begin{center}
\begin{tabular}{| c || c | c | c | c | c |c|}
\hline \vspace{-12pt} & & & & & & \\
 \multirow{2}{*}{$W(\phi)$} & $\leq\phi^2$& $>\phi^2$    & $e^{\phi/\sqrt{6}}$      &
$e^\phi \phi^\beta$              & $> e^{\phi/\sqrt{6}}\phi^{\frac{1}{2}}$        & $\geq e^{2
\phi/\sqrt{6}}$   \\
 && $<e^{\phi/\sqrt{6}}$
 &  & {\footnotesize $0 < \beta \le \frac12$}  &
$<e^{2\phi/\sqrt{6}}$
 & 
 \\
\noalign{\hrule height 0.9pt}
$y_s$ & $\infty$  &\multicolumn{5}{c|}{finite}  \\
\hline
$z_s$ &  \multicolumn{4}{c|}{$\infty$} &  \multicolumn{2}{c|}{finite}  \\
\hline
mass & \multicolumn{2}{c|}{\multirow{2}{*}{continuous}} & continuous
& \multicolumn{3}{c|}{discrete} \\
\cline{5-7}
spectrum &  \multicolumn{2}{c|}{} & w/ mass gap  & $m_n \sim
n^{2\beta}$ & \multicolumn{2}{c|}{$m_n \sim n$} \\
\hline
consistent & \multicolumn{5}{c|}{\multirow{2}{*}{yes}} & \multirow{2}{*}{no} \\
solution & \multicolumn{5}{c|}{} &  \\
\hline
\end{tabular}
\end{center}
\caption{\it Spectra resulting from different asymptotic forms of the
superpotential. In the first row we give the asymptotic behavior of
$W(\phi)$, with the strength of the divergence increasing from left to
right ($>$ means ``diverges faster than'', etc). Second and third row
show the finiteness of $y_s$ and $z_s$, with the behavior changing at
$W\sim \phi^2$ and $W\sim e^{\phi/\sqrt{6}}\phi^\frac{1}{2}$ respectively. The
third row shows the spectrum, while in the last one we indicate the
consistency of the solution.}
\label{default}
\end{table}

%%%
%%%\section{On the consistency of soft-wall models}
%%%

\section{Constructing soft-wall models with a hierarchy}
\label{sec:constructingsoftwallmodels}
The main motivation of studying warped extra-dimensional problems is their property of generating large hierarchies with little fine-tuning. In this section we will describe the essential properties a soft-wall model, or more precisely the superpotential, must have in order to display this property. 

Let us consider a soft-wall model with a superpotential $W(\phi)$, and assume that $W$ is a monotonically increasing function of $\phi$, i.e.~$W'(\phi)>0$. The location of the singularity, and hence the size of the extra dimension, is given in proper coordinates by
\be
y_s=\int_{\phi_0}^\infty\frac{d\phi}{W'(\phi)}\,.
\ee
The integral is finite whenever $W$ diverges faster than $W\sim \phi^2$. However, the inverse volume $y_s^{-1}$ is, in general, not the 4D {KK} scale nor the mass gap as there might be a strong {AdS} warping near the {UV} brane. The {KK} scale 
is given by the inverse conformal volume $z_s^{-1}$ (when it is finite), calculated as
\be
z_s=\int_0^{y_s} e^{A(y)} dy\,.
\ee
It is easy to warp the geometry near the brane without affecting $y_s$ by adding a positive constant of $\mathcal O(k)$ to the superpotential, leading to
\be
A(y)\to A(y)+ky\,,\qquad{\rm for}\qquad W\to W+k\,
.
\label{trick}
\ee
Notice that $A(y)$ is a monotonically increasing function of $y$, such that 
\be
k z_s>e^{k y_s}\,.
\label{z_s}
\ee
One sees that the {KK} scale is warped down with respect to the compactification scale, a phenomenon well known in {RS} models with two branes~\cite{Randall:1999ee}. In order to obtain, e.g., the TeV from the Planck scale we need
\be
k y_s=\int_{\phi_0}^\infty \frac{k}{W'(\phi)}\,d\phi\simeq 37\,.
\label{kys2}
\ee
This is not hard to achieve in a natural manner. In our model, Eq.~(\ref{ourW}), it works so well because the exponential behavior that was introduced for large values of $\phi$ is also valid at $\mathcal O(1)$ negative values and dominates the integral, leading to Eq.~(\ref{kys}). 

Moreover, there are many cases where $z_s$ is infinite, even though $y_s$ is finite. There can still be mass gaps or even a discrete spectrum, but $z_s$ is clearly inadequate to characterize the energy levels. One such example is the case $W=ke^\phi$ that leads to a mass gap. Let us be slightly more general and consider the class of superpotentials of type SW2 \footnote{Refer to Sec.~\ref{sec:aselfstabilizedsoftwallmodel} for the explicit construction of stabilized soft-wall models of type SW1.}
\be
W(\phi)= k\,e^{\phi}(\phi-\phi_1)^\beta\,,
\ee
with $\phi_1<\phi_0$.
This superpotential is monotonically increasing for $\beta\geq 0$ and has infinite $z_s$ for $\beta\leq\frac{1}{2}$, so we will assume $0\leq\beta\leq\frac{1}{2}$. The volume $y_s$ is approximately 
\be
ky_s \simeq e^{-\phi_0/\sqrt{6}}\,,
\label{ys2}
\ee
so, again, $k y_s$ is (mildly) exponentially enhanced when $|\phi_0|=\mathcal O(1-10)$, $\phi_0<0$. In order to estimate the spectrum, we need the asymptotic behavior of the warp factor in conformally flat coordinates. For large $z$, it is given by
\be
A(z)\simeq (\rho z)^\frac{1}{1-2\beta}\,
,
\label{metricz}
\ee
where $\rho=\mathcal O(y_s^{-1})$. \footnote{Metrics of the form Eq.~(\ref{metricz}) have been studied in detail in Refs.~\cite{Batell:2008me,Batell:2008zm}.} The coordinate change is given by
\be
z(y)=\int e^{A(y)} dy\,.
\ee
Using our trick of adding warping while keeping $y_s$ unchanged, Eq.~(\ref{trick}), we see that near the singularity 
\be
z(y)\to z_w(y)\simeq z(y)e^{k y_s}\,.
\ee
On the other hand, adding the warping leaves $A(y)$ nearly unchanged near the singularity (adding a constant $k y_s$ to infinity makes no difference).
Therefore,  demanding $A_w(y)\simeq A(y)$ near $y=y_s$ leads to
\be
[\rho z(y)]^{1/(1-2\beta)}=[\rho_w z_w(y)]^{1/(1-2\beta)}
=[\rho_w z(y)e^{k y_s}]^{1/(1-2\beta)}
,
\label{eq:warpeddownrho}
\ee
and hence
\be
\rho_w=\rho e^{-k y_s}\simeq\frac{e^{-k y_s}}{y_s}\,.
\ee
Combining this with Eq.~(\ref{ys2}) we find a strong suppression of  $\rho_w/k$ resulting just from $\mathcal O(1)$ numbers. The quantity $\rho_w$ sets the scale for the {KK} spectrum in this case, and more explicitly the spectrum is approximated as%
\be
m_n\simeq\rho_w\, n^{2\beta}\,,
\label{mn}
\ee
which can be obtained with a WKB approximation as in Sec.~\ref{subsec:classificationofspectra}. We see that $\rho_w$ indeed sets the scale of the 4D masses, which are hence parametrically suppressed with respect to $k$. The complete superpotential that accomplishes a hierarchy and leads to the spectrum Eq~(\ref{mn}) is
\be
W(\phi)=k(1+e^{\phi/\sqrt{6}}[\phi-\phi_1]^\beta)\,.
\label{super}
\ee

At this point, we already have a recipe of how to construct superpotentials that are consistent background solutions and that accomplish the stabilization of the hierarchy and feature a specific {KK} spectrum.  In a first step, one chooses the asymptotic (i.e.~large $\phi$) behavior of $W$. This will determine the asymptotic form of the spectrum.

 In a 
second step, one completes $W$ for smaller values of $\phi$ in such a way as to accomplish a mild hierarchy of the proper distance $y_s$ with respect to the fundamental 5D scale $k$, given by the simple relation Eq.~(\ref{kys2}). Notice that many of the interesting spectra require some kind of exponential behavior at large $\phi$, such that this region does not contribute at all to $k y_s$. 

Let us conclude this section by noting that there are certainly other ways to obtain the mild hierarchy $k y_s$, including moderate fine-tunings of parameters.
What is completely generic, though, is the fact that adding warping as in Eq.~(\ref{trick}) leaves $k y_s$ manifestly unchanged but suppresses the masses by an additional warp factor $e^{k y_s}$.

\section{A self-stabilized soft-wall model} 
\label{sec:aselfstabilizedsoftwallmodel}
Let us now apply the results of the previous sections an consider an explicit class of stabilized soft-wall models. That is, models with a single {UV} brane, a singularity at finite $y=y_s$ and with a mass scale hierarchically smaller than the Planck scale without fine-tuning of parameters. We will in addition require these models to behave as $\mathrm{AdS}_5$ near the {UV} brane. 

Following the recipe of Sec.~\ref{sec:constructingsoftwallmodels}, we know that these properties can be obtained, e.g., from the simple superpotential of the SW2 type
\begin{equation}
W(\phi)=6 k(1+e^{\nu\phi/\sqrt{6}})\,,
\label{ourW}
\end{equation}
where $k$ is some arbitrary dimensionful constant of the order of the 5D Planck scale, and $\nu<2$. The background solution can be easily obtained using Eqs.~(\ref{eq:EOMsuperpotential1},~\ref{eq:EOMsuperpotential2}) and reads
\begin{align}
A(y)&=k y-\frac{1}{\nu^2}\log\left(1-\frac{y}{y_s}\right)
\label{eq:backgroundA}
,\\
\phi(y)&=-\frac{\sqrt{6}}{\nu}\log[\nu^2k(y_s-y)]\,.
\label{eq:backgroundphi}
\end{align}
At the point $y=y_s$ we encounter a naked curvature singularity and for $y\ll y_s$, i.e.~near the boundary at $y=0$, the
geometry is $\mathrm{AdS}_5$.

The bulk potential which corresponds to the superpotential (\ref{ourW}) is given by
\be
V(\phi)=(3 k^2\nu^2-12 k^2)e^{2\nu\phi/\sqrt{6}}-24 k^2e^{\nu\phi/\sqrt{6}}-12 k^2
.
\label{potential}
\ee
\begin{itemize}
\item
For $\nu\leq 2$ the potential is bounded from above. More precisely it satisfies the condition
\be
V(\phi[y])\leq V(\phi_0)
,
\label{condition}
\ee
necessary for the corresponding bulk
geometry to support finite temperature in the form of black hole
horizons, a characteristic feature of physical
singularities~\cite{Gubser:2000nd}, as we commented in Sec.~\ref{subsec:solutionswithspacetimesingularities}. Moreover for $\nu<2$, as we have
seen in the previous section, the {EOM}s are satisfied at
the singularity and there is no need to resolve it.

\item
For $\nu>2$ the {EOM}s are not satisfied at the
singularity and the latter would need to be resolved to fine-tune to
zero the four-dimensional cosmological constant. Finally the potential
is not bounded from above and finite temperature is not supported in the dual theory.
\end{itemize}

\begin{figure}[t]
\centering
\begin{psfrags}
\psfrag{absphi0}[tc][tc]{ $\vert \phi_0 \vert$}%
\psfrag{Boxkys30}[cc][cc]{ $k y_s = 30$}%
\psfrag{log10rhok}[bc][bc]{ $\log_{10} ( \rho/k)$}%
\psfrag{Nu1}[cc][cc]{ $\nu =1$}%
\psfrag{Nu2}[cc][cc]{ $\nu =2$}%
\psfrag{Nu}[tc][tc]{ $\nu $}%
\psfrag{S11}[tc][tc]{ $1$}%
\psfrag{S121}[tc][tc]{ $1.2$}%
\psfrag{S141}[tc][tc]{ $1.4$}%
\psfrag{S151}[tc][tc]{ $1.5$}%
\psfrag{S161}[tc][tc]{ $1.6$}%
\psfrag{S181}[tc][tc]{ $1.8$}%
\psfrag{S21}[tc][tc]{ $2$}%
\psfrag{S251}[tc][tc]{ $2.5$}%
\psfrag{S31}[tc][tc]{ $3$}%
\psfrag{S351}[tc][tc]{ $3.5$}%
\psfrag{S41}[tc][tc]{ $4$}%
\psfrag{W0}[cr][cr]{ $0$}%
\psfrag{Wm12}[cr][cr]{ $-10$}%
\psfrag{Wm1342}[cr][cr]{ $-13.4$}%
\psfrag{Wm1362}[cr][cr]{ $-13.6$}%
\psfrag{Wm1382}[cr][cr]{ $-13.8$}%
\psfrag{Wm1422}[cr][cr]{ $-14.2$}%
\psfrag{Wm142}[cr][cr]{ $-14$}%
\psfrag{Wm1442}[cr][cr]{ $-14.4$}%
\psfrag{Wm152}[cr][cr]{ $-15$}%
\psfrag{Wm22}[cr][cr]{ $-20$}%
\psfrag{Wm252}[cr][cr]{ $-25$}%
\psfrag{Wm32}[cr][cr]{ $-30$}%
\psfrag{Wm51}[cr][cr]{ $-5$}%
\psfrag{x0}[tc][tc]{ $0$}%
\psfrag{x11}[tc][tc]{ $1$}%
\psfrag{x2}[tc][tc]{ $0.2$}%
\psfrag{x4}[tc][tc]{ $0.4$}%
\psfrag{x6}[tc][tc]{ $0.6$}%
\psfrag{x8}[tc][tc]{ $0.8$}%
\psfrag{y0}[cr][cr]{ $0$}%
\psfrag{y11}[cr][cr]{ $1$}%
\psfrag{y2}[cr][cr]{ $0.2$}%
\psfrag{y4}[cr][cr]{ $0.4$}%
\psfrag{y6}[cr][cr]{ $0.6$}%
\psfrag{y8}[cr][cr]{ $0.8$}%
\includegraphics{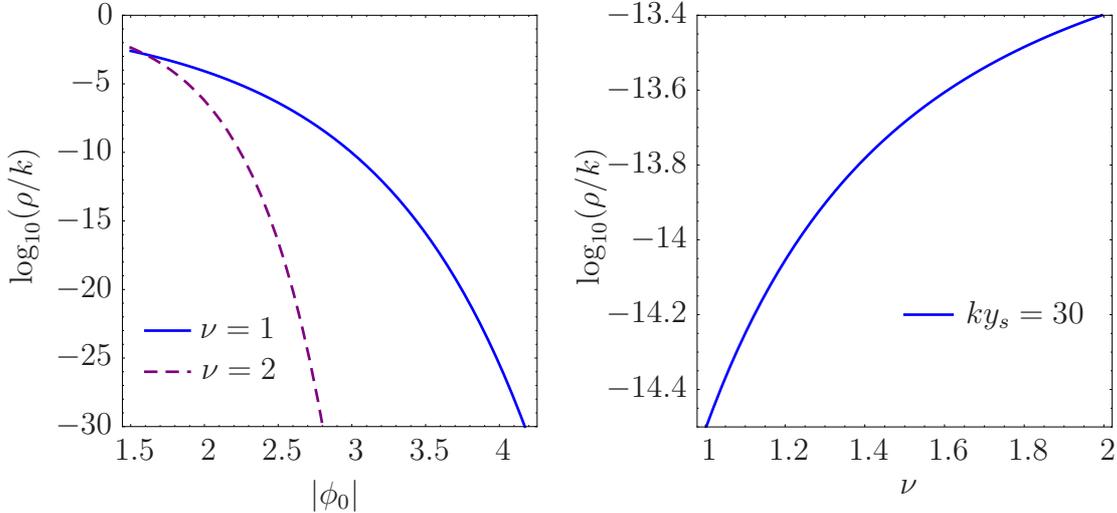}
\end{psfrags} \caption{\it Plot of $\log_{10}(\rho/k)$ as a function of $|\phi_0|$ for $\nu=1$ and $\nu=2$ [left panel], and as a function of $\nu$ for $k y_s = 30$  (this value will be used in following plots) [right panel].
}
\label{FT}
\end{figure}

The location of the singularity depends exponentially on the brane value of $\phi$,
\be
k y_s=\frac{1}{\nu^2} e^{-\nu \phi_0/\sqrt{6}}.
\label{kys}
\ee
We know from Sec.~\ref{sec:constructingsoftwallmodels} that the relevant mass scale for the 4D
spectrum is not the inverse volume, but rather a ``warped down'' quantity as in Eq.~\eqref{eq:warpeddownrho}. In the case of study, it is convenient to define
\be
\rho\equiv k(k y_s)^{-1 / \nu^2 }e^{-k y_s}\,,
\label{eq:rhosoftwalls}
\ee
which, as we will shortly see, is a quantity that reflects well the scale of {KK} excitations. All we need in order to create the electroweak hierarchy is thus
$\phi_0<0$ but otherwise of order unity.  This can be achieved with a
fairly generic brane potential, for instance by choosing a suitable $\lambda(\phi)$ such that the {BC} of Eq.~\eqref{eq:BC2} is satisfied
for our superpotential.\footnote{In order to satisfy the other {BC}, 
Eq.~\eqref{eq:BC1}, we still need a fine-tuning, for instance by adding a
$\phi$ independent term to $\lambda(\phi)$. This is precisely the
tuning of the 4D CC discussed above that of course has nothing to do
with the electroweak hierarchy we want to explain here.}  
For negative $\phi_0$, the ratio of scales $k/\rho$ exhibits a double exponential behavior
\be
\log\, \frac{k}{\rho}\sim \frac{e^{\nu(-\phi_0)/\sqrt{6}}}{\nu^2}   +\dots
,
\label{rhok}
\ee
 and we can create a huge hierarchy with very little fine-tuning. In Fig.~\ref{FT} we
plot $\rho/k$ as a function of $|\phi_0|$ for different values of $\nu$ and also as a function of $\nu$ for a fixed value $k y_s=30$ which generates a hierarchy of about fourteen orders of magnitude.

A comment about the motivation for choosing this particular superpotential is in order here. Its
particular form, Eq.~(\ref{ourW}), guarantees full analytic control over our
solution, providing a relatively simple framework to explore some of the interesting properties of soft-wall models, while introducing only one parameter, $\nu$,  that can be used to control the departure of our model from the {AdS} solution, which is recovered at $\nu= \infty$.

\subsection{The mass spectra}
From what we discussed in Sec.~\ref{sec:the4dspectrum}, and in particular having a look at Table~\ref{default} we can already advance that this set of models feature a discrete solution for $\nu>1$, a continuous spectrum with mass gap for $\nu=1$, and a continuous spectrum without mass gap for $\nu<1$. Let us, however, explicitly solve the fluctuation {EOM}s for our model in order to study the mass spectrum with more detail. 

Let us begin by finding an explicit expression for the relation between $z$ and $y$. From Eq.~\eqref{eq:relationzy} and for $\nu>0$ we easily find 
\be
\rho(z-z_0)=%(k y_s)^{1/\nu^2-1}\left[
\Gamma(1-1/\nu^2,ky_s-ky)-\Gamma(1-1/\nu^2,ky_s)%     \right]
\,,
\label{rel}
\ee
where $z_0$ corresponds to the location of the {UV} brane that we assume to be at $z_0=1/k$ and $\Gamma(a,x)$ is the incomplete gamma function. Since we are taking $e^{k y_s}\gg1$ and hence $k/\rho\gg 1$ we can approximate $\Gamma(1-1/\nu^2,k y_s)\simeq \rho/k$ and (\ref{rel}) simplifies to
\be
\rho z\simeq\Gamma(1-1/\nu^2,ky_s-ky)\,. \label{relsimp}
\ee
For $\nu>1$ the singularity at $y_s$ translates into a singularity at $z_s$ given by
\be
\rho z_s\simeq% 
\Gamma(1-1/\nu^2)\,.
\label{relsing}
\ee
For $0<\nu\leq1$ the singularity at $y_s$ translates into a singularity at $z_s\to\infty$. 

Let us now inspect the graviton and radion fluctuations for the case of study.

\subsubsection{The graviton}
\begin{figure}[t]
\begin{psfrags}
\psfrag{S1k}[tc][tc]{$1/k$}%
\psfrag{Szs}[tc][tc]{$z_s$}%
\psfrag{W0}[cr][cr]{ $0$}%
\psfrag{Wmg}[cr][cr]{$m_g$}%
\psfrag{WV0}[cr][cr]{$V_0$}%
\psfrag{z}[tc][tc]{$z$}%

\centering
\subfigure[$0<\nu<1$]{
\includegraphics{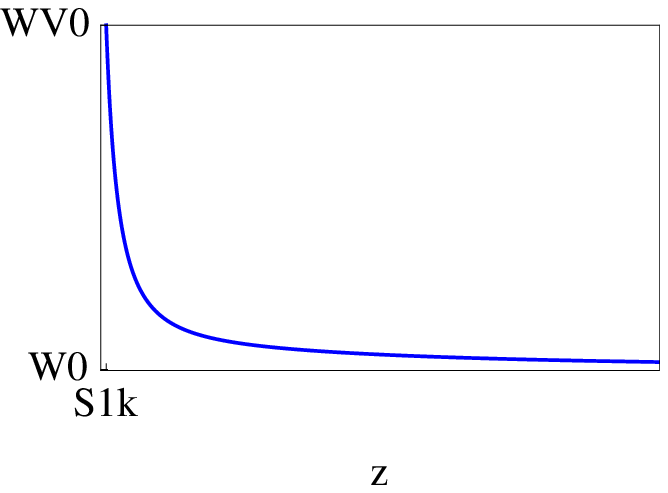}
\label{fig:potential<1}
}
\subfigure[$\nu=1$]{
\includegraphics[scale=1]{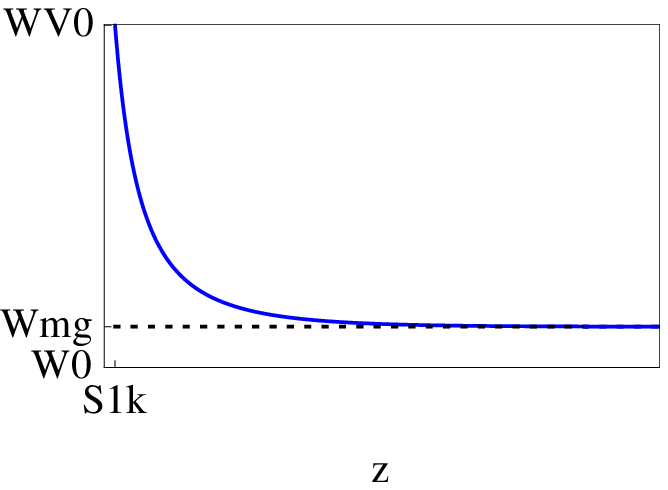}
\label{fig:potential=1}
}
\subfigure[$1<\nu<\sqrt{5/2}$]{
\includegraphics{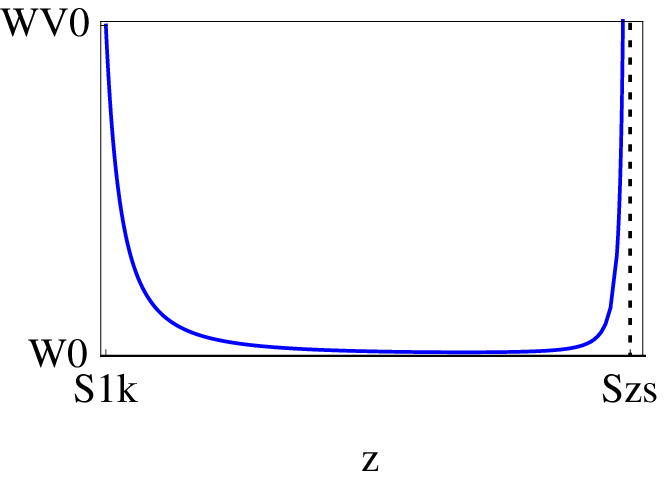}
\label{fig:potential>1}
}
\subfigure[$\nu>\sqrt{5/2}$]{
\includegraphics{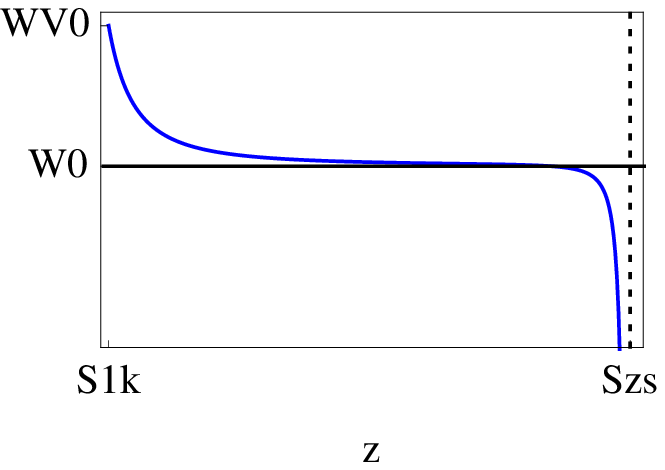}
\label{fig:potential>5/2}
}
\caption{\it Behavior of $V_{h}(z)$ for different values of $\nu$. Here, $V_0\equiv V_h(1/k)$. For the radion, $V_{F}(z)$ has the same behavior with the exception that Fig.~\ref{fig:potential>1} applies for all $\nu>1$.  }
\label{fig:potential}
\end{psfrags}
\end{figure}
From \eqref{GSchroedinger} we know that, after the redefinition $\tilde{h}(z) = e^{-3A(z)/2} h(z)$, we can express the fluctuation {EOM}s as 
\begin{equation}
 - {\tilde{ h}}''(z) +  V_{h}(z) \tilde h(z) = m^2 \tilde h(z) \,,
 \label{eq:GEOMy2}
\end{equation}
where $V_{h}(z)$ is given by \eqref{eq:Vhschroedinger}. In the case of study, it is only possible to obtain an analytic expression for $V_h(z)$ in the $y$-frame, where it reads
\begin{equation}
  V_{h} (z[y]) = \frac{3 e^{-2 k y} \left(1-\frac{y}{y_s}\right)^{\frac{2}{\nu^2}} \left[\,5 \nu^4 k^2  (y-y_s)^2
 -10 \nu ^2 k
   (y-y_s) - 2 \nu^2 +5\,\right]}{4 \nu ^4 (y-y_s)^2}\, .
 \label{GVhat}
\end{equation}
It is however possible to invert numerically the coordinate change \eqref{rel}, and so to plot \eqref{GVhat}. Its behavior for different values of $\nu$ is shown in Fig.~\ref{fig:potential}. One can distinguish three possible situations~\footnote{Similar potentials were considered in Ref.~\cite{Freedman:1999gk}.}:
\begin{itemize}
 \item $\nu<1$ [Fig.~\ref{fig:potential<1}] In this case $z$ extends to infinity where $ V_{h} \rightarrow 0$. The mass spectrum is continuous from $m=0$, i.~e.~without a mass gap. However, conformal symmetry is broken due to the occurrence of the scale $y_s$.
 \item $\nu=1$ [Fig.~\ref{fig:potential=1}] $z$ also extends to infinity but $ V_{h} \rightarrow (9/4)\rho^2$. This leads to a continuous spectrum with a mass gap $m_{\mathrm{g}} = (3/2) \rho$. 
 \item $\nu>1$ [Figs.~\ref{fig:potential>1} and \ref{fig:potential>5/2}] $z_s$ is finite and thus the mass spectrum is discrete. The potential diverges at $z_s$ changing sign at $\nu^2 = 5/2$, but this does not have observable consequences in the mass spectrum as we will see. 
\end{itemize}
Equations \eqref{GEOMy} and \eqref{GSchroedinger} do not have analytic solutions.  However, for $\nu>1$ one can find approximations for the wavefunction in the regions near the brane and near the singularity.  Let us first consider the region near the brane ($ky \simeq 0$). Assuming $k y_s \gg 1$ the potential \eqref{GVhat} is approximated as
\begin{equation}
  V_{h} \vert_{y\simeq 0} \simeq \frac{15\, k^2}{4} e^{-2 k y} \simeq \frac{15}{4} \frac{1}{z^2} \,, 
 \label{GVhatapproxy0}
\end{equation}
where the coordinate change is given by \eqref{relsimp}, which is approximated for $\nu > 1$ as
\begin{equation}
 k z  \simeq  e^{ky} \,.
 \label{coordapprox0}
\end{equation}
One can see that \eqref{GVhatapproxy0} corresponds to an {AdS} metric. With this approximated potential, Eq.~\eqref{eq:GEOMy2} is solved by
\begin{equation}
 \tilde h(z) \vert_{z\simeq z_0} = c_{1} \sqrt{k z} J_2 (m z ) + c_{2} \sqrt{k z} Y_{2} (m z) \,.
\end{equation}
The two coefficients can be determined by the normalization and the {BC} \eqref{Gconditionszs} at $z_0$, i.e.
\begin{equation}
\left.e^{-3 A(z)/2} \left( \dot{\tilde h} (z) + \frac32 \dot{A}(z) \tilde h(z) \right) \right\vert_{z_0,z_s} 
 = 0 \,,
\label{eq:GBCy2}
\end{equation}
which yields
\begin{equation}
 \frac{c_2}{c_1} = -\frac{J_1(m/k)}{Y_1(m/k)} \sim \left( \frac{m}{k} \right)^2 \simeq 0\,,
\end{equation}
since we expect the first mass modes to be of order $m \simeq (z_s - z_0)^{-1}$, and in our approximation $k (z_s - z_0) \gg 1$. %%

Let us now move on to consider the region next to the singularity ($y\simeq y_s$). In this case the potential is approximated by
\begin{equation}
  V_{h} \vert_{y\simeq y_s} \simeq \frac{3(5 - 2\nu^2)}{4 \nu^4} 
  \frac{ \rho^2 }{ [k(y_s-y)]^{2-2/\nu^2}} 
\simeq
\frac{3(5 - 2\nu^2)}{4(1 - \nu^2)^2} \frac{ 1 }{ (z_s-z)^{2}}
\,,
\label{GVhatapprox1}
\end{equation}
where we have used that, for $\nu>1$, the coordinate change \eqref{rel} is approximated by
\begin{equation}
 \rho(z_s - z) = \frac{\nu^2}{\nu^2 -1} [k (y_s - y)]^{1-1/\nu^2} \,.
 \label{coordsingularity}
\end{equation}
With this approximation, Eq.~\eqref{GSchroedinger} yields the solution
\begin{equation}
 \tilde h (z) = c_J  \sqrt{m \Delta z} J_\alpha ( m \Delta z) + c_Y  \sqrt{ m \Delta z} Y_\alpha ( m \Delta z )
 \,,
 \label{GsolutionII}
\end{equation}
where
\begin{equation}
 \alpha = \frac{4-\nu^2}{2(\nu^2-1)}
\end{equation}
and $\Delta z \equiv z_s - z $. The two integration constants can be obtained by imposing the {BC} at the singularity and normalizability and by matching this solution to the solution for the intermediate region between the  brane and the singularity. Near the singularity \eqref{GsolutionII} behaves like
\begin{equation}
 \tilde h (z) 
 \sim
  c^{(1)}_J (\Delta z)^{3/(2\nu^2-2)} + c^{(2)}_J (\Delta z)^{(4\nu^2 -1)/(2\nu^2 -2)} + 
 c^{(1)}_Y (\Delta z)^{(2\nu^2 - 5)/(2\nu^2 - 2)} 
\,,
\label{Gsolutionzs}
\end{equation}
where numerical factors are being absorbed in the constants $c_i$. We have included the next to leading order in the expansion of $J_\alpha$ as we need it for computing the {BC}, which reads
\begin{align}
 e^{-3A/2} \left( \dot{\tilde h} (z) + \frac{3}{2} \dot A(z) \tilde h(z) \right) 
 \sim 
 \hat{c}^{(2)}_J (\Delta z)^{(\nu^2 + 2)/(\nu^2 -1) }
+ 
 \hat{c}^{(1)}_Y (\Delta z)^0 \,.
 \label{Gbczs}
\end{align}
Again numerical factors have been absorbed in $c'_i$. Note that the {BC} is only satisfied when $c_Y = 0$, and that this condition also ensures that the solution \eqref{Gsolutionzs} is normalizable when $\nu^2 < 2$.

The {BC}s provide the quantization of the mass eigenstates for $\nu>1$. In order to compute the mass spectrum for the graviton one should match the solutions at the ends of the space with a solution for the intermediate region. Unfortunately, for the parameter range we are interested in we do not have good analytic control for this region. However we can extract a generic property of the spectrum by looking at the potential Eq.~(\ref{GVhat}) and using the form of the coordinate transformation Eq~(\ref{relsimp}) to deduce that, assuming $e^{k y_s} \gg 1$, the potential has the form\footnote{This behavior can actually be seen in the limiting cases Eq.~(\ref{GVhatapproxy0}) and Eq.~(\ref{GVhatapprox1}).}
\be
V_h(z)=\rho^2\, v_h(\rho z)\,,
\label{scaling1}
\ee
where $v_h$ is some dimensionless function of the dimensionless variable $\rho z$. 
In other words we have eliminated the two scales $k, y_s$ in favor of the single scale $\rho$ given in Eq.~(\ref{eq:rhosoftwalls}). The spectrum is therefore of the form
\be
m_n(\nu,k,y_s)=\mu_n(\nu)\,\rho(\nu,k,y_s)\,,
\label{scaling2}
\ee
where the pure numbers $\mu_n$ only depend on the parameter $\nu$ but not on the parameters $k$ or $y_s$. 

\begin{figure}[t]
\centering
\begin{psfrags}
\psfrag{mRhoNu}[bc][bc]{ $m/\rho(\nu)$}%
\psfrag{Nu}[tc][tc]{ $\nu $}%
\psfrag{S11}[tc][tc]{ $1$}%
\psfrag{S121}[tc][tc]{ $1.2$}%
\psfrag{S141}[tc][tc]{ $1.4$}%
\psfrag{S161}[tc][tc]{ $1.6$}%
\psfrag{S181}[tc][tc]{ $1.8$}%
\psfrag{S21}[tc][tc]{ $2$}%
\psfrag{W0}[cr][cr]{ $0$}%
\psfrag{W122}[cr][cr]{ $12$}%
\psfrag{W12}[cr][cr]{ $10$}%
\psfrag{W142}[cr][cr]{ $14$}%
\psfrag{W21}[cr][cr]{ $2$}%
\psfrag{W41}[cr][cr]{ $4$}%
\psfrag{W61}[cr][cr]{ $6$}%
\psfrag{W81}[cr][cr]{ $8$}%
\includegraphics[width=0.7\textwidth]{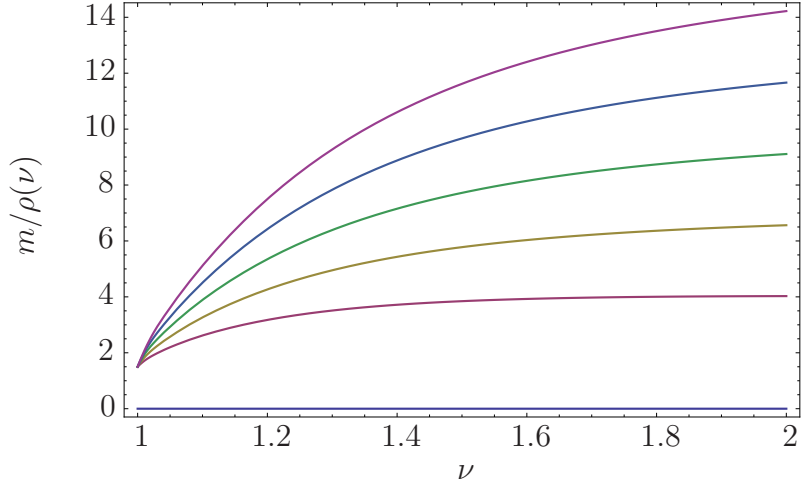}
\end{psfrags} 
\caption{\it Mass modes for the graviton, computed for $k y_s > 4$.\textsuperscript{\ref{footnotec}} The massless $(n=0)$ and the first 5 massive modes $(n=1,\ldots,5)$ are shown.}
\label{fig:massesG}
\end{figure}

\begin{figure}[t]
\centering
\begin{psfrags}
\psfrag{n0}[cc][cc]{ $\text{n=0}$}%
\psfrag{n1}[cc][cc]{ $\text{n=1}$}%
\psfrag{n2}[cc][cc]{ $\text{n=2}$}%
\psfrag{S0}[tc][tc]{ $0$}%
\psfrag{S11}[tc][tc]{ $1$}%
\psfrag{S2}[tc][tc]{ $0.2$}%
\psfrag{S4}[tc][tc]{ $0.4$}%
\psfrag{S6}[tc][tc]{ $0.6$}%
\psfrag{S8}[tc][tc]{ $0.8$}%
\psfrag{SqrtBoxBoxA}[Bc][Bc]{ $\sqrt{z_s} \, \tilde{h}(z)$}%
\psfrag{SqrtBoxBox}[cr][cr]{ $\sqrt{z_s} \, \tilde{h}(z)$}%
\psfrag{W0}[cr][cr]{ $0$}%
\psfrag{W11}[cr][cr]{ $1$}%
\psfrag{W151}[cr][cr]{ $1.5$}%
\psfrag{W5}[cr][cr]{ $0.5$}%
\psfrag{Wm11}[cr][cr]{ $-1$}%
\psfrag{Wm151}[cr][cr]{ $-1.5$}%
\psfrag{Wm5}[cr][cr]{ $-0.5$}%
\psfrag{zzsA}[Bc][Bc]{ $z/z_s$}%
\psfrag{zzs}[tc][tc]{ $z/z_s$}%
\includegraphics[width=0.7\textwidth]{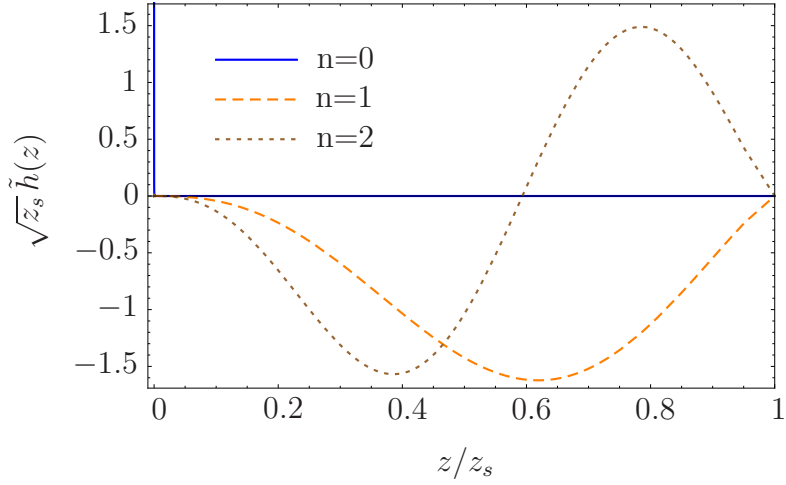}
\end{psfrags} 
\caption{\it {KK} graviton profiles in the $z$ frame for $k y_s = 30$ and $\nu=3/2$, using the normalization $\int dz \, \tilde h^2 =1$. The massless mode ($n=0$) is peaked near the brane. The two first massive modes ($n=1,2$) are also shown. The zero mode becomes more peaked near the brane in comparison to the massive modes as $ky_s$ increases.}
\label{fig:gprofile}
\end{figure}

Moreover one can find an expression for the spacing of the mass eigenstates by approximating the potential as an infinite well, which is valid for $m^2 \gg  V_{h}$. The result of this approximation is
\begin{equation}
 \Delta m \simeq   \frac{\rho\, \pi}{\Gamma\left( 1 - 1/\nu^ 2 \right)}=
\frac{\pi}{z_s} \,.
 \label{GDeltam}
\end{equation}
Note that the mass spectrum is linear ($m_n \sim n$), and that as one approaches $\nu=1$
\begin{equation}
  \lim_{\nu \to 1} \ \Delta m=0 ,
% \stackrel[\nu\rightarrow 1]{}{\longrightarrow} 0 ,
\end{equation}
recovering the expected continuous spectrum at this value (for $\nu<1$ the spectrum is continuous too, since \eqref{GDeltam} is only valid for $\nu>1$). The numerical result for the mass eigenvalues is shown in Fig.~\ref{fig:massesG} where these behaviors can be observed. Some profiles for the graviton computed numerically using the {EOM} \eqref{eq:GEOMy2} and the {BC}s \eqref{eq:GBCy2} are shown in Fig.~\ref{fig:gprofile}.

\footnotetext{Numerically one finds that the scaling property in Eq.~\eqref{scaling2} ceases to be valid for $k y_s \lesssim 3$, as discrepancies from this behavior become greater than $1\%$.  \label{footnotec}}

\subsubsection{The radion}
For the radion we make the redefinition $\tilde{F}(z) = e^{-3A(z)/2} F(z)/\dot{\phi}(z)$, after which we get the Schroedinger-like equation \eqref{Fschroedinger}
\begin{equation}
 -\ddot{\tilde F}(z) +  V_{F} (z) \tilde F(z) = m^2 \tilde F(z) \,,
 \label{Fschroedinger2}
\end{equation}
where, in the $y$-frame, the potential $V_F(z)$, defined in Eq.~\eqref{FVhatz}, is given by
\begin{equation}
  V_{F} (z[y]) = \frac{ e^{-2 k y} \left(1- \frac{y}{ y_s}\right)^{\frac{2}{\nu^2}} 
 \left[
 3 \nu^4 k^2  (y-y_s)^2
 + (-6\nu^2 + 8 \nu^4) k   (y-y_s)
 + 6\nu^2 + 3
 \right]}{4 \nu ^4 (y-y_s)^2}\,.
 \label{FVhat}
\end{equation}
This potential has similar form to the graviton potential, and the three situations presented above also apply for the radion (with the same mass gap for $\nu=1$). A difference is that this potential does not change the sign of divergence for $\nu>1$, although this does not have any observable consequences, as we have already seen.

\begin{figure}[t]
\centering
\begin{psfrags}
\psfrag{mRhoNu}[bc][bc]{   $m/\rho(\nu)$}%
\psfrag{Nu}[tc][tc]{   $\nu $}%
\psfrag{S11}[tc][tc]{   $1$}%
\psfrag{S121}[tc][tc]{   $1.2$}%
\psfrag{S141}[tc][tc]{   $1.4$}%
\psfrag{S161}[tc][tc]{   $1.6$}%
\psfrag{S181}[tc][tc]{   $1.8$}%
\psfrag{S21}[tc][tc]{   $2$}%
\psfrag{W0}[cr][cr]{   $0$}%
\psfrag{W1252}[cr][cr]{   $12.5$}%
\psfrag{W12}[cr][cr]{   $10$}%
\psfrag{W152}[cr][cr]{   $15$}%
\psfrag{W251}[cr][cr]{   $2.5$}%
\psfrag{W51}[cr][cr]{   $5$}%
\psfrag{W751}[cr][cr]{   $7.5$}%
\includegraphics[width=0.7\textwidth]{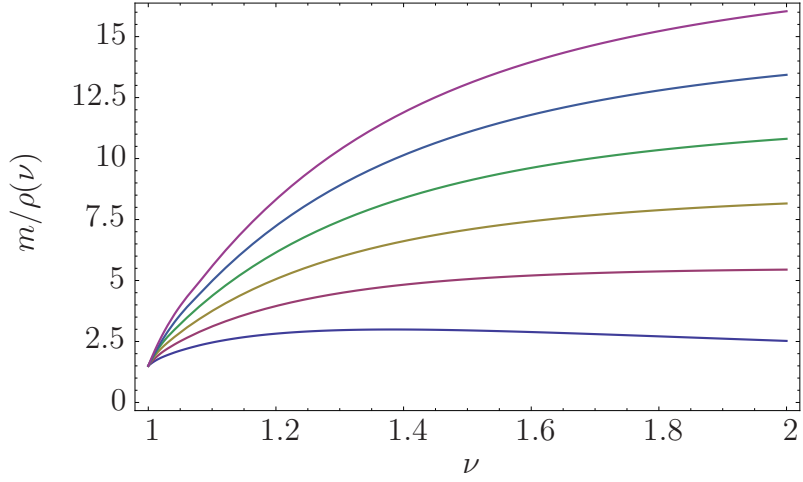}
\end{psfrags} 
\caption{\it Mass modes for the radion, computed for values of $k y_s > 4$ \textsuperscript{\ref{footnotec}}. The first 6 massive modes ($n=0,\ldots,5$) are shown.}
\label{fig:massesR}
\end{figure}

\begin{figure}[t]
\centering
\begin{psfrags}
\psfrag{n0}[cc][cc]{   $\text{n=0}$}%
\psfrag{n1}[cc][cc]{   $\text{n=1}$}%
\psfrag{n2}[cc][cc]{   $\text{n=2}$}%
\psfrag{S0}[tc][tc]{   $0$}%
\psfrag{S11}[tc][tc]{   $1$}%
\psfrag{S2}[tc][tc]{   $0.2$}%
\psfrag{S4}[tc][tc]{   $0.4$}%
\psfrag{S6}[tc][tc]{   $0.6$}%
\psfrag{S8}[tc][tc]{   $0.8$}%
\psfrag{SqrtBoxBoxA}[Bc][Bc]{   $\sqrt{z_s} \, \tilde{\varphi}(z)$}%
\psfrag{SqrtBoxBox}[cr][cr]{   $\sqrt{z_s} \, \tilde{\varphi}(z)$}%
\psfrag{W0}[cr][cr]{   $0$}%
\psfrag{W11}[cr][cr]{   $1$}%
\psfrag{W21}[cr][cr]{   $2$}%
\psfrag{Wm11}[cr][cr]{   $-1$}%
\psfrag{Wm21}[cr][cr]{   $-2$}%
\psfrag{zzsA}[Bc][Bc]{   $z/z_s$}%
\psfrag{zzs}[tc][tc]{   $z/z_s$}%
\includegraphics[width=0.7\textwidth]{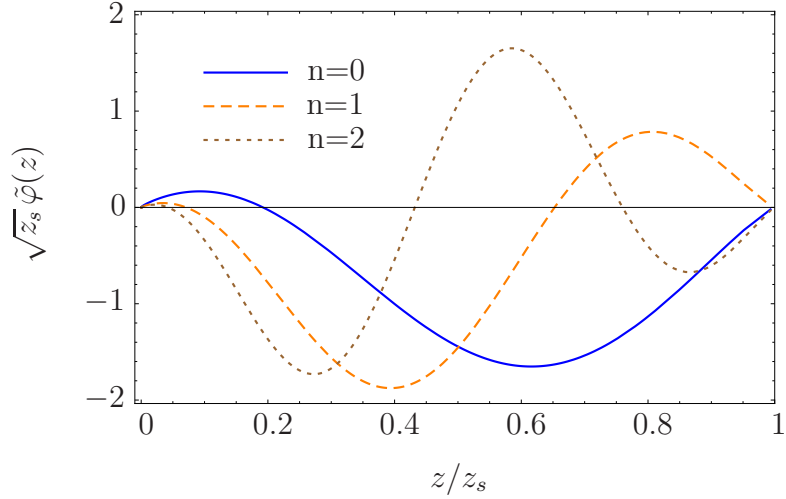}
\end{psfrags} 
\caption{\it {KK} profiles of the rescaled scalar fluctuations $\tilde\varphi(z)$ for
$k y_s = 30$ and $\nu=3/2$, using the normalization $\int dz \, \tilde \varphi^2 =1$.  The three first massive modes ($n=0,1,2$) are shown.}
\label{fig:rprofile}
\end{figure}

Let us now proceed to find the approximation for the wavefunction near the {UV} brane. Taking $k y \simeq 0$ and $k y_s \gg 1 $ and using \eqref{coordapprox0}, \eqref{FVhat} is given by
\begin{equation}
  V_{F} \vert_{y \simeq 0} \simeq \frac{3\,k^2}{4} e^{-2 k y} \simeq \frac34 \frac1{z^2} \,, 
\end{equation}
and hence the solution to \eqref{Fschroedinger} is
\begin{equation}
 \tilde F (z) \vert_{z\simeq z_0} = c_1 \sqrt{k z} J_1(mz) + c_2 \sqrt{k z} Y_1 (m z) \,. 
\end{equation}
The coefficients $c_i$ are to be determined using the {BC}s at the brane \cite{Csaki:2000zn}. As an example, using the condition~\footnote{This condition holds for brane potentials satisfying $\partial^2 \lambda / \partial \phi^2 \gg 1$ \cite{Csaki:2000zn}.} $\varphi(y=0) = 0$  (which we will use for the numerical computation) yields
\begin{equation}
 \frac{c_2}{c_1} = - \frac{J_2(m/k)}{Y_2(m/k)} \simeq 0
\,.
\end{equation}

Next to the singularity, and using Eq.~\eqref{coordsingularity}, the potential is approximated by
\begin{equation}
  V_{F} \vert_{y\simeq y_s} \simeq \frac{6\nu^2 + 3}{4 \nu^4} \frac{\rho^2}{ [k(y_s-y)]^{2-2/\nu^2}} 
\sim
\frac{6 \nu^2 + 3}{4(1 - \nu^2)^2} \frac{ 1 }{ (z_s-z)^{2}}
\,,
\end{equation}
that gives the solution
\begin{equation}
 \tilde F (z) = c_J  \sqrt{m \Delta z} J_\alpha ( m \Delta z) + c_Y  \sqrt{ m \Delta z} Y_\alpha ( m \Delta z ) \,,
 \label{FsolutionII}
\end{equation}
with
\begin{equation}
 \alpha = \frac{2+\nu^2}{2\nu^2 -2} \,.
\end{equation}
The behavior of this solution near the singularity is 
\begin{equation}
 \tilde F (z)  
\sim
  c^{(1)}_J  (\Delta z)^{(2\nu^2 +1)/(2\nu^2 -2)} + c^{(2)}_J (\Delta z)^{(6\nu^2 -3)/(2\nu^2 - 2)}   + 
 c^{(1)}_Y (\Delta z)^{-3/(2\nu^2 -2)}
\,.
\label{Rsolutionzs}
\end{equation}
The rescaled scalar fluctuation $\tilde\varphi(z) = e^{-3A(z)/2}\varphi(z)$, related to $\tilde F(z)$ by Eq.~\eqref{Ftildevarphi}, behaves near the singularity as 
\begin{equation}
\tilde{\varphi}(z) \sim  
 \hat{c}^{(1)}_J (\Delta z)^{3/(2\nu^2 -2)}  + 
 \hat{c}^{(1)}_Y (\Delta z)^{-(2\nu^2 + 1)/(2\nu^2 -2)}   \,,
\end{equation}
to which we have to apply the normalizability condition \eqref{Fnorm}, i.e.
\begin{equation}
\int_{z_0}^{z_s} dz \tilde{\varphi}^2(z)  < \infty\,,
\end{equation}
and the {BC} \eqref{Fboundary},
\begin{equation}
0 =  e^{-3A(z)} \dot{\varphi}(z) \sim
 c^{\prime (2)}_J (\Delta z)^{(\nu^2+2)/(\nu^2 -1)}  + 
 c^{\prime\prime (1)}_Y (\Delta z)^{(-2\nu^2 + 2)/(\nu^2 - 1)} \,.
\end{equation}
Again, the condition $c_Y = 0$ is sufficient to ensure both the fulfillment of the {BC}s and the normalizability. The scaling of the mass eigenvalues Eq~(\ref{scaling2}) and the 
approximation \eqref{GDeltam} for the spacing of the mass modes also holds for the radion. The numerically obtained values for the masses are shown in Fig.~\ref{fig:massesR}. In comparison to the graviton mass modes of Fig.~\ref{fig:massesG}, note that the first mode for the radion is lighter than the first massive mode of the graviton. This can be understood recalling that the radion does not have a zero mode.{\footnote{One can in fact show that for $\nu\rightarrow \infty$, which we can only take if we resolve the singularity, the lightest mode tends to be massless, corresponding to the radion profile in a RS2 model, as described in Ref.~\cite{Charmousis:1999rg}.}
Some profiles of the scalar fluctuations of the field $\tilde\varphi$ are shown in Fig.~\ref{fig:rprofile}.

\section{Applications of soft-wall models}
\label{sec:applicationsofsoftwallmodels}

As we already stated in the beginning of this chapter, there are a number of phenomenological applications of soft-wall models that have been already proposed, although most of them are still worth of future research in detail. In particular, the possibility of having naturally stabilized soft-wall models lets us describe phenomenology at the TeV scale or lower without incurring in a fine-tuning problem. 

To begin with, the SW2 case \eqref{eq:WSW2} for $\beta=1/4$ is particularly useful to provide a holographic description of QCD. When $\beta=1/4$ the spectrum behaves as $m_n^2\simeq \rho_w^2 n$, which corresponds to the linear Regge trajectory spectrum for the mesons, and has been shown to be appropriate for AdS/QCD models as in Ref.~\cite{Batell:2008zm}. In this case one would obtain the linear confinement behavior of e.g.~$\rho$-mesons by considering an additional piece in our action $\int d^5 x \sqrt{-g}\, e^{-\frac{1}{2}\phi} \mathcal{L}_\mathrm{mesons}$. The fact that asymptotically $A(z)\sim\phi(z)\sim z^2$ guarantees that the resonances of the vector mesons follow the same linear law as the ones for the scalars and tensors.

Another interesting case is the SW1 class \eqref{eq:WSW1} when $\nu=1$. In this case the mass spectrum of fields propagating in the bulk is a continuum above a mass gap, which can easily be set at a scale of $\mathcal O(TeV)$. This continuum (endowed with a given conformal dimension) can interact with {SM} fields propagating in the {UV} brane as operators of a CFT, where the conformal invariance is explicitly broken at a scale given by the mass gap, and can model and describe the unparticle phenomenology. This could also be applied to {EWSB}, as in the unHiggs theory of Ref.~\cite{Stancato:2008mp}, were the Higgs is embedded in such 5D background.

In the SW1 case for the parameter range $1>\nu>2$ the spectrum is quite similar to the RS1 model \cite{Randall:1999vf}, with some particularities. The most interesting feature is the fact that, at values of $\nu$ close to 1 the spectrum mimics a continuum with a mass gap (see e.g.~Fig.~\ref{fig:massesR}). In this case, if the first mode is at the TeV scale, several {KK} modes could be found with masses within the LHC energy range. Provided that the {KK} modes couple strongly enough to {SM} matter, this would provide a very particular signature of this kind of models, since in the standard RS1 models only one or two modes can be accommodated in the TeV range (i.e.~the density of states is higher than in RS1).

\subsection{Electroweak symmetry breaking}
\label{subsec:ewsbsw}
The fact that the SW1 case for $1>\nu>2$ is similar to RS1 makes it natural to try to apply these models to {EWSB}, the original motivation for RS1. The built-in stabilization mechanism of soft-wall models seems to hint towards a solution of the hierarchy problem, reducing even more the need for fine-tuning, provided that a {EWSB} mechanism can be described in a soft-wall background. 

One needs to take care of the electroweak precision observables when considering this possibility, and it was indeed shown in Ref.~\cite{Falkowski:2008fz} that the $S$ parameter could be reduced in soft-wall models featuring a custodial symmetry. What is more, applying the results on {EWSB} we will describe in the remaining of this thesis to soft-wall models indicates that the $T$ parameter could be reduced as well \cite{Cabrer:2011fb}. 

However, when one tries to solve the hierarchy problem in warped models, the Higgs needs to be localized in the {IR} brane or towards it \cite{Randall:1999vf}. In the soft-wall case, without an {IR} brane the Higgs can only be in {UV} brane or in the bulk. The former case does not solve the hierarchy problem, since we need to set a very small vacuum expectation value for the Higgs, which will be necessarily fine-tuned. 

One possibility would be to consider a bulk Higgs localized towards the {IR}. However, we still need some mechanism to localize the Higgs and thus trigger {EWSB}. In general, and as we will see in the following chapters, it is not possible to fix a {IR}-localized background for a scalar from {BC}s at the {UV} brane, unless we again introduce a fine-tuning. The only way-out would be to use some nonlinear dynamics on the {UV} brane or in the bulk, but a scalar potential that satisfies the required properties and preserves calculability turns out to be very difficult to construct. 

We will return to the question of {EWSB} using soft walls in Sec.~\ref{sec:commentssw}, when we will discuss on these difficulties with more detail. For the remaining of this Thesis, and in order to describe {EWSB}, we will choose to abandon the soft-wall models. However, we will insist in using singular metrics in a two-brane setup, i.e.~introducing a brane (a hard wall) before the singularity is reached. In fact, we will shortly see how some very interesting properties can be obtained by the use of singular metrics.

%%%%%%%%%%%%%%%%%%%%%%%%%%%%%%%%%%%%%%%%

\chapter{Electroweak Symmetry Breaking with a Bulk Higgs}
\chaptermark{EWSB with a Bulk Higgs}

\label{chap:ewsbbulkhiggs}

Having decided to abandon soft-wall models to describe {EWSB}, we will now focus on the construction of warped models with two branes and general metrics that break electroweak symmetry. In this chapter we will study some of the generalities of these models, and we will be focusing on those that have a Higgs boson propagating in the bulk of the extra dimension. In particular, we will consider a 5D extension of the {SM}, where the electroweak sector will be propagating in the bulk of the extra dimension, while the fermions will be located on the {UV} brane for simplicity.

When gauge bosons propagate in the 5D bulk their {KK} excitations can contribute to the Electroweak Precision Observables (EWPO).\footnote{In general, we would also need to consider the bounds coming from the fermionic sector. However, placing all fermions on the {UV} brane conveniently removes these constraints.} In order to construct realistic models, these contributions will have to be contrasted with all the {SM} Electroweak Precision Tests (EWPT)~\cite{Nakamura:2010zzi}. Since the {KK} modes decouple when they are heavy, {EWPT} will translate into lower bounds on their masses. If these bounds are much larger than the TeV scale, the theories will suffer of a ``little hierarchy" problem, which means that a certain amount of fine-tuning will be needed to stabilize the weak masses in the effective theory below the {KK} scale. Furthermore, when bounds on the {KK} scale are larger than a few TeV, the models become phenomenologically unappealing, for they are outside the LHC range.

In general, the precision observable responsible for the stronger constraints on 5D warped models is the T parameter \cite{Huber:2000fh,Davoudiasl:2009cd}, as the contributions to it are volume enhanced. In order to suppress these contributions, it was proposed to enlarge the gauge symmetry in the bulk by embedding the hypercharge in a extended gauge group $SU(2)_R \times U(1)_{B-L}$~\cite{Agashe:2003zs}. In this case, {KK} resonances preserve the \emph{custodial symmetry}  $SU(2)_V\subset SU(2)_L\times SU(2)_R$ after electroweak breaking, which protects the T parameter from large tree-level corrections. The relevant bounds arise then from the $S$ parameter and turn out to be $\mathcal O(3)$~TeV for the {RS} model. However, an extra discrete left--right symmetry is needed to keep under control volume enhanced corrections to the $Z\bar bb$ coupling~\cite{Agashe:2006at}. Another way of reducing the $T$ parameter in the absence of an extra custodial symmetry is introducing large IR brane kinetic terms~\cite{Davoudiasl:2002ua}. However, since IR brane radiative corrections are expected to be small this effect relies on unknown UV physics, which prevents calculability in the low energy effective theory. 

In this chapter, which relies in the results first presented in Refs.~\cite{Cabrer:2010si,Cabrer:2011fb}, we will study {EWSB} in models with arbitrary metrics and bulk Higgs profiles \emph{without} introducing this extra custodial symmetry. Our aim is to prepare the grounds for constructing models that do not need this custodial symmetry and have $\mathcal{O}$~(TeV) bounds, based on the use of generalized metrics and propagating the Higgs in the bulk. While here we will only present the generalities, we will construct models of this kind in Chapter~\ref{chap:noncustodial}. Finally, we will apply the results we obtain to the {RS} model, and we will see how a heavy bulk Higgs eases the {SM} naturalness problem and lowers the bounds on the masses of {KK} modes, as shown in Ref.~\cite{Cabrer:2011vu}.

\section{A 5D Standard Model}
\label{sec:5DSM}

We will consider the {SM} propagating in a warped 5D space with an arbitrary metric given by [as in Eq.~\eqref{eq:metricfirstdefinition}]
\begin{equation}
ds^2 = e^{-2A(y)} dx^\mu dx^\nu \eta_{\mu \nu} + dy^2
\end{equation}
and two flat branes localized at $y=0$ and $y=y_1$, at the edges of a finite $S^1/\mathbb{Z}_2$ interval. The dynamics of the gravitational system are described by the action
\begin{equation}
S = \frac{M_5^3}{2} \int d^5 x \sqrt{-g} R + S_5 \,,
\end{equation}
where $M_5$ is the 5D Planck scale and $S_5$ is the piece that contains the SM fields propagating in the bulk [see Eq.~\eqref{5Daction}]. The 4D (reduced) Planck mass $M_{\mathrm{Pl}} = 2.4 \cdot 10^{18}~\mathrm{GeV}$ is related to $M_5$ by
\begin{equation}
M_{\mathrm{Pl}} = M_5^3 \int e^{-2A(y)} dy \,.
\label{eq:MPlanck4D}
\end{equation}

The 5D $SU(2)_L\times U(1)_Y$ gauge bosons will be defined as $W^i_M(x,y)$, $B_M(x,y)$, where $i=1,2,3$ and $M=0,1,2,3,5$. In the weak basis, these gauge bosons will be given by $A_M^\gamma(x,y)$, $Z_M(x,y)$ and $W^{\pm}_M(x,y)$, corresponding respectively to the 5D photon and Z and W bosons. 

Finally, we will introduce a 5D version of the {SM} Higgs as
\be
H(x,y)=\frac{1}{\sqrt 2}e^{i \chi(x,y)} \left(\begin{array}{c}0\\h(y)+\xi(x,y)
\end{array}\right)
\,,
\label{Higgs}
\ee
where the matrix $\chi(x,y)$ contains the three 5D {SM} Goldstone fields $\vec\chi(x,y)$, see Eq.~(\ref{chimatrix}). In this chapter we will consider the Higgs 5D background $h(y)$, as well as the metric $A(y)$, to be arbitrary functions. 

We will consider the 5D action (in units of the 5D Planck scale) for the gauge fields, the Higgs field $H$ and other possible scalar fields of the theory, generically denoted as $\phi$, as 
\begin{eqnarray}
S_5&=&\int d^4x dy\sqrt{-g}\left(-\frac{1}{4} \vec W^{2}_{MN}-\frac{1}{4}B_{MN}^2-|D_M H|^2-\frac12(D_M \phi)^2
-V(H,\phi)
\right)\nonumber\\
&-&\sum_{\alpha}\int d^4x dy \sqrt{-g}\,(-1)^\alpha\,2\,\lambda^\alpha(\phi,H)\delta(y-y_\alpha)
\,,
\label{5Daction}
\end{eqnarray}
where $V(H,\phi)$ is the 5D scalar potential. $\lambda^\alpha(\phi,H)$ are the 4D brane potentials (with $\alpha=0,1$ respectively for the UV and IR branes). They are related to the background solution by
\begin{equation}
A'(y_\alpha) = \left. \frac23 \lambda^\alpha (\phi,H) \right\vert_{y=y_\alpha}, 
~~~~
\phi'(y_\alpha) =  \left. \frac{\partial \lambda^\alpha}{\partial \phi} \right\vert_{y=y_\alpha}.
\label{eq:BCAphi}
\end{equation}

From here on we will assume that $V(H,\phi)$ is quadratic in $H$. {EWSB} will be triggered on the {IR} brane. We thus choose the $H$-dependent part of the brane potentials as
\begin{align}
\lambda^0(\phi_0 , H)&=M_0 |H|^2 \,,
\label{eq:boundpot0}
\\
-\lambda^1(\phi_1 , H)&=-M_1 |H|^2+\gamma |H|^4 \,,
\label{eq:boundpot1}
\end{align}
where we denote by $\phi_\alpha$ the VEV of $\phi$ at the branes denoted by $\alpha=0,1$. 

\subsection{The gauge sector}
Let us start by considering the gauge sector of our theory. We can construct the 4D effective action out of the 5D action of Eq.~(\ref{5Daction}) by making the expansion in {KK} modes
\begin{equation}
A_\mu(x,y)= \frac{1}{\sqrt{y_1}} \sum_n a_\mu(x) f^{(n)}_A(y) \,,
\end{equation}
where $A=A^\gamma,Z,W^{\pm}$. Each of the functions $f^{(n)}_A$, that correspond to the $n$-th {KK} mode, satisfies the {EOM}\footnote{Some details on the derivation of the fluctuation {EOM}s can be found in App.~\ref{fluctuationsgauge}.}
\be
m_{f_A}^2 f_A+(e^{-2A}f^{\prime}_A)'-M_A^2 f_A=0 
\ ,
\label{eq:EOMf}
\ee
where the superscript $(n)$ has been removed to simplify the notation. The functions $f_A(y)$ are normalized as
\begin{equation}
\int_0^{y_1}f_A^2(y)dy=y_1
\end{equation}
and satisfy the Neumann {BC}s (which are required for having a light mode)
\begin{equation}
\left. f_A^{\prime}(y)\right|_{y=0,y_1}=0 \,.
\end{equation}
We have defined the 5D $y$-dependent gauge boson masses as
\be
M_W(y) \equiv \frac{g_5}{2} h(y)e^{-A(y)}
\ ,
\qquad M_Z(y) \equiv \frac{1}{c_W} M_W(y)
\ ,
\qquad M_\gamma(y)
\equiv 0
\ .
\label{masas1}
\ee
where $g_5$ and $g'_5$ are the 5D $SU(2)_L$ and $U(1)_Y$ couplings respectively, and 
\begin{equation}
c_W=\frac{g_5}{\sqrt{g_5^2+g_5'^2}} \,.
\end{equation}

In general, Eq.~\eqref{eq:EOMf} will not have analytic solutions. However, we can find an approximation for the mass of lightest mode of \eqref{eq:EOMf}, that we will denote by $m_A \equiv m_{f_A^0}$, in the cases where the breaking is small, i.e.~when the lightest mode is much lighter than the subsequent {KK} modes ($m_A \ll m_{f_A^n}$). In the limit where no breaking occurs, there will be a zero-mass mode with a constant profile. Expanding around this limit we can write 
\begin{equation}
f_A(y) = 1 - \delta_A + \delta f_A(y) \,,
\end{equation}
where
\begin{eqnarray}
 \delta f_A(y) &=& \int_0^y dy'\,e^{2A(y')} \int_0^{y'} dy''\left[ M_{A}^2(y'') - m_{A,0}^2 \right]   
\,,
 \\
 \delta_A&=& \frac{1}{y_1}\int_0^{y_1}  dy\, \delta f_A(y)
\,,
 \\
 m_{A,0}^2 &=&  \frac{1}{y_1}\int_0^{y_1} {M^2_{A}(y)} dy 
 \label{eq:mgaugezero}
\,,
 \end{eqnarray}
and where $m_{{A,0}}$ is a zeroth order approximation for the zero mode mass $m_A$.
Including the first order deviations from a constant profile we obtain for the light mode mass the refined expression  
\be
 m^2_{A} =m_{A,0}^2-\frac{1}{y_1}\int_0^{y_1}dy\, M^2_{A}(y)\left[\delta_A-\delta f_A(y)   \right]
 \,.
\label{masa1}
\ee
Following our definitions Eq.~\eqref{masas1}, we obtain for the photon $m_\gamma = 0$, while for the massive gauge bosons $m_Z$ and $m_W$ have to be matched to the physical values. 

Only the lightest mass eigenvalue will be significantly affected by the breaking, so we simplify our notation by defining
\be
m_{n}=m_{f_A^{n}}\,,\qquad f^{n}(y)=f^{n}_A(y)\,,
\ee
for the heavier modes ($n\geq 1$). In particular, masses and wave functions of the $n\geq 1$ KK excitations of the $W$ and $Z$ bosons as well as photon and gluons coincide, the splitting being negligible for all purposes in the models we will consider in this thesis. We can also find an approximation for the profiles of the heavy modes, expressed in conformal coordinates as
\begin{equation}
f^{(n)}(z) \simeq z \left[ Y_0 (m_n z_0 ) J_1 (m_n z) - J_0(m_n z_0) Y_1(m_n z) \right] \,,
\end{equation}
where $J_\alpha$ ($Y_\alpha$) are the Bessel functions of the first (second) kind and order $\alpha$.

On top of the modes from $A_\mu(x,y)$ there will also be pseudoscalar fluctuations $\eta_A(x,y)$ arising  from the $A_5$ -- $\chi$ sector (see App.~\ref{fluctuationsgauge} and Ref.~\cite{Falkowski:2008fz} for details). For each broken gauge symmetry there is a massive tower of such pseudoscalars.  The {EOM} and {BC} of the 5D wavefunctions $\eta_A(y)$ are given by [Eqs.~(\ref{eq.eta}--\ref{b.c.D})]
\begin{gather}
 m_{\eta_A}^2\eta_A+\left[M_A^{-2}\left(e^{-2A}M_A^2\eta_A \right)'  \right]'
 -M_A^2\eta_A=0 \,,
 \label{eq:EOMpseudoscalar}
\\
\left.%e^{-4A}
\eta_A(y)\right|_{y=0,y_1}=0\,.
\end{gather}
In the limit of vanishing {EWSB} these equations unify with their counterpart for the Higgs fluctuation  [Eqs.~(\ref{eq:EOMxi}--\ref{eq:BCxi})] to form complex doublets. Hence, the splitting for finite breaking is expected to be small (i.e.~proportional to the mass of the light Higgs).

\subsection{The Higgs sector}
\label{subsec:higgssector}
Let us now consider the Higgs sector. From Eq.~(\ref{Higgs}) and the 5D action (\ref{5Daction}) we can write the {EOM} for the background $h(y)$ as 
\begin{equation}
h''(y)-4A'h'(y)-\frac{\partial V}{\partial h}=0\,,
\label{eq:EOMh}
\end{equation}
while the {EOM}s at the branes read
\begin{equation}
h'(y_\alpha)=\left.\frac{\partial\lambda^\alpha}{\partial h}\right|_{y=y_\alpha}.
\label{eq:BCh}
\end{equation}
For the fluctuations $\xi(x,y)$, defined in Eq.~\eqref{Higgs}, and after proceeding similarly to the previous section, we obtain that the wavefunctions $\xi^{(n)}(y)$ satisfy the bulk {EOM} and {BC}s
\begin{gather}
\xi''(y)-4A'\xi'(y)-\frac{\partial^2 V}{\partial h^2}\xi(y)+m_H^2 e^{2A}\xi(y)=0\,,
\label{eq:EOMxi}
\\
\frac{\xi'(y_\alpha)}{\xi(y_\alpha)}=\left.\frac{\partial^2 \lambda^\alpha}{\partial h^2}\right|_{y=y_\alpha} .
\label{eq:BCxi}
\end{gather}
The functions $\xi(y)$ are normalized as
\begin{equation}
\int_0^{y_1} e^{-2A(y)} \xi^2(y) dy = y_1  \,.
\label{eq:normalizationxi}
\end{equation}

With our choice of the boundary potentials, Eqs.~(\ref{eq:boundpot0}--\ref{eq:boundpot1}), the {UV} {BC}s for the background and the fluctuations are the same. Hence, since we are considering a quadratic bulk Higgs potential, for $m_H=0$ the Higgs wave function $\xi(y)$ ($n=0$) is proportional to $h(y)$. For a small Higgs mass this will still be a good approximation to the exact wavefunction. This means that the 5D VEV will be carried almost entirely by the zero mode. Let us therefore simplify the discussion by considering an effective theory by writing 
\begin{equation}
H(x,y)=\sqrt k \, \frac{\mathcal H(x)h(y)}{h(y_1)}\,, 
\end{equation}
where we have introduced the UV scale $k$ to account for the correct dimension of $\mathcal H$. Now we can calculate the effective Lagrangian for the mode $\mathcal H(x)$ to find
\be
\mathcal L_{\rm eff}=- Z e^{-2A(y_1)}\,|D_\mu \mathcal  H|^2-
e^{-4A(y_1)}\left[\left(\frac{h'(y_1)}{h(y_1)}-M_1\right)k|\mathcal H|^2+\gamma k^2|\mathcal H|^4\right]\,,
\label{Leff}
\ee
where 
\be
Z=
k\int_0^{y_1}dy\frac{h^2(y)}{h^2(y_1)}e^{-2A(y)+2A(y_1)}\,.
\label{Z}
\ee

Several things can be learned from the effective Lagrangian of Eq.~(\ref{Leff}). 
The warp factors have the same effect as in the usual {RS} compactification with a Higgs localized on the {IR} brane: they red-shift  all mass scales in the IR. The quantity $Z$ is an additional wave function renormalization depending on both the 
gravitational and Higgs backgrounds.
In Sec.~\ref{sec:EWPO} we will see that a sizable $Z$ can reduce the electroweak precision observables.\footnote{In particular, we will see that the $T$ parameter is suppressed by two powers of $Z$, the $S$ parameter by just one, while the $Y$ and $W$ are unaffected by $Z$}

Minimizing the potential in $\mathcal L_{\rm eff}$ one finds the condition
\be
|\langle\mathcal H\rangle|^2=\frac{1}{2\gamma k}\left(M_1-\frac{h'(y_1)}{h(y_1)}\right)\,,
\ee
and hence the physical Higgs mass in the {EWSB} minimum is
\be
m_H^2
=\frac{2}{kZ}\left(M_1-\frac{h'(y_1)}{h(y_1)}\right)\rho^2
\ .
\label{mH}
\ee
Here the {UV} and {IR} scales, $k$ and $\rho$, are related by\footnote{This is a generalization of the RS model, where $\rho=k e^{-ky_1}$. Note that the definition of $\rho$ we use here is different to the one used in soft-wall models, for Eq.~\eqref{rho} would not make sense due to the lack of an IR brane}
\be
\rho\equiv k e^{-A(y_1)}\,.
\label{rho}
\ee
Let us now rederive Eq.~(\ref{mH}) in a different way.
The relation $\xi(y) \propto h(y)$  (exact for $m_H=0$) 
can be corrected  to $\mathcal O(m_H^2)$. Making an expansion of the {EOM} (\ref{eq:EOMxi}) and imposing the normalization \eqref{eq:normalizationxi} we obtain the wavefunction
\be
\xi(y)=\sqrt{\frac{ky_1}{Z}}\frac{h(y)}{h(y_1)}e^{A(y_1)}\left[1-m_H^2\left(\int_0^y e^{2A}\frac{\Omega}{\Omega'}
+\int_0^{y_1} e^{2A}\frac{\Omega}{\Omega'}(\Omega-1)\right)\right]\,
\ ,
\label{xiH}
\ee
where $\Omega$ is a function defined as
\be
\Omega(y)=\frac{\int_0^y h^2(y')e^{-2A(y')} \, dy'}{\int_0^{y_1}h^2(y')e^{-2A(y')} \,dy'} \,.
\label{Omega}
\ee
The true value of $m_H$ (and hence the validity of this expansion) is of course determined by the {BC}s given in Eq.~(\ref{eq:BCxi}). From Eq.~(\ref{xiH}) it follows that
\begin{equation}
\frac{\xi'(y_1)}{\xi(y_1)}  %= 3 \frac{h'(y_1)}{h(y_1)} - 2 M_1
=\frac{h'(y_1)}{h(y_1)}-k Z\frac{m_H^2}{\rho^2}
\label{BCxi}
,
\end{equation}
while from Eq.~(\ref{eq:EOMh}) and (\ref{eq:EOMxi}) we also have
\begin{gather}
\frac{h'(y_1)}{h(y_1)}=M_1-\gamma h^2(y_1)\,,
\\
\frac{\xi'(y_1)}{\xi(y_1)}=M_1-3\gamma h^2(y_1)=
3\frac{h'(y_1)}{h(y_1)}-2M_1\,.
\label{eq:BChiggsB}
\end{gather}
Combining Eqs.~(\ref{BCxi}--\ref{eq:BChiggsB}) one recovers Eq.~(\ref{mH}) for the light Higgs mass, as predicted from the effective theory. 
Moreover, notice that the {BC}s are universal for all modes and hence Eq.~(\ref{BCxi}) can be used to express the BC for the whole Higgs KK tower in terms of the Higgs mass ($m_H$) rather than the information about the boundary ($M_1$,\ $\gamma$).

We are finally interested in knowing the strength of the coupling of the light Higgs mode to the $W$ and $Z$ bosons and its KK modes, since this will determine how well perturbative unitarity is maintained. In fact, if the Higgs is light enough, we expect only small corrections to the SM coupling. Using the definition of the $WW\xi_n$ coupling, 
\be
h_{WW\xi_n}=\frac{g}{y_1}\int_0^{y_1} dy\,e^{-A(y)}M_A(y)f_0^2(y)\xi_n(y) \,, 
\ee
and the wave function (\ref{xiH}) one can deduce that
\be
h_{WW \xi_0}=h_{WW H}^{SM}\left[1-\mathcal O(m_H^2/m_{KK}^2,m_W^2/m_{KK}^2)\right]
,
\ee
so the coupling of the light to the gauge bosons only differs mildly to the {SM} result. The conclusion is then that a light Higgs unitarizes the theory in a similar way to the SM Higgs.

\subsubsection{Achieving a light Higgs}

Let us now make a few comments on the amount of fine-tuning required to have light modes in the Higgs and  gauge boson sectors. For this purpose, we will analyze the effective {SM} Lagrangian, which can be written as
\be
\mathcal L_{SM}=-\left|\mathcal D_\mu H_{SM}\right|^2+\mu^2|H_{SM}|^2-\lambda |H_{SM}|^4 \,,
\ee
where the SM Higgs field $H_{SM}(x)$ and the SM parameters $\mu^2$ and $\lambda$ are related to 5D quantities by 
\begin{eqnarray}
H_{SM}(x)&=& \sqrt{Z}e^{-A(y_1)} \mathcal H(x)\,,\\
\mu^2&=&(kZ)^{-1}\left(M_1-\frac{h'(y_1)}{h(y_1)}  \right)\; \rho^2\label{muSM}\,,\\
\lambda&=&\frac{\gamma k^2}{Z^2}\label{gammaSM}\,,
\end{eqnarray}
and from where the expressions for the Higgs mass (\ref{mH}) and the gauge boson masses (\ref{eq:mgaugezero}) easily follow from the usual SM relations. 

The required amount of fine-tuning at the tree-level in the 5D parameters is summarized in Eq.~(\ref{muSM}), where we see that, depending on the value of $\rho$ and $Z$, we eventually have to fine-tune the boundary mass $M_1$ with respect to $h'(y_1)/h(y_1)$ in order to obtain $\mu\sim 100$~GeV. For instance, a no-fine-tuning condition would imply that the parameter
\be
\frac{1}{k} \left( M_1-\frac{h'(y_1)}{h(y_1)} \right)=\frac{m_H^2}{\rho^2}\; \frac{Z}{2}\,,
\label{higgsnoft}
\ee
should be of  $\mathcal O(1)$, while a smaller value would imply some amount of fine-tuning\footnote{E.g.~a value of 0.1 (0.01) would amount to a 10\% (1\%) fine-tuning, and so on}. 

Controlling this fine-tuning is required when constructing particular models, and we will take care of this when presenting our model in Chapter~\ref{chap:noncustodial}. However, we can already advance some generalities. In fact, when constructing new models we will be interested in having a sizable $Z$, as it will reduce the contribution to {EWPO} and hence allow for a lower $\rho$. Moreover, we see here that the parameter $\mu$, or equivalently the Higgs mass, is further reduced with respect to $\rho$ by a factor $1/\sqrt{Z}$, which in turn reduces the required amount of fine-tuning in $\left[M_1-h'(y_1)/h(y_1)\right]$. The point is that, in general, trying to lower the contributions to {EWPO} seems to work in the same direction of reducing the fine-tuning of the Higgs mass, thus allowing for a lighter Higgs. 

\subsection{The metric fluctuations}
\label{radionEWSB}

To conclude this section, we will now analyze the tensorial and scalar fluctuations of the metric, which correspond respectively to the graviton and the radion. We have already analyzed these fluctuations in a soft-wall context (see Sec.~\ref{sec:the4dspectrum}), although now we will of course need to consider of the existence of two branes. 

Recall from Sec.~\ref{sec:the4dspectrum} that the metric and scalar fluctuations can be parametrized as
\begin{gather}
 ds^2 = e^{-2A(y) - 2F(x,y)}  [\eta_{\mu\nu}+ h_{\mu\nu}(x,y)] dx^{\mu} dx^{\nu}  + \left[ 1 + J(x,y) \right]^2dy^2  \,, 
 \label{eq:metricfluct}
 \\
 \phi(x,y) = \phi(y) + \varphi(x,y) \,,
 \label{eq:phifluct}
\end{gather}
and that the the tensor fluctuations satisfy the {EOM} and {BC}
\be
h_{\mu\nu}'' - 4A' h_{\mu\nu}' + e^{2A} m^2 h_{\mu\nu} = 0\,,\qquad
h_{\mu\nu}'(y_\alpha)=0\,.
\label{graviton}
\ee
The three scalars $F,\ J,\ \varphi$ are not independent, and their {EOM}s are given by~\eqref{Feqmotion}, recall
\begin{gather}
 F'' - 2 A' F' - 4 A'' F - 2 \frac{\phi''}{\phi'} F' + 4 A' \frac{\phi''}{\phi'} F = - m^2 e^{2A} F \,,
 \label{eq:radionEOM}\\
 \phi'(y) \varphi(y) =  F'(y) - 2 A'(y) F(y) \,, 
\label{eq:radionconstraint} 
 \\
 J(y) = 2F(y) \,.
 \label{eq:Feqmotion2}
\end{gather}
The boundary condition for the scalar at the {UV} and {IR} branes is given in Ref.~\cite{Csaki:2000zn} as
\be
\bigl[-\varphi'+\lambda''_\alpha \varphi+\lambda'_\alpha F\bigr]_{y=y_\alpha}=0\,,
\label{eq:radionBC2}
\ee
with the boundary potentials $\lambda_\alpha$ evaluated on the background. Notice that $\lambda_\alpha''(\phi_\alpha)$ is a constant that can be chosen at will without changing the corresponding BC for $\phi$, which only depends on $\lambda_\alpha'(\phi_\alpha)$. In order to decouple the boundary condition, one usually takes the limit of large $\lambda''_{\alpha}(\phi_\alpha)$, in which case $\varphi$ is frozen at the boundary and Eq.~(\ref{eq:radionBC2}) together with the constraint (\ref{eq:radionconstraint}) implies $(e^{-2A}F)'|_{y_\alpha}=0$. To be more general, we will for the moment not specify the value of $\lambda_\alpha''(\phi_\alpha)$, which remains as a free parameter of the theory. 

\subsubsection{The mass of the radion} 

In a slice of pure AdS$_5$ the radion is massless, and no other scalar modes are present. In the dual theory, the radion is the Goldstone mode of the breaking of scale invariance in the IR. Adding a stabilizing scalar field corresponds to an explicit breaking of conformal invariance by some relevant operator, so the radion becomes a pseudo-Goldstone field and acquires a mass. 
In most cases studied in the literature, the deformation of AdS by the scalar field is small, the radion remains light and its mass can be computed perturbatively~\cite{Csaki:2000zn}. However, when the deformation of AdS is large (as in the models we will consider later on in this thesis) one could expect the radion to be heavy. To be prepared, we will now derive a general approximation of the radion mass in models with general metrics.

Using Eq.~(\ref{eq:radionconstraint}), as well as the bulk {EOM} (\ref{eq:Feqmotion2}), we can rewrite the {BC} as\footnote{Here we are using the background {EOM}s and {BC}s. See Eqs.~(\ref{eq:eom1}--\ref{eq:eom3}) and subsequent equations for their explicit form.}
\be
\left[ m^2 F+\hat M_\alpha(e^{-2A}F)'\right]_{y=y_\alpha}=0\,,
\label{radiongen}
\ee
where we have defined the effective brane mass parameter%~\footnote{In terms of the superpotential method used in Sec.~\ref{model} one has $\phi''_\alpha/\phi'_\alpha=W''(\phi_\alpha)$ such that one can express the mass $\hat M_\alpha=\lambda''_\alpha(\phi_\alpha)-W''(\phi_\alpha)$. }
\be
\hat M_\alpha%=\lambda_\alpha''-W''
=\lambda''_\alpha-\frac{\phi_\alpha''}{\phi_\alpha'}\,.
\ee
It will also be convenient to recast the bulk {EOM} Eq.~(\ref{eq:Feqmotion2}) into the form
\be
\left(e^{2A}(A'')^{-1}[e^{-2A}F]'\right)'+(m^2e^{2A} (A'')^{-1}-2)F=0\,.
\label{eq:FEOMnewform}
\ee
The new system of Eqs.~(\ref{radiongen}--\ref{eq:FEOMnewform}) only depends on the background metric $A$ and the two free parameters $\hat M_\alpha$. Notice that the quantity $A''(y)$ is a measure of the back reaction and goes to zero in the AdS limit. Therefore, for $A''(y)=0$ the expected zero mode $F(y)=e^{2A}$ appears. For a small radion mass $m_0$ we can expand around this mode to obtain the first perturbation
\be
m_0^2=\frac{2\int e^{2A}}{\int e^{4A}(A'')^{-1}+(\hat M_0A_0'')^{-1}-e^{4A_1}(\hat M_1A_1'')^{-1}}\,.
\label{mrad}
\ee
Unless $\hat M_0$ is fine-tuned to zero, the second term in the denominator can always be neglected. The third term can however be important and hence the radion mass will depend on it. 
Since we are expanding in the dimensionful parameter $m^2$, the region of validity of this expansion is not so clear. 
To better judge on its convergence, one can compute the subleading correction $\delta m_0^2$ to Eq.~(\ref{mrad}). One finds 
\be
\frac{\delta m_0^2}{m_0^2}=-2\int_0^{y_1}e^{2A}\cdot\int_0^{y_1}e^{-2A}A''
\,\chi^2
\label{NLO}
\ee
with the function $\chi(y)$ defined as
\be
\chi(y)=\frac{\int_0^y e^{2A}}{\int_0^{y_1}e^{2A}}-
\frac{\int_0^y e^{4A}(A'')^{-1}}{\int_0^{y_1}e^{4A}(A'')^{-1}-e^{4A_1}(A''_1\hat M_1)^{-1}}\
.
\ee
The important observation is that, even if $A''$ is large, the correction Eq.~(\ref{NLO}) can be small if the function $\chi(y)$ is small.  For instance, consider the limit $\hat M_1\to \infty$. Then $\chi(y)$ is the difference of two functions that monotonically increase from zero to one. Therefore this function is always smaller than one and, in particular, it vanishes at $y=y_1$ where $A''(y)$ in Eq.~(\ref{NLO}) is expected to be largest. 
%For the model to be introduced in Sec.~\ref{model}, $\delta m_0^2/m_0^2$ is always negligible even for radion masses that one would naively consider to be large (i.~e.~of the order of the gauge boson's KK mass). On the other hand the approximation is slightly worse for finite values of $\hat M_1$ and of course breaks down completely when $\hat M_1$ is fine tuned to cancel the first term in the denominator of Eq.~(\ref{mrad}).

Whenever $\delta m_0^2/m_0^2$ is small we expect the couplings of the radion
to be very well approximated by using the leading order wave function 
\be
F(x,y)\simeq e^{2A(y)}r(x)\ .
\ee
This includes, but is not limited to, the case of very light radion and small back-reaction.

\section{Oblique Electroweak Precision Observables}
\label{sec:EWPO}

As advanced in the beginning of this chapter, the construction of models of {EWSB} requires the {EWPO} to be matched with the constraints coming from the experimental {EWPT}. In fact, the impact of the  KK modes on the EWPO will depend crucially on how strongly the former couple to the Higgs currents. Writing the relevant piece of the effective Lagrangian as 
\begin{equation}
\mathcal{L} = \sum_n \alpha_n \left( g W_\mu^n j_\mu^L + g' B_\mu^n j_\mu^Y \right)
\end{equation}
we can express the couplings as 
\begin{equation}
\alpha_n = \frac{k}{Z} \int_0^{y_1} dy \frac{h^2(y)}{h^2(y_1)} e^{-2A(y) + 2A(y_1)} f_n(y) \,,
\label{eq:alphan}
\end{equation}
where $Z$ was defined in Eq.~\eqref{Z}. We can readily see how enhanced $Z$ factors will reduce these couplings, provided that the integrals stay approximately constant.

We would now like to quantify the contributions of the KK modes to the EWPO with more detail. For this reason, in this section we will derive closed expressions for the contributions of these models to the {EWPO} in arbitrary backgrounds with arbitrary Higgs profiles.\footnote{Similar work has been performed in Refs.~\cite{Delgado:2007ne,Falkowski:2008fz,Falkowski:2009uy,Round:2010kj}} In particular, here we will parametrize the {EWPO} contributions in terms of the standard four parameters ($S$, $T$, $Y$ and $W$), introduced in Ref.~\cite{Barbieri:2004qk}, which are an extension of the Peskin-Takeuchi parameters \cite{Peskin:1991sw} that are more adequate for the kind of models considered in this thesis.

It is well known~\cite{Kennedy:1988sn} that the deviations to electroweak precision measurements will be encoded in the momentum dependence of the propagators of the electroweak gauge bosons. 
Recall that we are assuming that all the fermions are localized on the UV brane.
In this case, the coupling of the gauge bosons to the fermions are given by the brane values of the 5D gauge fields. Therefore,  
 need to calculate their inverse brane-to-brane propagators. The precision observables can be obtained from these quantities using an holographic method~\cite{Barbieri:2004qk}, which we will perform shortly. Equivalently, one could integrate out the KK modes to obtain effective dimension-six operators involving the fermions and the Higgs. This second method is particularly useful when considering more general settings, such as models with bulk fermions. We will thus make use of it in Chap.~\ref{chap:fermionsbulk}.

\subsection{Holographic method}
\label{STholo}
We will first use the holographic method to compute {EWPO}. For that purpose, we will need to compute the brane-to brane propagator, so let us define the quantity 
\be
P(p^2,y)=e^{-2A(y)}\frac{f_A'(p^2,y)}{f_A(p^2,y)}\,,
\ee
where the holographic profile $f_A(p^2,y)$ satisfies a similar {EOM} as the gauge boson profile (\ref{eq:EOMf}), that is,
\be
\left(e^{-2A}f'_A(p^2,y)\right)^\prime=(M_A^2+p^2)f_A(p^2,y) \,.
\ee
where here a prime ($^\prime$) denotes derivation with respect to $y$. From this it follows that $P(p^2,y)$ satisfies the differential equation and {BC}
\begin{gather}
P'+e^{2A}P^2=p^2+m_{A,0}^2\omega(y)
\,,
\label{prop1}
\\
P(p^2 , y_1)=0 \,,
\label{eq:BCP}
\end{gather}
where
\begin{equation}
\omega(y)=\frac{M_A^2(y)}{m_{A,0}^2}\,
\label{eq:omega}
\end{equation}
and we have introduced
\be
m_{A,0}^2=\frac{1}{y_1}\int_0^{y_1}M_A^2(y)\,,
\label{eq:mA0}
\ee 
which coincides with the zeroth order approximation for the gauge boson [see Eq.~\eqref{eq:mgaugezero}]. The function $\omega(y)$ is a distribution normalized to $y_1$. For a Higgs localized at the {IR} or {UV} brane, this function becomes a $\delta$ function supported at the respective boundary.

We will now solve Eq.~(\ref{prop1}) in a series expansion in powers of $p^2$ and $m_{A,0}^2$ and finally compute the inverse brane-to-brane propagator
\be
\Pi_A(p^2)=\frac{1}{y_1}P(p^2, 0) \,,
\ee
from which the precision observables can be extracted. The expansion should converge well when the precision observables are small, since the suppression scale is expected to be at the TeV scale and both the momentum and the mass $m_{A,0}$ are small compared to that scale.

Solving Eq.~(\ref{prop1}) order by order one finds
\begin{align}
P_0'+e^{2A}P^2_0&=0\,,\\
P_1'+2e^{2A}P_0P_1&=p^2+m_{A,0}^2\omega\,,\\
P_2'+e^{2A}(P_1^2+2P_0P_2)&=0\,.
\end{align}
where the subindex denotes the order of both $p^2$ and $m_{A,0}^2$ used when doing the expansion. Enforcing now the boundary condition \eqref{eq:BCP} at each order one easily finds the solution
\begin{align}
P_0(p^2,y)&=0\,,\\
P_1(p^2,y)&=-p^2(y_1-y)-m_{A,0}^2 y_1 \left[1-\Omega(y)\right]\,,\\
P_2(p^2,y)&=\int_y^{y_1} e^{2A(y)}\left(p^2(y_1-y')+m_{A,0}^2y_1\left[1-\Omega(y)\right]
\right)^2 dy' \,,
\end{align}
where $\Omega(y)$ is defined as in Eq.~(\ref{Omega}), recall
\begin{equation}
\Omega(y)=\frac{\int_0^y h^2(y')e^{-2A(y')} \, dy'}{\int_0^{y_1}h^2(y')e^{-2A(y')} \,dy'} \,.
\end{equation}
$\Omega(y)$ is monotonically increasing from $\Omega(0)=0$ to $\Omega(y_1)=1$. In the case of an IR brane localized Higgs it is actually a step function and in particular it vanishes identically in the bulk, $\Omega=0$. 

We end up with the simple expression for the inverse brane-to-brane propagator
\begin{align}
\Pi_A(p^2)=-p^2-m_{A,0}^2+y_1
\int_0^{y_1} e^{2A}\left[p^2\left(1-\frac{y}{y_1}\right)+m_{A,0}^2(1-\Omega)\right]^2+\cdots \,,
\label{prop2}
\end{align}
where the dots denote terms of higher order in $p^2$ and $m_{A,0}^2$. This is the quantity from which one can compute all electroweak precision observables related to effective operators of up to dimension six.

All the precision observables can be very easily calculated by applying the above $\Pi_A$'s to various gauge bosons. There are three experimental input parameters (usually referred to as $\epsilon_{1,2,3}$~\cite{Altarelli:1990zd}) that are commonly mapped to the three Peskin-Takeuchi ($S$, $T$, $U$) parameters~\cite{Peskin:1991sw}. However, in models with a gap between the electroweak and new physics scales the $U$ parameter is expected to be small since it corresponds to a dimension-eight operator. On the other hand, there are dimension six operators such as $(\partial_\mu B_{\nu\sigma})^2$ which in some models can have sizable coefficients and contribute to the $\epsilon_i$. Therefore, it was suggested in Ref.~\cite{Barbieri:2004qk} to consider an alternative set, defined by the $T$, $S$, $Y$ and $W$ parameters, as a more adequate basis for models of new physics. 

The oblique parameters are defined as
\begin{align}
\alpha T&=
  m_W^{-2}\left[c_W^2 \Pi_{Z}(0)-\Pi_{W}(0)\right]\,,
  \label{eq:Tprop}
  \\
\alpha S&=
4s_W^2c_W^2\left[\Pi'_{Z}(0)-\Pi_{\gamma}'(0)\right]\,,
\\
2m_W^{-2}Y&=
s_W^2\Pi_{Z}''(0)+c_W^2\Pi_{\gamma}''(0),,
\\
2m_W^{-2}W&=
c_W^2\Pi_{Z}''(0)+s_W^2\Pi_{\gamma}''(0)\,,
\label{eq:Wprop}
\end{align}
where $\alpha$ is the electromagnetic gauge coupling defined at the $Z$-pole mass. The 4D gauge couplings are defined as $g^2=g_5^2/y_1$ and $g'^2=g_5'^2/y_1$.
The quantities $Y$ and $W$ are expected to be relevant whenever the $\mathcal O(p^4)$ in $\Pi(p^2)$ terms cannot be neglected. In theories with a Higgs mode $H$ of mass $m_H\ll m_{\rm KK}$ one can relate $T$, $S$, $Y$, and $W$ to the 
 coefficients of the dimension six operators 
\be
|H^\dagger D_\mu H|^2\,,\qquad
H^\dagger W_{\mu\nu}HB^{\mu\nu}\,,\qquad
(\partial_\rho B_{\mu\nu})^2\,,\qquad (D_\rho W_{\mu\nu})^2
\label{ops}
\ee
in the effective low energy Lagrangian, respectively. 

Using Eq.~(\ref{prop2}) we can calculate the coefficients (\ref{eq:Tprop}--\ref{eq:Wprop}), and we obtain
\begin{align}
\alpha T&=s_W^2m_Z^2y_1\int_0^{y_1} e^{2A}(1-\Omega)^2 dy
=s_W^2 m_Z^2\, \frac{I_2}{\rho^2}\, \frac{ky_1}{Z^2}\,\,,
\label{eq:Texpr}
\\
\alpha S&=8s_W^2c_W^2 m_Z^2 \int_0^{y_1} e^{2A}\left(y_1-y\right)(1-\Omega) \, dy
=8 s^2_W c^2_W m_Z^2\, \frac{I_1}{\rho^2} \,\frac{1}{Z} \,,
\label{eq:Sexpr}
\\
Y&=\frac{c_W^2m_Z^{2}}{y_1}
\int_0^{y_1} e^{2A}\left(y_1-y\right)^2 dy
=c_W^2 m_Z^2\, \frac{I_0}{\rho^2}\,\frac{1}{ky_1}\,,
\label{eq:Yexpr}
\\
W&=Y\,,
\label{eq:Wexpr}
\end{align}
where we have used the identity
\be
Z \left[ 1-\Omega(y) \right]= \int_y^{y_1}dy'\,\frac{h^2(y')}{h^2(y_1)}e^{-2A(y')+2 A(y_1)} \equiv u(y)
\ee
and we have defined
\be
I_n=k^{3}\int_0^{y_1} (y_1-y)^{2-n}u^n(y)e^{2A(y)-2A(y_1)} .
\label{notacion}
\ee
The dimensionless integrals $I_n$ (for $n=0,1,2$) are expected to be of the same order. In particular, one expects $I_n/\rho^2=\mathcal O(1/m^2_{KK})$, as one can derive these expressions from integrating out the KK modes, as we will do in Chap.~\ref{chap:fermionsbulk}. 

The dependence of the {EWPO} on the $Z$ factors can actually be very easily understood: in Sec.~\ref{sec:5DSM} we saw that they appear in the low-energy effective Lagrangian (\ref{Leff}), so that $\sqrt{Z}$ could be interpreted as a wave function renormalization for the Higgs. Therefore, the powers of $Z$ in Eqs.~(\ref{eq:Texpr}--\ref{eq:Wexpr}) arise in front of the operators in (\ref{ops}) by canonically normalizing the Higgs field. 

Having a look at Eqs.~(\ref{eq:Texpr}--\ref{eq:Wexpr}) we can see that $T$ is enhanced by a volume factor, and therefore we expect it to be the leading observable. $S$ carries no power of the volume and it is thus the next to leading observable. On the other hand, the $W$ and $Y$ parameters are suppressed with an additional volume factor compared to $S$ (two compared to $T$). Therefore, for the majority of models the bounds coming from $T$ and $S$ will be the dominant ones, although the contribution of $W$ and $Y$ might not be negligible in some cases. The $T$ and $S$ parameters are expected to become of the same order when 
\be
\frac{ky_1}{Z} \frac{I_1}{I_2}=8c_W^2\approx 6.2\,.
\ee
Given that the volume is usually $ky_1\sim \mathcal{O}(30-35)$ (the amount required to generate a large enough hierarchy), only moderate values of $Z$ are required for this. In the RS model with a bulk Higgs one expects $Z<0.5$.

\parasep

%%%%%

In order to conclude this section, let us comment on the finiteness of the radiative corrections to the oblique parameters, which we only computed at tree-level. Since we are not including a gauged custodial symmetry, it is possible to include in the Lagrangian the term $|H^\dagger D_\mu H|^2$, which violates custodial symmetry. This is the first term in \eqref{ops}, and contributes to the T parameter. Therefore, in principle the T parameter could be {UV} sensitive and thus not computable. However, the authors of Ref.~\cite{Carmona:2011ib} found that the loop contributions to this operator (and all operators of dimension six) are finite when the Higgs propagates in the bulk, and hence that the oblique parameters are effectively calculable in this kind of models. Furthermore, the finite radiative corrections are expected to be subdominant when the coupling of the Higgs to the {KK} states is suppressed, and in general it will suffice to consider only the tree-level contribution.

\section{Application to the Randall-Sundrum model}
\label{sec:ApplicationtoRS}

Let us now apply the results of Sec.~\ref{sec:EWPO} to a generalization of the RS2 model which includes a bulk Higgs.\footnote{A similar analysis is performed in Refs.~\cite{Huber:2000fh,Round:2010kj}} That is, we will consider the model described in Sec.~\ref{sec:5DSM} with the $AdS_5$ metric, described by $A(y) = ky$.

We will also need to choose a particular form for the Higgs 5D background, so let us assume an exponential profile given by
\begin{equation}
h(y) = h(y_1) e^{a k (y - y_1)} \,,
\end{equation}
where $a$ is a parameter that controls the location of the Higgs field, with larger values corresponding to locating the Higgs closer to the {IR} brane. In the holographic dual, the quantity $a$ corresponds to the dimension of the Higgs condensate and we need to demand $a \gtrsim 2$ in order to solve the hierarchy problem (see Sec.~\ref{sec:hierarchy} for a detailed discussion on that). Having fixed the metric and the Higgs profile, the $T$, $S$, $Y$ and $W$ parameters can be readily computed from Eqs.~(\ref{eq:Texpr}--\ref{eq:Wexpr}), yielding
\begin{align}
\alpha T_{RS}&= s^2_W \frac{m_Z^2}{\rho^2}(ky_1)
\frac{(a-1)^2}{a(2a-1)}+\dots
,
\label{eq:TRS}
\\
\alpha S_{RS}&= 2 s_W^2c_W^2
\frac{m_Z^2}{\rho^2}\frac{a^2-1}{a^2}+\dots
,
\label{eq:SRS}
\\
\alpha Y_{RS}= \alpha W_{RS} &= \frac14 c_W^2 \frac{m_Z^2}{\rho^2} \frac{1}{ky_1} +\dots
,
\label{eq:YRS}
\end{align}
where the ellipses indicate subleading corrections in the large volume $ky_1$ and, in this case, the parameter $\rho$ just reads \eqref{rho}
\begin{equation}
\rho = k e^{-k y_1} \,.
\end{equation}
Note how the $T$~parameter is volume enhanced, that the $S$~parameter is not and that the $Y$ and $W$~parameters are volume suppressed, as advanced before. 

These parameters need to be compared with a {SM} fit. Assuming that only the $T$ and $S$ parameters are relevant, which is expected due to the others being volume suppressed, for a reference Higgs mass of $m_H^{ref} = 117$~GeV we read from Ref.~\cite{Nakamura:2010zzi} 
\begin{equation}
T = 0.07 \pm 0.08 \,, ~~~ S = 0.03 \pm 0.09 \,,
\label{eq:fit}
\end{equation}
with a correlation between $T$ and $S$ of 87\%. 

In order to translate Eqs.~(\ref{eq:TRS}--\ref{eq:YRS}) into bounds on the {KK} masses, we need to know how the mass of the {KK} modes relates to $\rho$. We will consider now the gauge bosons, described by Eq.~\eqref{eq:EOMf}. Solving this equation numerically for the {RS} model we find that the mass of the first {KK} mode is given by
\begin{equation}
m_{KK} \simeq 2.4 \rho \,.
\end{equation}
Now comparing the expressions (\ref{eq:TRS}--\ref{eq:YRS}) with the experimental data \eqref{eq:fit} will yield bounds on the {KK} masses. These bounds come from the most dominant $T$~parameter, for it is volume enhanced, and turn out to be quite strong. They are plotted in Fig.~\ref{fig:mHvsaRS} (continuous line), where we can see how, for a Higgs localized on the {IR} brane (which translates into the $a\to\infty$ limit) we get a $95\%$~{CL} bound $m_{KK}>12$~TeV.\footnote{This case is also studied in Ref.~\cite{Casagrande:2008hr}} However, for a bulk Higgs the bounds can be alleviated. In particular, for a delocalized Higgs with $a=2.1$ they are lowered by a factor $\sqrt{3}$, which leads to $m_{KK}>7.3$~TeV. This bound is still quite large for phenomenological purposes, but we will shortly see how it can be improved by considering a heavier Higgs.

Finally, for completeness, let us analyze what happens in presence of the custodial protection \cite{Agashe:2003zs}. In this case an extra symmetry~ $SU(2)_R\times U(1)_{B-L}$ is gauged, and there is a residual custodial symmetry which protects the $T$~parameter. Now the bounds will come from the $S$ parameter (see the dashed line in Fig.~\ref{fig:mHvsaRS}), and the $95\%$~{CL} experimental data translate into the restriction $m_{KK}>4.8$~TeV for a localized Higgs and into $m_{KK}>4.2$~TeV for a delocalized Higgs with $a=2.1$. 

\begin{figure}[t]
\centering
\begin{psfrags}
\input{figs/mHvsaRS-psfrag.tex}
\includegraphics[width=0.7\textwidth]{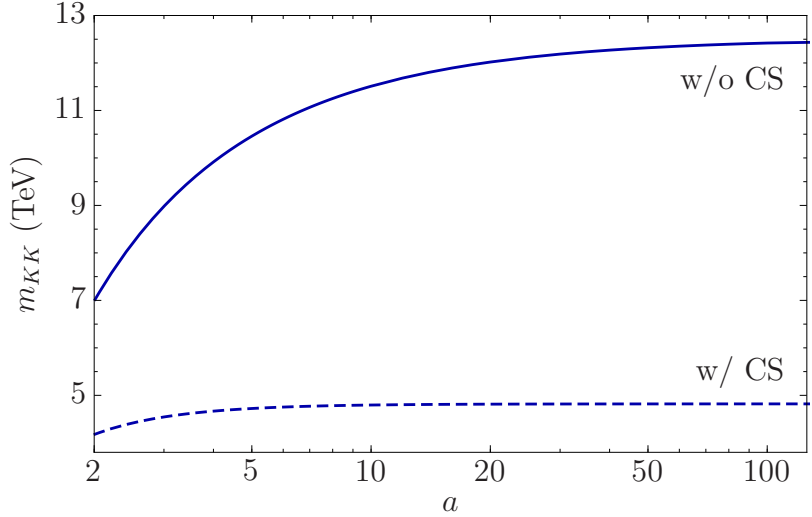}
\end{psfrags} \caption{\it 95\% {CL} lower bounds on the mass of the first {KK} mode of the electroweak gauge bosons, as a function of $a$, for the {RS} model with $A(y_1) = 35$ and a reference Higgs mass $m_H^{ref} = 117$~GeV, with and without an extra Custodial Symmetry (CS). The bounds arise from the experimental bounds on the oblique parameters. The $T$~parameter dominates in the case without CS while the $S$~parameter gives the bound in the case with CS (where we set $T=0$). 
}
\label{fig:mHvsaRS}
\end{figure}

\subsection{A heavy bulk Higgs}
\label{subsec:aheavybulkhiggs}

Let us now consider increasing the mass of the bulk Higgs boson, following the results of Ref.~\cite{Cabrer:2011vu}. Our motivation is driven by the fact that a heavy Higgs boson can be used to soften the {SM} naturalness problem (see e.g.~Refs.~\cite{Peskin:2001rw,Barbieri:2006dq}). Recall
% TODO: Did we announce that before?
 that, in the {SM}, the coupling of the top quark to the Higgs generates a one-loop shift on the Higgs mass which behaves as $\Delta m_H^2\sim (3/4\pi^2)\Lambda^2$, which translates into a sensitivity to the cutoff 
\begin{equation}
\delta_H = \frac{3}{4\pi^2} \frac{\Lambda^2}{m_H^2} \,.
\label{eq:sensitivity}
\end{equation}
Asking for naturalness in the theory, i.e.~$\delta_H \sim 1$, requires $\Lambda\sim 3.6\, m_H$, which in turn implies for a light Higgs that the UV physics should be around the corner at LHC (e.g.~for $m_H\sim 115$~GeV, $\Lambda\sim 400$~GeV) while for a heavy Higgs the UV physics can be at much higher scales (e.g.~for $m_H\sim 600$~GeV, $\Lambda\sim 2.2$~TeV). In view of the negative results on new physics searches at LEP2, and the increasing bounds imposed by ongoing LHC searches, it is thus interesting to consider models with heavy Higgs masses to alleviate this tension (also known as the little hierarchy problem). The {EWPT}, when applied to the {SM}, point towards to a light Higgs. However, we will now see that the {RS} model can easily accommodate heavy Higgs bosons without incurring in contradiction with the experimental results.

Moreover, the construction of the 5D {SM} model also favors heavy Higgs bosons. In fact, recall that the mass of the Higgs boson in the bulk is given by Eq.~\eqref{mH}, which in the {RS} model establishes the relation 
\begin{equation}
\frac{m_H^2}{\rho^2} \propto  \left( \frac{M_1}{k} - a \right) 
\,,
\end{equation}
where $M_1$ is the {IR}-brane mass term for the Higgs.  We can see that a certain fine-tuning is required to keep $m_H$ small, and thus larger Higgs masses will be favored.

Let us now analyze the bounds on $m_{KK}$ for different Higgs boson masses. We will need to know the dependence of the bounds on $T$ and $S$ with respect to the Higgs mass. The one-loop contribution to these parameters of a SM Higgs boson with a mass $m_H$, normalized to its values at the reference Higgs mass $m_{H,r}$, is given by
\begin{equation}
\Delta S=\frac{1}{2\pi} \left[ g_S(m_H^2/m_Z^2)-g_S(m_{H,r}^2/m_Z^2) \right] \,,
\label{eq:DeltaSlargeH}
\end{equation}
where
\begin{equation}
g_S(u)=\int_0^1 dx\, x(5x-3)\log(1-x+u x) \,, 
\end{equation}
and~\cite{Veltman:1976rt}
\begin{equation}
\Delta T=\frac{-3}{16\pi s^2_W}\left[ g_T(m_H^2/m_Z^2)-g_T(m_{H,r}^2/m_Z^2) \right]  \,,
\end{equation}
where
\begin{equation}
g_T(u)=y\frac{\log c^2_W-\log u}{c^2_W-u}+\frac{\log u}{c_W^2(1-u)}\,.
\label{eq:DeltaTlargeH2} 
\end{equation}
In the limit where the Higgs masses are much larger than $m_Z$ one recovers the approximate behavior in Ref.~\cite{Peskin:1991sw}. Now we can translate the experimental bounds of Eq.~\eqref{eq:fit}, evaluated at $m_H = 117$~GeV, to different Higgs masses.

\begin{figure}[t]
\centering
\begin{psfrags}
	\input{figs/ellipsesRS-psfrag.tex}
	\includegraphics[width=0.6\textwidth]{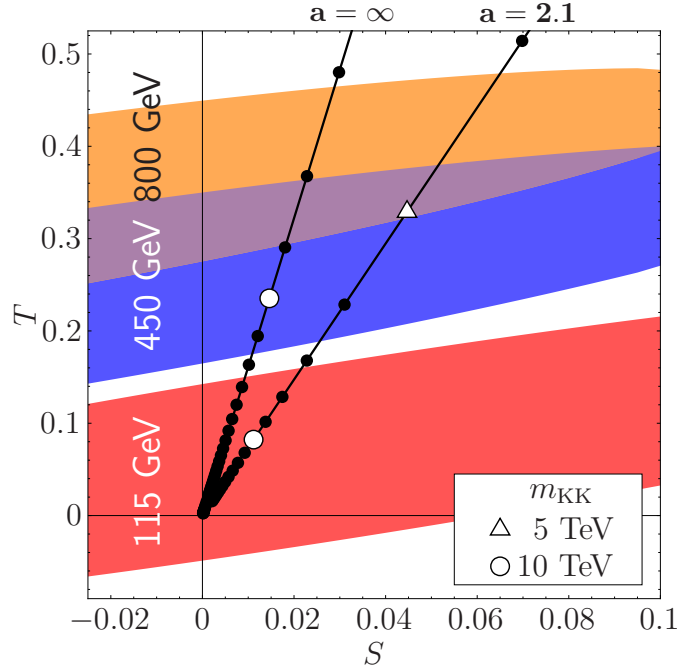}
\end{psfrags}
\caption{\label{fig:ellipses} \it 95\% CL regions in the $(S,T)$ plane for the {RS} model with different values of the Higgs mass. We show two rays corresponding to a {IR} localized ($a=\infty$) or a delocalized bulk Higgs (with $a=2.1$). Dot spacing is 1 TeV. Increasing values of $m_{KK}$ correspond to incoming fluxes.}
\end{figure}
In Fig.~\ref{fig:ellipses} we show the 95\% CL ellipses in the $(S,T)$ plane for different values of the Higgs mass $m_H=115,\, 450,\, 800$~GeV. The solid lines are obtained from the expressions for $S$ and $T$ [Eqs.~(\ref{eq:TRS}--\ref{eq:SRS})], and the dots correspond to different values of $m_{KK}$ and the dot spacing is 1 TeV. The values of $m_{KK}$ increase as the dots get closer to the origin. In this way the lower and (possibly) upper bounds on $m_{KK}$ can be read from the plot for the considered values of $m_H$. Note also how considering a custodial protection mechanism (which sets $T=0$) becomes in conflict with the experimental data when considering a heavier Higgs. 

\begin{figure}[t]
\centering
\begin{psfrags}
	\input{figs/boundsa-psfrag.tex}
	\includegraphics[width=0.7\textwidth]{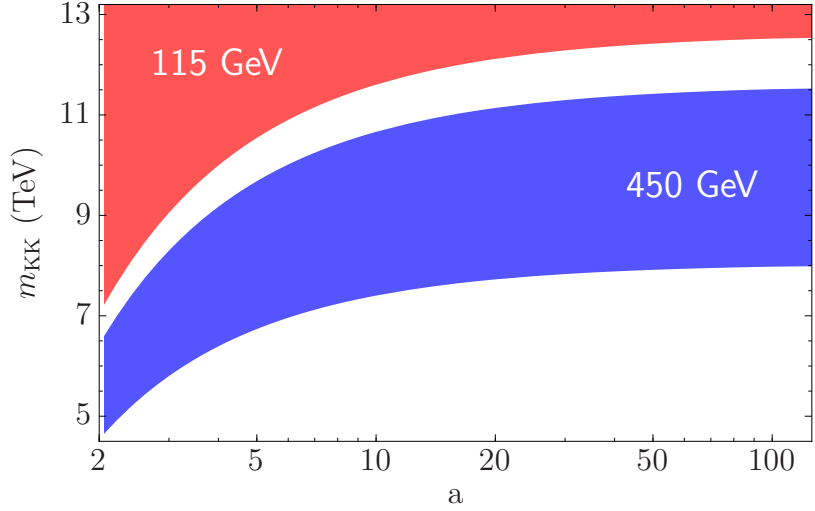}
\end{psfrags}
\caption{\label{fig:boundsa} \it 95\% CL regions in the $(a,m_{KK})$ plane for the {RS} model with a bulk Higgs and different values of the Higgs mass.}
\end{figure}
In Fig.~\ref{fig:boundsa} we show the 95\% CL allowed regions in the $(a,m_{KK})$ plane and different values of the Higgs mass. We can see from the plot how {RS} favors a heavy Higgs delocalized in the bulk. In particular, the 95\% CL window for a localized Higgs field with e.g.~$m_H=450$ GeV is 8.0~TeV~$\lesssim m_{KK}\lesssim 11.6$~TeV, while for a Higgs with $a=2.1$ the window is 4.6~TeV~$\lesssim m_{KK}\lesssim 6.6$~TeV. 

We also need to set bounds on the Higgs mass from theoretical considerations. As stated in the beginning of this section, we already know that solving the naturalness problem favors a heavy Higgs. However, a too massive Higgs can become in conflict with the bounds from perturbativity. In order to quantify this, we will follow the criterion of Ref.~\cite{Hambye:1996wb} on the beta-function of the Higgs quartic coupling, $\beta_\lambda$. The condition is that the two-loop corrections to $\beta_\lambda$ are less than $50\%$ of the one-loop correction ($\beta_\lambda^{(2)} = 0.5 \beta_\lambda^{(1)}$). 

\begin{figure}[t]
\centering
\begin{psfrags}
	\input{figs/boundsRS-psfrag.tex}
	\includegraphics[width=0.7\textwidth]{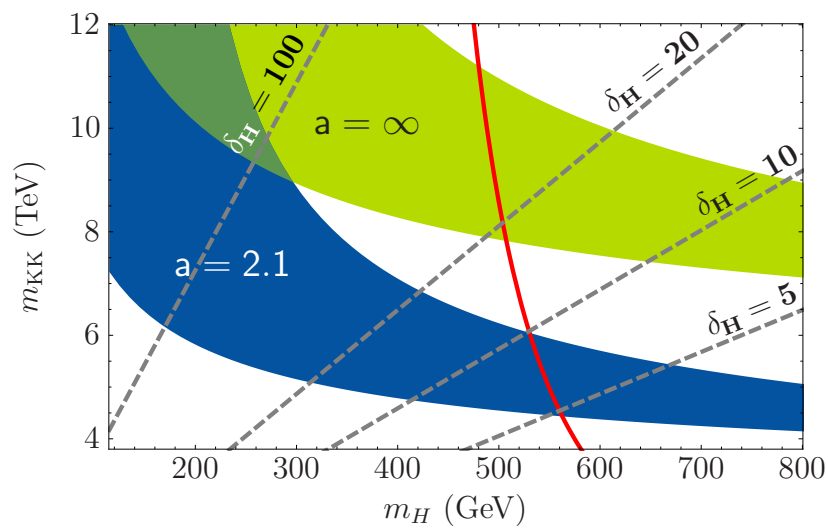}
\end{psfrags}
\caption{\label{fig:boundsRS} \it 95\% CL regions in the $(m_H,m_{KK})$ plane for RS with a localized ($a=\infty$) and  bulk ($a=2.1$) Higgs field. Dashed lines correspond to sensitivity $\delta_H=100$ (1\% fine-tuning),  $\delta_H=20$ (5\%), $\delta_H=10$ (10\%) and $\delta_H=5$ (20\%), for $\Lambda\sim m_{KK}$. The solid line is the perturbativity bound; the region to its right is excluded.}
\end{figure}
In Fig.~\ref{fig:boundsRS} we show the 95\% CL regions in the $(m_H,m_{KK})$ plane along with the perturbativity bound (solid line, the region on its right is excluded). We also show different contours for the sensitivity $\delta$ (see Eq.~\eqref{eq:sensitivity}).  We can see that having smaller values of $\delta$, without being in contradiction with the perturbativity bound, favors a delocalized Higgs. In particular, for a localized Higgs boson ($a=\infty$) the 95\% CL lower bound on the mass of gauge KK modes imposes $m_{KK}> 7.8$~TeV, which corresponds to $m_H< 510$~GeV, while the sensitivity satisfies $\delta_H>20$ ($<5\%$ fine-tuning). For a delocalized bulk Higgs, on the other hand, the bound is $m_{KK}> 4.4$~TeV with $m_H< 560$~GeV, and the sensitivity can be as low as $\delta_H>5$ ($<20\%$ fine-tuning).

The conclusion is that the {RS} model favors a heavy bulk Higgs, the further away from the {IR} brane the better, in order to satisfy the experimental and perturbativity bounds and solve the {SM} naturalness problem. However, the bounds on {KK} masses are still of order $m_{KK} \gtrsim 4$~TeV, which is low enough to be within the LHC range and solve the little hierarchy problem. In the following chapter we are going to see how we can construct a new class of models, based in a deformation of the AdS metric, that lower these bounds while preserving naturalness.

%%%%%%%%%%%%%%%%%%%%%%%%%%%%%%%%%%%%%%%%

\chapter{A Non-Custodial Warped Model}
\label{chap:noncustodial}

In the previous chapter, we have seen how propagating a heavy Higgs in the bulk of the {RS} model is enough to lower the bounds from {EWPT} on the masses of new non-{SM} excitations, without requiring the introduction of an additional gauge symmetry to protect the custodial symmetry. However, the improvement of the bounds is only large enough when the Higgs is heavy, and even in this case the bounds are not extraordinarily low (i.e.~for a Higgs of $560$~GeV, very close to the perturbativity bound, we have $m_{KK} > 4.6$~TeV). Therefore, it is certainly fair to study possible alternatives that could allow us to get better than that.

In this chapter, based on the results published in Refs.~\cite{Cabrer:2010si,Cabrer:2011fb,Cabrer:2011mw,Cabrer:2011vu}, we will present a model that consists in replacing the {RS} metric by a asymptotically {AdS} metric, generated by a bulk scalar field as in the Goldberger-Wise mechanism, along with considering a Higgs that propagates in the bulk. The metric will be inspired by the soft-wall model described in Sec.~\ref{sec:aselfstabilizedsoftwallmodel}, and therefore it will feature a singularity at a certain proper distance. However, we will consider a model with two branes, placing the singularity outside the physical interval, but nearby the {IR} brane so that the deformation it produces can have sizable effects.

The choice of the model has its roots in some previous results found in the literature. First, it was shown in Ref.~\cite{Huber:2000fh} that propagating the Higgs in the bulk of the {RS} model without custodial symmetry leads to a reduction of the $S$ and $T$ parameters (and therefore of the {EWPT} bounds). This  amounts to a sizable reduction of $\sim 46\%$ in the bounds of new {KK} states, although the resulting bound results in a the bound $m_{KK} \gtrsim 7~\mathrm{TeV}$, too large to present an alternative to custodial symmetry. Second, in Ref.~\cite{Falkowski:2008fz} it was shown how using soft-wall metrics in presence of the custodial symmetry leads to an extra suppression of the $S$ parameter with respect to the {RS} case. There were, however, negative results in the literature regarding the reduction of the $T$ parameter in models with generalized metrics when the Higgs is localized on the {IR} brane, as shown in Refs.~\cite{Delgado:2007ne,Archer:2010hh}.  

We will shortly see how with our model, based on a generalized metric and a bulk Higgs, we can reduce both the $T$ and $S$ parameters with respect to {RS}, to the extent that no custodial symmetry is necessary to have moderate bounds on the new {KK} states. This will allow us to construct a pure 5D {SM} without the presence of custodial symmetry (as presented in Chap.~\ref{chap:ewsbbulkhiggs}). In fact, we will show how this reduction of the bounds can be traced back to a large wavefunction renormalization for a light Higgs mode. We will finally find explicit values for these bounds and consider the cases of a light and heavy Higgs.

\section{Construction of the model}
\label{sec:model}

We will consider a 5D Standard Model, as introduced in Sec.~\ref{sec:5DSM} with a metric given by $A(y)$ and delimited by two branes. The $SU(2)_L \times U(1)_Y$ gauge bosons, the Higgs field $H$ and an additional scalar $\phi$ propagate in the bulk of the extra dimension, and the 5D action is given by Eq.~\eqref{5Daction}. 

The bulk scalar $\phi$ will source our non-trivial metric, and will also act as a Goldberger-Wise field stabilizing the distance between the two branes.\footnote{We will neglect the backreaction of the Higgs field and treat it as an external scalar, subject to the gravitational and $\phi$ scalar background. This working hypothesis will be justified later for our particular model.}  In order to find solutions to the gravitational background, we will follow the superpotential method described in Sec.~\ref{sec:scalargravitybackground}, and we will write our scalar potential similarly to Eq.~\eqref{eq:superpotential1} as 
\begin{equation}
V(\phi,H) = \frac12 \left( \frac{\partial W(\phi)}{\partial \phi} \right)^2 - \frac13 W^2(\phi) + M^2(\phi) \vert H \vert^2  \,.
\label{VW}
\end{equation}
Using this ansatz the background {EOM} can be written as simple first-order differential equations
\be
A'(y) = \frac{1}{6} W(\phi(y))\,,\quad
\phi'(y) = \partial_\phi W(\phi) \,,
 \label{metricEOM}
\ee
whose solutions will enter the usual second order linear equation for the 5D {VEV} $h(y)$ [see Eqs.~\eqref{Higgs} and \eqref{eq:EOMh}]
\be
h''(y)-4A'(y)h'(y)-M^2[\phi(y)]\, h(y)=0\,.
\label{higgsbg}
\ee
The boundary conditions for $A(y)$, $\phi(y)$ and $h(y)$ depend on the explicit form of the boundary potentials $\lambda^\alpha(\phi,h)$ [see Eqs.~\eqref{eq:BCAphi} and \eqref{eq:BCh}]. In our case the brane potentials are described by Eqs.~(\ref{eq:boundpot0}--\ref{eq:boundpot1}) and the {BC}s for $h(y)$ read
\begin{align}
\frac{h'(y_0)}{h(y_0)} &= M_0
\label{eq:BChiggs0}
\\
\frac{h'(y_1)}{h(y_1)} &= M_1 - \gamma h^2(y_1) 
\label{eq:BChiggs1}
\end{align}

\subsection{The scalar-gravity sector}
Our model will be inspired by the soft-wall model described in Sec.~\ref{sec:aselfstabilizedsoftwallmodel}. We will therefore consider the superpotential [Eq.~\eqref{ourW}]
\be
W_\phi(\phi)=6k(1+b e^{\nu\phi/\sqrt{6}})\,,
\label{superp}
\ee
where $\nu$ and $b$ are arbitrary parameters. This superpotential leads to the background configuration [Eqs.~(\ref{eq:backgroundA}--\ref{eq:backgroundphi})]
\begin{align}
A(y)&=ky-\frac{1}{\nu^2}\log\left(1-\frac{y}{y_s}\right)
\,,
\label{A}
\\
\phi(y)&=-\frac{\sqrt{6}}{\nu}\log[\nu^2 k(y_s-y)]
\label{phi}
\,.
\end{align}
The metric presents a singularity at $y=y_s$, although it will be hidden by a brane located at a certain position $y_1 < y_s$. Let us now discuss on the stabilization of these distances. 

We assume that the brane dynamics $\lambda^\alpha_\phi$ fixes the values of the field \mbox{$\phi=(\phi_0,\,\phi_1)$} on the UV and IR branes respectively. The inter-brane distance $y_1$, as well as the location of the singularity at $y_s\equiv y_1+\Delta$ and the warp factor $A(y_1)$, are related to the values of the field $\phi_\alpha$ at the branes by the following simple expressions:
\begin{eqnarray}
ky_1&=&\frac{1}{\nu^2}\left[e^{-\nu \phi_0/\sqrt{6}}-e^{-\nu \phi_1/\sqrt{6}}  \right], \quad k\Delta=\frac{1}{\nu^2}e^{-\nu \phi_1/\sqrt{6}}
\ ,
\nonumber\\
A(y_1)&\simeq&k y_1+\frac{1}{\nu}(\phi_1-\phi_0)/\sqrt{6}
\ ,
\end{eqnarray}
which shows that the required large hierarchy can naturally be 
fixed with values of the fields $\phi_1\gtrsim \phi_0$, $\phi_0<0$ and $\mathcal O(1)$ in absolute value.\footnote{Note that the soft-wall configuration described in Sec.~\ref{sec:aselfstabilizedsoftwallmodel} corresponds to the limit $\phi_1\gg 1,\, y_1\to y_s$.} 
Note that due to its exponential dependence on $\phi_1$, $\Delta$ can be small or, in other words, the {IR} brane can naturally be located very close to the singularity.

Let us now have a look at the 5D curvature. The curvature radius $L(y)$, related to the 5D curvature $R(y)$ by
\begin{equation}
L(y) = \sqrt{\frac{-20}{R(y)}} \,,
\end{equation}
is given in our case by [see Eq.~\eqref{eq:curvature}]
\begin{equation}
k L(y) = \frac{ \nu^2 k(y_s - y) }{\sqrt{1 - 2 \nu^2/5 + 2 \nu^2k(y_s - y) +  \nu^4k^2(y_s - y)^2}}\,.
\label{curvature-res}
\end{equation}
Near the UV brane we have $kL_0\sim 1$,\footnote{This shows that $k$ is approximately the inverse AdS curvature radius.}
and $k L(y)$ remains close to unity in most of the interbrane distance. Near the IR brane $L$ can get small due to the spurious singularity at $y=y_s$. The behavior of $L(y)$ in the IR is shown in Fig.~\ref{fig0}. $L(y)$ is a monotonically decreasing function when $\nu \le \sqrt{5/2}$. When $\nu > \sqrt{5/2}$ the curvature radius possesses a minimum (correspondingly the curvature presents a maximum) and eventually the curvature changes sign before the singularity. The quantity $k L(y)$, or better its maximum value, is a measure of the deviation of our model from pure {RS}. Not surprisingly, the deviation from {RS} is larger as $\nu$ and $\Delta$ increase.
\begin{figure}[tb]
\centering
\begin{psfrags}
\input{figs/L1-psfrag.tex}
\includegraphics[width=0.67\textwidth]{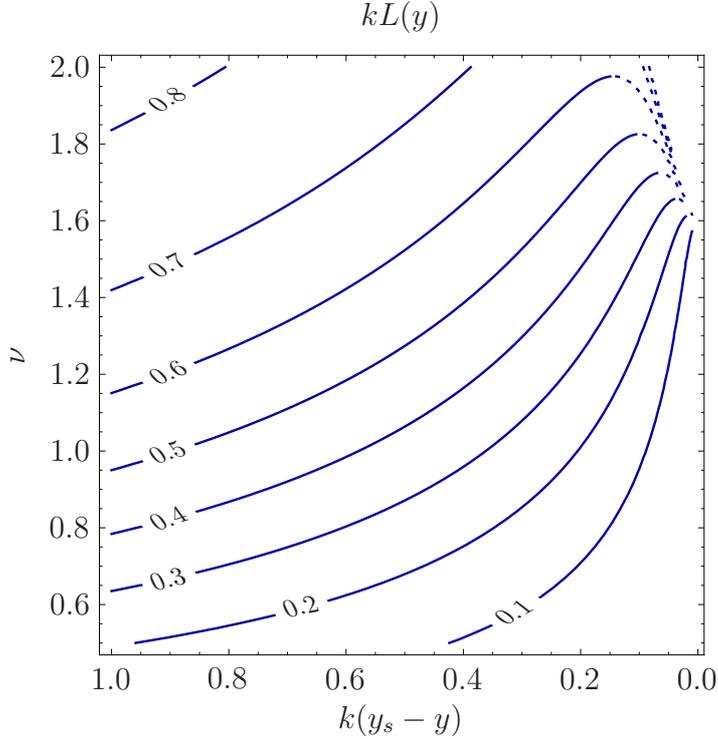}
\end{psfrags}
\caption{\it Contour levels of $k L(y)$ as a function of $\nu$ and $k(y_s - y)$. For $\nu > \sqrt{5/2}$, $L(y)$ presents a minimum close to the singularity. Since we demand $L(y)$ to be a monotonic function we will constrain ourselves outside the region after this minimum is reached (dotted lines).}
\label{fig0}
\end{figure}

\subsection{The Higgs sector}
For the Higgs bulk mass term we will choose
\be
M^2(\phi)=ak\left[ak-\frac{2}{3}W(\phi)\right]\,.
\label{eq:M2}
\ee
where $a$ is an arbitrary real parameter. In the next section we will see that $a$ is constrained by the hierarchy problem, restricting us to values $a>2$. The choice Eq.~(\ref{eq:M2}) ensures that one linearly independent solution to Eq.~(\ref{higgsbg}) is given by a simple exponential. Certainly other choices are possible, which lead to similar results. We will comment on some of them in Sec.~\ref{sec:hierarchy}.

Using the superpotential formalism to define the $\phi$ potential amounts to some fine-tuning among the different coefficients of the bulk potential, unless they are protected by some underlying 5D supergravity~\cite{DeWolfe:1999cp}. The quadratic Higgs term, which is generated by (\ref{VW}), can be written as
$k^2[a(a-4)-4 a b e^{\nu\phi/\sqrt{6}} ]|H|^2$
and the coefficients of the two operators $|H|^2$ and $e^{\nu\phi/\sqrt{6}}|H|^2$ can be considered as independent parameters.\footnote{Of course the coefficients of the operators not involving  the Higgs field remain fine-tuned as we are using the superpotential formalism to fix them.} However, since the parameter $b$ can be traded by a global shift in the value of the $\phi$ field, or in particular by a shift in its value at the UV brane $\phi_0$, for simplicity we will fix its value to $b=1$ hereafter.

Having fixed the background we can write the general solution to Eqs.~(\ref{higgsbg}--\ref{eq:BChiggs1}) and the BCs as
\be
h(y)= h_0 e^{aky}\left[ 1 + \left( \frac{M_0}{k} - a \right) \int^y e^{-2(a-2) k y'}\left(1-\frac{y'}{y_s}\right)^{-\frac{4}{\nu^2}}\right] \,,
\label{eq:solutiontoh}
\ee
where $h_0=h(0)$ and $M_0$ is the {UV} brane mass term in \eqref{eq:BChiggs0}. 

As we saw in Chap.~\ref{chap:ewsbbulkhiggs}, a reduction of the $S$ and $T$ parameters can occur provided that the $Z$ factors are sizable. Recall from Eq.~\eqref{Z} that the $Z$ factor is defined as
\be
Z=
k\int_0^{y_1}dy\frac{h^2(y)}{h^2(y_1)}e^{-2A(y)+2A(y_1)}\,.
\label{eq:Zrepeat}
\ee
From this expression follows that, in order to obtain a large $Z$, we would like to keep the exponential solution (which corresponds to the non-singular solution in the SW limit). In the next section we will see that this imposes some restrictions on the parameter space which will have a simple holographic interpretation.

\begin{figure}[p]
\centering
\begin{minipage}[c]{0.19\textwidth}
~~~~~~~~\textbf{(a)}
\end{minipage}
\begin{minipage}[c]{0.8\textwidth}
 \begin{psfrags}
  \input{figs/Z1-psfrag.tex}
  \includegraphics[width= 0.83\textwidth]{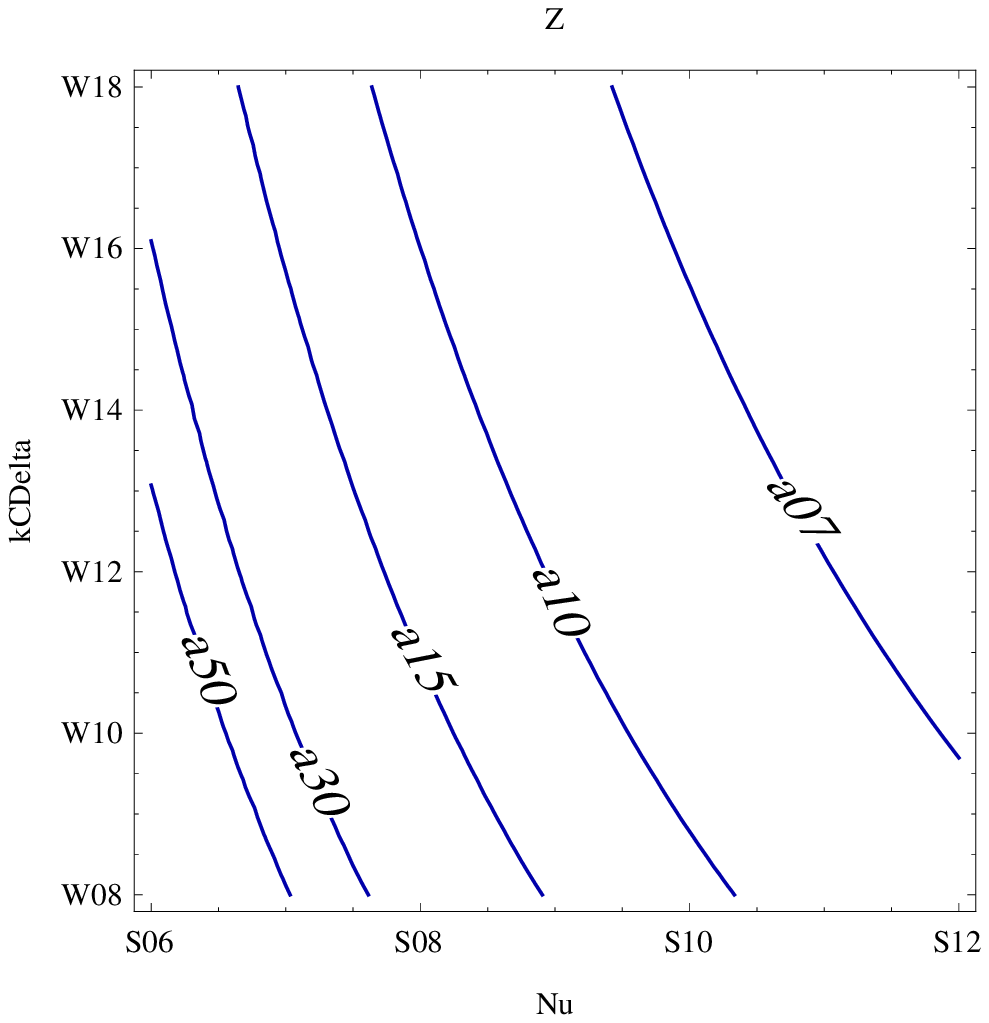}
 \end{psfrags}
\end{minipage}
\\ \vspace{5mm}
\begin{minipage}[c]{0.19\textwidth}
~~~~~~~~\textbf{(b)}
\end{minipage}
\begin{minipage}[c]{0.8\textwidth}
 \begin{psfrags}
  \input{figs/Z2-psfrag.tex}
  \includegraphics[width= 0.83\textwidth]{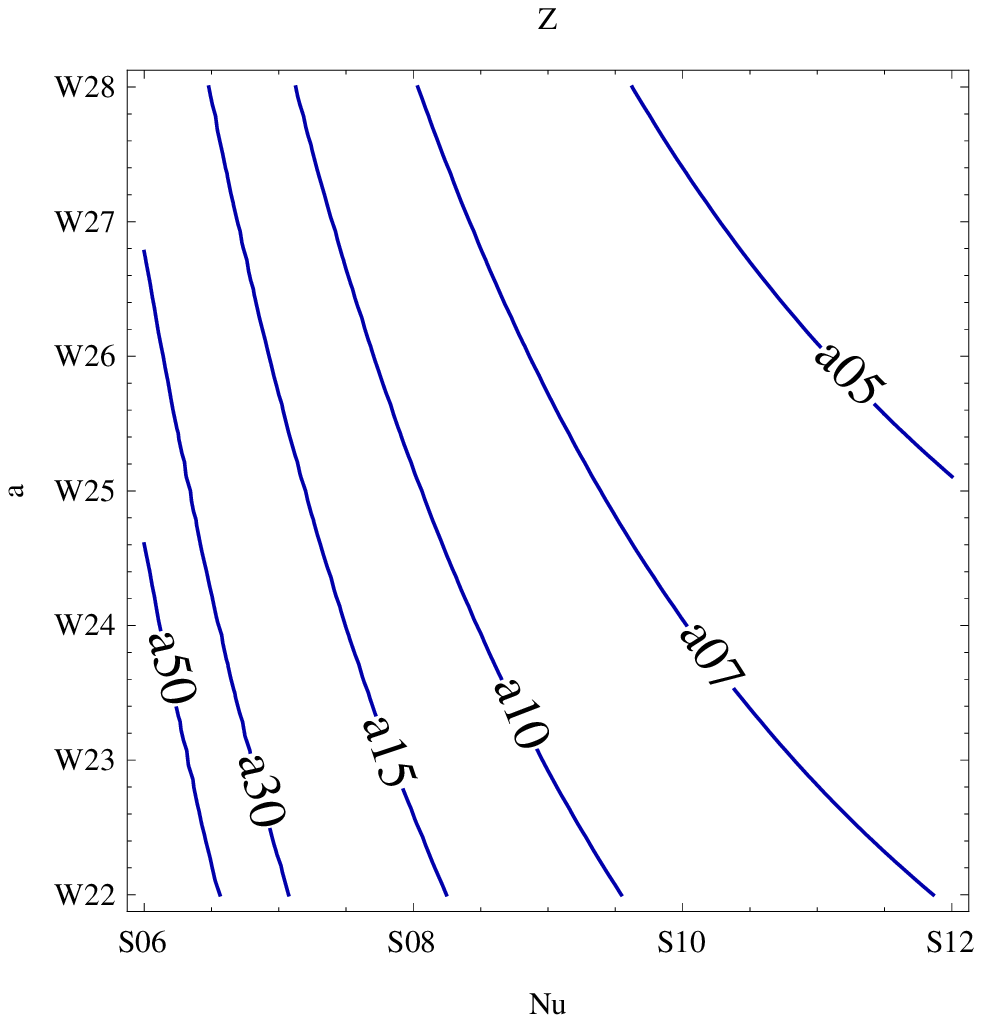}
 \end{psfrags}
\end{minipage}
\caption{\it Contour levels of $Z$ for $A(y_1)=35$: {\bf (a)} as a function of $\nu$ and $k\Delta$ for $a=2.2$; and {\bf (b)} as a function of $\nu$ and $a$ for $k\Delta=1$. }
\label{fig1}
\end{figure}
From Eq.~\eqref{eq:Zrepeat}, and considering a Higgs profile of the form $h(y)\sim e^{aky}$, we can evaluate the $Z$ factors in terms of the parameters $(a,\nu,\Delta)$, while we will fix the total warp factor as $A(y_1)=35$. To see the dependence of $Z$ on the various parameters we plot in Fig.~\ref{fig1} the contour levels of $Z$ in the planes ($\Delta$,$\nu$) and ($a$,$\nu$).

%%%%%%%%%%%%%%%%%%%%%%%%%%%%%%%%%%%%%%%%%%%%%%%

\section{The Hierarchy Problem}

\label{sec:hierarchy}

As we have found at the end of the last section, the $Z$ factors can become large for an exponential Higgs profile if the parameter $a$ is ``small". From a holographic dual point of view this can be translated to a small dimension for the Higgs condensate in the IR. This raises the question to which extent this reintroduces the hierarchy problem that we claimed to have solved by Higgs compositeness. 

In the context of RS models this question has been discussed in Ref.~\cite{Luty:2004ye}, whose main lines we follow essentially here. In the RS case, $A(y)=ky$, the Higgs background solution to Eqs.~(\ref{higgsbg}--\ref{eq:BChiggs1}) is given by 
\be
h(y)=h_0\left(\frac{M_0/k+a-4}{2(a-2)}e^{a k y}
-\frac{M_0/k-a}{2(a-2)}e^{(4-a) k y}\right) \,.
\label{RSsoln}
\ee
The observation in RS is that for $a>2$ no fine-tuning is necessary in order to keep only the first term, since near the IR brane (where EWSB occurs) the second term is always irrelevant. On the contrary, for $a<2$, the second term would be dominating and one needs to fine-tune $M_0/k=a$ in order to maintain the solution $h(y)\sim e^{aky}$. This fact has a simple holographic interpretation: since $\dim(\mathcal O_H)=a$ the hierarchy problem is solved by compositeness of the Higgs for $a>2$, but not for $a<2$ (see Ref.~\cite{Luty:2004ye} for a more detailed discussion).

In our case the situation is similar. Again for $a<2$ the solution $h(y)\sim e^{a k y}$  will  be fine-tuned due to the exponential enhancement in the integrand. However, now one has to be careful even for $a>2$. Let us rewrite the solution in Eq.~\eqref{eq:solutiontoh} as
\begin{equation}
h(y) = h_0 e^{aky} \bigg[
1 + (M_0/k -a) \left[ F(y) - F(0) \right]
\bigg]
,
\label{eq:hyF}
\end{equation}
where
\begin{equation}
F(y) =  e^{-2(a-2) k y_s} y_s \left[ -2(a-2) k y_s \right]^{-1 + 4/\nu^2} \Gamma \left[ 1 - \frac{4}{\nu^2} , -2(a-2) k( y_s - y) \right].
\end{equation}
Note that $F(y)$ is defined as a complex function but its imaginary part cancels in \eqref{eq:hyF} leading to a real solution. One should view $F(y)$ as the generalization of $F_{\rm RS}(y)=e^{-2(a-2)ky}$ in the RS case.

Similarly to the AdS case, in order to keep the exponential solution without the need of a fine-tuning we must require the function $F(y)$ to be small. Since $F$ is a monotonically increasing function of $y$ it will be enough to inspect $F(y_1)$. In order to quantify this let us define 
\begin{equation}
\delta \equiv \left\vert F(y_1) \right\vert ,
\label{definitiondelta}
\end{equation}
which will be a measure of the fine-tuning required in $(M_0/k - a)$ in order to keep the exponential solution. In particular the absence of fine-tuning requires roughly $\delta \lesssim \mathcal O(1)$. $\delta$ is a decreasing function of $a$, $\nu$ and $\Delta$, so we need to impose a lower bound on $a=a_0(\nu,\Delta)$ below which one would need to fine-tune $M_0/k\simeq a$ in order to keep the simple exponential solution that improves the EWPT. The behavior of $\delta$ as a function of $a$ and $\nu$ with $k\Delta=1$ is plotted in Fig.~\ref{fig2}.
\begin{figure}[p]
\centering
\begin{minipage}[c]{0.19\textwidth}
~~~~~~~~\textbf{(a)}
\end{minipage}
\begin{minipage}[c]{0.8\textwidth}
 \begin{psfrags}
  \input{figs/delta1-psfrag.tex}
  \includegraphics[width= 0.83\textwidth]{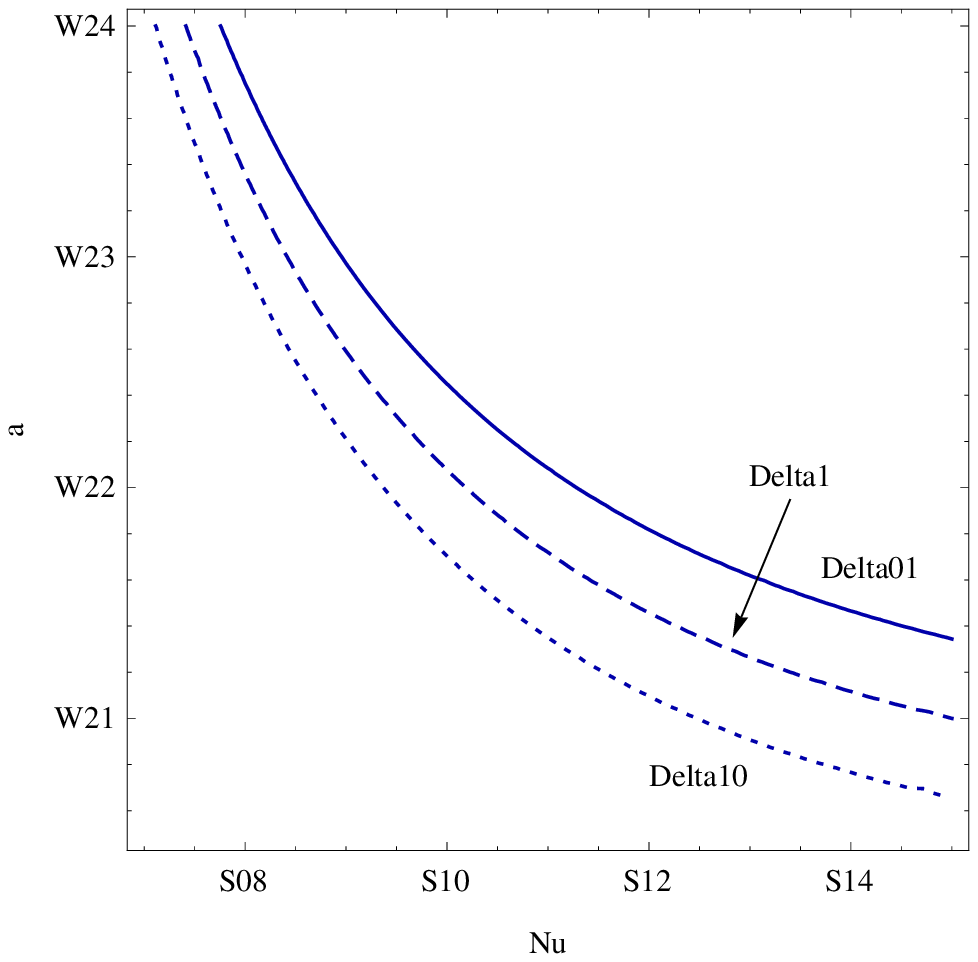}
 \end{psfrags}
\end{minipage}
\\ \vspace{5mm}
\begin{minipage}[c]{0.19\textwidth}
~~~~~~~~\textbf{(b)}
\end{minipage}
\begin{minipage}[c]{0.8\textwidth}
 \begin{psfrags}
  \input{figs/delta2-psfrag.tex}
  \includegraphics[width= 0.83\textwidth]{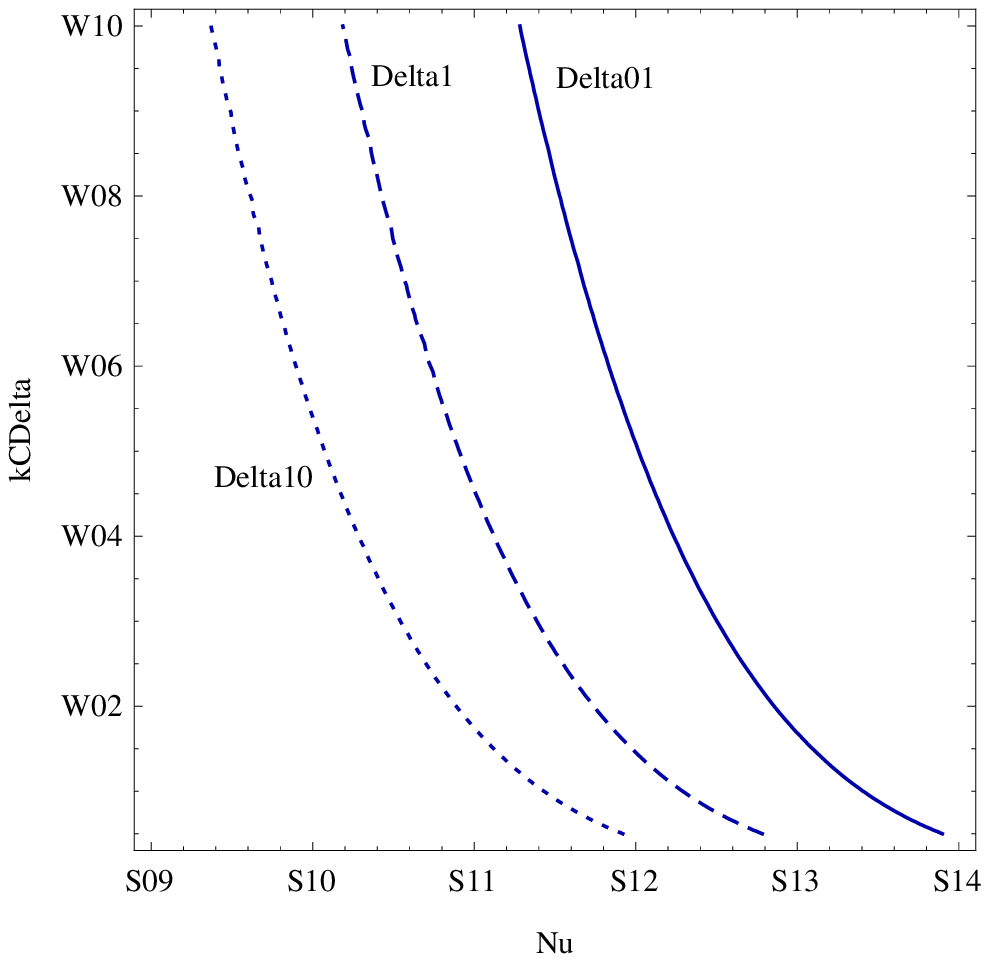}
 \end{psfrags}
\end{minipage}
\caption{\it Contour levels of $\delta$ for $A(y_1) = 35$: \textbf{(a)} as a function of $\nu$ and $a$ for $k\Delta=1$; and \textbf{(b)} as a function of $\nu$ and $k\Delta$ for $a=2.2$.}
\label{fig2}
\end{figure}

We can see that the lower bound on $a$ lies (depending on the value of $\nu$) a little above $a=2$, but not by much in the shown parameter range. One can then reinterpret this as stating that for a given $a>2$ there will be a curve $\Delta(\nu)$, as it is shown in Fig.~\ref{fig2}, below which keeping the exponential solution amounts to a fine-tuning. In particular it will be inconsistent to blindly take the limit $\Delta\to 0$.

There is here again a simple holographic interpretation.\footnote{See Refs.~\cite{Vecchi:2010aj} for related ideas.} The dimension of the Higgs condensate corresponding to the solution $h(y)\sim e^{aky}$ depends on $y$. Since the renormalization group (RG) scale is given by the warp factor we have
\be
\dim(\mathcal O_H)=\frac{h'}{h\,A'}=\frac{a}{1+\frac{1}{k(y_s-y)\nu^2}}\,.
\ee
Starting in the UV with some $\dim(\mathcal O_H)>2$, as required to avoid the fine-tuning and solve the hierarchy problem by a composite Higgs, the Higgs mass term $|\mathcal O_H|^2$ will have dimension $\dim(|\mathcal O_H|^2)=2\dim(|\mathcal O_H|)>4$,\footnote{We use the fact that in the large $N_c$ limit operator products become trivial.}  and will be an irrelevant operator becoming more and more suppressed along the RG flow. However, following the RG flow further, the theory departs from the conformal fixed point, $\dim(\mathcal O_H)$ decreases and there will be a critical RG scale $\mu_c$ at which $\dim(\mathcal O_H)<2$. As a consequence, $|\mathcal O_H|^2$ will become a relevant operator and will start increasing again. 

As long as this happens far enough, near the IR, there is no concern as, at the scale $\mu_c$, the mass term is really small and there is simply not enough RG time for it to become large enough before EWSB occurs. On the other hand, a low dimension of the condensate is essential to generate sizable wavefunction renormalizations for the light Higgs mode that will eventually allow us to suppress the $S$ and $T$ parameters. 

%%%%%%%%%%%%%%%%%%%%%%%%%%%%%%%%%%%%%%%%%%%%%%%

\section{Bounds from precision observables}
\label{results} 

\begin{figure}[t]
\centering
\begin{psfrags}
	\input{figs/alpha-psfrag.tex}
	\includegraphics[width= 0.7\textwidth]{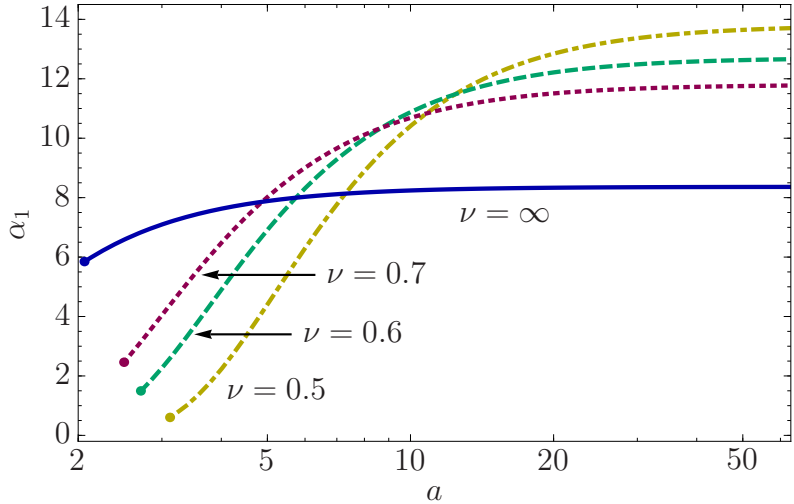}
\end{psfrags}
\caption{
\label{fig:coupling}  
\it Plot of the coupling $\alpha_1$ as a function of $a$ for $\nu=0.5,0.6,0.7$ and $\infty$ (RS), with $k\Delta=1$.  Lines end at $a=a_0$ where $\delta=0.1$.}
\end{figure}
We have seen how we can obtain small values of $Z$ for different values of our model parameters. This  reduction of $Z$ also reduces the coupling of the {KK} modes to the Higgs currents. In Fig.~\ref{fig:coupling} we plot the coupling of the first KK mode of the gauge boson to the Higgs current, $\alpha_1$, as defined in Eq.~\eqref{eq:alphan}, as a function of $a$ and for different values of $\nu$ (for $\Delta=1$). We can see how the coupling decreases as a combined effect of having a small $a$ (less localized Higgs field) and small $\nu$ (departure from {AdS} in the {IR}. This, we expect, will reduce the bounds from {EWPT} on the masses of the {KK} states, as we are now going to check.
   
We will now compute the explicit bounds on the lightest new states that appear in our model, namely the lightest ($n=1$) KK modes corresponding to the fields that propagate in the bulk. We will be applying the results of Chap.~\ref{chap:ewsbbulkhiggs} and, as we did there,for simplicity we will restrict ourselves to the case in which the fermions are localized on the {UV} brane. Recall that the strongest constraints on the masses of new {KK} modes in the theory will be given by the dominant $S$ and $T$ parameters, which can be computed from Eqs.~(\ref{eq:Texpr}--\ref{eq:Wexpr}).

It will be convenient to fix the  Planck-weak hierarchy by setting the warp factor $A(y_1)=35$, which imposes a functional relation $y_1 = y_1 (\Delta,\nu)$ so that $y_1$ increases with $\Delta$ and $\nu$. This functional relation is shown in Fig.~\ref{fig:y1}. We can see that $y_1$ gets closer to $35$ as $\nu$ or $\Delta$ increase, as we are approaching the RS limit.
\begin{figure}[t]
\centering
\begin{psfrags}
\input{figs/y1-psfrag.tex}
\includegraphics[width=0.67\textwidth]{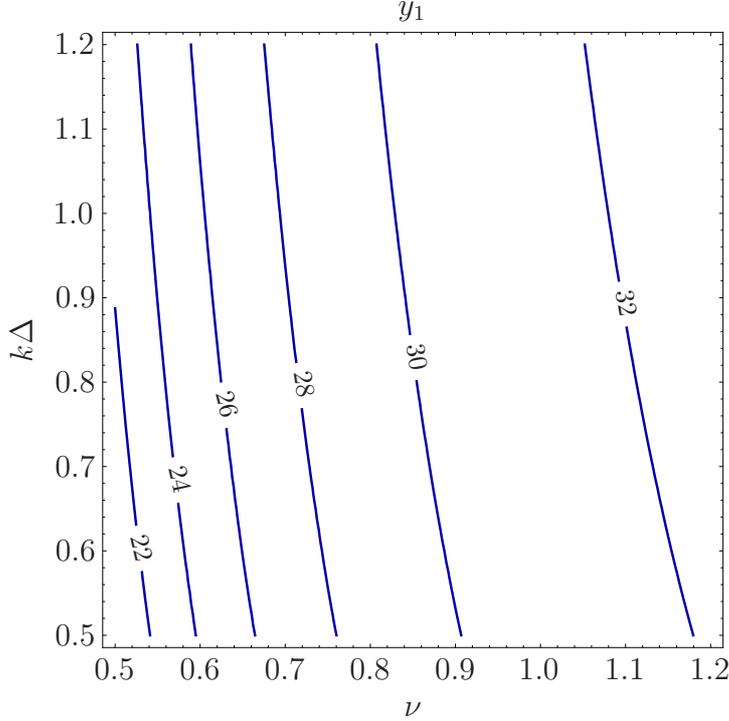}
\end{psfrags}
\caption{\it Contour levels of $y_1(\nu,\Delta)$, having fixed $A(y_1)=35$.}
\label{fig:y1}
\end{figure}

Moreover,  in order to account for the model to solve the hierarchy problem, as discussed in Sec.~\ref{sec:hierarchy}, we will trade $a$ for the fine-tuning parameter $\delta$, defined in Eq.~\eqref{definitiondelta} and shown in Fig.~\ref{fig2} (we are choosing the minimum value $a=a_0(\nu,\Delta)$). The Higgs solution is free of fine-tuning when $\delta \lesssim \mathcal{O}(1)$, and we will therefore use in the following plots the safe value $\delta=0.1$. Increasing $\delta$ would allow for smaller values of $a$ and hence lower bounds. However, an increase from $\delta=0.1$ to $\delta=1$ only amounts to decreasing the bounds in $\lesssim 3\%$ (see Ref.~\cite{Cabrer:2011fb}).

We are therefore left with three parameters: $\nu$, $\Delta$ and $\delta$. We can now compute the contribution of the {KK} of our model to the $T$ and $S$ parameters [Eqs.~(\ref{eq:Texpr}--\ref{eq:Sexpr})] and compare them with the experimental data \cite{Nakamura:2010zzi} in order to set a bound on $\rho$. We will for now consider a reference Higgs mass of $m_H=115$~GeV, although in the end of this chapter we will study different values for it. We will also assume that the contribution to $Y$ and $W$ is negligible, which will be true in most of the parameter space. We will discuss on the validity of this approximation later on, in Sec.~\ref{sec:constraining}.

Once we have found a bound on $\rho$, we need to compute the relation between the masses of the different {KK} modes and $\rho$. We will first consider the gauge bosons, which are described by Eq.~\eqref{eq:EOMf}. This equation needs to be solved numerically for each value of $\nu$ and $\Delta$. The ratio $m_{KK}/\rho$ is a monotonically decreasing function of both $\nu$ and $\Delta$; an approximate fit for this relation when $k\Delta=1$ and $\nu\gtrsim 0.8$ is provided by
\begin{equation}
\frac{m_{KK}}{\rho} \approx 2.4 + \frac{1.7}{\nu^2} \,,
\end{equation}
which, of course, coincides with the {RS} result when $\nu\rightarrow\infty$.

\begin{figure}
\centering
\begin{psfrags}
\input{figs/bounds-psfrag.tex}
\includegraphics[width=0.7\textwidth]{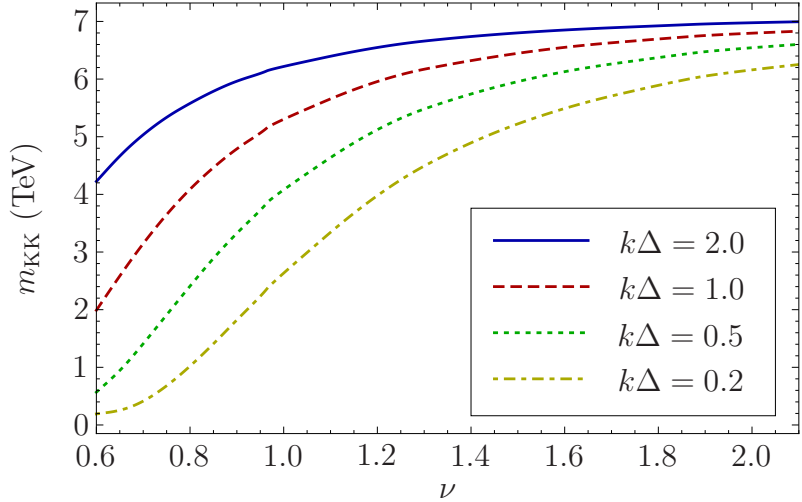}
\end{psfrags}
\caption{\it 95\%~C.L. bounds on the mass of the first KK mode of the gauge boson as a function of $\nu$ and for different values of $\Delta$, for $A(y_1) = 35$ and $\delta=0.1$. In this plot we are taking into account only the contribution from $S$ and $T$ parameters. }
\label{fig:bounds}
\end{figure}
We show in Fig.~\ref{fig:bounds} the $95\%$ {CL} bounds for the first {KK} mode of the gauge bosons, as a function of $\nu$ and $\Delta$, and for $\delta=0.1$. We can see that lowering both $\nu$ or $\Delta$ results in lower bounds on the mass of the {KK} masses. In fact, it seems that we can easily reach bounds of $\mathcal{O}($TeV$)$ or even lower, although we first need to consider other physical observables that might constrain our parameter range, as we will do in Sec.~\ref{sec:constraining}.

Finally, in Fig.~\ref{fig:boundsallnu} we also present the bounds on {KK} masses for the different fields living in the bulk, and compare them with the bounds on gauge boson {KK} modes. In particular, we present the first heavy {KK} mode mass of the physical Higgs [described by Eq.~\eqref{eq:EOMxi}] and pseudoscalar [Eq.~\eqref{eq:EOMpseudoscalar}], for the graviton [Eq.~\eqref{graviton}] and for the radion [Eq.~\eqref{eq:radionEOM}]. The fact that the Higgs and the pseudoscalar are almost degenerate follows from the fact that their {EOM}s unify in the limit of vanishing {EWSB}; indeed we can only see a slight difference in their masses when the $\rho$ scale becomes low enough compared to the {EWSB} scale. 
\begin{figure}
\centering
\begin{psfrags}
\input{figs/boundsallnu-psfrag.tex}
\includegraphics[width=0.7\textwidth]{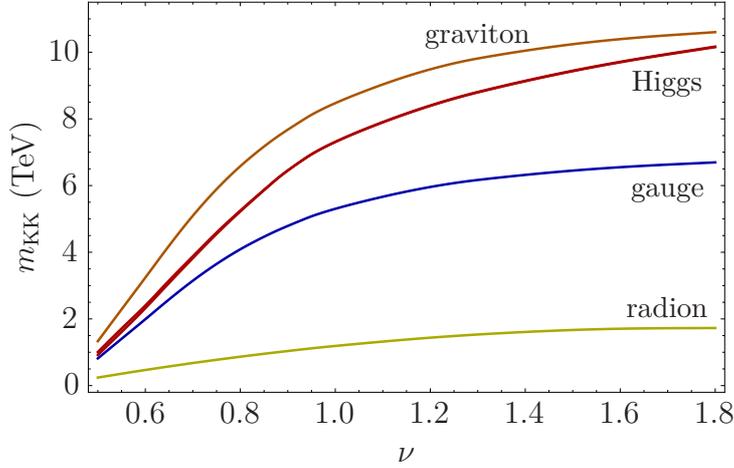}
\end{psfrags}
\caption{\it 95\%~C.L. bounds on the mass of the first KK mode of the different fields in the bulk as a function of $\nu$ and for $\Delta=1$, $\delta=0.1$ and $A(y_1)=35$. The curve for the Higgs includes the pseudoscalar field, the bound of which is around $1\%$ larger for small values of $\nu$.}
\label{fig:boundsallnu}
\end{figure}

The fact that lowering $\nu$ or $\Delta$ leads to lower bounds on $m_{KK}$ is an expected result, since a smaller $\nu$ or $\Delta$ corresponds to a larger $Z$, as can be read from  Fig.~\ref{fig1}. A larger $Z$ factor does in fact trigger:
\begin{itemize}
\item
A decreased coupling of the KK modes to the Higgs currents $\alpha_n$, which in turns leads to a decreasing dependence of $S$ and $T$ with respect to $\rho$ [Eqs.~(\ref{eq:Texpr}--\ref{eq:Sexpr})].
\item
A smaller angle defined by $\vartheta=\tan(T/S)$ and, consequently, using the large correlation between the $S$ and $T$ parameters, a path which approaches the major axis of the $95\%$ CL ellipse allowing for yet lower bounds on $m_{KK}$.
\end{itemize}
These two effects are exhibited in Fig.~\ref{fig:ellipsesnu}, where we show, in the ($T$,$S$) plane, the different paths corresponding to different values of $\nu$ (for $\Delta=1$) and the corresponding values of $T$ and $S$ for different values of $m_{KK}$. These are plotted along with the $95\%$ {CL} regions for different values of the Higgs mass (up to now we are considering $m_H=115~GeV$). For $\nu=\infty$ (the RS limit) the rays go mainly along the $T$ axis, while for smaller values of $\nu$ the rays get a longer path before getting of the ellipse.

\begin{figure}[t]
\centering
\begin{psfrags}
	\input{figs/ellipsesnu-psfrag.tex}
	\includegraphics[width=0.6\textwidth]{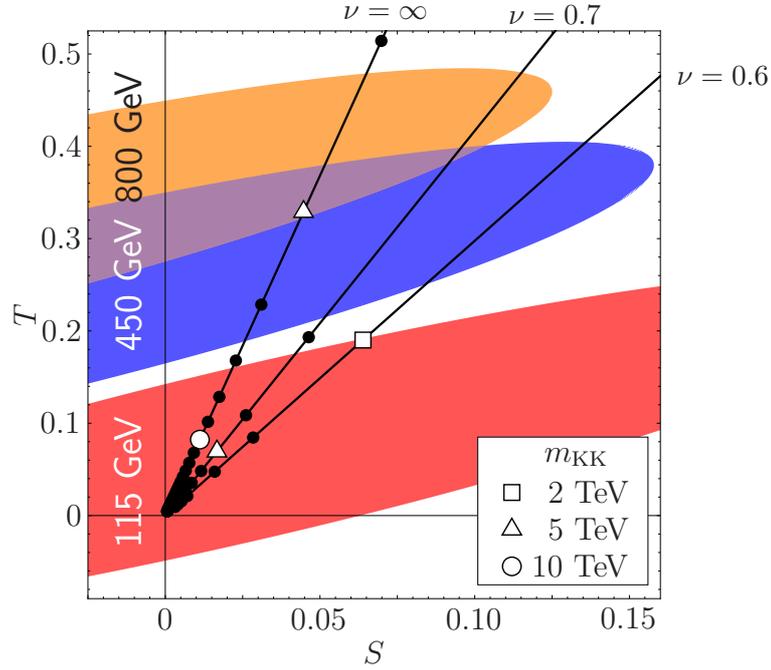}
\end{psfrags}
\caption{\label{fig:ellipsesnu} \it 95\% CL regions in the $(S,T)$ plane with different values of the Higgs mass. The rays correspond to different values of $\nu$ and $\Delta=1$. Dot spacing is 1 TeV. Increasing values of $m_{KK}$ correspond to incoming fluxes.}
\end{figure}

\subsection{On the Higgs mass}
Let us now analyze two things regarding the Higgs mass. We will first study how the bounds behave when increasing the Higgs mass, and later we will analyze the fine-tuning required to have a light Higgs. As we will shortly see, a larger Higgs mass yields lower bounds  on $m_{KK}$. On the other hand, there is no need to increase $m_{H}$ in our model, since the fine-tuning needed to achieve a light Higgs will turn out to be smaller than in {RS}.

%%%%%%
%%%%%%
\subsubsection{A heavy bulk Higgs}
Let us now consider the case of a heavy bulk Higgs, as we did in Sec.~\ref{subsec:aheavybulkhiggs} for {RS}. In Fig.~\ref{fig:ellipsesnu} we have already plotted the $95\%$~{CL} ellipses for different Higgs masses, using the relations in Eqs.~(\ref{eq:DeltaSlargeH}--\ref{eq:DeltaTlargeH2}). 

In fact, a quick look at Fig.~\ref{fig:ellipsesnu} reveals how considering larger Higgs masses allows for lower bounds on $m_{KK}$, although at the same time it might impose higher bounds. Moreover, a small value of $\nu$ can be excluded for certain values of $m_H$, as is the case of $\nu=0.6$ for $m_H=800$~GeV that can be clearly seen in the figure.

\begin{figure}[t]
\centering
\begin{psfrags}
	\input{figs/boundsMOD-psfrag.tex}
	\includegraphics[width=0.7\textwidth]{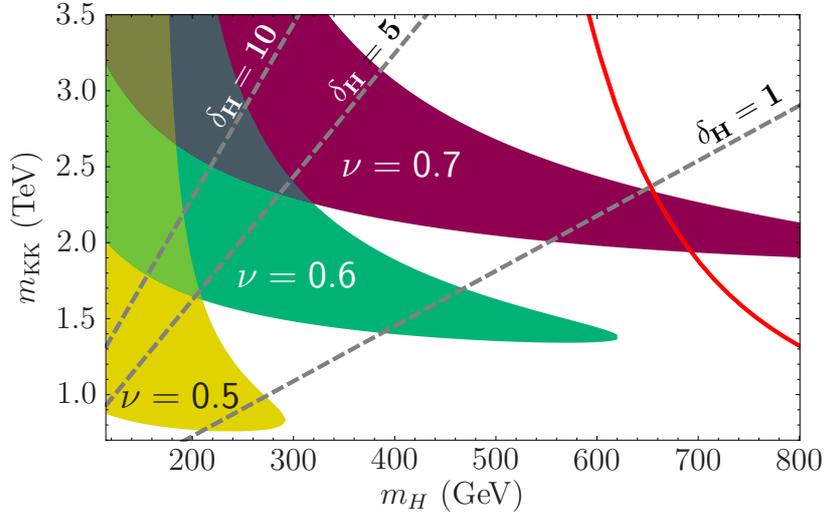}
\end{psfrags}
\caption{
\label{fig:boundsMOD}  
\it 95\% CL regions in the $(m_H,m_{KK})$ plane for 
model $\nu=0.7,\,0.6,\, 0.5$ and $k\Delta=1$, with $A(y_1)=35$ and $\delta=0.1$.  The dashed lines correspond to sensitivity $\delta_H=1$ (no fine-tuning), $\delta_H=5$ (20\%) and $\delta_H=10$ (10\%). The solid line corresponds to the perturbativity bound; the region to its right is excluded.}
\end{figure}
in Fig.~\ref{fig:boundsMOD} we show the 95\% CL allowed regions in the $(m_H,m_{KK})$ plane for various values of the parameters. The solid line is the perturbativity bound, as defined in Sec.~\ref{subsec:aheavybulkhiggs}, and the region on its right is excluded. The dashed lines correspond to different values of the Higgs mass sensivity $\delta_H$, defined in Eq.~\eqref{eq:sensitivity}.  

From Figs.~\ref{fig:ellipsesnu}--\ref{fig:boundsMOD} we see that, in general, a lower value of $\nu$ implies lowering the range of allowed values of $m_{KK}$. For example, for $m_H=450$ GeV the 95\% CL window for $m_{KK}$ is $2.1\, {\rm TeV}\lesssim m_{KK}\lesssim 2.9\, {\rm TeV}$ when $\nu=0.6$, while it becomes $1.4\, {\rm TeV}\lesssim m_{KK}\lesssim 1.7\, {\rm TeV}$ when $\nu=0.7$. Moreover, given a value of $\nu$ implies an upper bound on the Higgs mass; the overall bound is in fact $m_H\lesssim 750$~GeV. This shows that, for our model, a heavy Higgs field can be consistent with KK-modes accessible at LHC energies, and the measurement of the Higgs mass at LHC should constrain the model parameters.

%%%%%%
%%%%%%
\subsubsection{Fine-tuning in the Higgs sector}
\begin{figure}[t]
\centering
\begin{psfrags}
\input{figs/higgsftnu-psfrag.tex}
\includegraphics[width=0.7\textwidth]{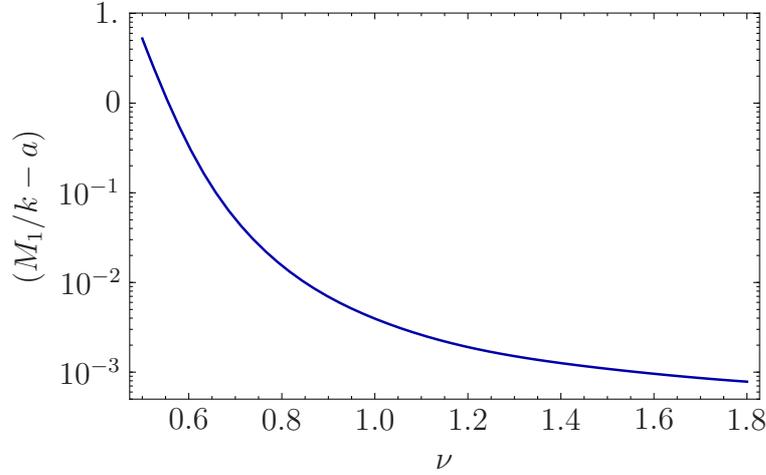}
\end{psfrags}
\vspace{2mm}
\caption{\it $(M_1/k - a)$ for a light Higgs mode with a mass of $120~\mathrm{GeV}$, as a function of $\nu$ and for $\Delta=1$, using the bounds on $\rho$ that can be read from Fig.~\ref{fig:bounds}, where $\delta = 0.1$ and $A(y_1) = 35$. This quantity can be understood as the amount of fine-tuning required to have a Higgs with this mass.}
\label{fig:higgsft}
\end{figure}

As we saw in Sec.~\ref{subsec:higgssector}, having a light Higgs requires some amount of fine-tuning. Recall that the zero-mode of the Higgs field has a mass given by Eq.~(\ref{mH}), that is,
\be
m_H^2
=\frac{2} {Z}\left(M_1/k-a \right)\rho^2
\,,
\label{mHfinal}
\ee
where $M_1$ is IR the brane Higgs mass term. The fine-tuning required to have a light Higgs can be quantified by studying the quantity $M_1/k-a$, so that when it is of $\mathcal O(1)$ there is no fine-tuning, for $\mathcal O(0.1)$ there is a 10\% fine-tuning, and so on and so forth. It is clear that having a heavy Higgs does not require a fine-tuning, although we have seen that a heavy Higgs constrains our model. However, in our model there are two separate effects which favor a light Higgs, as can be seen from Eq.~(\ref{mHfinal}): the suppression in the required value of $\rho$ and the enhancement in the value of $Z$. 

We can quantify this fine-tuning by plotting the prefactor $M_1/k-a$ as a function of $\nu$ (for $\Delta=1$), as shown in Fig.~\ref{fig:higgsft} for a Higgs with mass $m_H=115$~GeV. There we can see that for values of $\nu \lesssim 0.7$ (where we find bounds $m_{KK}\gtrsim 1-2$~TeV) there is no fine-tuning, while in the RS case the fine-tuning would amount to more than 0.1\%. Therefore, our model removes the fine-tuning problem related to obtaining a light Higgs.

%%%%
\subsection{Constraining the parameter space}
\label{sec:constraining}
At this point it is obvious that we would like to choose values for $\nu$ and $\Delta$ as small as possible in order to get the lowest possible bounds on $m_{KK}$. However, the question of which are the minimum value of these parameters that we can take arises. In fact, there are some factors that we need to take into account, let us now study them.

\subsubsection{The $W$ and $Y$ parameters}

To begin with, in our analysis we have assumed that the contribution of the new states to the $W$ and $Y$ parameters is negligible. However, one should be concerned if they might become too large in part of the parameter space. Recall that the expressions for $W$ and $Y$ are given in Eqs.~(\ref{eq:Wexpr}--\ref{eq:Yexpr}). These parameters are volume suppressed but, in fact, the region were the lowest bounds are obtained also corresponds to where the volume $ky_1$ is relatively small (see Fig.~\ref{fig:y1} and Fig.~\ref{fig:bounds}, or also Tab.~\ref{tablevalues}). Also, they correspond to the four-fermion effective operators generated by the exchange of KK-gauge bosons, so it is natural to question whether they remain small enough in the range of small $\nu$ and $\Delta$.

\begin{figure}[t]
\centering
\begin{psfrags}
\input{figs/WYnu-psfrag.tex}
\includegraphics[width=0.7\textwidth]{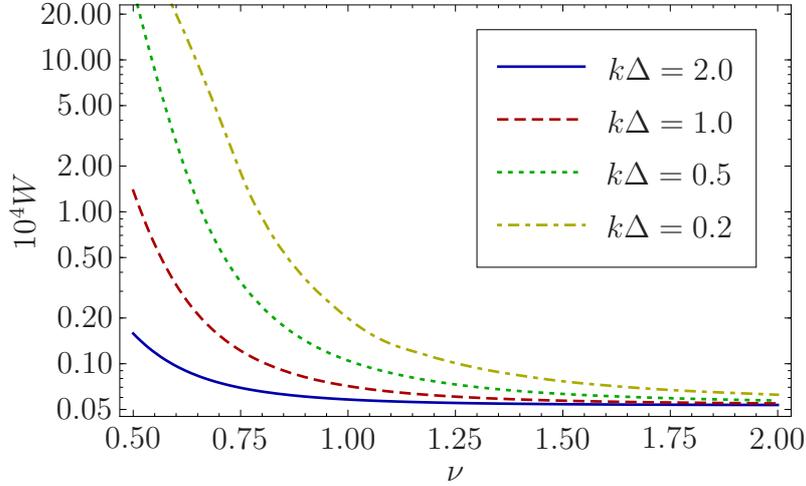}
\end{psfrags}
\caption{\it Plot of the observables $W=Y$ as a function of $\nu$ and for different values of $\Delta$, using the bounds on $\rho$ that can be read from Fig.~\ref{fig:bounds}, where $A(y_1) = 35$ and $\delta=0.1$. }
\label{fig:wynu}
\end{figure}
In Fig.~\ref{fig:wynu} we plot the values of $W=Y$ as a function of $\nu$ for different values of $\Delta$. We should ask these values to be well below the experimental bounds. A fit to all observables for a light Higgs yielded~\cite{Barbieri:2004qk} $W = Y\lesssim 10^{-3}$ at 95\% CL. Therefore, we can see that in most of the shown parameter space the values of $W$ are well below this bound, although they can become in conflict with it when $\nu$ or $\Delta$ becomes too small. For example, when $k\Delta=0.2$ we hit the experimental bound for $\nu \lesssim 0.65$, although we could in principle reach smaller values of $\nu$ for a smaller $\Delta$.

\subsubsection{Gravitational scales}

We also need to check whether the perturbativity in the 5D gravity theory is kept under control. In fact, it will be the case as long as $M_5 L(y) \gtrsim 1$ in all the space, where $M_5$ is the 5D Planck scale, which can be be deduced from Eq.~\eqref{eq:MPlanck4D}. Actually, it turns out that $M_5 \neq k$, so we also have to check that the hierarchy between $M_5$ and $k$ does not grow too large.

Let us now consider this point having a look on the involved scales in our theory. For every set of variables $(\nu, \Delta, a)$, the EWPO fix a lower bound on the parameter $\rho$. On the other hand, since we have fixed the total warp factor by $A(y_1)=35$, it turns out that, for every value of $\rho$, $k$ is fixed as $k=e^{35}\rho$.~\footnote{Notice that this procedure is purely operational. We could as well have fixed $k$ (or even the volume $ky_1$) and have considered different warp factors for every case. Physics should not depend on the chosen procedure.}  Considering the minimal lower bounds on $\rho$ provided by the {EWPT} we can obtain the corresponding values of $M_5$ and $k$. We plot in Fig.~\ref{fig:M5k} these two scales (along with the corresponding values of $\rho$, see right axis) as a function of $\nu$ for $\Delta=1$. In the top axis of the figure we show the minimum value of $kL(y)$ ($kL_{\mathrm{min}}$) that corresponds to each value of $\nu$ (see Fig.~\ref{fig0}).\footnote{In parameter range shown in Fig.~\ref{fig:M5k} the minimum value of $kL(y)$ satisfies $kL_{\mathrm{min}}=kL(y_1)$}
\begin{figure}[p!]
\centering
\begin{psfrags}
\input{figs/M5k-psfrag.tex}
\includegraphics[width=0.7\textwidth]{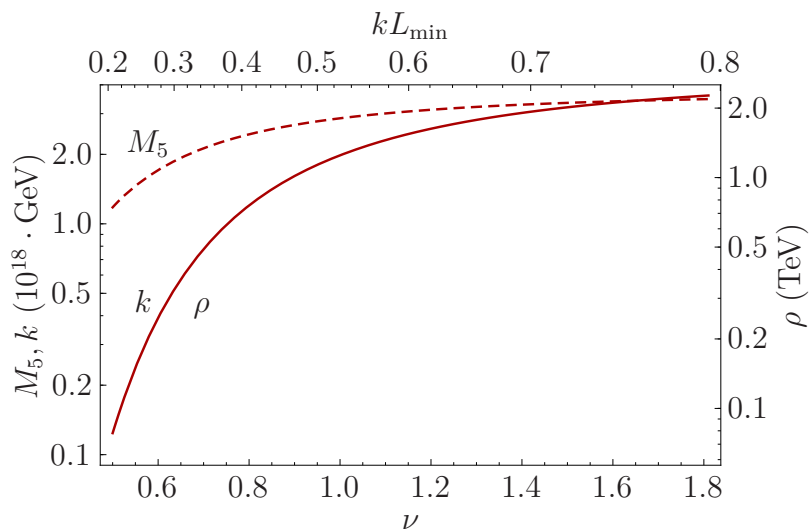}
\end{psfrags}
\caption{\it $M_5$ and $k$  as a function of $\nu$ and for $\Delta=1$, using the bounds on $\rho$ that can be read from Fig.~\ref{fig:bounds}, where $\delta = 0.1$ and $A(y_1) = 35$. On the right axis we show the value of $\rho$ in TeV. On the top axis the value of $k L_\mathrm{min}$ corresponding to $\nu$ can be read. }
\label{fig:M5k}
\end{figure}
\begin{table}[p!]
\centering
\begin{tabular}{c|c}
$k \Delta$ & $\nu$  \\
\hline
2.0 & $>0.40$ \\
1.0 & $>0.49$  \\
0.5 & $>0.64$  \\
0.2 & $>0.80$ \\
\end{tabular}
\caption{\it Minimum values of $\nu$, for different values of $\Delta$, such that the ratio between the 5D Planck scale and $k$ satisfies $M_5/k < 10$.}
\label{tab:minnu}
\end{table}
From Fig.~\ref{fig:M5k} we can see that, when the minimum curvature radius $kL_{\mathrm{min}}$ decreases by a factor of $\sim 5$, $\rho$ decreases by a factor $\sim 20$ while $M_5$ only decreases by a factor $\sim 2$. The parameter $M_5 L_{\mathrm{min}} =(M_5/k) kL_{\mathrm{min}}$, which controls perturbativity in the 5D gravity theory, thus \emph{increases} by a factor $\sim 2$. This counter-intuitive result comes from the fact that the suppression in the curvature radius is overcompensated by the suppression in the value of $k$. This produces a small hierarchy (one order of magnitude) between $k$ and $M_5$. Lowering $kL_1$ further would translate into a larger hierarchy which in turn would translate into a subsequent amount of fine-tuning. In order to be more explicit about this, in Tab.~\ref{tab:minnu} we show the requirements on $\nu$ for different values of $\Delta$ such that $M_5/k < 10$. The bounds here are more stringent than the previous requirements from the $W$ and $Y$ parameters, and therefore they provide a rough estimate of the minimum values of $\nu$ and $\Delta$ that can be allowed for.

One should keep in mind that the scale $k$ is really a free parameter of the theory that only comes out here as a prediction because we have fixed the volume by the condition $A(y_1)=35$ throughout our analysis. Slightly increasing the volume does not change $S$ and $T$, and we could have used this freedom to fix $k$ such that e.g.~$M_5/k\sim 5$  and hence $M_5 L_{\mathrm{min}}\sim 1$. 
Let us also mention that the volume is generally reduced in our model since the deformation of AdS is positive, $A(y_1)>ky_1$ (see e.g.~Tab.~\ref{tablevalues}). However, it is not the main effect in improving $T$ and moreover the hierarchy is fixed by $A(y_1)$, and not $y_1$.\footnote{Of course by relaxing the hierarchy requirement one can easily lower the bounds from EWPT. The simplest examples are the so-called ``little RS models"~\cite{Davoudiasl:2008hx} which solve the hierarchy problem up to scales much below the Planck mass. For instance in these models the requirement $T\sim S$ for an IR localized Higgs (a bulk Higgs with $a\simeq 2$) provides the volume condition $ky_1\sim 4c^2_W$ ($ky_1\sim 9c^2_W$), which translates into mass stabilization up to scales of $\sim 30 $ TeV ($\sim 1200$ TeV).}

\parasep

To conclude this section, in Tab.~\ref{tablevalues} we provide for reference some benchmark points for our model, where we display the explicit parameters $\nu,\ \Delta$, $a$ and $k y_1$ along with the minimum curvature radius $kL_{\mathrm{min}}$, the product  $M_5 L_{\mathrm{min}}$, $Z$ and the bounds on $\rho$ and the different kinds of fields living in the bulk.
\begin{table}[h]
\centering
\begin{tabular}{ | c | c | c | c || c | c | c || c | c | c | c | c | }
\hline
  $\nu$ 
& $k\Delta$ 
& $a$ 
& $ky_1$ 
& $ \displaystyle \!\! \rule[-1.2em]{0pt}{3.2em} kL_{\mathrm{m}} \!\!$  
& $\!\!M_5L_{\mathrm{m}}\!\!$
& $Z$
&$\displaystyle\!\!\frac{\rho}{\mathrm{TeV}}\!\!$
 & $\displaystyle\!\!\frac{m^{\mathrm{KK}}_{\mathrm{g.b.}}}{\mathrm{TeV}}\!\!$ 
 & $\displaystyle\!\!\frac{m^{\mathrm{KK}}_{\mathrm{H}}}{\mathrm{TeV}}\!\!$ 
 &  $\displaystyle\!\!\frac{m^{\mathrm{KK}}_{\mathrm{grav}}}{\mathrm{TeV}}\!\!$ 
 & $\displaystyle \!\!\frac{m_{\mathrm{rad}}}{\mathrm{TeV}} \!\!$
\\ \hline \hline
 0.48 & 1.0 & 3.2& 22&0.2   & 2.0 & 6.6 & 0.072 & 0.82 & 0.92 & 1.3 & 0.28 
\\ \hline

 0.55 & 1.3 & 2.8  & 25&0.3  & 1.1  & 2.1 & 0.32 & 2.4 & 2.9 & 4.0 & 0.84 

\\ \hline

 0.64 & 1.6 & 2.5 & 28&0.4   & 0.86 & 1.2 & 0.70 & 4.0 & 4.9 & 6.5 & 1.3

\\ \hline

 0.73 & 1.7 & 2.4 & 30&0.5   & 0.78 & 0.86 & 1.1 & 5.2 & 6.6 & 8.4 & 1.6

\\ \hline

 $\infty$ & $\infty$ & 2.1 & 35&1   & 0.79 & 0.47 & 3.1 & 7.5 & 12 & 12 & 0

\\ \hline 
\end{tabular}
\caption{\it Values of different relevant quantities for different points, sorted by their value of $L_{\mathrm{min}}$ (abbreviated as $L_{\mathrm{m}}$) , for $\delta= \nobreak 0.1$ and $A(y_1)=35$. The stabilization mechanism in our model disappears when we take the RS limit, leading to vanishing radion mass as expected.
} 
\label{tablevalues}
\end{table}

%%%%%%%%%%%%%%%%%%%%%%%%%%%%%%%%%%%%%%%%%%

\section{The soft-wall limit}
\label{sec:commentssw}

As we stated in the beginning of this chapter, the metric in Eq.~(\ref{A}) has a curvature singularity at a point $y_s>y_1$, beyond the IR brane and outside the physical interval. In fact, in the absence of an {IR} the singularity would become naked and we would be in front of a soft-wall model (in fact the SW2 model  described in Chap.~\ref{chap:softwalls}).

It is then natural to ask what would happen if we removed the {IR} brane and considered a true soft-wall background. We briefly discussed about this possibility in Sec.~\ref{subsec:ewsbsw}. Moreover, from what we have seen, it can be expected that the $T$ and $S$ parameters could be greatly reduced in a soft-wall setup, making this possibility worth of study. Also, recall from Chap.~\ref{chap:softwalls} that soft walls show some other interesting features such as a greater variety of KK spectra: in particular the density of states above the first KK excitation is typically higher than in two-brane models (and can even be continuous) and could lead to very interesting collider signatures.

A soft-wall model can be recovered in our model by taking the limit $\Delta\to 0$. However, we cannot take this limit blindly.\footnote{Another minor consideration when taking the soft-wall limit is related to our fixing of the Planck-weak hierarchy. In our analysis we have kept $\rho/k= e^{-A(y_1)}$ fixed, but this quantity vanishes when we take the limit $y_1 \to y_s$. Therefore we should fix the Planck-weak hierarchy differently, e.g.~by keeping fixed $\rho/k= e^{-ky_s}(ky_s)^{-\frac{1}{\nu^2}}$, where we have used the definition of $\rho$ used in Chap.~\ref{chap:softwalls}.}
First, because of the lower bounds on $\nu$ that might arise from the considerations of Sec.~\ref{sec:constraining}. Second, we have to check that our soft-wall background is consistent, according to the condition on the superpotential expressed in \eqref{criterion}, which requires in our case that $\nu > 2$. And finally, a soft-wall model with the metric in Eq.~(\ref{A}) only presents a mass gap when $\nu\geq 1$. Therefore, from these last two requirements the parameter space is reduced to, at least, $1 \leq \nu < 2$.

However, if we intend to implement {EWSB} there is a more important factor to be accounted for. Without an {IR} brane and with a bulk Higgs we need to find an alternative location where to trigger {EWSB}. The only sensible {IR} boundary condition for the Higgs profile in the singular background is to demand regularity of the solution. With a linear bulk {EOM} and linear {UV} boundary conditions a nontrivial profile can only arise at the price of a fine tuning: satisfying the boundary conditions at the {UV} brane fixes a certain linear combination of the two bulk solutions (up to an overall normalization). This solution will in general not be regular at $y_s$, so the only solution is the trivial one $h(y)\equiv 0$ and {EWSB} does not occur. This does not come as a surprise, as we have lost the {IR} brane with its {EWSB} potential. 

Therefore, the only possibility to obtain a nontrivial profile without an {IR} brane is a fine-tuning of some parameters in the Lagrangian. One can see how this fine-tuning arises in the smooth limit $y_1\to y_s$, following the logics of Sec.~\ref{sec:hierarchy}. For a fixed value of $a>2$ there will be a finite $\Delta=y_s-y_1>0$ at which the singular solution (initially suppressed by $e^{-2(a-2)ky}$) resurfaces due to the presence of the singular factor in Eq.~(\ref{eq:hyF}). In this case, a fine-tuning of the UV mass $M_0$ is needed in order to select the regular solution. This is the original hierarchy problem. Clearly, in the exact SW limit ($\Delta=0$) one needs $M_0=a$ precisely, and therefore an infinite fine-tuning.

The only way out is to trigger EWSB by nonlinear dynamics on the UV brane~\cite{Batell:2008me} or in the bulk. In the former case, $h(0)$ (the VEV of the Higgs field at the UV brane) is unrelated to the IR scale generated by the warping.\footnote{In fact $h(0)$ has to be fine tuned to a very small number even though this can be made technically natural as long as the UV brane localized Higgs mass term is Planckian.}
Breaking in the bulk requires to go beyond a simple quadratic potential in the Higgs field and cannot be tackled analytically. Things could be simplified by integrating over the soft wall to create an effective IR brane at $y=y_1< y_s$, along the lines of Ref.~\cite{vonGersdorff:2010ht}. Form factors arise on the IR brane that mimic the characteristic SW spectra, while the bulk potential near the singularity will generate a nontrivial brane potential that could be used to trigger EWSB~\cite{vonGersdorff:2010ht}. It is however not straightforward to engineer a bulk potential $V(\phi,h)$ that can be totally neglected in the reduced bulk $y<y_1$ and whose only effect is the effective IR brane tension. Although it can be expected that some of the findings of this section can be translated to the case of a genuine bulk, breaking a quantitative statement is hard to be settled in the absence of a precise and calculable model.

%%%%%%%%%%%%%%%%%%%%%%%%%%%%%%%%%%%%%%%%

\chapter{Fermions in the Bulk}
\label{chap:fermionsbulk}

In Chap.~\ref{chap:ewsbbulkhiggs} we studied how to break EWSB using generalized models of extra dimensions where the Higgs and gauge bosons propagate in the bulk. Later, in Chap.~\ref{chap:noncustodial} we applied those results to a concrete model where the contributions to EWPT are greatly reduced. While those constructions were quite general, we chose to locate the fermions on the UV brane, as this greatly simplified calculations. This leads to the question of what would change in our conclusions if we allowed the fermions to propagate in the bulk. 

Besides our aim to consider the most general possible setup, there are a number of reasons why it is interesting to study the propagation of fermions in the bulk. One of them is the increased strength of the interactions between the SM fermions and the new KK states that arise in warped models, as the coupling between bulk and UV-brane fields are strongly suppressed. This leads to a much greater range of interesting new phenomenology to be tested at the LHC but, on the other hand, it might also carry stronger constraints from experimental data.

But the strongest motivation for propagating the fermions in the bulk is to address the SM \emph{flavor puzzle}, which is the failure of the SM to explain why the masses of the fermions span 5 orders of magnitude with no apparent structure. The RS model can address that by localizing the different kinds of fermions at different positions further or closer to the IR brane, so that their couplings with the IR-localized Higgs boson get respectively decreased or enhanced%(see Sec.~\ref{subsec:addressingflavorpuzzle})
. However, it is well known that the different fermion localization generates non-oblique observables. Mainly, it modifies the $Z b \bar{b}$ coupling, as well as flavor changing neutral current (FCNC) and CP violating dimension-six operators, which might impose further constraints on the KK scale.  

In this chapter, based of Ref.~\cite{Cabrer:2011qb}, we will study this mechanism to explain the SM flavor structure, but we will allow the Higgs to propagate in the bulk (as in Chap.~\ref{chap:ewsbbulkhiggs}) and consider both RS and the model described in Chap.~\ref{chap:noncustodial}, defined by the metric of Eq.~(\ref{A}), i.e.~
\begin{equation}
A(y)=ky-\frac{1}{\nu^2}\log\left(1-\frac{y}{y_s}\right) \,.
\label{ourmetric}
\end{equation}
 In doing so, we will consider 5D Yukawa couplings of $\mathcal{O}(1)$ with no particular structure\footnote{The absence of a 5D Yukawa hierarchy is also referred to as \emph{anarchy}.} and choose specific mass terms for each kind of fermions in order to fix their locations. We will also get bounds on $m_\mathrm{KK}$ from the $Z b \bar{b}$ coupling and flavor violating operators, and we will shortly see how the model \eqref{ourmetric} leads to milder constraints when compared to RS. 

Notice that the issue of non-oblique EWPOs in the model (\ref{ourmetric}) has recently been addressed in Ref.~\cite{Carmona:2011ib}.  Although we employ different  fermion profiles, we find similar bounds from the $Z b \bar{b}$ coupling, with slightly more optimistic bounds in the fully anarchic case. Moreover, our analysis is quite different and complementary to the one done in~\cite{Carmona:2011ib}, as we perform a democratic scan over possible 5D Yukawa couplings deducing the bulk masses needed to reproduce the observed masses and mixings. In that way we are able to associate a probability to a certain choice of bulk masses and hence quantify the fine tuning to achieve a given KK scale that leads to agreement with experimental constraints. The improvement of flavor/$CP$ bounds with modified metrics has recently been noted in the context of soft wall models~\cite{Atkins:2010cc}. Here we show that a similar improvement can be obtained in the hard wall setup, and again we quantify the amount of fine tuning needed to obtain a satisfactory bound on the KK scale.

The plan of this chapter goes as follows. In Sec.~\ref{low} we describe the low energy 4D theory obtained after integrating out the KK weak gauge bosons, the KK gluons and the KK fermions. In Sec.~\ref{sec:EWPOfermions} we give explicit expressions of oblique and non-oblique EWPOs for arbitrary metrics, and in Sec.~\ref{sec:flavorfermions} we do the same for flavor and CP violating observables. We complete our general analysis in Sec.~\ref{quarks}, where we present an approximation of the quark mass eigenstates and mixing angles, assuming a left-handed hierarchy more general than other existing approximations in the literature. Finally, in Sec.~\ref{sec:applicationflavor} we apply these results to the RS model and the non-custodial model introduced in Chap.~\ref{chap:noncustodial}.

\section{The low-energy effective theory}
\label{low}

In this section we will present general expressions for oblique and non-oblique corrections, as well as FCNC operators, for arbitrary profiles for the metric, the Higgs and the fermions. We will first integrate out the KK modes of the weak gauge bosons, which will be relevant for EWPOs, in Sec.~\ref{KKweak}. Dimension six operators generated from integration of the KK modes of gluons will be considered in Sec.~\ref{KKgluons} and those obtained from integration of the KK modes of fermions in Sec.~\ref{secKKfermions}.

\subsection{Integrating out the KK weak gauge bosons}
\label{KKweak}

Let us consider a general gauge interaction of the form
\begin{equation}
g_5 \int d^5 x \sqrt{g} A_M(x,y) \left[ \sum_\psi J_\psi^M (x,y) + J_h^M (x,y) \right] \,,
\end{equation}
where $J_h$ stands for the Higgs current and $J_\psi$ for the fermion currents, and where $\psi = Q, u, d, L, \ell$ runs over the SM fermions before EWSB. For simplicity, we are omitting here the adjoint gauge indexes and assuming an implicit sum over them. The EOMs for the EW gauge bosons are then
\be
D^\mu F^A_{\mu\nu}+ J^{A}_{h\, \nu}+\sum_\psi J^{A}_{\psi\, \nu}=0\,.
\label{eom}
\ee

We would like to integrate out the KK modes of the gauge bosons in order to obtain the coefficients of the effective Lagrangian with dimension-six operators 
\be
\mathcal L_{\rm eff}=\mathcal L_{\rm SM}+
 \frac{1}{2}\alpha_{hh}\,J_h\cdot J_h+\sum_\psi\alpha_{h\psi}\, J_h\cdot J_\psi
+\frac{1}{2}\sum_{\psi,\psi'} \alpha_{\psi\psi'}\,J_\psi\cdot J_{\psi'}\,.
\label{eff}
\ee
In other words, we would like to compute the diagrams in Fig.~\ref{diagramas}. 
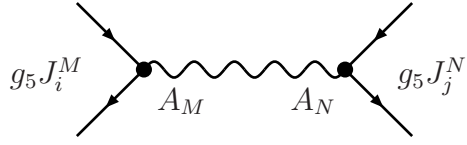
\begin{figure}[t]
\begin{center}
%\SetScale{.5}
\begin{picture}(65,80)(35,-20)
\SetWidth{1.}
\Vertex(25,25){3}
\Vertex(100,25){3}
\ArrowLine(0,50)(25,25)
\ArrowLine(25,25)(0,0)
\Photon(25,25)(100,25){3}{5}
\ArrowLine(125,50)(100,25)
\ArrowLine(100,25)(125,0)
\put(30,10){$A_M$}
\put(80,10){$A_N$}
\put(-25,20){$g_5 J_i^M$}
\put(120,20){$g_5J_j^N$}
\end{picture}
\vspace{-1cm}
\end{center}
\caption{\it Diagram contributing to the effective Lagrangian.}
\label{diagramas}
\end{figure}
To this end we need to evaluate 
\be
\alpha_{X X'}= \frac{1}{y_1^3}\int_0^{y_1}
 dy\,dy'\,\omega_X(y)\,\omega_{X'}(y')\,G(y,y')
 \,,
 \label{eq.alpha}
\ee
where $G(y,y')$ is the 5D gauge boson propagator at zero 4D momentum (with any possible zero modes subtracted), and where we have defined
\begin{align}
\omega_h(y) &= e^{-2A(y)} \psi^2_0(y) \,,
\\
\omega_\psi(y) &= e^{-3A(y)} \psi^2_0(y) \,.
\end{align}

The precise definition of the propagator $G(y,y')$ and explicit expressions for it are given in App.~\ref{app:propagators}. Using the result given in Eq.~\eqref{GNN},\footnote{Recall that gauge bosons have Neumann-Neumann BCs.} we obtain  
\be
\alpha_{XX'}=y_1\int e^{2A}\left(\Omega_X-\frac{y}{y_1}\right)\left(\Omega_{X'}-\frac{y}{y_1}\right)\,,
\ee
where
\be
\left\lbrace 
\begin{array}{l}
\displaystyle \Omega_{h} = \int_0^y dy'\,e^{-2A(y')}\,\xi_0^2(y')\,
\vspace{1mm}
\\
\vspace{1mm}
\displaystyle \Omega_{\psi} = \int_0^y dy'\,e^{-3A(y')}\,\psi_0^2(y')\,
\end{array}
\right.
\,,
\label{Omegahpsi}
\ee
and the wavefunctions correspond to zero-modes. The normalization conditions on the wavefunctions imply $\Omega_X(y_1)=1$.

We will rewrite the Lagrangian \eqref{eff} as
\be
\mathcal L_{\rm eff}=\mathcal L_{\rm SM}+\mathcal L_{\rm oblique}+\mathcal L_{\rm non-oblique}\,,
\ee
with
\bea
\mathcal L_{\rm oblique}&=&
\frac{1}{2}\hat\alpha_{hh}\,J_h\cdot J_{h}
+\,\hat\alpha_{hg}\, [D_\mu F^{\mu\nu}\cdot J_{h\,\nu}]
+\frac{1}{2}\hat\alpha_{gg} \, [D_\mu F^{\mu\nu}]^2 \,,
\label{oblique}
\\
\mathcal L_{\rm non-oblique}&=&
\sum_{\psi} \hat\alpha_{h\psi}\,J_h\cdot J_{\psi}
+\frac{1}{2}\sum_{\psi,\psi'} \hat\alpha_{\psi\psi'}\, J_{\psi}\cdot J_{\psi'}\,,
\label{non-oblique1}
\eea
which can be shown to be physically equivalent to Eq.~(\ref{eff}) using Eq.~(\ref{eom})  and requiring the  conditions 
\bea
\alpha_{hh}&=&\hat \alpha_{hh}-2\hat \alpha_{hg}+\hat\alpha_{gg}\,,\\
\alpha_{h\psi}&=&\hat \alpha_{h\psi}-\hat\alpha_{hg}+\hat\alpha_{gg}\,,\\
\alpha_{\psi\psi'}&=&\hat\alpha_{\psi\psi'}+\hat\alpha_{gg}\,.
\eea
hold. 

We can now transform away the non-oblique corrections for near UV-localized fermions (mostly elementary) such as first and second generation leptons (which have $\Omega\approx 1$), so all the new physics for them is encoded in the oblique parameters.  
We will refer to this basis as the ``oblique basis" and use it from now on. In order to achieve $\hat\alpha_{h\psi}\approx 0$ and $\hat \alpha_{\psi\psi'}\approx 0$ for the elementary fields
a good choice is
\begin{align}
\hat\alpha_{hg}&=y_1\int e^{2A}(1-\Omega_h)\left(1-\frac{y}{y_1}\right)\,,
\\
\hat\alpha_{gg}&=y_1\int e^{2A}\left(1-\frac{y}{y_1}\right)^2  \,,
\end{align}
which leads to
\begin{align}
\label{alphahat}
\hat\alpha_{hh}&=\alpha_{hh}+2\hat\alpha_{hg}-\hat\alpha_{gg}=y_1\int e^{2A}(1-\Omega_h)^2\,,
\\
\label{alphahatB}
\hat \alpha_{h\psi}&=\alpha_{h\psi}+\hat \alpha_{hg}-\hat\alpha_{gg}
  =y_1\int e^{2A}\left(\Omega_h-\frac{y}{y_1}\right)\left(\Omega_\psi-1\right)\,,
  \\
  \label{alphahatC}
\hat\alpha_{\psi\psi'}&=\alpha_{\psi\psi'}-\hat\alpha_{gg}
  =y_1\int e^{2A}\left[\left(\Omega_\psi-\frac{y}{y_1}\right)\left(\Omega_{\psi'}-\frac{y}{y_1}\right)-\left(1-\frac{y}{y_1}\right)^2\right]\,.
\end{align}
It is clear from Eqs.~(\ref{alphahat}--\ref{alphahatC}) that, for fermions strictly localized on the UV brane ($\Omega_\psi=1$), $\hat\alpha_{h\psi}$ and $\hat\alpha_{\psi\psi'}$ vanish. Consequently fermions that are only near UV localized will still have strongly suppressed non-oblique corrections. 

The oblique Lagrangian \eqref{oblique} gives rise to the $T$, $S$, $W$ and $Y$ parameters, while  the first term of the non-oblique Lagrangian \eqref{non-oblique1}  contributes to modified $Z$ and $W$ couplings to fermions. These effects will be described in Sec.~\ref{sec:EWPOfermions}. On the other hand, the second term of \eqref{non-oblique1} generates flavor-violating four-fermion operators although, as we will see in Sec.~\ref{sec:flavorfermions}, they are subleading with respect to those induced by the KK gluons.

%%%%%%%%%%%%%%%

\subsection{Integrating out the KK gluons}
\label{KKgluons}

Let us now have a look at the gluons. Integrating out their KK modes we obtain
\be
\mathcal L=\frac{g_s^2}{2}\sum_{\psi,\psi'} \alpha_{\psi\psi'}\, 
\bar\psi \gamma^\mu\lambda^a\psi\, \bar\psi'\gamma^\mu\lambda^a\psi'\,,
\ee
where here $\psi$ runs over the quarks ($u_{L,R},d_{L,R}$) and $\lambda^a$ are the $SU(3)$ matrices normalized to $\tr \lambda^a \lambda^b=\frac{1}{2}\delta^{ab}$.

Using appropriate spinor and $SU(3)$ identities, we can rewrite this as
\begin{equation}
\begin{split}
\mathcal L=&\frac{g_s^2}{2}\sum_{q,q'}\left[
\gamma_{q_L,q'_L}^{ij,k\ell}\left(-\frac{1}{6}\,
\bar q_L^i\gamma^\mu q_L^j\ \bar q'^k_L\gamma^\mu q'^\ell_L+\frac{1}{2}
\bar q_L^i\gamma^\mu q'^\ell_L\ \bar q'^k_L\gamma^\mu q^j_L\right)
+L\to R\right.
\\
&\left.+2\,\gamma_{q_L,q'_R}^{ij,k\ell}\left(
\bar q^i_Lq_R'^\ell\ \bar q'^k_Rq^j_L-\frac{1}{3}\,
\bar q^{i\,\alpha}_Lq_R'^{\ell\,\beta} \bar q'^{k\,\beta}_R q^{j\,\alpha}_L
\right)\right]\,,
\end{split}
\label{gKK}
\end{equation}
where $\alpha$ and $\beta$ are color indexes,\footnote{We suppress color indexes whenever they are contracted in the same way as the spinor indexes.} $q$ and $q'$ run over $u,d$, and we have defined
\be
\gamma_{q_\chi,q'_{\chi'}}^{ij\,k\ell}=y_1\int e^{2A}\left(\Omega_{q_\chi}^{ij}-\frac{y}{y_1}\delta^{ij}\right)\left(\Omega_{q'_{\chi'}}^{k\ell}-\frac{y}{y_1}\delta^{k\ell}\right)\,,
\ee
with the hermitian matrices $\Omega$ defined as
\be
\Omega_{q_\chi}^{ij}=(V_{q_\chi}\Omega^{diag}_{q_\chi} V_{q_\chi}^\dagger)^{ij} ,\quad \chi=L,R\,.
\label{OmegaMatrix}
\ee

The Lagrangian (\ref{gKK}) will give rise to the leading flavor violating effects, as we will see in detail in Sec.~\ref{sec:flavorfermions}.

%%%%%%%%%%%%%%%%%%%%%%%%%

\subsection{Integrating out the KK fermions}
 \label{secKKfermions}
 
We will now consider the fermion action~\cite{MertAybat:2009mk}
\begin{equation}
\begin{split}
S=& \int dy\,e^{-3A}\left(i\bar\psi_L\, /\hspace{-.22cm}\partial\,\psi_L
+i\bar\psi_R\, /\hspace{-.22cm}\partial\,\psi_R\right)
\\
&+
e^{-4A}\,\left(
\bar\psi_R\psi_L'-2A'\,\bar\psi_R\psi_L
-M_\psi(y)\bar\psi_R\psi_L+{\rm h.c.}\right)\,,
\end{split}
\label{A1}
\end{equation}
where $\psi=(\psi_L,\psi_R)^T$ (which runs over $Q,u,d,L,\ell$) is a 5D (Dirac) fermion and, for the sake of generality, we have allowed the bulk mass to depend on $y$.

Defining new wave functions  
\be
\psi_{L,R}(y)=e^{2A(y)} \hat\psi_{L,R}(y)\,,
\ee
we can rewrite the Dirac equation as 
\be
m \hat \psi_{L,R}=e^{-A}(M_\psi\pm \partial_y)\hat\psi_{R,L}\,.\label{Dirac2}
\ee
For the BC we take either $\psi_L=0$ or $\psi_R=0$ at both branes.  Then there is a zero mode with profile
\be
 \hat\psi^0_{L,R}(y)=\frac{e^{- Q_{\psi}(y)}}{ \left({\int_0^{y_1} e^{A-2Q_{\psi}}}\right)}\,,\qquad \hat\psi^0_{R,L}(y)\equiv 0\,,
\ee
where 
\begin{equation}
Q_{\psi}(y)=\mp\int_0^y M_\psi(y') dy'\,.
\label{eq:Qpsi}
\end{equation}
The upper sign corresponds to left-handed zero modes (i.~e.~$SU(2)$ doublets $\psi=Q,L$) and the lower one to right-handed zero modes (i.e.~$SU(2)$ singlets $\psi=u,d,\ell$).
The function $\Omega_{\psi}$ defined in Eq.~(\ref{Omegahpsi}) can then be written as
\be
\Omega_{\psi}(y)=\frac{U_{\psi}(y)}{U_{\psi}(y_1)}\,,\qquad U_{\psi}(y)=\int_0^y e^{A(y')- 2Q_{\psi}(y')} dy'\,.
\label{Omegafermion}
\ee
The quark Yukawa coupling is then
\be
Y_{ij}^{q}=\hat Y_{ij}^q\frac{\displaystyle \int_0^{y_1} h\,e^{-Q_{Q^i_L}-Q_{q^j_R}}}
  {\displaystyle \left(
    \int_0^{y_1} e^{-2A}h^2\int_0^{y_1} e^{A-2Q_{Q^i_L}} \int_0^{y_1} e^{A-2Q_{q^j_R}}
  \right)^{\frac{1}{2}}}\,,
  \label{eq.Y4DY5D}
\ee
where $q=(u,\, d)$. Here $\hat Y_{ij}^q$ are 5D Yukawa couplings with mass dimension $-\frac{1}{2}$.

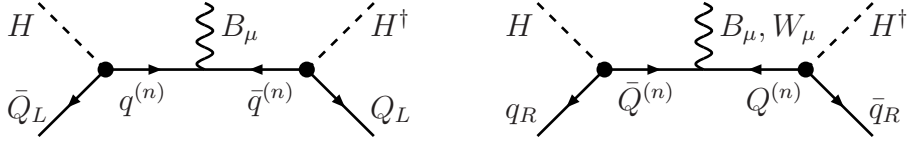
\begin{figure}[t]
\begin{center}
%\SetScale{.5}
\begin{picture}(65,80)(35,-20)
\SetWidth{1.}
\Vertex(25,25){3}
\Vertex(100,25){3}
\ArrowLine(25,25)(0,0)
\DashLine(0,50)(25,25){3}
\ArrowLine(25,25)(62,25)
\ArrowLine(100,25)(62,25)
\DashLine(125,50)(100,25){3}
\ArrowLine(100,25)(125,0)
\Photon(62,25)(62,50){3}{3}
\put(30,10){$ q^{(n)}$}
\put(78,10){$\bar q^{(n)}$}
\put(-12,38){$H$}
\put(124,38){$H^\dagger$}
\put(-12,7){$\bar Q_L$}
\put(124,7){$ Q_L$}
\put(68,38){$B_\mu$}
\end{picture}
%
%\SetScale{.5}
\hspace{4cm}
\begin{picture}(65,80)(35,-20)
\SetWidth{1.}
\Vertex(25,25){3}
\Vertex(100,25){3}
\ArrowLine(25,25)(0,0)
\DashLine(0,50)(25,25){3}
\ArrowLine(25,25)(62,25)
\ArrowLine(100,25)(62,25)
\DashLine(125,50)(100,25){3}
\ArrowLine(100,25)(125,0)
\Photon(62,25)(62,50){3}{3}
\put(30,10){$\bar Q^{(n)}$}
\put(78,10){$ Q^{(n)}$}
\put(-12,38){$H$}
\put(124,38){$H^\dagger$}
\put(-12,7){$ q_R$}
\put(124,7){$\bar q_R$}
\put(68,38){$B_\mu, W_\mu$}
\end{picture}

\vspace{-1cm}
\end{center}
\caption{\it Integrating out the KK modes of the singlets (left) and doublets (right). Notice that the zero mode is not included in the internal line.}
\label{KKfermion}
\end{figure}

%%%%%%%%%
We would now like to integrate out the KK modes of the quarks. In particular we are interested in the diagrams shown in Fig.~\ref{KKfermion}. Notice that we can neglect these contributions if the external quarks are near UV localized. This is because, unlike the coupling of a UV localized fermion to gauge KK modes, 
the coupling of a UV localized fermion to the Higgs zero mode and a KK fermion does not go to a universal constant but rather to zero. 
Since we are primarily interested in corrections to the $Z\bar bb$ coupling we will focus on the down sector (the up sector works analogously with $H\to i\sigma_2 H^*$). One obtains the effective 4D Lagrangian of zero modes
\begin{equation}
\begin{split}
 \mathcal L'_{\rm non-oblique} =&i \beta^{d_L}_{i\ell}\, \bigl[\bar Q_L^i\,H\bigr]\, / \!\!\!\! D\,\bigl[ H^\dagger\, Q^\ell_L\bigr]
+i\beta^{d_R}_{i\ell}
\, \bigl[\bar d_R^i\,H^\dagger\bigr]\, / \!\!\!\! D\,\bigl[ H\, d^\ell_R\bigr]
\\
&\supset \beta^{d_L}_{i\ell}\frac{v^2}{4}\frac{g}{c_w}\bar d_L^i \gamma^\mu Z_\mu\,d_L^\ell
-\beta^{d_R}_{i\ell}\frac{v^2}{4}\frac{g}{c_w}\bar d_R^i \gamma^\mu Z_\mu\,d_R^\ell
\, ,
\end{split}
\label{Zbb2}
\end{equation}
where in the second line we have also used the Dirac equation. 
Using the results of App.~\ref{app:fermions} we can express the couplings $\beta$ as
\begin{align}
\beta^{d_L}_{i\ell}&=
\sum_jY_{ij}^{d}Y_{\ell j}^{d*}
\int_0^{y_1}
e^{2A}
\,
(\Omega'_{d_R^j})^{-1}
(\Gamma^d_{\ell j}-\Omega_{d^j_R})(\Gamma^d_{ij}-\Omega_{d_R^j})\,,
\label{betaINI}
\\
\beta^{d_R}_{i\ell}&=
\sum_jY_{ji}^{d^*}Y_{j\ell}^{d}
\int_0^{y_1}
e^{2A}
\,
(\Omega'_{d_L^j})^{-1}
(\Gamma^d_{j\ell}-\Omega_{d^j_L})(\Gamma^d_{ji}-\Omega_{d_L^j})\,,
\label{betaFIN}
\end{align}
where we have defined
\be
\Gamma^d_{ij}(y)=\frac{\displaystyle \int_0^{y}h\,e^{-Q_{Q_L^i}-Q_{d_R^j}}}{\displaystyle \int_0^{y_1}h\,e^{-Q_{Q_L^i}-Q_{d_R^j}}}\,,
\ee
which is the cumulative distribution of the physical 4D down-type Yukawa coupling along the fifth dimension. Thus, $\Gamma$ monotonically increases from zero to one; if any of the three fields (Higgs, $Q_L^i$ or $d_R^j$) is UV (IR) localized we can take the limit $\Gamma^d_{ij}\to 1$ ($\Gamma^d_{ij}\to 0$).\footnote{Using this simple limit we have checked that we obtain the same result as quoted in Ref.~\cite{Casagrande:2008hr} for an IR-brane localized Higgs.}

The non-oblique Lagrangian (\ref{Zbb2}) will contribute with a significant amount to the $Z\bar bb$ coupling, as we will describe in detail in Sec.~\ref{sec:EWPOfermions}.

%%%%%%%%
%%%%%%%%
%%%%%%%%

\section{Electroweak Precision Observables}
\label{sec:EWPOfermions}

In this section we will consider the relevant pieces of the effective Lagrangians we just obtained after integrating bulk KK modes. The general procedure to evaluate the effect of oblique and non-oblique EWPOs in the presence of New Physics is described in detail in Ref.~\cite{Burgess:1993vc}, and we will just present here some final results for a general warped 5D model with a bulk Higgs.

\subsection{Oblique corrections}
Let us start by considering the oblique corrections, described in terms of the $(T,S,W,Y)$ parameters. These can be obtained from Eq.~(\ref{oblique}), the oblique part of the effective Lagrangian that results from integrating the KK weak gauge bosons. Let us rewrite the relevant piece of this Lagrangian as\footnote{The ellipsis denotes operators not relevant to the oblique precision observables.}
\begin{align}
\begin{split}
\mathcal L_{\rm }=\mathcal L_{\rm SM}
&+\frac{g_5'^2}{2} \hat{\alpha}_{hh} |H^\dagger D_\mu H|^2
- g_5g_5' \hat{\alpha}_{hg} H^\dagger W_{\mu\nu}HB^{\mu\nu}
\\
&+\hat{\alpha}_{gg} \frac{ y_1}{2}(\partial_\rho B_{\mu\nu})^2
+\hat{\alpha}_{gg} \frac{ y_1}{2} (D_\rho W_{\mu\nu})^2+\dots
\,,
\end{split}
\end{align}
from which we can read the precision observables\footnote{Besides the equality $Y=W$ one can also derive another relation $(\alpha S)^2\leq x (\alpha T) Y$ with $x=64 c_w^2s_w^2\approx 11.2$. This follows from Eqs.~(\ref{eq:Texpr2}--\ref{eq:Wexpr2}) by use of the Cauchy-Schwarz inequality.} 
\begin{align}
\alpha T&=s_W^2m_Z^2 \hat{\alpha}_{hh}\,,
\label{eq:Texpr2}
\\
\alpha S&=8s_W^2c_W^2 m_Z^2 \hat{\alpha}_{hg} \,,
\\
Y&=c_W^2m_Z^{2} \hat{\alpha}_{gg}
 \,,
 \\
W&=Y\,.
\label{eq:Wexpr2}
\end{align}
Note how this result coincides with the one obtained in Sec.~\ref{STholo}, where we used a holographic method instead.

\subsection{Non-oblique corrections}
%Following Ref.~\cite{Burgess:1993vc}, we diagonalize and canonically normalize the kinetic terms of the gauge bosons, and afterwards express the SM parameters $\tilde{e}$, $\tilde{s}_w$ and $\tilde{m}_Z$ that appear in the Lagrangian in terms of the physically measured ones ($e$, $s_w$ and $m_Z$). 
% TODO: UNDERSTAND & WRITE DETAILS!?!?!?!?!

Let us now focus on the non-oblique operators. In particular, the first term of the non-oblique Lagrangian (\ref{non-oblique1}) contributes to modified $Z$ and $W$ couplings to fermions. The same holds true for the operator given in Eq.~(\ref{Zbb2}). We can then extract the contributions to the vertex corrections by going to the electroweak vacuum.\footnote{
Notice, in particular that from Eq.~(\ref{eom}) one has
$
\Box A_\mu^A = -J_{h\,\mu}^A+\dots.
$
Hence, after EWSB, $ J_h=-\mathcal M^2 A$ where $\mathcal M^2$ is the gauge boson mass matrix and one can directly evaluate the product with the fermion currents.}

Diagonalizing the physical Yukawa couplings with the rotation matrices $V_{d_\chi}$, the $Z\bar qq$ couplings receive the corrections 
\begin{align}
\delta g_{q^{1,2}_{L,R}}&=\frac{g^{SM}_{q_{L,R}}}{2} \left(\alpha T+\frac{Y}{c_w^2}\right)
-Q^{em}_{q}\,\frac{1}{c_w^2-s_w^2}\left(\frac{\alpha S}{4}-c_w^2 s_w^2\,\alpha T-s_w^2\,Y\right)\,,%\quad (q=u,d)
\\
\delta g_{b_{L,R}}&=%\delta g_{d_{L,R}}
\frac{g^{SM}_{d_{L,R}}}{2} \left(\alpha T+\frac{Y}{c_w^2}\right)
+\frac{1}{3}\,\frac{1}{c_w^2-s_w^2}\left(\frac{\alpha S}{4}-c_w^2 s_w^2\,\alpha T-s_w^2\,Y\right)
+\delta\tilde g_{b_{L,R}}\,,
\\
\delta\tilde g_{b_L}&\equiv \delta\tilde g^{33}_{d_L}=\left(-g^{SM}_{d_{L}}\, m_Z^2\, 
\hat\alpha_{h,d^i_{L}} \delta_{i\ell}+\frac{v^2}{4}\beta^{d_L}_{i\ell}\right)
V^{3i}_{d_L}V_{d_L}^{*3 \ell}\,,
\\
\delta \tilde g_{b_R}&\equiv \delta\tilde g^{33}_{d_R}=\left(-g^{SM}_{d_{R}}\, m_Z^2\, 
\hat\alpha_{h,d^i_{R}} \delta_{i\ell}-\frac{v^2}{4}\beta^{d_R}_{i\ell}\right)
V^{3i}_{d_R}V_{d_R}^{*3 \ell}\,,
\label{deltagoverg}
\end{align}
where 
\begin{equation}
g_{q_L}^{SM}=T^3_{q}-Q^{em}_q\,s_w^2 \,, ~~~ g^{SM}_{q_R}=-Q^{em}_qs_w^2 \,,
\end{equation}
and the integrals $\hat\alpha_{h\psi}$ and $\beta_{ij}^\psi$ have been given in Eqs.~(\ref{alphahat}) and (\ref{betaINI}--\ref{betaFIN}). The tilde here denotes an explicit vertex correction coming from the non-oblique operators.

The most relevant observable that we can extract from these results is the deviation of the $Z b \bar{b}$ coupling from the SM result. In particular, we should compute the effect of the non universal $Z b \bar{b}$ coupling to the partial width
\be
R_b=\frac{\Gamma(Z\to \bar b b)}{\Gamma(Z\to \bar q q)} \,.
\ee
For that matter, we write
\be
R_b=R_b^{SM}+\left.\left(\sum_{q\neq t}\frac{\partial R_b^{tree}}{\partial g_{q_\chi}}\delta g_{q_\chi}\right)\right|_{g_{q_\chi}^{SM}}\,,
\ee
where 
\be
R_b^{tree}=\frac{g_{b_L}^2+g_{b_R}^2}{\sum_{q\neq t}(g_{q_L}^2+g_{q_R}^2)}\,,\qquad
R_b^{SM}=0.21578\,.
\ee
Only the left handed bottom has both oblique and non-oblique corrections, while all other couplings only have corrections coming from the oblique parameters, see Eq.~(\ref{deltagoverg}). This should be compared to the experimental value \cite{Nakamura:2010zzi}
\be
R_b^{exp}=0.21629\pm 0.00066\,,
\label{expRb}
\ee
which will translate into a lower bound on the KK scale.

%%%%%%%%%%%%%%%%%%%%%%%%%%%%%%%%%%%%%%%%%%%%%%%%%%%%%%%%%%%%%%%%%%%%%%
%%%%%%%%%%%%%%%%%%%%%%%%%%%%%%%%%%%%%%%%%%%%%%%%%%%%%%%%%%%%%%%%%%%%%%

\section{Flavor Violation}
\label{sec:flavorfermions}

The dominant flavor violation comes from the KK gluons, in particular the off-diagonal elements in Eq.~(\ref{gKK}). Additionally, the second term of the non-oblique Lagrangian (\ref{non-oblique1}) generates flavor violating four-fermion operators. However, the corresponding effects will be subleading with respect to those induced by the KK gluons. The details about the four-fermion operators are included in App.~\ref{4f}.

Following standard convention~\cite{Bona:2007vi,Isidori:2010kg} we parametrize the most constraining $\Delta F=2$ Lagrangian as\footnote{The minus signs in the first two operators reflect our convention for the metric, $\eta^{\mu\nu}=\diag(-+++)$.}
\bea
- \mathcal L_{sd}^{\Delta F=2}=
\mathcal H_{sd}^{\Delta F=2}&=&
-C_1^{sd} (\bar d_L\gamma^\mu s_L)^2-\tilde C_1^{sd}(\bar d_R\gamma^\mu s_R)^2\nn\\
  &&+\, C_4^{sd}(\bar d_L s_R)(\bar d_R s_L)
  +C_5^{sd}(\bar d_L^\alpha s_R^\beta)(\bar d_R^\beta s_L^\alpha)\,.
  \label{Wilson}
\eea
In full analogy we can write similar operators by replacing $sd\to uc$ or $bd$.
We can use the results of Sec.~\ref{KKgluons} to write the coefficients explicitly as
\bea
C_1^{sd}&=&\frac{g_s^2\, y_1}{6}\int e^{2A}(\Omega_{d_L}^{12})^2\label{C1}\,,\\
\tilde C_1^{sd}&=&\frac{g_s^2\, y_1}{6}\int e^{2A}(\Omega_{d_R}^{12})^2\label{C1tilde}\,,\\
C_4^{sd}&=&-g_s^2\, y_1\int e^{2A}(\Omega_{d_L}^{12}\Omega_{d_R}^{12})
\label{C4}\,,\\
C_5^{sd}&=&\frac{g_s^2\, y_1}{3}\int e^{2A}(\Omega_{d_L}^{12}\Omega_{d_R}^{12})\,,\label{C5}
\eea
where $\Omega^{12}_{d_\chi}$ has been defined in Eq.~(\ref{Omega}).
Notice that using the unitarity of the mixing matrices we can write
\be
\Omega_{d_L}^{12}=
  (\Omega_{d_L}^2-\Omega_{d_L}^1)V_{d_L}^{12}V_{d_L}^{*22}
 +(\Omega_{d_L}^3-\Omega_{d_L}^1)V_{d_L}^{13}V_{d_L}^{*23}\,,
\ee
and similarly for $L\to R$. 

The coefficients $C_i$ are related to flavor violating and/or $CP$ violating observables~\cite{Isidori:2010kg} from where they get upper bounds. Those will translate into lower bounds on the KK scale of the model, as we will see shortly.

%%%%%%%%%%%%%%%%%%%%%%%%%%%%%%%%%%%%%%%%%%%%%%%%%%%%%%%%%%%%%%%%%%%%%%%%
%%%%%%%%%%%%%%%%%%%%%%%%%%%%%%%%%%%%%%%%%%%%%%%%%%%%%%%%%%%%%%%%%%%%%%%%
%%%%%%%%%%%%%%%%%%%%%%%%%%%%%%%%%%%%%%%%%%%%%%%%%%%%%%%%%%%%%%%%%%%%%%%%

\section{Quark masses and mixing angles}
\label{quarks}

In this section we will introduce explicit quark zero mode profiles and fit the parameters in the 5D Lagrangian to the observed quark masses, mixing angles and $CP$-violating phase. 

First of all, we need to choose an explicit mass term for the fermions or. equivalently, a function $Q_\psi (y)$ as defined in Eq.~(\ref{eq:Qpsi}). We will make the choice 
\begin{equation}
Q_\psi(y)=c_\psi A(y)\,,
\label{choice}
\end{equation}
which coincides with that used in RS models where $Q^{RS}_\psi=c_\psi ky$.\footnote{Of course one could also use for a general model $Q_\psi=c_\psi ky$. See e.g.~Ref.~\cite{Carmona:2011ib}.} The mass term that leads to our choice \eqref{choice} can be achieved if the stabilizing field $\phi$ couples to the fermions. In particular, if we use the superpotential method (see Sec.~\ref{sec:model}) we can achieve (\ref{choice}) by choosing $M_\psi(\phi)=\mp c_\psi W(\phi)$.\footnote{The upper sign refers to 5D fermions with left-handed zero modes and the lower sign to those with right-handed zero modes.} In the end, this choice leads to fermion zero modes given by
\begin{equation}
\psi(y) = e^{(2 - c_\psi)A(y)} \,.
\end{equation}

In this case, the function $U_\psi (y)$, defined in Eq.~\eqref{Omegafermion}, simply reads
\begin{equation}
U_\psi(y)=\int_0^y \exp\left[(1-2c_\psi)A(y')\right]\,.
\end{equation}
The relation between 4D and 5D Yukawa couplings given in Eq.~\eqref{eq.Y4DY5D} is now 
\be
Y_{ij}^q=\hat Y_{ij}^q \ F(c_{Q^i_L},c_{q^j_R})\,,
\ee
where the function $F$ is defined as
\begin{equation}
F(c_L,c_R)=\frac{\displaystyle \int_0^{y_1} h\,e^{-(c_{L}+c_{R})A}}
  {\displaystyle \left[
    \int_0^{y_1} e^{-2A}h^2\int_0^{y_1} e^{ (1-2 c_{L})   A} \int_0^{y_1} e^{(1-2 c_{R})A}
  \right]^{\frac{1}{2}}}\ .
\end{equation}
We note the following properties of the fermion profiles and the function $F$:
\begin{itemize}
\item
For any strength of the metric deformation, fermions $\psi$ are IR (UV) localized for $c_\psi<\frac{1}{2}$ ($c_\psi>\frac{1}{2}$). This is thus the same situation as in the RS model. Notice also that this choice of profile corresponds, in the dual theory, to the special case of constant anomalous dimension, i.e.~the fermionic operators are not significantly disturbed by the presence of the CFT deformation.
\item
The larger the deformation of the AdS background, the larger is the portion of the parameter space $(c_L, c_R)$ where the function $F(c_L,c_R)$ is enhanced. Consequently,
 the coefficients $c_\psi$  can be slightly larger for the same 5D Yukawa coupling in order to reproduce the same (physical) 4D Yukawa coupling. This in turn leads to a weaker coupling of the fermions to the KK modes of the gauge fields. 

 Alternatively, for fixed values of the coefficients $c_\psi$ and 4D Yukawa couplings, the 5D Yukawa couplings (and correspondingly the couplings of fermion KK-modes to the fermion and Higgs zero modes in Fig.~\ref{KKfermion}) are decreased with respect to their values in the AdS background, leading to a softening of the bounds on the value of $m_{KK}$, as we will see in Sec.~\ref{sec:boundsEWPOfermions}. 

%\item
%The function $F$ develops a maximum at a value $c_L=c_R\lesssim 1/2$ for large deformations \cite{Carmona:2011ib} due to the fact that the Higgs profile $h(y)e^{-A(y)}$ develops a maximum near the IR brane, and the overlap decreases again for very IR localized fermions.
\end{itemize}
This enhancement of the function $F(c_L,c_R)$ for a background with a large AdS deformation is illustrated in Fig.~\ref{yukawaenhancement} for the model (\ref{ourmetric}) with $k\Delta=1$ and $\nu=0.5$.
\begin{figure}[t]
\begin{center}
\begin{psfrags}
\input{figs/F-psfrag.tex}
\includegraphics[width=0.5\textwidth]{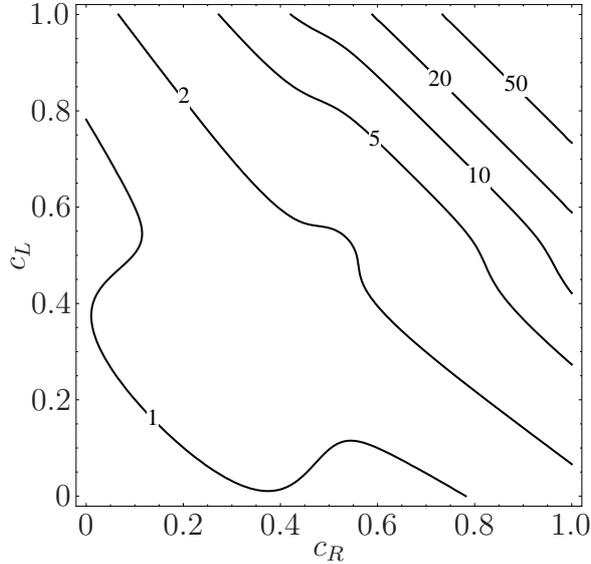}
\end{psfrags}
\end{center}
\caption{\it The function $F(c_L,c_R)$ for $k\Delta=1,\ \nu=0.5$ normalized to the corresponding RS value. One can see that for a wide portion of the parameter space it leads to an enhancement of the 4D Yukawa couplings with respect to RS. %{\bf eventually we will make a nicer plot....}
}
\label{yukawaenhancement}
\end{figure}

Let us now consider the quark mass-squared matrices
\be
\M^{q_L}=\frac{v^2}{2}\,  Y^{q} Y^{q\,\dagger},\qquad q=u,d
\ee
where $v=246$ GeV is the VEV of the Higgs field. Unitary matrices should be introduced to diagonalize the matrices $\M^{q_L}$ as
\be
\M_{diag}^q=V^{q_L}\,\M^{q_L}\,V^{q_L\,\dagger}\,.
\label{rotaciones}
\ee
Next let us write expressions for the eigenvalues and mixing angles of the hierarchical Yukawa couplings. In the following we will just make two reasonable assumptions:
\begin{itemize}
\item
First we will assume a left-handed hierarchy, i.e.
\be
Y^q_{1i}\ll Y^q_{2i}\ll Y^q_{3i} \,.
\label{LHhierarchy}
\ee
This will be the case whenever there is a mild hierarchy between the left-handed $c_\psi$, i.e.~$c_{Q_L^1}>c_{Q_L^2}>c_{Q_L^3}$.\footnote{We will comment later on about possible further simplifications that take place in case there is also a right handed hierarchy.}
\item
The only second assumption we are making is that the left-handed rotations show a similar hierarchy as the CKM matrix
\be
|V^{q_L}_{12}|\sim \epsilon\,,\quad |V^{q_L}_{23}|\sim \epsilon^2\,,\quad |V^{q_L}_{13}|\sim \epsilon^3\,,
\label{rotations}
\ee
where $\epsilon$ is of the order of the Cabbibo angle. 
This assumption is natural since otherwise the smallness of the CKM mixing angles would be a consequence of large cancellations. As it turns out it is also consistent with the assumption Eq.~(\ref{LHhierarchy}).
\end{itemize}

Owing to our assumption (\ref{rotations}), these unitary rotations can be given in a Wolfenstein-like parameterization as~\cite{Nakamura:2010zzi}
\be
V^{q_L}=\left(
\begin{array}{ccc}
1-\frac{1}{2}|V^{q_L}_{12}|^2&V^{q_L}_{12}&V^{q_L}_{13}\\
-V^{{q_L}*}_{12}&1-\frac{1}{2}|V^{q_L}_{12}|^2&V^{q_L}_{23}\\
(-V^{{q_L}}_{13}+V^{q_L}_{12}V^{q_L}_{23})^*&-V_{23}^{{q_L}*}&1
\end{array}
\right)\,,\qquad q=u,d
\label{matrices}
\ee
where terms of order $\mathcal O(\epsilon^4)$ have been neglected. The matrix $V^{q_L}$ contains three independent complex parameters and it is unitary to the considered order. The angles and eigenvalues are best expressed in terms of the quantities
\be
\widetilde \M^{q_L}_{ij}=\M^{q_L}_{ij}-\frac{\M^{q_L}_{i3}\M^{q_L}_{3j}}{\M_{33}}\,.
\ee
First, by demanding the off-diagonal terms in Eq.~(\ref{rotaciones}) to vanish we obtain the mixing angles
\begin{alignat}{2}
&\displaystyle V^{q_L}_{12} =-\frac{\widetilde \M^{q_L}_{12}}{\widetilde \M^{q_L}_{22}}\,, 
&\displaystyle V^{q_L}_{21} =\frac{\widetilde \M^{q_L}_{21}}{\widetilde \M^{q_L}_{22}}\,,
\label{angulosINI}
\\
&\displaystyle V^{q_L}_{23} =-\frac{\M^{q_L}_{23}}{\M^{q_L}_{33}}\,,
&\displaystyle V^{q_L}_{32} =\frac{\M^{q_L}_{32}}{\M^{q_L}_{33}}\,,
\\
&\displaystyle V^{q_L}_{13} =-\frac{\M^{q_L}_{13}}{\M^{q_L}_{33}}
+\frac{\widetilde\M^{q_L}_{12}\M^{q_L}_{23}}{\widetilde\M^{q_L}_{22}\M^{q_L}_{33}}\,, \qquad
&\displaystyle V^{q_L}_{31} =\frac{\M^{q_L}_{31}}{\M^{q_L}_{33}}\,,
\label{angulosFIN}
\end{alignat}
The mass eigenvalues are then obtained as:
\begin{align}
(m_3^q)^2&=\M_{33}^{q_L},
\label{masascompINI}
\\
(m_2^q)^2&=\widetilde \M_{22}^{q_L}\,,
\\
(m_1^q)^2&=\widetilde\M_{11}^{q_L}-\frac{\widetilde\M_{12}^{q_L}\widetilde\M_{21}^{q_L}}{\widetilde\M^{q_L}_{22}}\,,
\label{masascompFIN}
\end{align}
where we are using the notation $m_3^u=m_t$, $m_3^d=m_b$, and so on.
The comparison with CKM matrix ($V=V^{u_L} V^{d_L\dagger}$) elements leads to the relations
\begin{align}
V_{us}&=%V^{u_L}_{12}-V^{d_L}_{12}
   \hat V_{12}\,,
   \label{CKMcompINI}
   \\
V_{cb}&=%V^{u_L}_{23}-V^{d_L}_{23}
   \hat V_{23}\,,
   \\
V_{ub}&=%[-(V^{u_L}_{31}-V^{d_L}_{31})+V_{21}^{u_L}\,( V^{u_L}_{32}-V^{d_L}_{32})]^*
   (-\hat V_{31}+V_{21}^{u_L}\hat V_{32})^*\,,
   \\
%        &=&|(V^{d_L}_{13}-V^{u_L}_{13})-V_{23}^{d_L}\,( V^{d_L}_{12}-V^{u_L}_{12})|\nonumber\\
V_{td}&=%(V^{u_L}_{31}-V^{d_L}_{31})-V_{21}^{d_L}\,( V^{u_L}_{32}-V^{d_L}_{32})
   \hat V_{31}-V_{21}^{d_L}\hat V_{32}\,,
\label{CKMcompFIN}
\end{align}
where $\hat V=V^{u_L}-V^{d_L}$.
The CKM matrix defined this way does not obey the usual phase convention~\cite{Nakamura:2010zzi}. One can easily obtain the standard convention  by multiplying $V^{q_L}$ from the left with appropriate phases. One finds
\be
e^{i\delta}=\frac{V^*_{ub}V_{us} V_{cb}}{|V_{ub}V_{us}V_{cb}|}\,.
\ee
Alternatively we can write the Jarlskog invariant as
\begin{equation}
J=\operatorname{Im}\,( V^*_{ub}V^*_{sc}V_{us} V_{cb}) =-\operatorname{Im}\,(\hat V_{12}\hat V_{23}\hat V_{31})+|V_{cb}|^2
    \operatorname{Im}\,(V_{12}^{d_L}V^{u_L}_{21})\,.
   \label{Jarlskog}
\end{equation}

To summarize, given the complex 5D Yukawa couplings (two $3\times 3$ matrices with complex coefficients), there are nine constants $c_{Q^i_L},c_{u^i_R},c_{d^i_R}$ which should be adjusted to satisfy the mass relations (\ref{masascompINI}--\ref{masascompFIN}) and the experimental CKM matrix relations (\ref{CKMcompINI}--\ref{CKMcompFIN}) and (\ref{Jarlskog}). Note that, for a given model, not every set of 5D Yukawa couplings will admit a solution, which will in turn constrain our possible initial Yukawa couplings. In the following chapter we will apply this to two particular models: RS and the model \ref{ourmetric}.

\section[Application to RS and the non-custodial model]{Application to Randall-Sundrum and the non-custodial model}
\label{sec:applicationflavor}

Let us now apply the results we just found to the Randall-Sundrum model and to the non-custodial model \ref{ourmetric} described in Chap.~\ref{chap:noncustodial}. For that matter, we will first generate a set of solutions to a random distribution of 5D Yukawa couplings, and then we will compute the bounds imposed to each of the solutions by the Flavor and EWPO constraints. This way, we will be able study statistically the behavior of both models, which is arguably the most rigorous way to compare them.

\subsection{Generating a set of solutions}

We will start by considering a random sample of 5D Yukawa couplings (two complex $3\times 3$ matrices, $\hat Y^u_{ij}$ and $\hat Y^d_{ij}$). For each point of the sample, we will perform a $\chi^2$ fit to the experimental quark masses (measured at the KK scale~\cite{Csaki:2008zd}) from which we obtain the nine coefficients $c_\psi$ that minimize the $\chi^2$ function, using the results of Sec.~\ref{quarks}. We will accept those points that yield $\chi^2 \lesssim 4$ to \emph{both} the RS model and the model \eqref{ourmetric} with $\nu=0.5$ and $k\Delta = 1$.  For each set of Yukawas, this will give rise to two sets of the $c_\psi$ coefficients, one for RS and other for model \eqref{ourmetric}. This facilitates a direct comparison between the two models, since the 4D effective theories originate from the same set of 5D Yukawa couplings.

\begin{figure}[p!]
\centering
\input{figs/cPDF-psfrag.tex}
\includegraphics[width=0.8\textwidth]{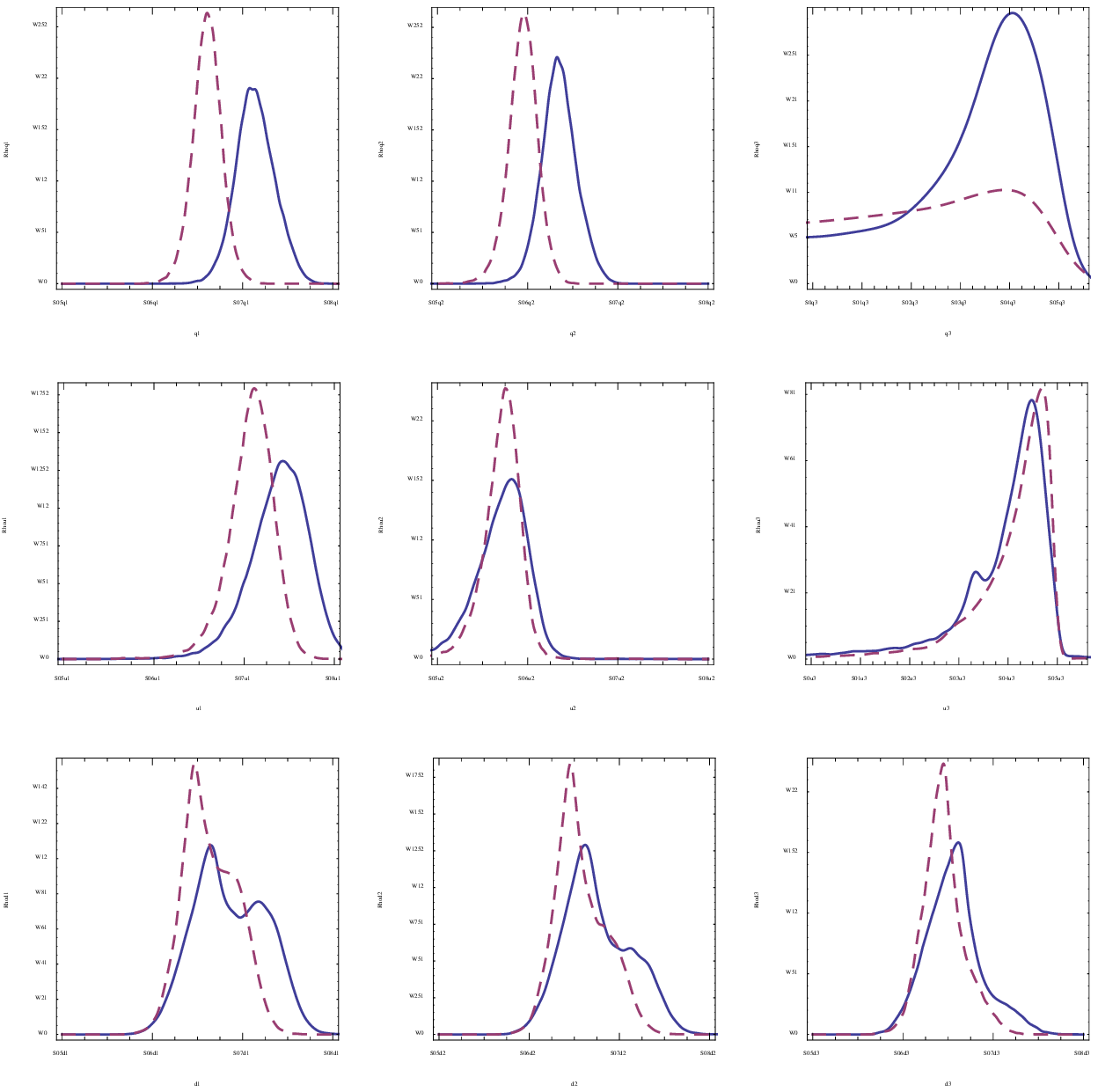}
\caption{\it The distribution of the $c$ parameters for the different %species of 
quarks and chiralities, for RS [dashed] and for model (\ref{ourmetric}) with $k\Delta=1$ and $\nu=0.5$ [solid]. Only the parameters $c_{(t,b)_L}$ and $c_{t_R}$ are IR localized ($c<0.5$). Notice the highly asymmetric forms of the corresponding distributions. } 
\label{figc}
\end{figure}
\begin{table}[p!]
\centering
\begin{tabular}{rl}
%%%
RS: 
&
\begin{tabular}{| l | l | l |}
\hline
$c_{(u,d)L} =0.66 \pm 0.02$ & $c_{(c,s)L} = 0.59 \pm 0.02  $ & $c_{(t,b)L} =  -0.11^{+0.45}_{-0.53} $
\\ \hline 
$c_{uR} = 0.71 \pm 0.02$ & $c_{cR} = 0.57 \pm 0.02$ & $c_{tR} =  0.42^{+0.05}_{-0.17}$%{{\bf*}}
\\ \hline
$c_{dR} = 0.66 \pm 0.03$ & $c_{sR} = 0.65 \pm 0.03 $ & $c_{bR} = 0.64 \pm 0.02$ 
\\ \hline
\end{tabular}
\\
\ & \ 
\\
 $%k\Delta=1,\,
 \nu=0.5$:
 &
\begin{tabular}{| l | l | l |}
\hline
$c_{(u,d)L} = 0.71 \pm 0.02 $ & $c_{(c,s)L} =  0.63 \pm 0.02$ & $c_{(t,b)L} =  0.31^{+0.13}_{-0.52}\phantom{-}$%{{\bf*}}
\\ \hline
$c_{uR} =  0.74 \pm 0.03$ & $c_{cR} =  0.57 \pm 0.03$ & $c_{tR} =  0.42^{+0.05}_{-0.11}$%{{\bf*}}
\\ \hline
$c_{dR} = 0.68 \pm 0.04 $ & $c_{sR} = 0.67 \pm 0.04$ & $c_{bR} =  0.66 \pm 0.03$
\\ \hline 
\end{tabular}
%
%%%
\end{tabular}

~

\caption{\it Medians and $1\sigma$ confidence intervals of the $c$ parameters corresponding to the different species of quarks and chiralities, for RS and for model (\ref{ourmetric}) with $k\Delta=1,\,\nu=0.5$. }
\label{tabc}
\end{table}

From these premises, we generated a sample of 40,000 points, having chosen for the 5D Yukawas flat prior distributions $1\leq |\sqrt{k}\hat Y^q_{ij}|\leq 4$ and $0\leq \arg(\hat Y^q_{ij})<2\pi$. The results of the fit are presented in Fig.~\ref{figc}, while the corresponding central values and $1\sigma$ confidence intervals of the $c_\psi$ parameters are listed in Tab.~\ref{tabc}.
As it is clear from the individual plots, the  $c_{\psi}$ are slightly larger for our model than in the RS model, as anticipated above.\footnote{We have restricted ourselves to the region $c_\psi>-1$ in order to avoid strongly IR localized fermions, which typically have stricter perturbativity bounds for the Yukawa couplings. }
This means that the couplings of the individual quarks to KK modes are more suppressed than in RS.\footnote{We have checked that for fixed $c$, the individual couplings of KK gauge bosons to fermion zero modes are of the same order for both models.}

An interesting fact that we find is that the $c_{d^i_R}$ are very much non-hierarchical. In fact only about 30\% of all points show the ``traditional" hierarchy $c_{d_R^1}>c_{d_R^2}>c_{d_R^3}$. 
As expected, our expressions (\ref{angulosINI}--\ref{angulosFIN}) and (\ref{masascompINI}--\ref{masascompFIN}) are much better approximations to the true angles and eigenvalues in these cases than the ones usually employed in the literature~\cite{Hall:1993ni}. Note that, in practice, we never need to have explicit expressions for the right-handed angles in terms of the Yukawa matrices, as the former do not enter in the fit.\footnote{For each data point obtained in the fit it is of course a simple matter to numerically find the right handed rotations.} 
On the other hand, the up-type sector will always be hierarchical, $c_{u_R^1}>c_{u_R^2}>c_{u_R^3}$ or 
$Y^u_{i1}\ll Y^u_{i2}\ll Y^u_{i3}$ equivalently, and we could have used the simpler expressions for the eigenvalues and angles described in App.~\ref{RH}. 
%
% TODO: Introduce App.~\ref{RH} earlier!!!!!!
%

%%%%%%%%%
%%%%%%%%%

\subsection{Bounds from electroweak precision observables}
\label{sec:boundsEWPOfermions}

Let us now find the bounds on the KK scale that arise from the electroweak precision observables, as described in Sec.~\ref{sec:EWPOfermions}. 

The analysis for oblique observables was already performed in Chap.~\ref{chap:ewsbbulkhiggs} for the RS model and in Chap.~\ref{chap:noncustodial} for model \eqref{ourmetric}. As we did there, we will choose the minimum value of $a$ consistent with solving the Hierarchy problem (see Sec.~\ref{sec:hierarchy}).

As for the non-oblique observables, we can draw on the results shown in Fig.~\ref{figc} to simplify our analysis. Recall that, as mentioned before, we can neglect the diagrams in Fig.~\ref{KKfermion} if the external quarks are near UV localized. Moreover, the contribution from the gauge KK modes is universal for near-UV localized modes and is summarized in the oblique parameters. As shown in Fig.~\ref{figc}, only the left handed top-bottom doublet and the right handed top singlet are near IR localized. We will therefore neglect the explicit correction $\delta \tilde g_{b_R}$. 

\begin{figure}[t]
\begin{center}
\includegraphics[width=0.48\textwidth]{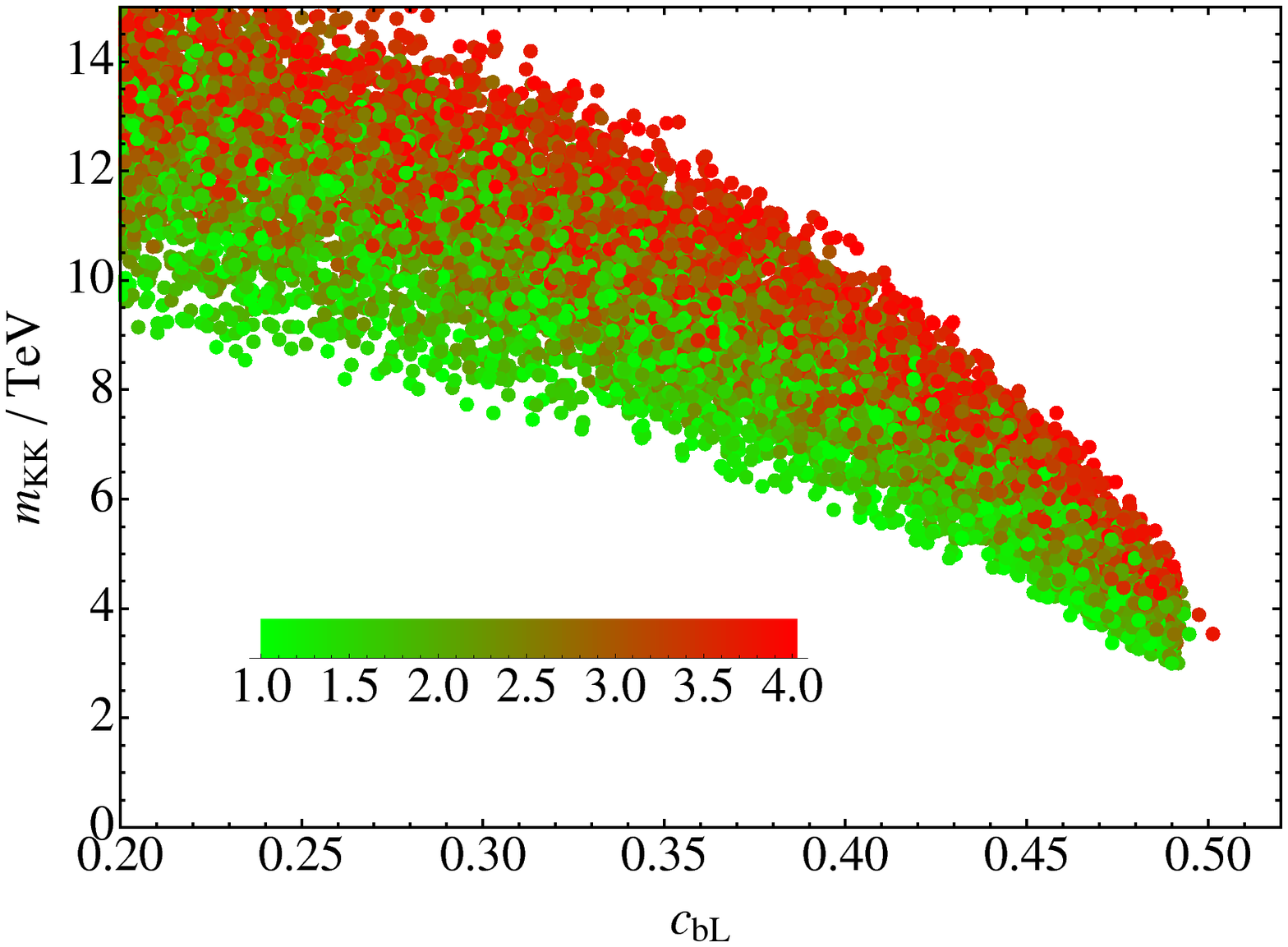}
~
\includegraphics[width=0.48\textwidth]{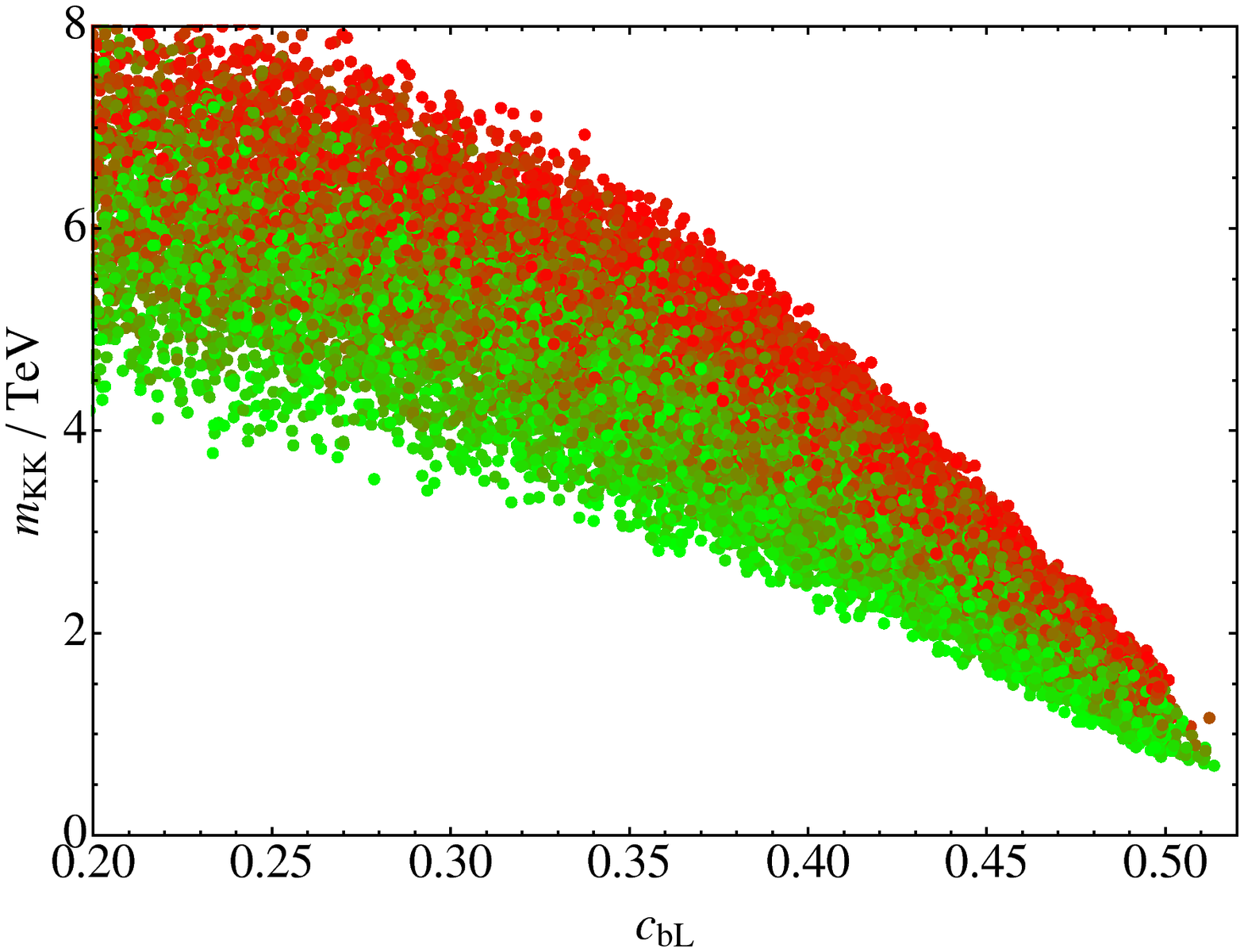}
\end{center}
\caption{\it The bounds (in TeV) implied by the experimental limits on $R_b$, as a function of $c_{Q_L^3}$. Left panel: RS model. Right panel: our model with $k\Delta=1$, $\nu=0.5$. We also display the dependency on the 5D bottom Yukawa:  the coloring interpolates between green (light gray) for $\sqrt{k}\hat Y^d_{33}=1$ to red (dark gray) for $\sqrt{k}\hat Y^d_{33}= 4$.}
\label{figZbb2}
\end{figure}

There are several factors which will influence the size of the non universal $Z\bar bb$ coupling:
\begin{itemize}
\item
The more UV localized the left handed top-bottom doublet, the more suppressed are its coupling to the KK modes of the gauge bosons and to those of the singlet quarks. Hence we expect a suppression of the contribution to $Z\bar b b$ for larger $c_{Q_L^3}$. 
\item
The smaller the 5D bottom Yukawa $\hat Y^d_{33}$, the more suppressed the Yukawa coupling of $b_L$ to the singlet KK modes appearing in the left panel of Fig.~\ref{KKfermion}, and hence the more suppressed is this contribution to $\delta \tilde g_{b_L}$.
\item
As shown in Chap.~\ref{chap:noncustodial}, the Higgs can become more decoupled from the IR in the deformed background, and this reduces both the coupling to KK gauge bosons and KK fermions. 
\end{itemize} 

All effects enumerated above are clearly visible in Fig.~\ref{figZbb2} where we present plots of the minimal KK scale required to sufficiently suppress the observable
 $R_b$ as computed in Eq.~(\ref{deltagoverg}) when compared to the experimental value \eqref{expRb}. In particular, the third effect above reduces the bounds (for fixed $c_{Q_L^3}$ and $\hat Y^d_{33}$) by roughly a factor of 2 when comparing the RS model to the model defined by Eq.~(\ref{ourmetric}) for $k\Delta=1$ and $\nu=0.5$.

Moreover, in Fig.~\ref{figZbb}  we have considered the probability distribution functions (PDF) and cumulative distribution functions (CDF) for the  lower bound on $m_{KK}$. The fact that the model with the modified background generally requires larger $c_{Q^3_L}$ (see Fig.~\ref{figc}) further pushes these distributions to lower KK scales, implying milder bounds.

\begin{figure}[p!]
\begin{center}
\begin{psfrags}
\input{figs/ZbbPDF-psfrag.tex}
\includegraphics[width=0.6\textwidth]{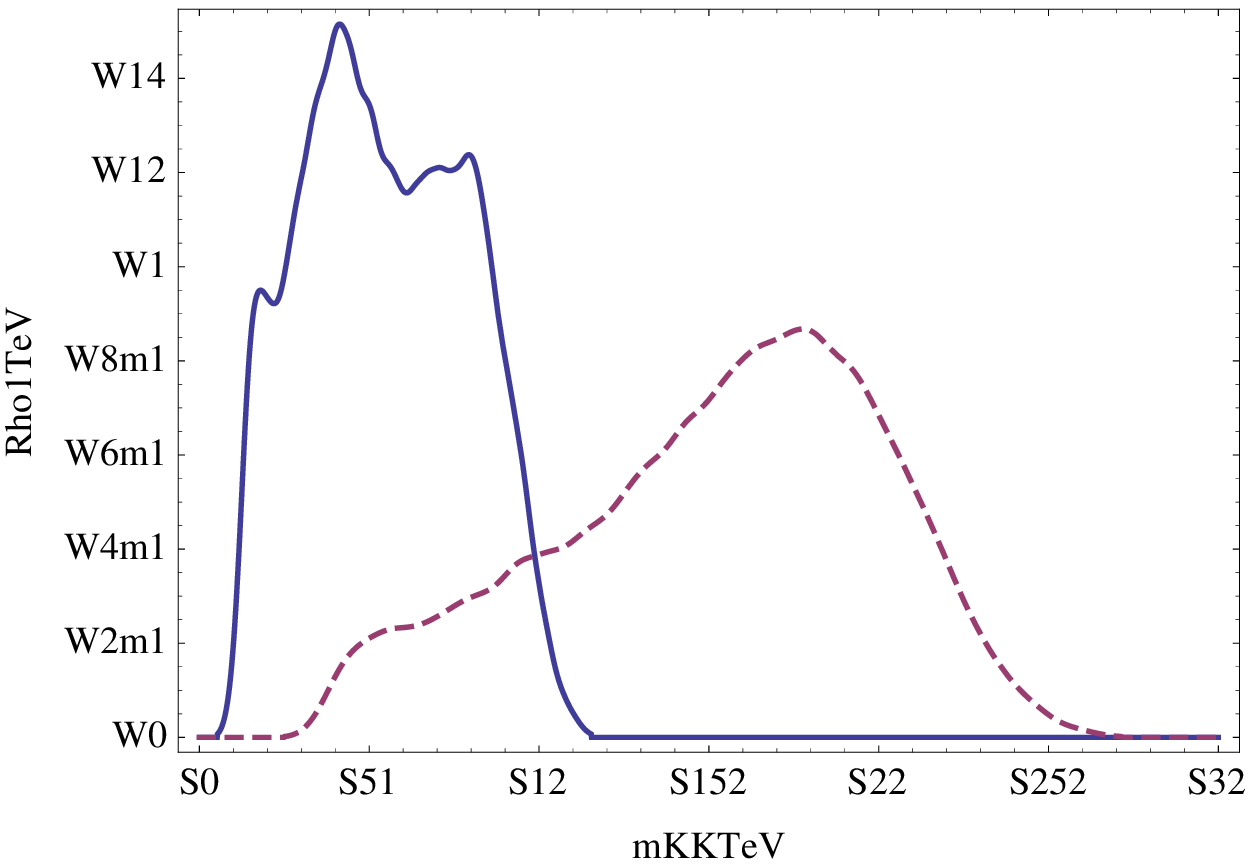}
\end{psfrags}
\\
\vspace{5mm}
\begin{psfrags}
\input{figs/ZbbCDF-psfrag.tex}
\includegraphics[width=0.6\textwidth]{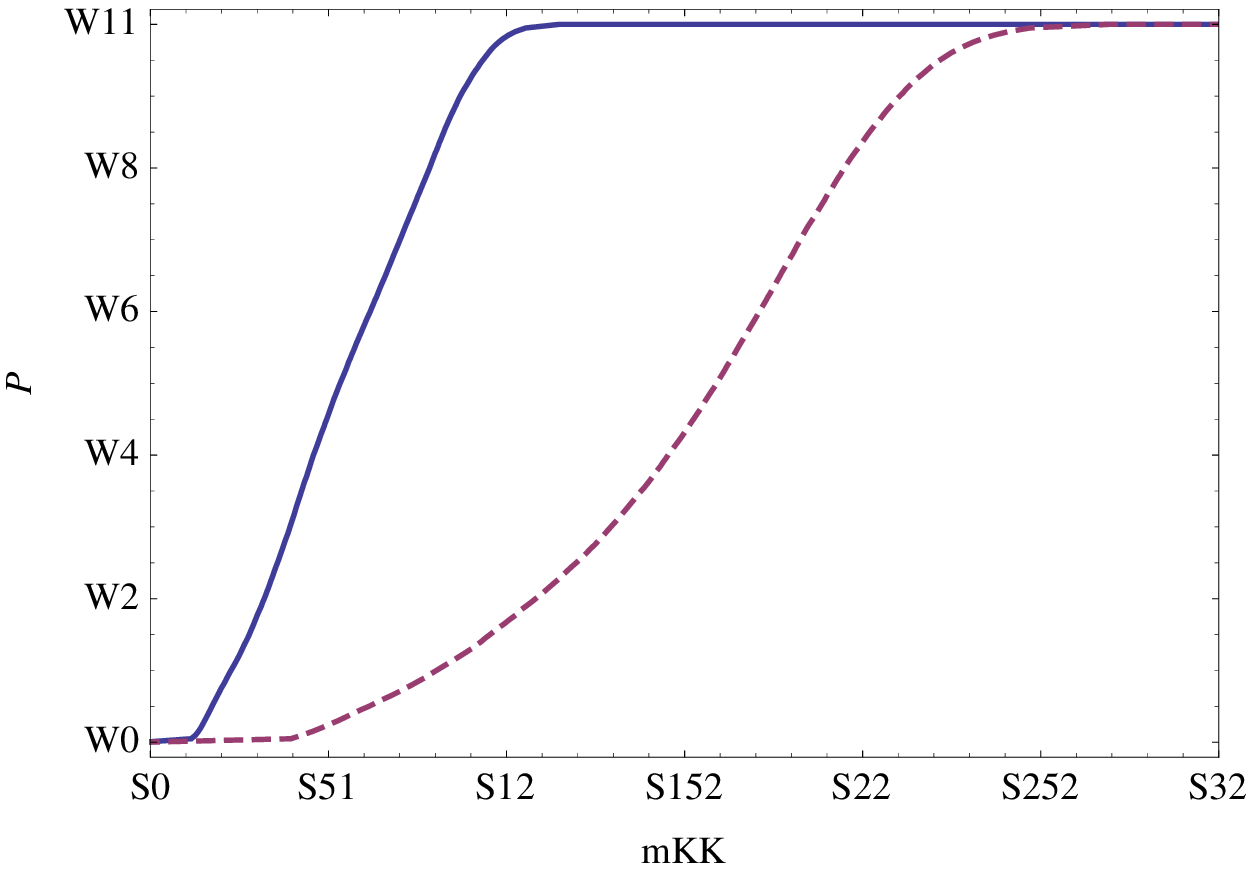}
\end{psfrags}
\end{center}
\caption{\it PDF (upper panel) and CDF (lower panel) for $m_{KK}$ from comparison with $R_b$. Dashed lines correspond to the RS model and solid lines to the model in Eq.~(\ref{ourmetric}) for $k\Delta=1$ and $\nu=0.5$.}
\label{figZbb}
\end{figure}
\begin{table}[p!]
\centering
\begin{tabular}{|c|ccc|ccc|}
\cline{2-7}
\multicolumn{1}{c|}{}&\multicolumn{3}{c|}{\textit{Probability for $m_{KK}$ below}}&\multicolumn{3}{c|}{\textit{Percentile}}\\ %\cline{2-7}
\multicolumn{1}{c|}{}&$ 3$ TeV &$5$ TeV &$10$ TeV&10\% &20\% &50\% \\
\hline
RS &0\% 			&2.4\% &17\% &8.0 TeV&11 TeV& 16 TeV\\
$\nu=0.5$ &18\% 	&46\% &98\% &2.3 TeV&3.2 TeV& 5.3 TeV\\
\hline
\end{tabular}
\caption{\it Left panel: Integrated probability for values of $m_{KK}$ below 3, 5 and 10 TeV from $R_b$ for RS (upper row) and the model in Eq.~(\ref{ourmetric}) for $k\Delta=1$ and $\nu=0.5$ (lower row). Right panel: 10th, 20th and 50th (median) percentiles for both models.}
\label{tabZbb}
\end{table}

For a given KK scale on the horizontal axis one can read off from the CDF (right panel of Fig.~\ref{figZbb}) the fraction of points consistent with such a scale for both models on
the vertical axis. This fraction is thus the probability that the KK scale is smaller than a given value.
Notice that it can also be viewed as the amount of fine tuning necessary to obtain a given bound. Conversely one can start from a given fraction (fine tuning) and read off the percentile on the horizontal axis for both models. 

In Tab.~\ref{tabZbb} we present some explicit numbers obtained from these distributions. As we can see from Tab.~\ref{tabZbb}, getting ``acceptable" bounds depends to a large extent on the amount of fine-tuning that we tolerate. For instance assuming a 20\%  (50\%) fine-tuning the lower bound is 11 TeV (16 TeV) for the RS model and 3.2 TeV (5.3 TeV) for the model with modified background.\footnote{It has previously been noted that one can fine-tune the fermion bulk-masses in RS in order to achieve $R_b$ in agreement with experiment~\cite{Djouadi:2006rk}. Our analysis shows that in the minimal anarchic RS model such a fine-tuning is sizable.}

We have also checked the dependence on the choice of the fermion bulk mass term. In particular Ref.~\cite{Carmona:2011ib} used a constant mass term $Q(y)=c k$. Although our analysis is quite different, we have verified the results in Ref.~\cite{Carmona:2011ib} qualitatively. In particular, for the anarchic case the bounds are slightly higher than the ones for the choice $Q(y)=cA(y)$ and show a stronger dependence on the 5D bottom Yukawa coupling. This indicates that the effect of the KK fermions is dominating for large $\hat Y^d_{33}$, which can easily be mitigated by lowering that coupling at the cost of a mild $\mathcal O(10)$ hierarchy  in the 5D Yukawa couplings \cite{Carmona:2011ib}.

Finally, we should mention that, to be fully consistent, one should consider a global fit of the EWPT data to the observables $S, T$ and $\delta\tilde g_{b_L}$ and also include possible loop corrections~\cite{Carmona:2011ib}. We will leave this to future work.

\subsection{Bounds from flavor and CP violation}

Using our sample of data points we can then compute the exact mixing matrices numerically necessary to find the coefficients $C_i$ defined in Sec.~\ref{sec:flavorfermions}. It turns out that the most constraining parameter is Im $C_4^{sd}$, which is related to the $CP$ violating observable in the $K$-system, $\epsilon_K$, and is bounded by~\cite{Isidori:2010kg}
\be
\left|\textrm{Im}\, C_4^{sd}\right|<2.6\times 10^{-11}\ \textrm{TeV}^{-2}\,.
\label{ImC4}
\ee

By using the expression for $C_4^{sd}$ provided in Eq.~(\ref{C4}) and comparing with the experimental bound in Eq.~(\ref{ImC4}) we obtain bounds on $m_{KK}$ for every data point,\footnote{A more refined procedure would be to link the Wilson coefficients in Eq.~(\ref{Wilson}) to the actual observables, in particular $\epsilon_K$ and $\Delta m_K$, and apply the direct experimental bounds, see e.g.~Ref.~\cite{Archer:2011bk}.} as we did in the previous section for the coupling $R_b$. The result is exhibited in Fig.~\ref{figC4}, where we show both the PDF and CDF for the distribution of points.

\begin{figure}[p!]
\begin{center}
\begin{psfrags}
\input{figs/C4PDF-psfrag.tex}
\includegraphics[width=0.6\textwidth]{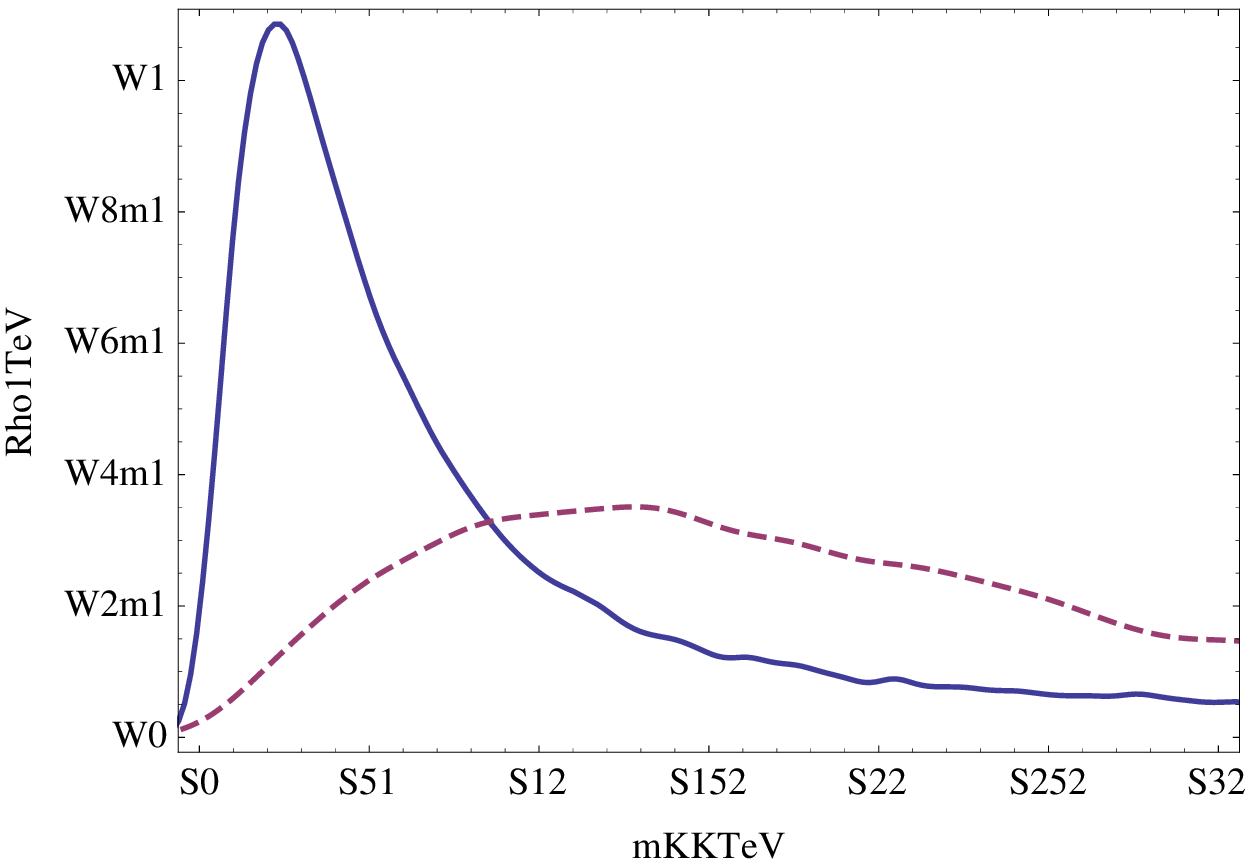}
\end{psfrags}
\\
\vspace{5mm}
\begin{psfrags}
\input{figs/C4CDF-psfrag.tex}
\includegraphics[width=0.6\textwidth]{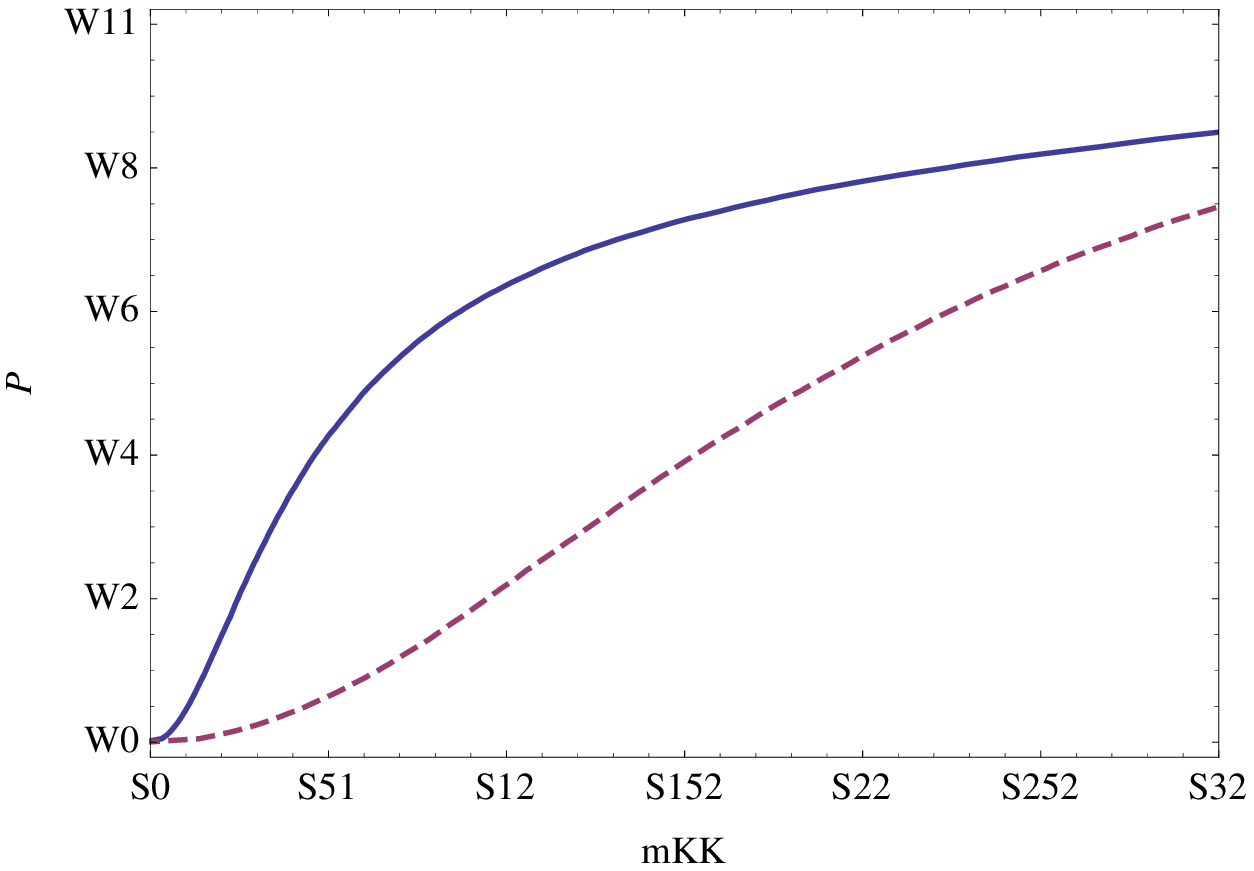}
\end{psfrags}
\end{center}
\caption{\it PDF (upper panel) and CDF (lower panel) for $m_{KK}$ from comparison with $\left|\textrm{Im} \,C_4^{sd}\right|$. Dashed lines correspond to the RS model and solid lines to the model in Eq.~(\ref{ourmetric}) for $k\Delta=1$ and $\nu=0.5$.}
\label{figC4}
\end{figure}
\begin{table}[p!]
\centering
\begin{tabular}{|c|ccc|ccc|}
\cline{2-7}
\multicolumn{1}{c|}{}&\multicolumn{3}{c|}{\textit{Probability for $m_{KK}$ below}}&\multicolumn{3}{c|}{\textit{Percentile}}\\
\multicolumn{1}{c|}{}&$3$ TeV &$5$ TeV &$10$ TeV&10\% &20\% &50\% \\
\hline
RS 			& 2.4\% &6.4\% &22\% & 6.5 TeV & 9.7 TeV & 19 TeV\\
$\nu=0.5$ 	& 26\% &43\% &64\%   & 1.6 TeV & 2.5 TeV & 6.2 TeV\\
\hline
\end{tabular}
\caption{\it Left panel: Integrated probability for values of $m_{KK}$ below 3, 5 and 10 TeV from Im $C_4^{sd}$ for RS (upper row) and the model in Eq.~(\ref{ourmetric}) for $k\Delta=1$ and $\nu=0.5$ (lower row). Right panel: 10th, 20th and 50th percentiles for both models.}
\label{tabC4}
\end{table}

A statistical analysis similar to that done in Sec.~\ref{sec:boundsEWPOfermions} can be performed here,
and in Tab.~\ref{tabC4} we present some explicit numbers obtained from these distributions. 
We can trace back the improvement in the bounds on $m_{KK}$ in the modified background model with respect to the RS model on the weakening of couplings of gauge KK modes to the first and second generation SM fermions, resulting in turn from the enhancement in the coefficients $c_\psi$. 
For instance, assuming a 20\% (50\%) fine-tuning the lower bound for the RS model is 9.7 TeV (19 TeV), while for the modified background model they are 2.5 TeV (6.3 TeV). The combined bounds will be much stronger, as we will shortly see. 

Finally, we should pay attention to the other coefficients: ${\rm Re\ }C^{sd}_4$, $C^{sd}_1$, $\tilde C^{sd}_1$ and $C^{sd}_5$. The bounds on ${\rm Re\ }C^{sd}_4$, coming mostly from $\Delta m_K$, are about one to two orders of magnitude weaker than for ${\rm Im\ }C^{sd}_4$. However, it is conceivable that the favorable points that allow a low KK scale could result from an accidental cancellation of the phase and hence the bounds from the real part turn out to dominate. We have verified that this is not the case and the bounds are not changed by taking into account the real part. Furthermore, notice that $C^{sd}_5=\frac{1}{3}C^{sd}_4$, and hence whenever $C^{sd}_4$ is suppressed so is $C^{sd}_5$ (the experimental constraints on the two quantities are comparable). The experimental constraints on the coefficients $C^{sd}_1$ and $\tilde C^{sd}_1$ are about two orders of magnitude weaker with again a similar suppression as $C_4^{sd}$. We thus do not expect any additional constraints from here either.

\subsection{Combined bounds}
\label{sec:combinedbounds}

Concerning non-oblique versus FCNC and $CP$ violating observables in both models the final comparison is as follows:
\begin{itemize}
\item
The bounds for the modified metric model are milder than those in the RS model. This can be clearly seen from Figs.~\ref{figZbb} and \ref{figC4} and from Tabs.~\ref{tabZbb} and \ref{tabC4}. The main origin of this improvement in the modified metric model with respect to the RS model can be traced back to the fact that because of the IR deformation of the metric fermions fitting the quark mass eigenvalues and CKM matrix elements are shifted towards the UV in the former model which produced a general suppression of effects in the observables. 
\item
The bounds from FCNC and $CP$ violating effective operators are stronger than those from non-oblique observables in both models. This is mainly due to the strong constraints on these operators, in particular from the $CP$ violating observable $\epsilon_K$. Of course we expect  the combined bonds to be stronger than those from the individual constraints.
\end{itemize}
\begin{figure}[p!]
\begin{center}
\begin{psfrags}
\input{figs/BothPDF-psfrag.tex}
\includegraphics[width=0.6\textwidth]{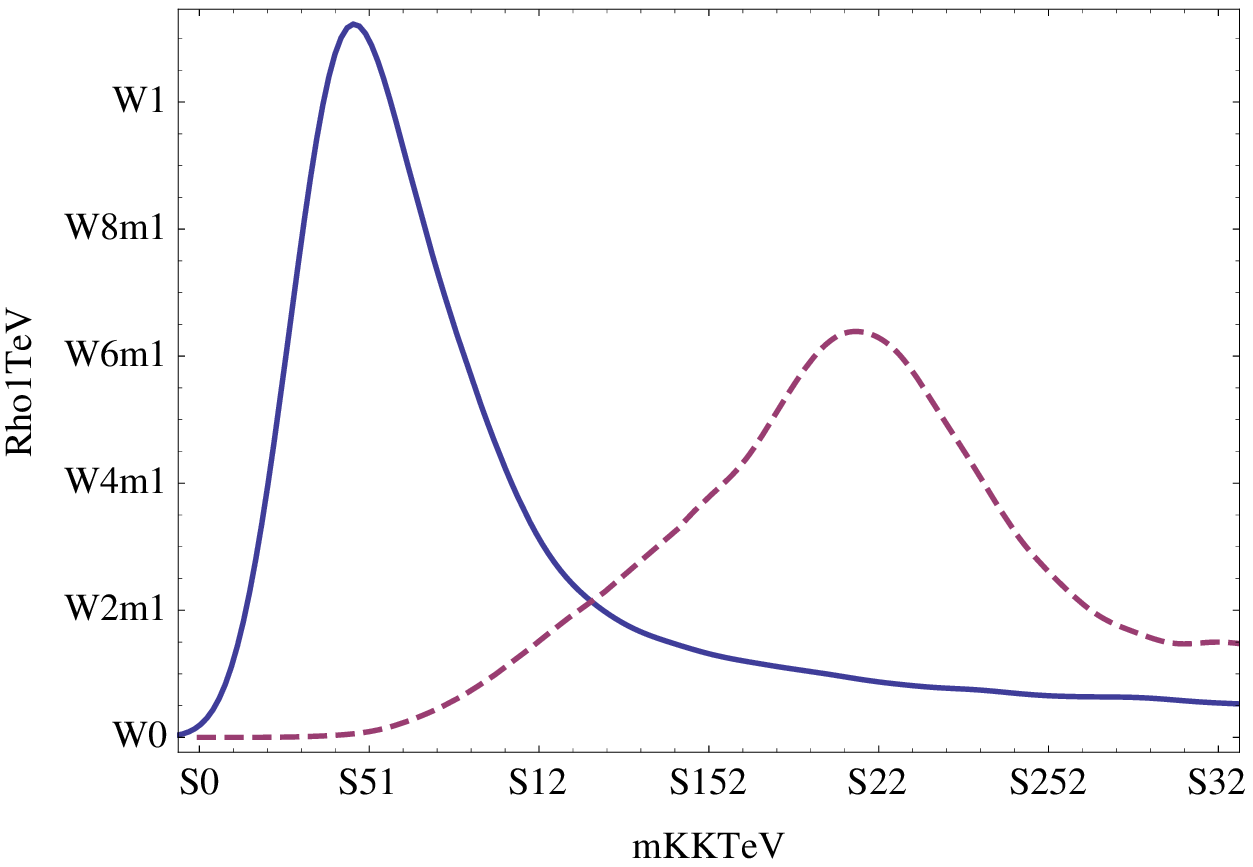}
\end{psfrags}
\\
\vspace{5mm}
\begin{psfrags}
\input{figs/BothCDF-psfrag.tex}
\includegraphics[width=0.6\textwidth]{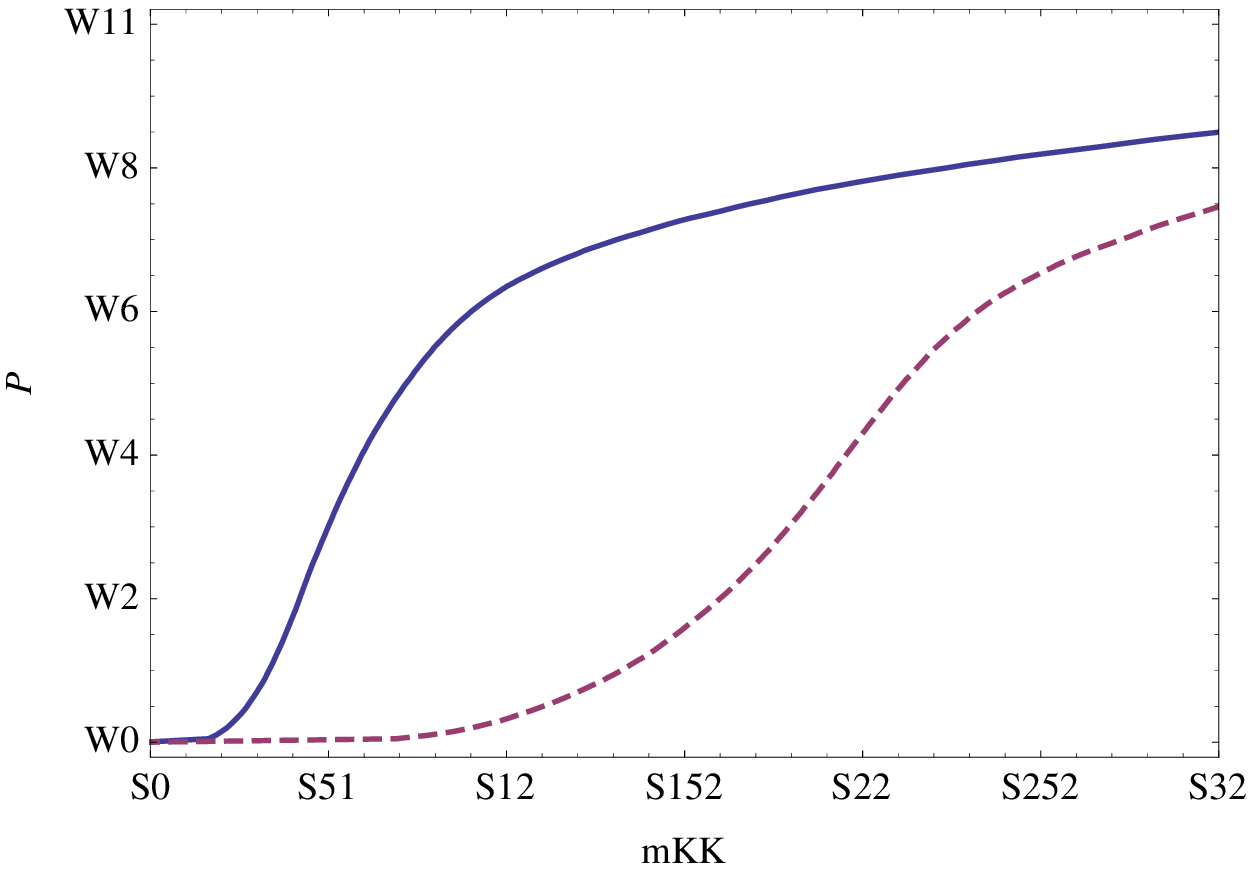}
\end{psfrags}
\end{center}
\caption{\it PDF (upper panel) and CDF (lower panel) for $m_{KK}$ from comparison with $\left|\textrm{Im} \,C_4^{sd}\right|$ and $R_b$. Dashed lines correspond to the RS model and solid lines to the model in Eq.~(\ref{ourmetric}) for $k\Delta=1$ and $\nu=0.5$.}
\label{figboth}
\end{figure}
\begin{table}[p!]
\centering
\begin{tabular}{|c|ccc|ccc|}
\cline{2-7}
\multicolumn{1}{c|}{}&\multicolumn{3}{c|}{\textit{Probability for $m_{KK}$ below}}&\multicolumn{3}{c|}{\textit{Percentile}}\\
\multicolumn{1}{c|}{}&$3$ TeV &$5$ TeV &$10$ TeV&10\% &20\% &50\% \\
\hline
RS &0\% &0\% &3.3\% &13 TeV&16 TeV& 21 TeV\\
$\nu=0.5$ &7.1\% &30\% &64\% &3.3 TeV&4.2 TeV& 7.2 TeV\\
\hline
\end{tabular}
 \caption{\it Left panel: Integrated probability for values of $m_{KK}$ below 3, 5 and 10 TeV from $R_b$ and Im $C_4^{sd}$ for RS (upper row) and the model in Eq.~(\ref{ourmetric}) for $k\Delta=1$ and $\nu=0.5$ (lower row). Right panel: 10th, 20th and 50th percentiles for both models.}
\label{tabboth}
\end{table}
In Fig.~\ref{figboth} we show the PDF and CDF distributions corresponding to the combined bounds from non-oblique observables and flavor/$CP$ violating effective operators. A similar statistical analysis to those presented for the individual contributions is done here and the results are presented in Tab.~\ref{tabboth}.
From there we can see that  assuming a 20\% (50\%) fine-tuning the lower bound for the RS model is 16 TeV (21 TeV) while for the modified background model they are 4.2 TeV (7.2 TeV). Then since the percentile is also a measure of the fine-tuning we can conclude that if we tolerate a fine tuning $\sim$10\%-20\% a KK-mass $\sim$ 3 TeV can be roughly acceptable.

Let us remark that the derived bounds can be considered the most conservative ones (i.e. the worst case scenario in the absence of further suppressions). In particular, the $Z b \bar{b}$ bounds can be improved if one allows for a moderate hierarchy in the 5D Yukawas, i.e.~by lowering the 5D bottom Yukawa. On the other hand, flavor bounds can be improved by including the effects of UV brane localized kinetic term for the gluon\cite{Csaki:2008zd}, or by invoking some flavor symmetries \cite{Santiago:2008vq,Bauer:2011ah}.

%%%%%%%%%%%%%%%%%%%%%%%%%%%%%%%%%%%%%%%%

\chapter{Concluding Remarks}
\label{chap:conclusion}

In the previous chapters, we have presented two different kinds of warped models: a self-stabilized soft-wall model (Chap.~\ref{chap:softwalls}) and a two-brane warped model for EWSB that does not require the introduction of a custodial symmetry (Chap.~\ref{chap:noncustodial}). While they share the same metric, their construction is very different and so are their phenomenology and possible applications. 

In this chapter we will briefly discuss, for each of these two classes of models, some aspects that have not been covered in the previous chapters but might be worth researching. We will also comment on the new research paths these models could open, and how they might provide new interesting results.

\section{Self-stabilized soft walls}

In Chap.~\ref{chap:softwalls} we studied the stabilization of soft walls, i.e.~4+1 dimensional geometries with 4D Poincar\'e invariance, that are only bounded by a single three-brane but that nevertheless exhibit a finite volume for the extra dimensional coordinate. That is achieved by replacing the second brane by a naked singularity at a finite proper distance. In particular, we saw how those soft walls arise in models with a single scalar field, and classified the type of models that can be realized as full solutions to the Einstein equations without destabilizing contributions at the singularities. 

Our main objective was to show how how to stabilize the position of the singularity at parametrically large values compared to the 5D Planck length.  We proposed a family of models that accomplishes this goal, which can be classified in function of the 4D spectra they can yield: continuous, continuous with a mass gap, and discrete. The 4D mass scale $\rho$, controlling the mass gap in the continuous case and the spacing in the discrete one, depends in a double exponential manner on the value of the scalar field at the brane [see Eq.~(\ref{rhok})] and can thus be naturally suppressed with respect to the 5D mass scale $k$ without fine-tuning. 

Depending on the class of models considered, there are a number of phenomenological applications of soft-walls that are beyond the scope of this thesis but are worth of future investigations. For the case when the spectrum is continuous and given by $m_n \sim n$ (i.e.~$1<\nu<2$ for a superpotential that behaves as $W(\phi) \sim e^{\nu \phi}$) these applications are common with two-brane models, like RS1, but with some peculiarities. However, as discussed in Sec.~\ref{sec:commentssw}, it seems difficult to achieve a model for EWSB that solves the hierarchy problem by using soft-wall models. Without an IR brane, there is no way to localize a mechanism to trigger EWSB, unless a fine-tuning is introduced at the UV brane. Finding a suitable way to implement EWSB in soft-wall models while retaining calculability and without introducing another kind of fine-tuning remains an open problem. 

Another interesting feature of soft-walls is the fact that the spectrum might present a spacing between KK modes much smaller than the mass of the first mode. This would mean that, if the first KK mode is within the range of the LHC, many other models could be found inside the LHC range. This would provide a striking signature of soft-wall models, since two-brane warped models do not exhibit this behavior. In particular, graviton (and radion) KK~modes can be produced and decay at LHC by their interaction with matter $\sim h_{\mu\nu} T^{\mu\nu}$, so they are expected to be produced through gluon annihilation \cite{Lillie:2007yh}.

In one class of soft-wall models (when the superpotential behaves as $W(\phi) \sim e^{\phi}$) the mass spectrum of fields propagating in the bulk is a continuum above a mass gap, which can be set to be of $\mathcal{O}$(TeV). This continuum (endowed with a certain conformal dimension) can interact with SM fields living in the UV brane as operators of a CFT, where conformal invariance is explicitly broken at a scale given by the mass gap, and can model and describe unparticle phenomenology \cite{Georgi:2007ek}. In particular, the authors of Ref.~\cite{Stancato:2008mp} embedded the Higgs in a soft-wall 5D background to describe an unHiggs theory. It would certainly be interesting to study this phenomenology in a stabilized model such as the one described here.

Finally, soft-wall models can be constructed so that the mass spectrum behaves as $m_n \sim \sqrt{n}$ (when $W(\phi) \sim e^{\phi} \phi^{1/4}$). This case is particularly interesting as it models the Regge trajectories for mesons. Therefore, it can be used to exploit the AdS/CFT correspondence for modeling QCD (a setup usually referred to as AdS/QCD), along the lines of Ref.~\cite{Karch:2006pv}. The stabilization mechanism we described opens the interesting possibility to study AdS/QCD in a model where the QCD scale can be naturally stabilized by the scalar field.

%%%%%

\section{Non-custodial warped models}

In Chap.~\ref{chap:ewsbbulkhiggs} we studied how to break electroweak symmetry in a two-brane 5D warped model when the Higgs field is propagating in the bulk of the extra dimension. We derived a series of general results valid for arbitrary metric and scalar backgrounds. These results include the effective theory of a light bulk Higgs, general expressions for electroweak observables $S$, $T$, $Y$ and $W$, and expressions for the mass of the radion. We identified a new contribution $Z$ to the wave function renormalization of the Higgs zero mode which can suppress the tree level Higgs mass and plays a major role in reducing the contributions to the $S$ and $T$ parameters. In holographic language the Higgs wave function renormalization $Z$ can be large if the dimension of the Higgs condensate decreases towards the IR, while staying sufficiently large in the UV to solve the hierarchy problem.  

Looking for a way to avoid the usual paradigm of custodial gauge symmetries, we introduced, in Chap.~\ref{chap:noncustodial}, a generalization of the RS model based on an asymptotically AdS (AAdS) metric and a Higgs propagating in the bulk, which allows us to achieve sizable values of $Z$ and therefore avoid the need for a custodial symmetry.  For that, one needs a strong deviation from conformality in the IR, parametrized by a large back reaction of the stabilizing field on the AdS background metric. In order to achieve this, we propose a metric with a curvature singularity, although we locate the IR brane so that the singularity remains outside the physical interval.

Our model is described by three input parameters: the quantities $\nu$ and $\Delta=y_s-y_1$ entering our metric, Eq.~(\ref{A}), and the parameter $a$, which corresponds to the UV dimension of the Higgs condensate. The location of the IR brane, $y_1$, has been fixed by imposing the Planck-EW hierarchy $A(y_1)=35$. The parameter $k$ [or equivalently $\rho=k e^{-A(y_1)}$] is then computed from requiring consistence with EWPT, which in turn sets bounds on the KK masses. We considered the tree-level contributions to the oblique EWPO and found that they are reduced when $a,\ \nu$ and $\Delta$ are lowered. 

There are a number of arguments that can be used to constrain our parameter space from theoretical considerations. First, for each fixed $\nu$ and $\Delta$ there is a minimum value of $a$ required to solve the hierarchy problem, as described in Sec.~\ref{sec:hierarchy}. Moreover, the requirement that a not too large hierarchy is introduced between $k$ and $\rho$ introduces a lower bound on $\nu$ and $\Delta$. In particular, for $\Delta=1$ we find $\nu \gtrsim 0.5$ and $a \gtrsim 3.1$, which translates into a tree-level bound $m_{\mathrm{KK}} \gtrsim 0.8~\mathrm{TeV}$ from oblique EWPOs and for a Higgs mass $m_H \approx 115~\mathrm{TeV}$. Conversely, the lowest bound that can be obtained for RS without custodial symmetry is about $7~\mathrm{TeV}$.

In Chap.~\ref{chap:fermionsbulk} we studied the propagation of fermions in the bulk in our non-custodial model as well as in RS. This lead to further constraints from non-oblique EWPOs (mainly from a modified $Z b \bar{b}$ coupling) and from flavor and CP violation. After an statistic analysis, we found that the non-custodial model can improve the bounds from these two effects about a factor $3$ or better (e.g.~the median combined bound for RS is $21~\mathrm{TeV}$ while for our model it is $7.2~\mathrm{TeV}$).

Therefore, the non-custodial model presented here is an attractive alternative, or extension, to the RS model, allowing for some interesting phenomenology in the LHC range, and therefore easing the little hierarchy problem that many models of New Physics feature. Moreover, the fact that it features a minimal 5D extension of the SM makes it, arguably, a simpler alternative than the introduction of additional gauge symmetries, which in turn add calculability complications.

There are many topics that have not been covered in this thesis but which are worth of future research. To begin with, one topic that deserves closer investigation is the phenomenology of the radion. One unequivocal prediction of our model is a heavy radion due to the large deviation of the metric from AdS in the IR. Fortunately the radion wave function is approximated very well by $F(y)=e^{2A(y)}$ due to the excellent accuracy of the leading approximation for the mass, Eq.~(\ref{mrad}). It should therefore be straightforward to work out its couplings to the other light fields and establish its possible signatures at the LHC. 

Another important point to keep in mind is that we have done all of our calculations at tree-level. The authors of Ref.~\cite{Carmona:2011ib} showed that the loop corrections to the EWPO are finite and are in fact expected to be small when compared to the tree-level result. However, in the case where fermions are in the bulk, the radiative corrections should depend to a large extent on the size of 5D Yukawa couplings, so that they are very model dependent. 

In fact, there are some $CP$ violating effects which appear only at the loop level and which we did not consider in Chap.~\ref{chap:fermionsbulk}. In particular, the one-loop contribution to the neutron electric dipole moment~\cite{Agashe:2004cp} due to non-removable Majorana phases would probably require some kind of flavor alignment, although a bulk Higgs should certainly alleviate the problem since it renders the one-loop diagram contributing to it finite.

Also, we should keep in mind that the construction presented in Chap.~\ref{chap:fermionsbulk} is far from being a complete theory of flavor. The anarchic solution to the flavor problem (more precisely, the values of the 5D Yukawa couplings and the localizing fermion coefficients $c_f$) should arise from a more fundamental theory. 

Finally, it is worth studying with detail the collider phenomenology of the non-custodial model. In particular, it would be interesting to find signals that could differentiate one particular model from general models of warped extra dimensions. In fact, the LHC data is already starting to probe a region of the parameter space allowed by the precision observables.  Let us hope that the LHC will shed some light on the quest for New Physics and that it will soon give us a hint about the fate of models of warped extra dimensions.

%[...]

%%%%%%%%%%%%%%%%%%%%%%%%%%%%%%%%%%%%%%%%

\appendix

\chapter{Gauge Fluctuations}
\label{fluctuationsgauge}

In this Appendix we will study the propagation of gauge bosons in the 5D bulk in the presence of the background provided by the Higgs field and the gravitational metric. The results we will find here are presented in Chapter~\ref{chap:ewsbbulkhiggs}, where a 5D copy of the Standard Model is described. We will first (in Sec.~\ref{sec:warming}) consider the case of an abelian theory, and we will generalize it later (in Sec.~\ref{sec:theSM}) to a non-abelian {SM}-like model.

\section{An abelian theory}
\label{sec:warming}
We will first analyze the gauge fixing in a spontaneously broken abelian 5D theory with a Higgs defined by
\be
H(x,y)=\frac{1}{\sqrt{2}}[h(y)+\xi(x,y)]e^{ig_5\chi(x,y)}
\ee
where $h(y)$ is the $y$-dependent Higgs field background, $\xi(x,y)$ the Higgs fluctuation and $\chi(x,y)$ the Goldstone fluctuation. The 5D action for the gauge field $A_M(x,y)$ and the Goldstone boson is given by
\be
S_5=\int d^4x dy\sqrt{-g}\left(-\frac{1}{4} F_{MN}F^{MN}-|D_M H|^2
\right)
\label{accion}
\ee
where $F_{MN}=\partial_MA_N-\partial_NA_M$, $D_MH=\partial_MH-ig_5A_MH$ and $g_5$ is the 5D gauge coupling with mass dimension $-1/2$. The mass dimension of the 5D fields $h$, $\xi$ and $A_M$ is $3/2$ and that of $\chi$ is $1/2$. The action (\ref{accion}) is invariant under 5D gauge transformations
\begin{eqnarray}
A_M(x,y)&\to& A_M(x,y)+\frac{1}{g_5}\partial_M \alpha(x,y)\nonumber\\
\chi(x,y)&\to&\chi(x,y)+\frac{1}{g_5}\alpha(x,y)
\end{eqnarray}

To quadratic order in the fluctuations $A_M$ and $\chi$,\footnote{We need to consider here only the fluctuations of fields $A_M$ and $\chi$ which mix to each other through the mechanism of electroweak breaking. The Higgs fluctuations $\xi$ will decouple from them and are considered in Sec.~\ref{sec:5DSM}.} the action (\ref{accion}) can be written as
\begin{eqnarray}
S_5&=&\int d^4xdy \biggl[-\frac{1}{4}(F_{\mu\nu})^2-\frac{1}{2} \,e^{-2A}(F_{\mu 5})^2
-\frac{1}{2}M_A^2(\partial_\mu\chi- A_\mu)^2  \biggr.\nonumber\\
&&-\frac{1}{2}M_A^2\,e^{-2A}\biggl.(\chi'- A_5)^2 \biggr] ,
\label{accion2}
\end{eqnarray}
where we have defined the $y$ dependent bulk mass 
\be
M_A(y)=g_5 h(y)e^{-A(y)} .
\ee
The bulk EOM's from action (\ref{accion2}) are
\begin{eqnarray}
\Box A_\mu+(e^{-2A}A'_\mu)'-M_A^2 A_\mu
+\partial_\mu\left\{ M_A^2\chi-(\partial^\nu A_\nu)-(e^{-2A}A_5)' \right\}&=&0\label{eq.A}\\
\Box A_5-\partial^\nu A'_\nu+M_A^2(\chi'-A_5)&=&0\label{eq.A5}\\
\Box \chi- \partial^\nu A_\nu+M_A^{-2}\left\{\left(M_A^2 e^{-2A}\right)(\chi'- A_5)
\right\}'&=&0\label{eq.chi}
\end{eqnarray}
and the boundary conditions are
\begin{eqnarray}
%e^{-2A}
 \left( \partial_\mu A_5 - A'_\mu \right) \vert_{y=0,y_1} = 0 \\
%e^{-4A} h' \vert_{y=0,y_1} = 0 \\
 %e^{-4A} 
 \left( \chi' -  A_5 \right) \vert_{y=0,y_1} = 0   
\label{b.c.}
\end{eqnarray}

We can gauge away the last term in Eq.~(\ref{eq.A}) by the gauge condition
\be
\partial^\mu A_\mu-M_A^2\chi+(e^{-2 A}A_5)'=0.
\label{gaugecond}
\ee
By making the ansatz (the dot product denotes an expansion in modes)
\be
A_\mu(x,y)=\frac{a_\mu(x)\cdot f(y)}{\sqrt{y_1}}
\label{ansatzA}
\ee
the EOM (\ref{eq.A}) becomes
\be
m_f^2 f+(e^{-2A}f')'-M_A^2 f=0 \,,
\label{eq.f}
\ee
where the functions $f(y)$ are normalized as
\be
\frac{1}{y_1}\int_0^{y_1}f^2(y)dy=1
\ee
and satisfy the boundary conditions
\be
%e^{-2A}
\left. f'\right|_{y=0,y_1}=0 \,.
\label{eq:boundaryf}
\ee

It is easy to see that the gauge condition remains invariant under the whole set of 5D gauge transformations
\be
\alpha(x,y)=\alpha(x)\cdot f(y) \,,
\ee
where $\alpha_n(x)$ are arbitrary 4D gauge transformations which are the remaining 4D invariances. A quick glance at the action, Eq.~(\ref{accion2}), shows that the Goldstone boson degree of freedom (which couples to $\partial^\mu A_\mu$ in the action) should be defined as
\be
G(x,y)=M_A^2 \chi-\left(e^{-2A} A_5\right)'
\ee
while the remaining degree of freedom is defined as~\footnote{The pseudoscalar modes $K_n$ are physical (they are in particular gauge invariant) and could play an important role in experimentally identifying the Higgs as a bulk field. Their equations of motion have been derived previously in Ref.~\cite{Falkowski:2008fz}.}
\be
K(x,y)=\chi'-A_5
\ee
which satisfy the decoupled EOM's [from Eqs.~(\ref{eq.A5}) and (\ref{eq.chi})]
\begin{eqnarray}
\Box G+(e^{-2 A}G')'-M_A^2 G&=&0\label{eq.G}\\
 \Box K+\left[M_A^{-2}\left(e^{-2A}M_A^2K \right)'  \right]'
 -M_A^2K&=&0\ . \label{eq.K}
\end{eqnarray}
Eqs.~(\ref{eq.G}) and (\ref{eq.K}) are satisfied by
\begin{eqnarray}
G(x,y)&=&\frac{m_f\,G(x)\cdot f(y)}{\sqrt{y_1}}\label{ansatzG}\\
K(x,y)&=&\frac{K(x) \cdot\eta(y)}{\sqrt{y_1}}\label{ansatzK}
\end{eqnarray}
where $f(y)$ satisfies Eq.~(\ref{eq.f}), and $\eta(y)$ satisfies the bulk EOM
\be
 m_\eta^2\eta+\left[M_A^{-2}\left(e^{-2A}M_A^2\eta \right)'  \right]'
 -M_A^2\eta=0
\label{eq.eta}
\ee
and the (Dirichlet) boundary conditions
\be
\left.%e^{-4A}
\eta\right|_{y=0,y_1}=0\ .
\label{b.c.D}
\ee
The normalization for $\eta$ will be fixed below.
Notice that in the limit $M_A\to 0$ there is no massless mode since the zero mode would have the (trivial) wave function, consistent with the boundary conditions, $\eta(y)\equiv 0$. Only massive KK modes do appear.

In the 4D theory the degrees of freedom are the gauge field $a_\mu(x)$ the Goldstone boson $G(x)$ and the gauge invariant scalar $K(x)$. They transform under the 4D gauge transformation $\alpha(x)$ as
\begin{eqnarray}
\delta_\alpha a_\mu(x)&=&\frac{1}{g}\partial_\mu\alpha(x)
\\
 \delta_\alpha G(x)&=&\frac{m_f}{g}\,\alpha(x)
 \\
\delta_\alpha K(x)&=&0
\end{eqnarray}
where the 4D gauge coupling is defined as $g=g_5/\sqrt{y_1}$.
It is easy to obtain the 4D effective Lagrangian upon integration of the $y$-coordinate in the action (\ref{accion2}) by using the decomposition 
\begin{eqnarray}
\sqrt{y_1}\,A_5(x,y)&=&\frac{1}{m_f}G(x)\cdot f'(y)-\frac{M_A^2}{m_\eta^2}\,K(x)\cdot \eta(y)\nonumber\\
\sqrt{y_1}\,\chi(x,y)&=&\frac{1}{m_f}G(x)\cdot f(y)-\frac{1}{m_\eta^2}M_A^{-2}\left(M_A^2e^{-2A}\eta(y)\right)'\cdot K(x).
\end{eqnarray}
In fact after integration over the $y$-coordinate and using the EOM (\ref{eq.f}) and (\ref{eq.eta}) one can write down the 4D Lagrangian as
\be
\mathcal L_{4D}=-\frac{1}{4}(\partial_\mu a_\nu-\partial_\nu a_\mu)^2-\frac{1}{2}(m_f a_\mu-\partial_\mu G)^2 
-\frac{1}{2}(\partial_\mu K)^2-\frac{1}{2}m_\eta^2 K^2
\label{modelag}
\ee
where we have fixed the normalization for the wave function $\eta$ as
\be
\frac{1}{y_1}\int_0^{y_1} M_A^2e^{-2A}\eta^2\,dy=m_\eta^2\,.
\ee
In Eq.~(\ref{modelag}) all the squares are to be understood as summations over modes.

%%%%%%%%%%%%
%\begin{comment}
Notice that although the EOM (\ref{eq.G}) and (\ref{eq.K}) are decoupled they arise from the coupled set (\ref{eq.A5}) and (\ref{eq.chi}) and as such the mass eigenvalues are common. Of course this does not mean that Eqs.~(\ref{eq.f}) and (\ref{eq.eta}) should have the same mass eigenvalues and the puzzle can be resolved by noticing that Eqs.~(\ref{eq.f}) and (\ref{eq.eta}) always admit the trivial solutions $f(y)\equiv 0$ and/or $\eta(y)\equiv 0$. In fact a solution $m_1^2$ and $f^{(1)}(y)$ from Eq.~(\ref{eq.f}) corresponds to the mass eigenfunctions $(f^{(1)}(y),\eta^{(1)}(y)\equiv 0)$ and the corresponding solution $m_2^2$ and $\eta^{(2)}(y)$ from Eq.~(\ref{eq.eta}) corresponds to the mass eigenfunctions $(f^{(2)}(y)\equiv 0,\eta^{(2)}(y))$. Then the effective Lagrangian corresponding to all the modes can be written as
\begin{eqnarray}
\mathcal L_{4D}&=&-\sum_{n_1}\left(\frac{1}{4}(\partial_\mu a_\nu^{(n_1)}(x)-\partial_\nu a_\mu^{(n_1)}(x))^2+\frac{1}{2}m^2_{n_1}(a_\mu^{(n_1)}(x))^2\right.\nonumber\\
&+&\left.\frac{1}{2}(\partial_\mu G^{(n_1)}(x))^2+m_{n_1}(\partial^\mu a_\mu^{(n_1)}(x)) G^{(n_1)}(x)\right)\nonumber\\
&-&\sum_{n_2}\left(\frac{1}{2}(\partial_\mu K^{(n_2)}(x))^2+\frac{1}{2}m^2_{n_2} (K^{(n_2)}(x))^2\right)
\end{eqnarray}
%\end{comment}
%%%%%%%%%%%%%

\section{The 5D Standard Model}
\label{sec:theSM}
The generalization to non-abelian gauge theories is straightforward. In particular in the $SU(2)\times U(1)$ {SM} the gauge and Higgs bosons are introduced in the usual way with a 5D action given by
\be
S_5=\int d^4x dy\sqrt{-g}\left(-\frac{1}{4} (F^i_{MN})^2-\frac{1}{4}(F_{MN}^Y)^2-|D_M H|^2
-V(\Phi,H)
\right) \,,
\ee
where the 5D Higgs field is written as
\be
H=\frac{1}{\sqrt 2}e^{i g_5 \chi} \left(\begin{array}{c}0\\h+\xi\end{array}\right)
\ee
and where the matrix $\chi$ only includes the coset and $g_5$ is the 5D $SU(2)_W$ coupling. The $\xi$ field will again decouple so we consider it separately.
Following the standard notation we have
\be
D_M=\partial_M-ig_5A_M\,,\qquad A_M=\left(
\begin{array}{cc}
s_wA^{em}_M+\frac{c_w^2-s_w^2}{2 c_w}Z_M&\frac{1}{\sqrt 2}W^+_M\\
\frac{1}{\sqrt 2}W_M^-&-\frac{1}{2c_w}Z_M
\end{array}
\right)
\ee
and, analogously,
\be
\chi=\left(
\begin{array}{cc}
\frac{c_w^2-s_w^2}{2 c_w}\chi_Z&\frac{1}{\sqrt 2}\chi_+\\
\frac{1}{\sqrt 2}\chi_-&-\frac{1}{2c_w} \chi_Z
\end{array}
\right) \,,
\label{chimatrix}
\ee
where the weak angle is defined as in the 4D theory, $t_w\equiv g'_5/g_5=g'/g$.
Expanding the Lagrangian to second order we obtain a straightforward generalization of the abelian case, Eq.~(\ref{accion2})
\be
\mathcal L=\mathcal L^{\gamma}+\mathcal L^{Z}+\mathcal L^W \,,
\ee
with
\begin{eqnarray}
\mathcal L^{\gamma}&=&-\frac{1}{4}(F^{\gamma}_{\mu\nu})^2-\frac{1}{2}e^{-2A}(F^{\gamma}_{\mu 5})^2
\,,\\
\mathcal L^{Z}&=&-\frac{1}{4}(F_{\mu\nu}^Z)^2-\frac{1}{2}e^{-2A}(F_{\mu5}^Z)^2-\frac{1}{2}M_Z^2(\partial_\mu\chi_Z-A_\mu^Z)^2\nonumber\\
&&-\frac{1}{2}e^{-2A}M_Z^2(\chi_Z'-A_5^Z)^2
\,,\\
\mathcal L^W&=&-\frac{1}{2}F_{\mu\nu}^+F_{\mu\nu}^--\frac{1}{2}e^{-2A}F_{\mu5}^+F_{\mu5}^--M_W^2(\partial_\mu\chi_+-A_\mu^+)(\partial_\mu\chi_--A_\mu^-)\nonumber\\
&&-e^{-2A}M_W^2(\chi_+'-A_5^+)(\chi_-'-A_5^-)
\,.
\end{eqnarray}
Here we have defined the 5D $y$-dependent gauge boson masses
\be
M_W(y)=\frac{g_5}{2} h(y)e^{-A(y)}\,,\qquad M_Z(y)=\frac{1}{c_w} M_W(y)\,,\qquad M_\gamma(y)\equiv 0
\label{masas2}
\ee

Now we should proceed as in the abelian case and define the mode expansion for the different gauge bosons $A_\mu(x,y)$ with profiles $f_A(y)$ ($A=W,Z,\gamma$) as in Eq.~(\ref{ansatzA}) and the corresponding pseudoscalars $K_A(x,y)$ with profiles $\eta_A(y)$ as in Eq.~(\ref{ansatzK}) which satisfy [Eqs.~(\ref{eq.f}) and (\ref{eq.eta})]
\begin{eqnarray}
m_{f_A}^2 f_A+(e^{-2A}f'_A)'-M_A^2 f_A&=&0 \label{eqfinalf}\\
 m_{\eta_A}^2\eta_A+\left[M_A^{-2}\left(e^{-2A}M_A^2\eta_A \right)'  \right]'
 -M_A^2\eta_A&=&0 
 \label{eqfinaleta}
\end{eqnarray}
where $M_A$ is defined in Eq.~(\ref{masas2}) and $m_{f_A}$ and $m_{\eta_A}$ the mass eigenvalues which are identified with the physical gauge boson masses.

%%%%
%%%%
%%%%
%%%%

\chapter{Gauge Boson Propagators}
\label{app:propagators}

In this appendix we compute the gauge boson propagators at zero momentum. In other words, we would like to compute
\be
G_{B_0B_1}(y,y')=\sum_{n\geq 1} \frac{f^n(y)\, f^n(y')}{m_n^2}\,,
\label{defG}
\ee
where $f_n$ are the wave functions of the gauge bosons and $B_\alpha=D,N$ denote Dirichlet or Neumann BC at the boundaries at $y=y_\alpha$:
\be
G_{DB_1}(0,y')=0\quad {\rm or}\quad G'_{NB_1}(0,y')=0\,,
\label{B0}
\ee
and
\be
G_{B_0D}(y_1,y')=0\quad {\rm or}\quad G'_{B_0N}(y_1,y')=0\,,
\label{B1}
\ee
respectively.
The sum excludes any zero mode (if present). 

The $f_n$ are the wave functions in the symmetric phase, and they satisfy
\be
(e^{-2A}f_n')'+m_n^2f_n=0\,.
\label{eomf}
\ee
Therefore, the propagators satisfy the EOM
\be
\partial_y\left[e^{-2A(y)}\partial_y G_{B_0B_1}(y,y')\right]=-y_1\delta(y-y')\,,
\label{eomG}
\ee
for $B_0B_1\neq NN$ and
\be
\partial_y\left[e^{-2A(y)}\partial_y G_{NN}(y,y')\right]=1-y_1\delta(y-y')\,,
\label{eomGNN}
\ee
in the case of Neumann-Neumann BC with zero mode subtracted.
These equations are easily derived from Eq.~(\ref{eomf}) using the completeness relation\footnote{Furthermore recall that our normalization reads $\int_0^{y_1}f_n^2=y_1$, in particular $f_0(y)=1$ in the $NN$ case.}
\be
\sum_{n\geq 0}f_n(y)f_n(y')=y_1\delta(y-y')\,.
\ee
Moreover, the boundary conditions have to be supplemented by the jump and continuity relations
\begin{align}
e^{-2A(y')} \left[ \partial_y G_{B_0B_1}(y'+\epsilon,y')- \partial_y G_{B_0B_1}(y'-\epsilon,y') \right] &=-y_1\,,
\\
G_{B_0B_1}(y'+\epsilon,y')-G_{B_0B_1}(y'-\epsilon,y')&=0\,.
\label{jump}
\end{align}

For $B_0B_1\neq NN$ the solutions are straightforward and read
\begin{align}
G_{DN}(y,y')&=y_1\int_0^{y_<}e^{2A}\,,
\\
G_{ND}(y,y')&=y_1\int_{y_>}^{y_1}e^{2A}\,,
\\
G_{DD}(y,y')&=y_1\frac{\displaystyle \left( \int_0^{y_<}e^{2A} \right)\left( \int_{y_>}^{y_1}e^{2A}  \right)}{\displaystyle \int_0^{y_1}e^{2A}}\,,
\label{GBB}
\end{align}
where $y_<$ ($y_>$) denotes the smaller (larger) of the pair $y,y'$. One can immediately verify that these are solutions to the system of Eqs.~(\ref{eomG}), (\ref{B0}), (\ref{B1}) and (\ref{jump}). 

The $NN$ case requires more care. One can always shift the solution by a $y'$ dependent constant: the bulk Eq.~(\ref{eomGNN}) is invariant under such a shift and so are the BC and the conditions Eq.~(\ref{jump}). After imposing symmetry in the interchange of $y$ and $y'$ (obvious from the definition Eq.~(\ref{defG})), one still has an undetermined $y'$ independent constant. In fact one, can immediately verify that
\be
G_{NN}(y,y')=\int_0^{y_<}d\hat y\,e^{2A(\hat y)}\hat y
+\int_{y_>}^{y_1}d\hat y\, e^{2A(\hat y)}(y_1-\hat y)+c
\ee
is a solution (symmetric under interchange of $y,y'$) to the system for arbitrary
constant $c$. To fix $c$, we impose that $G_{NN}(0,0)$ reduces to the brane to brane propagator computed in Eq.~(\ref{prop2}):
\be
G_{NN}(0,0)=-\lim_{p\to 0}\left(\frac{1}{\Pi(p^2)}-\frac{1}{p^2}\right)\,.
\ee
Notice that in Eq.~(\ref{prop2}) one can set $m_A$ to zero in the symmetric phase. We then find
\be
G_{NN}(0,0)=y_1\int_0^{y_1}d\hat{y}\,e^{2A(\hat y)}\left(1-\frac{\hat{y}}{y_1}\right)^2
\ee
which fixes $c$ uniquely. Finally, we end up with
\be
G_{NN}(y,y')=\int_0^{y_<}d\hat y\,e^{2A(\hat y)}\hat y
+\int_{y_>}^{y_1}d\hat y\, e^{2A(\hat y)}(y_1-\hat y)
-\frac{1}{y_1}\int_0^{y_1}e^{2A(\hat y)}\hat y(y_1-\hat y)\,.
\label{GNN}
\ee
%

%%%%%%%
%%%%%%%

\chapter{Fermion Propagators}
\label{app:fermions}

In this appendix we will provide a few more details concerning the procedure of integrating out fermionic KK modes, as done in Sec.~\ref{secKKfermions}. This parallels and generalizes the computation in Appendix \ref{app:propagators} for the gauge bosons. We will restrict ourselves to the case where the KK tower contains a zero mode that has to be subtracted, which is the most complicated case and the only relevant for this work. In fact, in order to evaluate the first diagram in Fig.~\ref{KKfermion} we need to compute
\be
\beta^{d_L}_{i\ell}=\sum_j\hat Y^d_{ij}\hat Y^{d*}_{\ell j}\sum_{n\neq 0}\int_0^{y_1}dy\,dy'
\,\frac{
 \left[\xi^0(y)\,\hat \psi^0_{Q_L^i}\!(y)\,\hat\psi^n_{d_R^j}\!(y)\right]
 \left[\xi^0(y')\,\hat \psi^0_{Q_L^\ell}\!(y')\,\hat\psi^n_{d_R^j}\!(y')\right]}{m_n^2}\,,
\label{betaexpl}
\ee
where $\xi^0(y)$ is the normalized Higgs zero mode wave function. The expression for $\beta^{d_R}_{i\ell}$ is completely analogous. The idea is now to first perform the sum over the KK modes and then the integrations.
Let us thus consider the equations of motion for the KK modes of a fermion,
\be
e^{2A} m_n^2 \hat \psi_n+e^A(\partial_y-Q')e^{-A}(\partial_y+Q')\hat \psi_n=0\,,
\qquad 
\left.\hat \psi_n'+Q'\hat\psi_n\right|_{y=0,y_1}=0\,,
\label{eomfermion1}
\ee
which follow from Eq.~(\ref{Dirac2}).
It will be convenient to factor out the zero mode as
\be
\hat\psi_n(y)=e^{-Q(y)}\chi_n(y)
\ee
which transforms Eq.~(\ref{eomfermion1}) into
\be
e^{A-2Q}\,m_n^2\,\chi_n+(e^{-A-2Q}\chi'_n)'=0\,,\qquad \left.\chi_n'\right|_{y=0,y_1}=0\,.
\label{eomfermion2a}
\ee
The completeness and orthonormality conditions in this basis read:
\be
\sum_{n=0}^{\infty} \chi_n(y)\chi_n(y')=e^{-A+2Q}\delta(y-y')\,,\qquad
\int_0^{y_1}e^{A(y)-2Q(y)}\chi_n(y)\chi_m(y)=\delta_{mn}\,. \label{completeness}
\ee
We now need to compute the sum
\be
G(y,y')=\sum_{n\neq 0}\frac{\chi_n(y)\chi_n(y')}{m_n^2}\,.
\ee 
Note that the zero mode has been excluded from the sum.
To compute $G(y,y')$, we integrate Eq.~(\ref{eomfermion2a}) twice
\be
\chi_n(y)=\chi_n(0)-m_n^2\int^y_0 e^{A(u)+2Q(u)} \int_0^{u}  e^{A(v)-2Q(v)}\chi_n(v)\,,
\ee
and use Eq.~(\ref{completeness}) to get
\be
G(y,y')
=\sum_{n\neq0}\frac{\chi_n(0)\chi_n(0)}{m_n^2}
-\int_0^{y_>}e^{A+2Q}(1-\Omega)
+\int_0^{y_<}e^{A+2Q}\Omega\,,
\ee
where $\Omega(y)$ has been defined in Eq.~(\ref{Omegafermion}), and $y_>$ ($y_<$) is the larger (smaller) of the pair $(y,y')$. We thus have reduced the problem of finding $G(y,y')$ to that of finding $G(0,0)$, which is the zero momentum limit of the (zero mode subtracted) brane-to-brane propagator. The latter can be written as 
\be
G(0,0;p^2)=-\frac{\chi(0,p^2)}{\chi'(0,p^2)}-\frac{\chi^2_0(0)}{p^2}\,
\ee
where $\chi(0,p^2)$ is the solution to
\be
-e^{A-2Q}p^2\chi+(e^{-A-2Q}\chi')'=0\,,\qquad \left.\chi'\right|_{y_1}=0\,.
\label{eomfermion2b}
\ee
(note that we do not impose a BC at $y=0$). One can easily derive an equation for $\chi'/\chi $ and solve it in a power series in $p^2$:
\be
\begin{split}
e^{-A(y)-2Q(y)}
\frac{\chi'(y,p^2)}{\chi(y,p^2)}=&-p^2\int_y^{y_1}e^{A(u)-2Q(u)}
\\
&+p^4\int_y^{y_1}e^{A(u)+2Q(u)}\left[\int_u^{y_1}e^{A(v)-2Q(v)}\right]^2+\dots
\end{split}
\ee
One ends up with
\be
G(0,0)=\int_0^{y_1}e^{A+2Q}(1-\Omega)^2\,,
\ee
and hence
\be
G(y,y')
=\int_0^{y_1}e^{A+2Q}(1-\Omega)^2
-\int_0^{y_>}e^{A+2Q}(1-\Omega)
+\int_0^{y_<}e^{A+2Q}\Omega\,.
\ee
Using this expression in Eq.~(\ref{betaexpl}) we arrive, after a series of partial integrations, at the quoted result Eqs.~(\ref{betaINI}--\ref{betaFIN}).

%%%%%%%%
%%%%%%%%

%%%%% TODO: BEGIN C&P %%%%%%

\chapter{Four-Fermion Terms from Electroweak KK Modes}
\label{4f}
In this appendix we explicitely write the four-fermion interactions that appear after integrating out the KK modes of weak gauge bosons. For the neutral currents the effective Lagrangian reads
\be
\mathcal L^{4f}_{NC}=\sum_{f_\chi,f'_\chi}
\delta^{k\ell,rs}_{f_\chi,f'_\chi}
(\bar f^k_\chi\gamma^\mu f^\ell_\chi) (\bar f'^r_{\chi'}\gamma_\mu f'^s_{\chi'})
\label{4fNC}\,,
\ee
where the constants $\delta$ are tensors in flavor space:
\be
\delta^{k\ell,rs}_{f_\chi,f'_\chi}=
%-\delta^{k\ell}\delta^{rs}c^{EW}_{f_\chi,f'_\chi}\hat\alpha_{gg}
%+
%(
%c^{\rm EW}_{f_\chi,f'_\chi}%+c^{\rm QCD}_{\chi,\chi'})
\frac{e^2}{2}\left(Q_fQ_{f'}+\frac{1}{s_W^2c_W^2}g^{SM}_{f_\chi}g^{SM}_{f'_\chi}\right)
\sum_{i,j}\hat\alpha_{f^i_\chi,f'^{j}_\chi}(V^{k i}_{f_\chi}
V^{*\ell i}_{f_\chi})(V^{rj}_{f'_\chi}V^{*sj}_{f'_\chi})\,.
\ee
%
%with 
%
%\be
%c^{\rm EW}_{f_\chi,f'_\chi}=\frac{e^2}{2}\left(Q_fQ_{f'}+\frac{1}{s_W^2c_W^2}g^{SM}_{f_\chi}g^{SM}_{f'_\chi}\right)\,,\qquad
%c^{\rm QCD}_{\chi\chi'}
%%=-\frac{g_s^2}{4}\left(\delta_{\chi\chi'}+\frac{1}{3}\right)
%\ee
%
%the last correction is the contribution from integrating out the KK gluons which is only present for quarks. 
Finally, integrating out the KK modes of the $W$ boson also leads to four-fermion terms, which we write explicitly as
\begin{equation}
\begin{split}
\mathcal L^{4 f}_{CC} &= \delta_{u^kd^\ell,d^r u^s} (\bar u^k_L\gamma^\mu d_L^\ell)  (\bar d_L^r\gamma^\mu u_L^s)%\nn\\
+\delta_{\nu^ke^\ell,e^r \nu^s} (\bar \nu^k_L\gamma^\mu e_L^\ell)  (\bar e_L^r\gamma^\mu \nu_L^s)
\\
&\phantom{=}+\left[\delta_{u^kd^\ell,e^r \nu^s} (\bar u^k_L\gamma^\mu d_L^\ell)  (\bar e_L^r\gamma^\mu \nu_L^s)+h.c.\right]\,,
\end{split}
 \label{4fCC}
\end{equation}
with
\begin{align}
\delta_{u^kd^\ell,d^r u^s}&=\frac{g^2}{2}\sum_{ij}\hat\alpha_{q_L^i,q^j_L}
   V_{u_L}^{ki}V_{d_L}^{*\ell i}V^{rj}_{d_L}V^{*sj}_{u_L}\,,
   \\
\delta_{\nu^ke^\ell,e^r \nu^s}&=\frac{g^2}{2}\sum_{ij}\hat\alpha_{\ell_L^i,\ell^j_L}
   V_{\nu_L}^{ki}V_{e_L}^{*\ell i}V^{rj}_{e_L}V^{*sj}_{\nu_L}\,,
   \\
\delta_{u^kd^\ell,e^r \nu^s}&=\frac{g^2}{2}\sum_{ij}\hat\alpha_{q_L^i,\ell^j_L}
   V_{u_L}^{ki}V_{d_L}^{*\ell i}V^{rj}_{e_L}V^{*sj}_{\nu_L}\,.
\end{align}

\chapter{Right-Handed Hierarchies}
\label{RH}

In Sec.~\ref{quarks} we gave expressions for the masses and left handed mixing angles in case there is a left handed hierarchy, $Y^q_{1i}\ll Y^q_{2i}\ll Y^q_{3i}$. This fact is well supported by experiment, given that the CKM mixing angles are hierarchical. There is no such analogous measurement for the right handed mixing angles. However, making the assumptions that we also have a right-handed hierarchy,
\be
Y^q_{i1}\ll Y^q_{i2}\ll Y^q_{i3}\,,
\label{RHhierarchy}
\ee
the expressions given in Sec.~\ref{quarks} simplify. Although the calculation is a bit tedious, the result is very simple: We just have to replace the mass-squared matrices by the Yukawas. Indeed,
by writing the expressions in Eqs.~(\ref{angulosINI}--\ref{angulosFIN}) and (\ref{masascompINI}--\ref{masascompFIN}) explicitly in terms of the 
Yukawa couplings and taking the limit Eq.~(\ref{RHhierarchy}) we obtain for the angles
\begin{alignat}{2}
&\displaystyle V^{q_L}_{12} =-\frac{\widetilde Y^q_{12}}{\widetilde Y^q_{22}}\,, 
&\displaystyle V^{q_L}_{21} =\left(\frac{\widetilde Y^{q}_{12}}{\widetilde Y^q_{22}}\right)^*\,,
\label{angulos2INI}
\\
&\displaystyle V^{q_L}_{23} =-\frac{Y^q_{23}}{Y^q_{33}}\,,
&\displaystyle V^{q_L}_{32} =\left(\frac{Y^q_{23}}{Y^q_{22}}\right)^*\,,
\\
&\displaystyle V^{q_L}_{13} =-\frac{Y^q_{13}}{Y^q_{33}}
+\frac{\widetilde Y^q_{12}Y^q_{23}}{\widetilde Y^q_{22}Y^q_{33}}\,, \qquad
&\displaystyle V^{q_L}_{31} =\left(\frac{Y^q_{13}}{Y^q_{33}}\right)^*\,,
\label{angulos2FIN}
\end{alignat}
and for the mass eigenvalues
\begin{align}
(m_3^q)^2&=\frac{v^2}{2}\,|Y^q_{33}|^2,
\label{masascomp2INI}
\\
(m_2^q)^2&=\frac{v^2}{2}\,|\widetilde Y_{22}^{q}|^2\,,
\\
(m_1^q)^2&=\frac{v^2}{2}\,|Y_{11}^{q}-\widetilde Y_{12}^{q}\widetilde Y_{21}^{q}/\widetilde Y^{q}_{22}|^2\,,
\label{masascomp2FIN}
\end{align}
where we have defined
\be
\widetilde Y^{q}_{ij}=Y^{q}_{ij}-\frac{Y^{q}_{i3}Y^{q}_{3j}}{Y_{33}}\,.
\ee
These results agree with the ones quoted in Ref.~\cite{Hall:1993ni}, whose authors considered real Yukawas. In the case of a right handed hierarchy, there is also an approximation to the right handed rotations . It can be obtained from Eqs.~(\ref{angulos2INI}--\ref{angulos2FIN}) by replacing $Y^q\to Y^{q\dagger}$, leading to expressions again in accordance with Ref.~\cite{Hall:1993ni}.

\clearpage
\addcontentsline{toc}{chapter}{Bibliography}
\small
\bibliographystyle{utphys}
\bibliography{thesis}

\providecommand{\href}[2]{#2}\begingroup\raggedright\begin{thebibliography}{10}

\bibitem{Cabrer:2009we}
J.~A. Cabrer, G.~von Gersdorff, and M.~Quiros, ``{Soft-Wall Stabilization},''
  \href{http://dx.doi.org/10.1088/1367-2630/12/7/075012}{{\em New J. Phys.}
  {\bf 12} (2010)  075012},
\href{http://arxiv.org/abs/0907.5361}{{\tt arXiv:0907.5361 [hep-ph]}}.
%%CITATION = 0907.5361;%%.

\bibitem{Cabrer:2010si}
J.~A. Cabrer, G.~von Gersdorff, and M.~Quiros, ``{Warped Electroweak Breaking
  Without Custodial Symmetry},''
  \href{http://dx.doi.org/10.1016/j.physletb.2011.01.058}{{\em Phys. Lett.}
  {\bf B697} (2011)  208--214},
\href{http://arxiv.org/abs/1011.2205}{{\tt arXiv:1011.2205 [hep-ph]}}.
%%CITATION = 1011.2205;%%.

\bibitem{Cabrer:2011fb}
J.~A. Cabrer, G.~von Gersdorff, and M.~Quiros, ``{Suppressing Electroweak
  Precision Observables in 5D Warped Models},''
  \href{http://dx.doi.org/10.1007/JHEP05(2011)083}{{\em JHEP} {\bf 05} (2011)
  083},
\href{http://arxiv.org/abs/1103.1388}{{\tt arXiv:1103.1388 [hep-ph]}}.
%%CITATION = 1103.1388;%%.

\bibitem{Cabrer:2011mw}
J.~A. Cabrer, G.~von Gersdorff, and M.~Quiros, ``{Warped 5D Standard Model
  Consistent with EWPT},'' \href{http://dx.doi.org/10.1002/prop.201100054}{{\em
  Fortschritte d. Physik} {\bf 59} (2011)  1135--1138},
\href{http://arxiv.org/abs/1104.5253}{{\tt arXiv:1104.5253 [hep-ph]}}.
%%CITATION = 1104.5253;%%.

\bibitem{Cabrer:2011vu}
J.~A. Cabrer, G.~von Gersdorff, and M.~Quiros, ``{Improving Naturalness in
  Warped Models with a Heavy Bulk Higgs Boson},''
  \href{http://dx.doi.org/10.1103/PhysRevD.84.035024}{{\em Phys. Rev.} {\bf
  D84} (2011)  035024},
\href{http://arxiv.org/abs/1104.3149}{{\tt arXiv:1104.3149 [hep-ph]}}.
%%CITATION = 1104.3149;%%.

\bibitem{Cabrer:2011qb}
J.~A. Cabrer, G.~von Gersdorff, and M.~Quiros, ``{Flavor Phenomenology in
  General 5D Warped Spaces},''
\href{http://arxiv.org/abs/1110.3324}{{\tt arXiv:1110.3324 [hep-ph]}}.
%%CITATION = 1110.3324;%%.

\bibitem{Hagiwara:2006jt}
K.~Hagiwara, A.~Martin, D.~Nomura, and T.~Teubner, ``{Improved predictions for
  g-2 of the muon and alpha(QED) (M**2(Z))},''
  \href{http://dx.doi.org/10.1016/j.physletb.2007.04.012}{{\em Phys.Lett.} {\bf
  B649} (2007)  173--179}, \href{http://arxiv.org/abs/hep-ph/0611102}{{\tt
  arXiv:hep-ph/0611102 [hep-ph]}}.

\bibitem{PhysRevD974}
{Weinberg, Steven}, ``{Implications of dynamical symmetry breaking},'' {\em
  {Phys. Rev. D}} {\bf {13}} ({1976})  {974--996}.

\bibitem{PhysRevD1667}
E.~Gildener, ``Gauge-symmetry hierarchies,''
  \href{http://dx.doi.org/10.1103/PhysRevD.14.1667}{{\em Phys. Rev. D} {\bf 14}
  (1976)  1667--1672}.

\bibitem{PhysRevD2619}
L.~Susskind, ``Dynamics of spontaneous symmetry breaking in the weinberg-salam
  theory,'' \href{http://dx.doi.org/10.1103/PhysRevD.20.2619}{{\em Phys. Rev.
  D} {\bf 20} (1979)  2619--2625}.

\bibitem{Randall:1999ee}
L.~Randall and R.~Sundrum, ``{A large mass hierarchy from a small extra
  dimension},'' \href{http://dx.doi.org/10.1103/PhysRevLett.83.3370}{{\em Phys.
  Rev. Lett.} {\bf 83} (1999)  3370--3373},
\href{http://arxiv.org/abs/hep-ph/9905221}{{\tt arXiv:hep-ph/9905221}}.
%%CITATION = HEP-PH/9905221;%%.

\bibitem{Randall:1999vf}
L.~Randall and R.~Sundrum, ``{An alternative to compactification},''
  \href{http://dx.doi.org/10.1103/PhysRevLett.83.4690}{{\em Phys. Rev. Lett.}
  {\bf 83} (1999)  4690--4693},
\href{http://arxiv.org/abs/hep-th/9906064}{{\tt arXiv:hep-th/9906064}}.
%%CITATION = HEP-TH/9906064;%%.

\bibitem{Donoghue:1992dd}
J.~Donoghue, E.~Golowich, and B.~R. Holstein, ``{Dynamics of the standard
  model},'' {\em Camb.Monogr.Part.Phys.Nucl.Phys.Cosmol.} {\bf 2} (1992)
  1--540.

\bibitem{Peskin:1995ev}
M.~E. Peskin and D.~V. Schroeder, ``{An Introduction to quantum field
  theory},''. Reading, USA: Addison-Wesley (1995) 842 p.

\bibitem{standardmodeldiagram}
Public domain, redrawn from
  \url{http://en.wikipedia.org/wiki/File:Elementary_particle_interactions.svg}.

\bibitem{Martin:1997ns}
S.~P. Martin, ``{A Supersymmetry Primer},''
\href{http://arxiv.org/abs/hep-ph/9709356}{{\tt arXiv:hep-ph/9709356}}.
%%CITATION = HEP-PH/9709356;%%.

\bibitem{Dimopoulos:1979es}
S.~Dimopoulos and L.~Susskind, ``{Mass Without Scalars},''
\href{http://dx.doi.org/10.1016/0550-3213(79)90364-X}{{\em Nucl. Phys.} {\bf
  B155} (1979)  237--252}.
%%CITATION = NUPHA,B155,237;%%.

\bibitem{Csaki:2003zu}
C.~Csaki, C.~Grojean, L.~Pilo, and J.~Terning, ``{Towards a realistic model of
  Higgsless electroweak symmetry breaking},''
  \href{http://dx.doi.org/10.1103/PhysRevLett.92.101802}{{\em Phys.Rev.Lett.}
  {\bf 92} (2004)  101802}, \href{http://arxiv.org/abs/hep-ph/0308038}{{\tt
  arXiv:hep-ph/0308038 [hep-ph]}}.

\bibitem{Davoudiasl:2009cd}
H.~Davoudiasl, S.~Gopalakrishna, E.~Ponton, and J.~Santiago, ``{Warped
  5-Dimensional Models: Phenomenological Status and Experimental Prospects},''
  \href{http://dx.doi.org/10.1088/1367-2630/12/7/075011}{{\em New J. Phys.}
  {\bf 12} (2010)  075011},
\href{http://arxiv.org/abs/0908.1968}{{\tt arXiv:0908.1968 [hep-ph]}}.
%%CITATION = 0908.1968;%%.

\bibitem{Csaki:2004ay}
C.~Csaki, ``{TASI lectures on extra dimensions and branes},''
  \href{http://arxiv.org/abs/hep-ph/0404096}{{\tt arXiv:hep-ph/0404096
  [hep-ph]}}.

\bibitem{Sundrum:2005jf}
R.~Sundrum, ``{Tasi 2004 lectures: To the fifth dimension and back},''
  \href{http://arxiv.org/abs/hep-th/0508134}{{\tt arXiv:hep-th/0508134
  [hep-th]}}.

\bibitem{Goldberger:1999wh}
W.~D. Goldberger and M.~B. Wise, ``{Bulk fields in the Randall-Sundrum
  compactification scenario},''
  \href{http://dx.doi.org/10.1103/PhysRevD.60.107505}{{\em Phys.Rev.} {\bf D60}
  (1999)  107505}, \href{http://arxiv.org/abs/hep-ph/9907218}{{\tt
  arXiv:hep-ph/9907218 [hep-ph]}}.

\bibitem{Huber:2000fh}
S.~J. Huber and Q.~Shafi, ``{Higgs mechanism and bulk gauge boson masses in the
  Randall-Sundrum model},''
  \href{http://dx.doi.org/10.1103/PhysRevD.63.045010}{{\em Phys. Rev.} {\bf
  D63} (2001)  045010},
\href{http://arxiv.org/abs/hep-ph/0005286}{{\tt arXiv:hep-ph/0005286}}.
%%CITATION = HEP-PH/0005286;%%.

\bibitem{Maldacena:1997re}
J.~M. Maldacena, ``{The large N limit of superconformal field theories and
  supergravity},'' \href{http://dx.doi.org/10.1023/A:1026654312961}{{\em Adv.
  Theor. Math. Phys.} {\bf 2} (1998)  231--252},
\href{http://arxiv.org/abs/hep-th/9711200}{{\tt arXiv:hep-th/9711200}}.
%%CITATION = HEP-TH/9711200;%%.

\bibitem{Gubser:1998bc}
S.~S. Gubser, I.~R. Klebanov, and A.~M. Polyakov, ``{Gauge theory correlators
  from non-critical string theory},''
  \href{http://dx.doi.org/10.1016/S0370-2693(98)00377-3}{{\em Phys. Lett.} {\bf
  B428} (1998)  105--114},
\href{http://arxiv.org/abs/hep-th/9802109}{{\tt arXiv:hep-th/9802109}}.
%%CITATION = HEP-TH/9802109;%%.

\bibitem{Witten:1998qj}
E.~Witten, ``{Anti-de Sitter space and holography},'' {\em Adv. Theor. Math.
  Phys.} {\bf 2} (1998)  253--291,
\href{http://arxiv.org/abs/hep-th/9802150}{{\tt arXiv:hep-th/9802150}}.
%%CITATION = HEP-TH/9802150;%%.

\bibitem{Contino:2003ve}
R.~Contino, Y.~Nomura, and A.~Pomarol, ``{Higgs as a holographic
  pseudo-Goldstone boson},''
  \href{http://dx.doi.org/10.1016/j.nuclphysb.2003.08.027}{{\em Nucl. Phys.}
  {\bf B671} (2003)  148--174},
\href{http://arxiv.org/abs/hep-ph/0306259}{{\tt arXiv:hep-ph/0306259}}.
%%CITATION = HEP-PH/0306259;%%.

\bibitem{Goldberger:1999uk}
W.~D. Goldberger and M.~B. Wise, ``{Modulus stabilization with bulk fields},''
  \href{http://dx.doi.org/10.1103/PhysRevLett.83.4922}{{\em Phys. Rev. Lett.}
  {\bf 83} (1999)  4922--4925},
\href{http://arxiv.org/abs/hep-ph/9907447}{{\tt arXiv:hep-ph/9907447}}.
%%CITATION = HEP-PH/9907447;%%.

\bibitem{Goldberger:1999un}
W.~D. Goldberger and M.~B. Wise, ``{Phenomenology of a stabilized modulus},''
  \href{http://dx.doi.org/10.1016/S0370-2693(00)00099-X}{{\em Phys. Lett.} {\bf
  B475} (2000)  275--279},
\href{http://arxiv.org/abs/hep-ph/9911457}{{\tt arXiv:hep-ph/9911457}}.
%%CITATION = HEP-PH/9911457;%%.

\bibitem{Chamblin:1999ya}
H.~A. Chamblin and H.~S. Reall, ``{Dynamic dilatonic domain walls},''
  \href{http://dx.doi.org/10.1016/S0550-3213(99)00520-9}{{\em Nucl. Phys.} {\bf
  B562} (1999)  133--157},
\href{http://arxiv.org/abs/hep-th/9903225}{{\tt arXiv:hep-th/9903225}}.
%%CITATION = HEP-TH/9903225;%%.

\bibitem{ArkaniHamed:2000eg}
N.~Arkani-Hamed, S.~Dimopoulos, N.~Kaloper, and R.~Sundrum, ``{A small
  cosmological constant from a large extra dimension},''
  \href{http://dx.doi.org/10.1016/S0370-2693(00)00359-2}{{\em Phys. Lett.} {\bf
  B480} (2000)  193--199},
\href{http://arxiv.org/abs/hep-th/0001197}{{\tt arXiv:hep-th/0001197}}.
%%CITATION = HEP-TH/0001197;%%.

\bibitem{Kachru:2000hf}
S.~Kachru, M.~B. Schulz, and E.~Silverstein, ``{Self-tuning flat domain walls
  in 5d gravity and string theory},''
  \href{http://dx.doi.org/10.1103/PhysRevD.62.045021}{{\em Phys. Rev.} {\bf
  D62} (2000)  045021},
\href{http://arxiv.org/abs/hep-th/0001206}{{\tt arXiv:hep-th/0001206}}.
%%CITATION = HEP-TH/0001206;%%.

\bibitem{Gubser:2000nd}
S.~S. Gubser, ``{Curvature singularities: The good, the bad, and the naked},''
  {\em Adv. Theor. Math. Phys.} {\bf 4} (2000)  679--745,
\href{http://arxiv.org/abs/hep-th/0002160}{{\tt arXiv:hep-th/0002160}}.
%%CITATION = HEP-TH/0002160;%%.

\bibitem{Forste:2000ps}
S.~Forste, Z.~Lalak, S.~Lavignac, and H.~P. Nilles, ``{A comment on self-tuning
  and vanishing cosmological constant in the brane world},''
  \href{http://dx.doi.org/10.1016/S0370-2693(00)00468-8}{{\em Phys. Lett.} {\bf
  B481} (2000)  360--364},
\href{http://arxiv.org/abs/hep-th/0002164}{{\tt arXiv:hep-th/0002164}}.
%%CITATION = HEP-TH/0002164;%%.

\bibitem{Csaki:2000wz}
C.~Csaki, J.~Erlich, C.~Grojean, and T.~J. Hollowood, ``{General properties of
  the self-tuning domain wall approach to the cosmological constant problem},''
  \href{http://dx.doi.org/10.1016/S0550-3213(00)00390-4}{{\em Nucl. Phys.} {\bf
  B584} (2000)  359--386},
\href{http://arxiv.org/abs/hep-th/0004133}{{\tt arXiv:hep-th/0004133}}.
%%CITATION = HEP-TH/0004133;%%.

\bibitem{Forste:2000ft}
S.~Forste, Z.~Lalak, S.~Lavignac, and H.~P. Nilles, ``{The Cosmological
  Constant Problem from a Brane-World Perspective},'' {\em JHEP} {\bf 09}
  (2000)  034,
\href{http://arxiv.org/abs/hep-th/0006139}{{\tt arXiv:hep-th/0006139}}.
%%CITATION = HEP-TH/0006139;%%.

\bibitem{Karch:2006pv}
A.~Karch, E.~Katz, D.~T. Son, and M.~A. Stephanov, ``{Linear Confinement and
  AdS/QCD},'' \href{http://dx.doi.org/10.1103/PhysRevD.74.015005}{{\em Phys.
  Rev.} {\bf D74} (2006)  015005},
\href{http://arxiv.org/abs/hep-ph/0602229}{{\tt arXiv:hep-ph/0602229}}.
%%CITATION = HEP-PH/0602229;%%.

\bibitem{Gursoy:2007cb}
U.~Gursoy and E.~Kiritsis, ``{Exploring improved holographic theories for QCD:
  Part I},'' \href{http://dx.doi.org/10.1088/1126-6708/2008/02/032}{{\em JHEP}
  {\bf 02} (2008)  032},
\href{http://arxiv.org/abs/0707.1324}{{\tt arXiv:0707.1324 [hep-th]}}.
%%CITATION = 0707.1324;%%.

\bibitem{Batell:2008zm}
B.~Batell and T.~Gherghetta, ``{Dynamical Soft-Wall AdS/QCD},''
  \href{http://dx.doi.org/10.1103/PhysRevD.78.026002}{{\em Phys. Rev.} {\bf
  D78} (2008)  026002},
\href{http://arxiv.org/abs/0801.4383}{{\tt arXiv:0801.4383 [hep-ph]}}.
%%CITATION = 0801.4383;%%.

\bibitem{Cacciapaglia:2008ns}
G.~Cacciapaglia, G.~Marandella, and J.~Terning, ``{The AdS/CFT/Unparticle
  Correspondence},''
  \href{http://dx.doi.org/10.1088/1126-6708/2009/02/049}{{\em JHEP} {\bf 02}
  (2009)  049},
\href{http://arxiv.org/abs/0804.0424}{{\tt arXiv:0804.0424 [hep-ph]}}.
%%CITATION = 0804.0424;%%.

\bibitem{Falkowski:2008yr}
A.~Falkowski and M.~Perez-Victoria, ``{Holographic Unhiggs},''
  \href{http://dx.doi.org/10.1103/PhysRevD.79.035005}{{\em Phys. Rev.} {\bf
  D79} (2009)  035005},
\href{http://arxiv.org/abs/0810.4940}{{\tt arXiv:0810.4940 [hep-ph]}}.
%%CITATION = 0810.4940;%%.

\bibitem{Georgi:2007ek}
H.~Georgi, ``{Unparticle Physics},''
  \href{http://dx.doi.org/10.1103/PhysRevLett.98.221601}{{\em Phys. Rev. Lett.}
  {\bf 98} (2007)  221601},
\href{http://arxiv.org/abs/hep-ph/0703260}{{\tt arXiv:hep-ph/0703260}}.
%%CITATION = HEP-PH/0703260;%%.

\bibitem{Falkowski:2008fz}
A.~Falkowski and M.~Perez-Victoria, ``{Electroweak Breaking on a Soft Wall},''
  \href{http://dx.doi.org/10.1088/1126-6708/2008/12/107}{{\em JHEP} {\bf 12}
  (2008)  107},
\href{http://arxiv.org/abs/0806.1737}{{\tt arXiv:0806.1737 [hep-ph]}}.
%%CITATION = 0806.1737;%%.

\bibitem{vonGersdorff:2010ht}
G.~von Gersdorff, ``{From Soft Walls to Infrared Branes},''
  \href{http://dx.doi.org/10.1103/PhysRevD.82.086010}{{\em Phys. Rev.} {\bf
  D82} (2010)  086010},
\href{http://arxiv.org/abs/1005.5134}{{\tt arXiv:1005.5134 [hep-ph]}}.
%%CITATION = 1005.5134;%%.

\bibitem{DeWolfe:1999cp}
O.~DeWolfe, D.~Z. Freedman, S.~S. Gubser, and A.~Karch, ``{Modeling the fifth
  dimension with scalars and gravity},''
  \href{http://dx.doi.org/10.1103/PhysRevD.62.046008}{{\em Phys. Rev.} {\bf
  D62} (2000)  046008},
\href{http://arxiv.org/abs/hep-th/9909134}{{\tt arXiv:hep-th/9909134}}.
%%CITATION = HEP-TH/9909134;%%.

\bibitem{George:2011gs}
D.~P. George and M.~Postma, ``{Avoiding the dangers of a soft-wall
  singularity},''
\href{http://arxiv.org/abs/1105.3390}{{\tt arXiv:1105.3390 [hep-th]}}.
%%CITATION = 1105.3390;%%.

\bibitem{Csaki:2000zn}
C.~Csaki, M.~L. Graesser, and G.~D. Kribs, ``{Radion dynamics and electroweak
  physics},'' \href{http://dx.doi.org/10.1103/PhysRevD.63.065002}{{\em Phys.
  Rev.} {\bf D63} (2001)  065002},
\href{http://arxiv.org/abs/hep-th/0008151}{{\tt arXiv:hep-th/0008151}}.
%%CITATION = HEP-TH/0008151;%%.

\bibitem{Gursoy:2007er}
U.~Gursoy, E.~Kiritsis, and F.~Nitti, ``{Exploring improved holographic
  theories for QCD: Part II},''
  \href{http://dx.doi.org/10.1088/1126-6708/2008/02/019}{{\em JHEP} {\bf 02}
  (2008)  019},
\href{http://arxiv.org/abs/0707.1349}{{\tt arXiv:0707.1349 [hep-th]}}.
%%CITATION = 0707.1349;%%.

\bibitem{Batell:2008me}
B.~Batell, T.~Gherghetta, and D.~Sword, ``{The Soft-Wall Standard Model},''
  \href{http://dx.doi.org/10.1103/PhysRevD.78.116011}{{\em Phys. Rev.} {\bf
  D78} (2008)  116011},
\href{http://arxiv.org/abs/0808.3977}{{\tt arXiv:0808.3977 [hep-ph]}}.
%%CITATION = 0808.3977;%%.

\bibitem{Freedman:1999gk}
D.~Z. Freedman, S.~S. Gubser, K.~Pilch, and N.~P. Warner, ``{Continuous
  distributions of D3-branes and gauged supergravity},'' {\em JHEP} {\bf 07}
  (2000)  038,
\href{http://arxiv.org/abs/hep-th/9906194}{{\tt arXiv:hep-th/9906194}}.
%%CITATION = HEP-TH/9906194;%%.

\bibitem{Charmousis:1999rg}
C.~Charmousis, R.~Gregory, and V.~A. Rubakov, ``{Wave function of the radion in
  a brane world},'' \href{http://dx.doi.org/10.1103/PhysRevD.62.067505}{{\em
  Phys. Rev.} {\bf D62} (2000)  067505},
\href{http://arxiv.org/abs/hep-th/9912160}{{\tt arXiv:hep-th/9912160}}.
%%CITATION = HEP-TH/9912160;%%.

\bibitem{Stancato:2008mp}
D.~Stancato and J.~Terning, ``{The Unhiggs},''
  \href{http://dx.doi.org/10.1088/1126-6708/2009/11/101}{{\em JHEP} {\bf 11}
  (2009)  101},
\href{http://arxiv.org/abs/0807.3961}{{\tt arXiv:0807.3961 [hep-ph]}}.
%%CITATION = 0807.3961;%%.

\bibitem{Nakamura:2010zzi}
{\bf Particle Data Group} Collaboration, K.~Nakamura {\em et al.}, ``{Review of
  particle physics},''
\href{http://dx.doi.org/10.1088/0954-3899/37/7A/075021}{{\em J. Phys.} {\bf
  G37} (2010)  075021}.
%%CITATION = JPHGB,G37,075021;%%.

\bibitem{Agashe:2003zs}
K.~Agashe, A.~Delgado, M.~J. May, and R.~Sundrum, ``{RS1, custodial isospin and
  precision tests},'' {\em JHEP} {\bf 08} (2003)  050,
\href{http://arxiv.org/abs/hep-ph/0308036}{{\tt arXiv:hep-ph/0308036}}.
%%CITATION = HEP-PH/0308036;%%.

\bibitem{Agashe:2006at}
K.~Agashe, R.~Contino, L.~Da~Rold, and A.~Pomarol, ``{A custodial symmetry for
  Z b anti-b},'' \href{http://dx.doi.org/10.1016/j.physletb.2006.08.005}{{\em
  Phys. Lett.} {\bf B641} (2006)  62--66},
\href{http://arxiv.org/abs/hep-ph/0605341}{{\tt arXiv:hep-ph/0605341}}.
%%CITATION = HEP-PH/0605341;%%.

\bibitem{Davoudiasl:2002ua}
H.~Davoudiasl, J.~L. Hewett, and T.~G. Rizzo, ``{Brane localized kinetic terms
  in the Randall-Sundrum model},''
  \href{http://dx.doi.org/10.1103/PhysRevD.68.045002}{{\em Phys. Rev.} {\bf
  D68} (2003)  045002},
\href{http://arxiv.org/abs/hep-ph/0212279}{{\tt arXiv:hep-ph/0212279}}.
%%CITATION = HEP-PH/0212279;%%.

\bibitem{Delgado:2007ne}
A.~Delgado and A.~Falkowski, ``{Electroweak observables in a general 5D
  background},'' \href{http://dx.doi.org/10.1088/1126-6708/2007/05/097}{{\em
  JHEP} {\bf 05} (2007)  097},
\href{http://arxiv.org/abs/hep-ph/0702234}{{\tt arXiv:hep-ph/0702234}}.
%%CITATION = HEP-PH/0702234;%%.

\bibitem{Falkowski:2009uy}
A.~Falkowski and M.~Perez-Victoria, ``{Electroweak Precision Observables and
  the Unhiggs},'' \href{http://dx.doi.org/10.1088/1126-6708/2009/12/061}{{\em
  JHEP} {\bf 12} (2009)  061},
\href{http://arxiv.org/abs/0901.3777}{{\tt arXiv:0901.3777 [hep-ph]}}.
%%CITATION = 0901.3777;%%.

\bibitem{Round:2010kj}
M.~Round, ``{Holographic Renormalisation and the Electroweak Precision
  Parameters},'' \href{http://dx.doi.org/10.1103/PhysRevD.82.053002}{{\em Phys.
  Rev.} {\bf D82} (2010)  053002},
\href{http://arxiv.org/abs/1003.2933}{{\tt arXiv:1003.2933 [hep-ph]}}.
%%CITATION = 1003.2933;%%.

\bibitem{Barbieri:2004qk}
R.~Barbieri, A.~Pomarol, R.~Rattazzi, and A.~Strumia, ``{Electroweak symmetry
  breaking after LEP-1 and LEP-2},''
  \href{http://dx.doi.org/10.1016/j.nuclphysb.2004.10.014}{{\em Nucl. Phys.}
  {\bf B703} (2004)  127--146},
\href{http://arxiv.org/abs/hep-ph/0405040}{{\tt arXiv:hep-ph/0405040}}.
%%CITATION = HEP-PH/0405040;%%.

\bibitem{Peskin:1991sw}
M.~E. Peskin and T.~Takeuchi, ``{Estimation of oblique electroweak
  corrections},''
\href{http://dx.doi.org/10.1103/PhysRevD.46.381}{{\em Phys. Rev.} {\bf D46}
  (1992)  381--409}.
%%CITATION = PHRVA,D46,381;%%.

\bibitem{Kennedy:1988sn}
D.~C. Kennedy and B.~W. Lynn, ``{Electroweak Radiative Corrections with an
  Effective Lagrangian: Four Fermion Processes},''
\href{http://dx.doi.org/10.1016/0550-3213(89)90483-5}{{\em Nucl. Phys.} {\bf
  B322} (1989)  1}.
%%CITATION = NUPHA,B322,1;%%.

\bibitem{Altarelli:1990zd}
G.~Altarelli and R.~Barbieri, ``{Vacuum polarization effects of new physics on
  electroweak processes},''
\href{http://dx.doi.org/10.1016/0370-2693(91)91378-9}{{\em Phys. Lett.} {\bf
  B253} (1991)  161--167}.
%%CITATION = PHLTA,B253,161;%%.

\bibitem{Carmona:2011ib}
A.~Carmona, E.~Ponton, and J.~Santiago, ``{Phenomenology of Non-Custodial
  Warped Models},''
\href{http://arxiv.org/abs/1107.1500}{{\tt arXiv:1107.1500 [hep-ph]}}.
%%CITATION = 1107.1500;%%.

\bibitem{Casagrande:2008hr}
S.~Casagrande, F.~Goertz, U.~Haisch, M.~Neubert, and T.~Pfoh, ``{Flavor Physics
  in the Randall-Sundrum Model: I. Theoretical Setup and Electroweak Precision
  Tests},'' \href{http://dx.doi.org/10.1088/1126-6708/2008/10/094}{{\em JHEP}
  {\bf 10} (2008)  094},
\href{http://arxiv.org/abs/0807.4937}{{\tt arXiv:0807.4937 [hep-ph]}}.
%%CITATION = 0807.4937;%%.

\bibitem{Peskin:2001rw}
M.~E. Peskin and J.~D. Wells, ``{How can a heavy Higgs boson be consistent with
  the precision electroweak measurements?},''
  \href{http://dx.doi.org/10.1103/PhysRevD.64.093003}{{\em Phys. Rev.} {\bf
  D64} (2001)  093003},
\href{http://arxiv.org/abs/hep-ph/0101342}{{\tt arXiv:hep-ph/0101342}}.
%%CITATION = HEP-PH/0101342;%%.

\bibitem{Barbieri:2006dq}
R.~Barbieri, L.~J. Hall, and V.~S. Rychkov, ``{Improved naturalness with a
  heavy Higgs: An alternative road to LHC physics},''
  \href{http://dx.doi.org/10.1103/PhysRevD.74.015007}{{\em Phys. Rev.} {\bf
  D74} (2006)  015007},
\href{http://arxiv.org/abs/hep-ph/0603188}{{\tt arXiv:hep-ph/0603188}}.
%%CITATION = HEP-PH/0603188;%%.

\bibitem{Veltman:1976rt}
M.~J.~G. Veltman, ``{Second Threshold in Weak Interactions},''
{\em Acta Phys. Polon.} {\bf B8} (1977)  475.
%%CITATION = APPOA,B8,475;%%.

\bibitem{Hambye:1996wb}
T.~Hambye and K.~Riesselmann, ``{Matching conditions and Higgs mass upper
  bounds revisited},'' \href{http://dx.doi.org/10.1103/PhysRevD.55.7255}{{\em
  Phys. Rev.} {\bf D55} (1997)  7255--7262},
\href{http://arxiv.org/abs/hep-ph/9610272}{{\tt arXiv:hep-ph/9610272}}.
%%CITATION = HEP-PH/9610272;%%.

\bibitem{Archer:2010hh}
P.~R. Archer and S.~J. Huber, ``{Electroweak Constraints on Warped Geometry in
  Five Dimensions and Beyond},''
  \href{http://dx.doi.org/10.1007/JHEP10(2010)032}{{\em JHEP} {\bf 10} (2010)
  032},
\href{http://arxiv.org/abs/1004.1159}{{\tt arXiv:1004.1159 [hep-ph]}}.
%%CITATION = 1004.1159;%%.

\bibitem{Luty:2004ye}
M.~A. Luty and T.~Okui, ``{Conformal technicolor},''
  \href{http://dx.doi.org/10.1088/1126-6708/2006/09/070}{{\em JHEP} {\bf 09}
  (2006)  070},
\href{http://arxiv.org/abs/hep-ph/0409274}{{\tt arXiv:hep-ph/0409274}}.
%%CITATION = HEP-PH/0409274;%%.

\bibitem{Vecchi:2010aj}
L.~Vecchi, ``{Technicolor at Criticality},''
  \href{http://dx.doi.org/10.1007/JHEP04(2011)127}{{\em JHEP} {\bf 04} (2011)
  127},
\href{http://arxiv.org/abs/1007.4573}{{\tt arXiv:1007.4573 [hep-ph]}}.
%%CITATION = 1007.4573;%%.

\bibitem{Davoudiasl:2008hx}
H.~Davoudiasl, G.~Perez, and A.~Soni, ``{The Little Randall-Sundrum Model at
  the Large Hadron Collider},''
  \href{http://dx.doi.org/10.1016/j.physletb.2008.05.024}{{\em Phys. Lett.}
  {\bf B665} (2008)  67--71},
\href{http://arxiv.org/abs/0802.0203}{{\tt arXiv:0802.0203 [hep-ph]}}.
%%CITATION = 0802.0203;%%.

\bibitem{Atkins:2010cc}
M.~Atkins and S.~J. Huber, ``{Suppressing Lepton Flavour Violation in a
  Soft-Wall Extra Dimension},''
  \href{http://dx.doi.org/10.1103/PhysRevD.82.056007}{{\em Phys. Rev.} {\bf
  D82} (2010)  056007},
\href{http://arxiv.org/abs/1002.5044}{{\tt arXiv:1002.5044 [hep-ph]}}.
%%CITATION = 1002.5044;%%.

\bibitem{MertAybat:2009mk}
S.~Mert~Aybat and J.~Santiago, ``{Bulk Fermions in Warped Models with a Soft
  Wall},'' \href{http://dx.doi.org/10.1103/PhysRevD.80.035005}{{\em Phys. Rev.}
  {\bf D80} (2009)  035005},
\href{http://arxiv.org/abs/0905.3032}{{\tt arXiv:0905.3032 [hep-ph]}}.
%%CITATION = 0905.3032;%%.

\bibitem{Burgess:1993vc}
C.~Burgess, S.~Godfrey, H.~Konig, D.~London, and I.~Maksymyk, ``{Model
  independent global constraints on new physics},''
  \href{http://dx.doi.org/10.1103/PhysRevD.49.6115}{{\em Phys.Rev.} {\bf D49}
  (1994)  6115--6147}, \href{http://arxiv.org/abs/hep-ph/9312291}{{\tt
  arXiv:hep-ph/9312291 [hep-ph]}}.

\bibitem{Bona:2007vi}
{\bf UTfit Collaboration} Collaboration, M.~Bona {\em et al.},
  ``{Model-independent constraints on $\Delta$ F=2 operators and the scale of
  new physics},'' \href{http://dx.doi.org/10.1088/1126-6708/2008/03/049}{{\em
  JHEP} {\bf 0803} (2008)  049}, \href{http://arxiv.org/abs/0707.0636}{{\tt
  arXiv:0707.0636 [hep-ph]}}.

\bibitem{Isidori:2010kg}
G.~Isidori, Y.~Nir, and G.~Perez, ``{Flavor Physics Constraints for Physics
  Beyond the Standard Model},'' {\em Ann.Rev.Nucl.Part.Sci.} {\bf 60} (2010)
  355, \href{http://arxiv.org/abs/1002.0900}{{\tt arXiv:1002.0900 [hep-ph]}}.

\bibitem{Csaki:2008zd}
C.~Csaki, A.~Falkowski, and A.~Weiler, ``{The Flavor of the Composite
  Pseudo-Goldstone Higgs},''
  \href{http://dx.doi.org/10.1088/1126-6708/2008/09/008}{{\em JHEP} {\bf 0809}
  (2008)  008}, \href{http://arxiv.org/abs/0804.1954}{{\tt arXiv:0804.1954
  [hep-ph]}}.

\bibitem{Hall:1993ni}
L.~J. Hall and A.~Rasin, ``{On the generality of certain predictions for quark
  mixing},'' \href{http://dx.doi.org/10.1016/0370-2693(93)90175-H}{{\em
  Phys.Lett.} {\bf B315} (1993)  164--169},
  \href{http://arxiv.org/abs/hep-ph/9303303}{{\tt arXiv:hep-ph/9303303
  [hep-ph]}}.

\bibitem{Djouadi:2006rk}
A.~Djouadi, G.~Moreau, and F.~Richard, ``{Resolving the $A_{FB}^b$ puzzle in an
  extra dimensional model with an extended gauge structure},''
  \href{http://dx.doi.org/10.1016/j.nuclphysb.2007.03.019}{{\em Nucl. Phys.}
  {\bf B773} (2007)  43--64},
\href{http://arxiv.org/abs/hep-ph/0610173}{{\tt arXiv:hep-ph/0610173}}.
%%CITATION = HEP-PH/0610173;%%.

\bibitem{Archer:2011bk}
P.~R. Archer, S.~J. Huber, and S.~Jager, ``{Flavour Physics in the Soft Wall
  Model},'' \href{http://arxiv.org/abs/1108.1433}{{\tt arXiv:1108.1433
  [hep-ph]}}.

\bibitem{Santiago:2008vq}
J.~Santiago, ``{Minimal Flavor Protection: A New Flavor Paradigm in Warped
  Models},'' \href{http://dx.doi.org/10.1088/1126-6708/2008/12/046}{{\em JHEP}
  {\bf 0812} (2008)  046}, \href{http://arxiv.org/abs/0806.1230}{{\tt
  arXiv:0806.1230 [hep-ph]}}.

\bibitem{Bauer:2011ah}
M.~Bauer, R.~Malm, and M.~Neubert, ``{A Solution to the Flavor Problem of
  Warped Extra-Dimension Models},''
\href{http://arxiv.org/abs/1110.0471}{{\tt arXiv:1110.0471 [hep-ph]}}.
%%CITATION = 1110.0471;%%.

\bibitem{Lillie:2007yh}
B.~Lillie, L.~Randall, and L.-T. Wang, ``{The Bulk RS KK-gluon at the LHC},''
  \href{http://dx.doi.org/10.1088/1126-6708/2007/09/074}{{\em JHEP} {\bf 09}
  (2007)  074},
\href{http://arxiv.org/abs/hep-ph/0701166}{{\tt arXiv:hep-ph/0701166}}.
%%CITATION = HEP-PH/0701166;%%.

\bibitem{Agashe:2004cp}
K.~Agashe, G.~Perez, and A.~Soni, ``{Flavor structure of warped extra dimension
  models},'' \href{http://dx.doi.org/10.1103/PhysRevD.71.016002}{{\em Phys.
  Rev.} {\bf D71} (2005)  016002},
\href{http://arxiv.org/abs/hep-ph/0408134}{{\tt arXiv:hep-ph/0408134}}.
%%CITATION = HEP-PH/0408134;%%.

\end{thebibliography}\endgroup

\newpage

\pagestyle{empty}
\ 

\end{document}